\documentclass[12pt]{book}
\pdfoutput=1
\usepackage{times}
\usepackage{fullpage}
\usepackage{mjz-titlepage}
\usepackage{jrs-fancyheadings}
\usepackage{epsfig}
\usepackage{amsmath}
\usepackage{supertabular}
\usepackage{dcolumn}
\usepackage{multirow}
\usepackage{rotating}
\usepackage[numbers]{natbib}
\usepackage[
backref,
pageanchor=true,
plainpages=false,
pdfpagelabels, % makes the status line page labels the same as latex ones
bookmarks,
bookmarksnumbered,
pdfborder=0 0 0,
pdfpagemode=UseOutlines]{hyperref}
\hypersetup{colorlinks=true, linkcolor=red}

\usepackage{color}

\usepackage{draftstamp}
\usepackage{subfigure}
\usepackage[letterpaper,dvips,twoside,nomarginpar]{geometry}
\usepackage{listings}
\usepackage{color}
\usepackage{xcolor}
\usepackage{xspace}
\usepackage{booktabs}
\usepackage{float}
\usepackage{multirow}
\usepackage{comment}

\usepackage{amsmath}
\usepackage{amssymb}
\usepackage{xspace}
\usepackage{pifont}
\newcommand{\bdc}{B$\Delta$I\xspace}
\newcommand{\bd}{B$+\Delta$\xspace}
\newcommand{\ignore}[1]{}

\newcommand{\carp}{CAMP} 
\newcommand{\insertionpolicy}{SIP} 
\newcommand{\mineviction}{MVE} 
\newcommand{\REM}[1]{}

\usepackage{tabularx,booktabs}

\newcolumntype{C}{>{\centering\arraybackslash}X} 
\newcolumntype{L}{>{\raggedright\arraybackslash}X}

\usepackage{lipsum}

\newcommand\blfootnote[1]{%
  \begingroup
  \renewcommand\thefootnote{}\footnote{#1}%
  \addtocounter{footnote}{-1}%
  \endgroup
}

\begin{document}

\frontmatter
%\include{definitions}

%initialize page style, so contents come out right (see bot) -mjz
\pagestyle{empty}

\title{\bf Practical Data Compression\\ for Modern Memory Hierarchies}
\author{\bf Gennady G. Pekhimenko}
\date{July 2016}
\Year{2016}
\trnumber{CMU-CS-16-116}

\committee{
Todd C. Mowry, Co-Chair \\
Onur Mutlu, Co-Chair \\
Kayvon Fatahalian \\
David A. Wood, University of Wisconsin-Madison\\
Douglas C. Burger, Microsoft\\
Michael A. Kozuch, Intel 
}

\support{ This research was sponsored by the National Science Foundation under
grant numbers CNS-0720790, CCF-0953246, CCF-1116898, CNS-1423172, CCF-1212962, CNS-1320531, CCF-1147397, and CNS- 1409723,
 the Defense Advanced Research Projects Agency, the Semiconductor Research Corporation, the Gigascale Systems Research Center,
Intel URO Memory Hierarchy Program, Intel ISTC-CC, and gifts from AMD, Google, IBM, Intel, Nvidia, Oracle, Qualcomm, Samsung, and VMware.
We also acknowledge the support through PhD fellowships from Nvidia, Microsoft, Qualcomm, and NSERC.}

\disclaimer{
The views and conclusions contained in this document are those of the author
and should not be interpreted as representing the official policies, either
expressed or implied, of any sponsoring institution, the U.S. government or any
other entity.  }

% copyright notice generated automatically from Year and author.
% permission added if \permission{} given.


\maketitle

%\begin{dedication}
  %For graduation.  How I love thee.  Hopefully someday I will attain
  %you.
%\end{dedication}

\pagestyle{plain} % for toc, was empty

%\begin{abstract}
\chapter*{Abstract}%
\addcontentsline{toc}{chapter}{\numberline{}Abstract}%

Although compression has been widely used for decades to reduce file sizes
(thereby conserving storage capacity and network bandwidth when transferring
files), there has been limited
use of hardware-based compression within modern memory
hierarchies of commodity systems. Why not?  
Especially as programs become increasingly
data-intensive, the capacity and bandwidth within the memory hierarchy
(including caches, main memory, and their associated interconnects) have
already become increasingly important bottlenecks.  If 
hardware-based data compression could be
applied successfully to the memory hierarchy, it could potentially relieve
pressure on these bottlenecks by increasing effective capacity, increasing
effective bandwidth, and even reducing energy consumption.  

In this thesis, we describe a new, practical approach to integrating 
hardware-based data
compression within the memory hierarchy, including on-chip caches, main memory,
and both on-chip and off-chip interconnects. This new approach is fast, simple,
and effective in saving storage space. A key insight in our approach is that
access time (including decompression latency) is critical in modern memory
hierarchies. By combining inexpensive hardware support with modest OS support,
our holistic approach to compression achieves substantial improvements in
performance and energy efficiency across the memory hierarchy. 
Using this new approach, we make several major contributions in this thesis.

First, we propose a new compression algorithm, \emph{Base-Delta-Immediate Compression} (\emph{\bdc}), that achieves high
compression ratio with very low compression/decompression latency. 
\bdc exploits the existing low dynamic range of values
present in many cache lines to compress them to smaller sizes using Base+Delta
encoding. 

Second, we observe that the compressed size of a cache block can be indicative of
its reuse. We use this observation to develop a new cache insertion policy for
compressed caches, the \emph{Size-based Insertion Policy} (\emph{SIP}), which uses the size
of a compressed block as one of the metrics to predict its potential future
reuse.  

Third, we propose a new main memory compression framework, \emph{Linearly
Compressed Pages} (\emph{LCP}), that significantly reduces the complexity 
and power cost of supporting main memory compression. 
We demonstrate that \emph{any} compression
algorithm can be adapted to fit the requirements of LCP, and  
that LCP can be efficiently integrated with the existing cache compression
designs, avoiding extra compression/decompression.

Finally, in addition to exploring compression-related issues and enabling practical solutions in modern
CPU systems, we discover new problems in realizing hardware-based compression
for GPU-based systems and develop new solutions to solve these problems.
%end{abstract}

%%% Local Variables: 
%%% mode: latex
%%% TeX-master: "main"
%%% End: 

\pagestyle{plain}

\chapter*{Acknowledgments}
\addcontentsline{toc}{chapter}{\numberline{}Acknowledgments}%
First of all, I would like to thank my advisers, Todd Mowry and Onur Mutlu, for
always trusting me in my research experiments, giving me enough resources and
opportunities to improve my work, as well as my presentation and writing
skills.

I am grateful to Michael Kozuch and Phillip Gibbons for being both my mentors
and collaborators. I am grateful to the members of my PhD committee: Kayvon
Fatahalian, David Wood, and Doug Burger for their valuable feedback and for
making the final steps towards my PhD very smooth.  I am grateful to Deb
Cavlovich who allowed me to focus on my research by magically solving all other
problems.  

I am grateful to SAFARI group members that were more than just lab
mates.  Vivek Seshadri was always supportive for my crazy ideas and was willing
to dedicate his time and energy to help me in my work.  Chris Fallin was a rare
example of pure smartness mixed with great work ethic, but still always had
time for an interesting discussion.  From Yoongu Kim I learned a lot about the
importance of details, and hopefully I learned something from his aesthetic
sense as well.  Lavanya Subramanian was my fellow cubic mate who showed me an
example on how to successfully mix work with personal life and how to be
supportive for others.  Justin Meza helped me to improve my presentation and
writing skills in a very friendly manner (as everything else he does).
Donghyuk Lee taught me everything I know about DRAM and was always an example
of work dedication for me.  Nandita Vijaykumar was my mentee, collaborator, and
mentor all at the same time, but, most importantly, a friend that was always
willing to help.  Rachata Ausavarungnirun was our food guru and one of the most
reliable and friendly people in the group.  Hongyi Xin reminded me about
everything I almost forgot from biology and history classes, and also taught me
everything I know now in the amazing field of bioinformatics.  Kevin Chang
and Kevin Hsieh were always helpful and supportive when it matters most.
Samira Khan was always available for a friendly chat when I really need it.
Saugata Ghose was my rescue guy during our amazing trip to Prague.  I also
thank other members of the SAFARI group for their assistance and support:
HanBin Yoon, Jamie Liu, Ben Jaiyen, Yixin Luo, Yang Li, and Amirali Boroumand.

Michelle Goodstein, Olatunji Ruwase and Evangelos Vlachos, senior PhD students,
shared their experience and provided a lot of feedback early in my career.  I am
grateful to Tyler Huberty and Rui Cai for contributing a lot to my research and
for being excellent undergraduate/masters researchers who selected me as a
mentor from all the other options they had.  

During my time at Carnegie Mellon, I
met a lot of wonderful people: Michael Papamichael, Gabe Weisz, Alexey Tumanov,
Danai Koutra and many others who helped and supported me in many different
ways. I am also grateful to people at PDL and CALCM groups for accepting me in their
communities.  

I am grateful to my internship mentors for making my work in
their companies mutually successful for both sides. At Microsoft Research, I had
the privilege to closely work with Karin Strauss, Dimitrios Lymberopoulos,
Oriana Riva, Ella Bounimova, Patrice Godefroid, and David Molnar. At NVIDIA
Research, I had the privilege to closely work with Evgeny Bolotin, Steve
Keckler, and Mike O'Connor. I am also grateful to my amazing collaborators
from Georgia Tech: Hadi Esmaeilzadeh, Amir Yazdanbaksh, and Bradley Thwaites. 

And last, but not least, I would like to acknowledge the enormous love and
support that I received from my family: my wife Daria and our daughter Alyssa,
my parents: Gennady and Larissa, and my brother Evgeny.

%\begin{comment}
%I love Jello.  This stuff has fed me from the day I proposed until the
%$day I defended.  Jello is my friend, my sustenance, my very being.

%Oh yeah.  My advisor is cool too.

%In addition, {\Large\bf Catherine Copetas} and {\Large\bf Sharon Burks}
%should be mentioned in a {\Large\bf large} font in everyone's
%Acknowledgements section, since they said so.  (It is a required part
%of the thesis formatting guidelines.)

%\end{comment}
\clearpage

%%% Local Variables: 
%%% mode: latex
%%% TeX-master: t
%%% End: 

\tableofcontents
\listoffigures
\listoftables

\mainmatter

%% Double space document for easy review:
%\renewcommand{\baselinestretch}{1.66}\normalsize

% Catherine Copetas says you should use at most 1.5 spacing in your 
% thesis.  It is preferable to single space (save the trees!):
%\renewcommand{\baselinestretch}{1.24}\normalsize

% The other requirements Catherine has:
%
%  - avoid large margins.  She wants the thesis to use fewer pages, 
%    especially if it requires colour printing.
%
%  - The thesis should be formatted for double-sided printing.  This
%    means that all chapters, acknowledgements, table of contents, etc.
%    should start on odd numbered (right facing) pages.
%
%  - You need to use the department standard tech report title page.  I
%    have tried to ensure that the title page here conforms to this
%    standard.
%
%  - Use a nice serif font, such as Times Roman.  Sans serif looks bad.
%
% Other than that, just make it look good...

\chapter{Introduction}

%This thesis totally blows away most other work~\citep{dummy:99}.
%\citet{dummy:99} is not so smart.

The recent Big Data revolution has had a transformative effect on many
areas of science and technology~\citep{nsf-big-data-press-release}.  Indeed,
a key factor that has made Cloud Computing attractive is the ability to
perform computation near these massive data sets.  As we look toward the
future, where our ability to capture detailed data streams from our
environment is only expected to increase, it seems clear that many
important computations will operate on increasingly larger data set sizes.

Unfortunately, data-intensive computing creates significant challenges for
system designers.  In particular, the large volume and flow of data places
significant stress on the capacity and bandwidth across the many layers
that comprise modern {\em memory hierarchies}, thereby making it difficult
to deliver high performance at low cost with minimal energy consumption.

\section{Focus of This Dissertation: Efficiency of the Memory Hierarchy } 
\begin{figure}[ht] 
\centering
\includegraphics[width=0.95\textwidth]{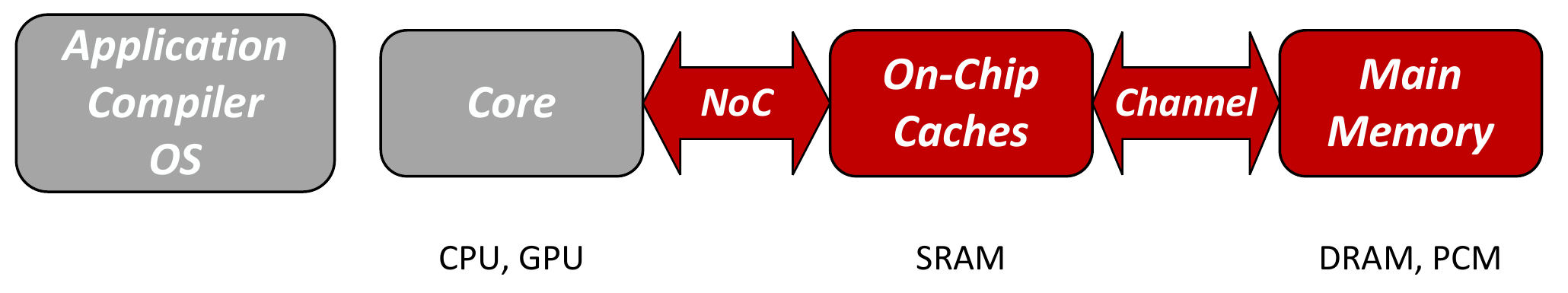} \caption{Data
compression from the core to the main memory. } 
\label{fig:full} \end{figure}

This
dissertation focuses on performance and energy efficiency of the modern memory
hierarchies. We observe that existing systems have significant redundancy in
the data (i) \emph{stored} in the memory hierarchies (e.g., main memory, on-chip
caches) and (ii) \emph{transferred} across existing communication channels (e.g.,
off-chip bus and on-chip interconnect).  Figure~\ref{fig:full} shows parts of
the system stack where we aim to apply data compression (in red/dark).

In this dissertation, we first propose a simple and fast yet efficient
compression algorithm that is suitable for on-chip cache compression. This
algorithm solves one of the key challenges for cache compression: achieving
low decompression latency, which is on the critical path of the execution.
%while maintaining high compression ratio. 
Then, we show that \emph{compressed
cache block size} is a new important factor when making cache replacement decisions
that helps to outperform state-of-the-art cache replacement mechanisms.

We then propose a new design for main memory
compression that solves a
key challenge in realizing data compression in main memory: the disparity between
how the data is stored (i.e., at a \emph{page} granularity) and how it is accessed (i.e., at a
\emph{cache line} granularity).

Finally, we show that bandwidth compression---both
on-chip and off-chip---can be efficient in providing high effective bandwidth
 in the context of modern GPUs (with more than a hundred real applications
evaluated). At the same time, we find that there is a new important problem
with bandwidth compression that makes it potentially energy
inefficient -- the significant increase in the number of \emph{bit toggles}
(i.e., the number of transitions between zeros and ones) that leads to an 
increase in dynamic energy. We provide an efficient solution to this problem. 

\subsection{A Compelling Possibility: Compressing Data throughout the Full Memory Hierarchy}

At first glance, {\em data compression} may seem like an obvious approach
to reducing the negative impacts of processing large amounts of data.  In
theory, if data compression could effectively reduce the size of the data
without introducing significant overheads, it would relieve pressure on
both the {\em capacity} of the various layers of the memory hierarchy
(including caches, DRAM, non-volatile memory technologies, etc.) as well as
the {\em bandwidth} of the communication channels (including memory buses, etc.)
that transfer data between these layers.  This in turn would allow system
designers to avoid over-provisioning these resources, since they could
deliver performance more efficiently as a function of system cost and/or
power budget.  Perhaps surprisingly, although forms of data compression
have been used for many years to reduce file system storage requirements
(e.g., by using {\tt gzip} to compress files), there has been little to no
use of compression within modern memory hierarchies.\footnote{The only real
  exception that we are aware of is IBM's MXT technology~\citep{MXT}, which
  was shipped in commercial products roughly 10 years ago, but which has
  not become widely adopted.} Why not?

\subsection{Why Traditional Data Compression Is Ineffective for Modern Memory Systems} 

Traditional file compression algorithms such as Lempel-Ziv~\citep{lz}
achieve high compression ratios by scanning through the file from the
beginning, building up a dictionary of common character sequences (which is
stored within the compressed file and used for decompression).  In the
context of storing files on disk, variations of Lempel-Ziv have been very
popular because files are often accessed as sequential streams, and because
the large decompression latencies are considered to be acceptable given
that (i) disk accesses are already slow, and (ii) saving as much
disk space as possible is typically a very high priority.

In contrast to accessing compressed files on disk, two things are
fundamentally different when a processor accesses data (via loads and
stores) within its memory hierarchy: (i) {\em latency} is extremely
critical, and (ii) data is commonly {\em accessed randomly} (rather than
sequentially).  Because processor performance is so sensitive to memory
access latency, it is critical that the {\em decompression latency} must be
as small as possible when accessing compressed data within the memory
hierarchy.  Otherwise, system designers and users will quickly become
disenchanted with memory compression if it costs them significant
performance.  Ideally, if decompression latency is small enough,
compression within the memory hierarchy should actually {\em improve
  performance} by improving cache hit rates and reducing bandwidth-related
stalls.  The fact that main memory is randomly accessed creates additional
challenges, including {\em locating} (as well as decompressing) arbitrary
blocks of data efficiently, plus achieving significant compression ratios
without being able to use Lempel-Ziv's approach of building up dictionaries
over large access streams.

\section{Related Work}
Several prior works have proposed different mechanisms to improve the
efficiency of the memory hierarchy to provide (i) higher capacity, (ii) higher
bandwidth, (iii) lower latency, and (iv) higher energy efficiency. In this
section, we summarize some of the approaches that are related to our work. We
summarize those works based on their high-level insight and compare them with
the mechanisms proposed in this thesis. 

\subsection{3D-Stacked DRAM Architectures} One of the major limitations of the
existing DRAM-based memories is their limited off-chip bandwidth. One way to
overcome this limitation is by vertically stacking multiple DRAM chips that
provide wider IO interfaces, and hence increase the available off-chip
bandwidth to improve performance.  Many recent works have proposed designs and
architectures based on this idea (e.g., ~\cite{jedec-wideio2,jedec-hbm,
jedec-hbm,lee-isscc14,hmc10,hmc11}) to get higher off-chip bandwidth, or
to utilize 3D-stacked memory's higher capacity as a cache
(e.g.,~\cite{black-micro08,loh-isca08,loh-micro09,woo-hpca10}).  These designs
are largely orthogonal to the ideas proposed in this thesis, and hence can be
used together.

\subsection{In-Memory Computing}
Processing in memory (PIM) has been previously (e.g., \cite{LogicInMemory,NON-VON,EXECUBE,Terasys,computationalRAM,IRAM,ActivePages,FlexRAM2,FlexRAM})
and more recently (e.g., \cite{rowclone,GS-DRAM,AndOrDRAM,LazyPIM,PointerChasing,ContRunAhead,SchedPIM,Milad1,PIM2,PIM3}) explored to perform
computation near the data to reduce the off-chip bandwidth bottleneck improving
both the performance and energy efficiency.  More recently the idea of PIM have
been actively explored again in the context of 3D-stacked memory (e.g.,
\cite{Ahn1,Ahn2,Akin,Babarinsa,NDA,Gao1,BBSync,in-memory1,Gao2,TOM,LazyPIM,SchedPIM}).
These prior works might require (i) programmer effort to
map regular computation and data to PIM, or (ii) significant increase in the overall
cost of the system and/or cost-per-bit of the modern DRAM.
The mechanisms proposed in this dissertation are also applicable to systems that
perform in-memory computation.
% that has a very
%high-density process that is not good for building high-speed PIM logic.

\subsection{Improving DRAM Performance}
Many prior works look at different ways to improve the efficiency of modern
DRAM architectures by either reducing the average access latency (e.g.,~\cite{lee-hpca2013,lee-hpca2015,rowclone,malladi-isca2012,ChangKHGHLLPKM16}) 
or enable higher parallelism within the DRAM itself (e.g.,~\cite{salp,Chang1}).
The approaches used by these work include (i) exploiting DRAM heterogeneity (e.g., Tiered-Latency DRAM~\cite{lee-hpca2013}), Dynamic Asymmetric Subarray~\cite{DynamicAsymmetricSubarray},
Low-Cost Interlinked Subarrays~\cite{lisa}), (ii) improving DRAM parallelism~\cite{salp,Chang1},
(iii) exploiting variation in DRAM latency (e.g., Adaptive Latency DRAM~\cite{lee-hpca2015}, ChargeCache~\cite{chargecache}),
(iv) smarter refresh and scheduling mechanisms (e.g.,~\cite{ESKIMO,raidr,Chang1,Avatar,Liu2,RAPID}), and
(v) more intelligent memory scheduling and partitioning 
algorithms (e.g.,~\cite{parbs,stfm,TCM,ATLAS,Ebrahimi1,hps-tr,bliss,DASH,ASM,MISE,sch2,sch3,sch4,sch5,sch6,part1,part2,part3,part4}). 
Many of these techniques can significantly improve DRAM performance (in terms of latency and energy efficiency),
but are not capable of providing higher effective off-chip bandwidth or higher effective DRAM capacity
by exploiting the existing redundancy in the data itself. The ideas in this dissertation can be exploited
in conjunction with many of these techniques, e.g., intelligent memory scheduling.  

\subsection{Fine-grain Memory Organization and Deduplication}
Several different proposals aim to improve memory performance by changing its page-granularity organization
(e.g., fine-grain memory deduplication~\cite{HICAMP}, fine-grain virtual page management~\cite{overlays}).
The proposed frameworks usually require significant changes to the existing virtual page organization 
that frequently leads to a significant increase in the cost. The techniques proposed in this thesis are much
less radical in the way they affect the higher levels of the systems stack.
The key difference with the deduplication approach~\cite{HICAMP} is that data redundancy is exploited
at a much finer granularity (e.g., 1--4 byte vs. 16--64 byte), hence much higher compression ratios are possible
for many applications. Our techniques are complementary to fine-grain virtual page management works (e.g.,~\cite{overlays}).

\subsection{Data Compression for Graphics}
Data compression is a widely used technique in the specialized area of texture compression~\cite{LDR,floating,bufferCompression}
used in modern GPUs. These approaches have several major limitations.
First, compressed textures are usually read-only that is not acceptable for many applications.
Second, compression/decompression latency is quite significant that limits applicability of these
algorithms to latency-insensitive applications. Our work is targeted towards more general-purpose workloads
where it is difficult to customize the compression algorithm to very specialized characteristics found in
graphics processing. 

\subsection{Software-based Data Compression}
Several mechanisms were proposed to perform memory compression in software
(e.g., in the compiler~\cite{PointerComp}, in the operating system~\cite{vm-compression}) for various modern
operating systems (e.g., Linux~\cite{linux}, MacOS~\cite{macos}, Windows~\cite{windows}, AIX~\cite{aix}). 
While these techniques can be quite efficient in reducing applications' memory footprint, their major limitation
is very slow (usually software-based) decompression. This limits these mechanisms to compressing
only ``cold'' pages (e.g., swap pages).

\subsection{Code Compression}
Compression was successfully applied not only to the application data, but also to the code itself~\cite{instr0,
instr1,instr2,instr3,instr4,instr5,instr6,instr7,instr8,instr9,instr10}. 
The primary goal in these works was usually to reduce the program footprint (especially in the context of
embedded devices).% that is similar to some of the goals when data compression is applied for applications' data.
The reduced footprint can allow for more instructions to be stored in the instruction caches, and hence
reduce the number of instruction cache misses, which, in turn, improves performance.
In this dissertation, we do not specialize for code compression. Instead, our goal is to enable
general data compression. Hence, the key difference between these prior works on code compression with the designs proposed in this dissertation 
is in the compression algorithms themselves: 
code compression algorithms are usually significantly tuned for a specific input -- instructions, 
and usually not effective for data compression.

\subsection{Hardware-based Data Compression}
Hardware-based data compression received some
attention in the past (e.g., \cite{fvc,MXT,fpc,reetu1,c-pack,MMCompression}), but unfortunately proposed general-purpose designs were
not practical either due to unacceptable compression/decompression latency or high design complexity and high overhead to support variable
size blocks after compression. In this thesis, we will show how to overcome these challenges in several practical designs across
the whole memory hierarchy. We will provide comprehensive quantitative comparisons to multiple previous
state-of-the-art works on hardware-based data compression (e.g., ~\cite{fpc,c-pack,zca,fvc,MMCompression,MXT}).

\section{Thesis Statement: Fast and Simple Compression \\ throughout the Memory Hierarchy} 

The key insight in our approach is that (i) {\em decompression latency} and
(ii) {\em simplicity of design} are far more critical than {\em compression
  ratio} when designing a compression scheme that is effective for modern
memory systems (in contrast to traditional file compression techniques
aimed at disk storage).  We have identified simple and effective
mechanisms for compressing data in on-chip caches (e.g., by exploiting {\em
  narrow dynamic ranges}) and in main memory (e.g., by adopting a common
compression ratio for all cache blocks within a page) that achieve
significant compression ratios (roughly a factor of two in most cases)
while adding minimal access latency overhead~\cite{bdi,LCP,camp,toggles-hpca}.  The
simplicity of our proposed mechanisms enables elegant solutions for
dealing with the practical challenges of how on-chip caches and main memories
are organized in modern systems.

The ultimate goal of this research is to validate the following thesis:
\begin{quote}
\textbf{\em It is possible to develop a new set of designs for data compression within modern memory
  hierarchies that are fast enough, simple enough, and effective enough in
  saving storage space and consumed bandwidth such that the resulting improvements in performance, cost,
  and energy efficiency will make such compression designs attractive to
  implement in future systems.}
\end{quote}

The hope is to achieve this goal through the following new mechanism:
\begin{quote}
{\em Data compression hardware (along with appropriate operating system
  support) that (i) efficiently achieves significant compression ratios with
  negligible latencies for locating and decompressing data, and (ii)
  enables the seamless transfer of compressed data between all memory
  hierarchy layers. }
\end{quote}

%The new compression framework should
%relieve pressure on both the {\em capacity} of the various layers of the
%memory hierarchy (including caches, DRAM, non-volatile memory technologies,
%etc.) as well as the {\em bandwidth} of the interconnects (including memory
%buses, etc.)  that transfer data between these layers.  This in turn would
%allow system designers to avoid over-provisioning these resources, enabling
%them to choose more favorable tradeoffs between performance, cost, and
%power. 
As a result of this, future computer systems would be better suited to the
increasingly data-intensive workloads of the future.

\section{Contributions}
This dissertation makes the following contributions.
\begin{enumerate}

\item We propose a new compression algorithm (\bdc) that achieves a high
compression ratio. \bdc exploits the existing low dynamic range of values
present in many cache lines to compress them to smaller sizes using Base+Delta
encoding. \bdc yields itself to a very low latency decompression pipeline
(requiring only a masked vector addition). To our knowledge, no prior work
achieved such low latency decompression at high compression ratio. \textbf{Chapter 3} describes \bdc
implementation and its evaluation in more detail.

\item We observe that the compressed size of a cache block can be indicative of
its reuse. We use this observation to develop a new cache insertion policy for
compressed caches, the Size-based Insertion Policy (SIP), which uses the size
of a compressed block as one of the metrics to predict its potential future
reuse. We introduce a new compressed cache replacement policy, Minimal-Value
Eviction (MVE), which assigns a value to each cache block based on both its
size and its reuse and replaces the set of blocks with the smallest value.  Both
policies are generally applicable to different compressed cache designs (both
with local and global replacement) and can be used with different compression
algorithms.  \textbf{Chapter 4} describes our proposed design, Compression-Aware
Management Policies (CAMP = MVE + SIP) in detail.

\item We propose a new compression framework (LCP) that solves the problem of
efficiently computing the physical address of a compressed cache line in main
memory with much lower complexity and power consumption than prior
proposals. We demonstrate that \emph{any} compression
algorithm can be adapted to fit the requirements of LCP, and  
that LCP can be efficiently integrated with existing cache compression
designs (\textbf{Chapter 7}), avoiding extra compression/decompression.
\textbf{Chapter 5} provides detailed implementation and evaluation of this framework.

%\item We comprehensively evaluate the use of data
%  compression to alleviate the memory bandwidth bottleneck in modern
%  GPUs. Our evaluations across a wide variety of GPGPU applications
%  show that data compression with simple hardware-based compression
%  algorithms (e.g., \bdc) on average (1) reduces memory bandwidth
%  by 2.1X, (2) improves performance by 42\%, and (3)
%  reduces overall system energy by 22\%. We provide more details and analysis
%  in \textbf{Chapter 6}.

\item We observe that hardware-based bandwidth compression applied to
on-chip/off-chip communication interfaces poses a new challenge for system
designers: a potentially significant increase in the bit toggle count as a
result of data compression. Without proper care, this increase can lead to
significant energy overheads when transferring compressed data that was not
accounted for in prior works. We propose 
a set of new mechanisms to address this new challenge: Energy
Control and Metadata Consolidation.
We provide a detailed analysis and evaluation of a large
spectrum of GPU applications that justify (i) the usefulness of
data compression for bandwidth compression in many real
applications, (ii) as well as the existence of the bit toggle problem for
bandwidth compression, and (iii) effectiveness of our
new mechanisms to address bit toggle problem, in \textbf{Chapter 6}.
  
\end{enumerate}

%%% Local Variables: 
%%% mode: latex
%%% TeX-master: "main"
%%% End: 

\chapter{Key Challenges for Hardware-Based Memory Compression}

%%%%%%%%%%%%%%%%%%%%%%%%%%%%%%%%%%%%%%%%%%%%%%%%%%%%%%%%%%%%%%%%%%%%%%%%%%%%%
\label{chap:challenges}
There are two major factors that limit
the current use of data compression in modern memory hierarchies:
(i) the increase in access latency due to compression/decompression and 
(ii) supporting variable data size after compression.
In this chapter, we discuss these major factors and how they affect
the possibility of applying data compression at different levels of the
memory hierarchy.

\section{Compression and Decompression Latency}
\subsection{Cache Compression}
%Cache compression is a promising technique to increase on-chip cache capacity
%and to decrease on-chip and off-chip bandwidth usage. Unfortunately, directly
%applying well-known compression algorithms (usually implemented in software)
%leads to high hardware complexity and unacceptable decompression/compression
%latencies, which in turn can negatively affect performance.  
In order to make
cache compression practical, we have to answer the following
 key question: what is the
right compression algorithm for an on-chip memory hierarchy?

The conventional wisdom is usually to aim for the highest possible compression
ratio. This is usually achieved by using existing software-based compression
algorithms that work by finding common subsets of data and storing them only
once (i.e., dictionary-based compression), and then simplifying these algorithms
so that they can be implemented in hardware.  Instead of following this
conventional path, another option is to prioritize simplicity of the compression
algorithm over its efficiency (i.e., compression ratio).  In summary,
the major challenge is to balance the compression/decompression {\em speed} (decompression latency
is especially important, because it is on the execution critical path) and {\em
simplicity} (no complex or costly hardware changes), while still being {\em effective}
(having good compression ratio) in
saving storage space. 

\subsection{Main Memory}
For main memory, compression/decompression latency is still an important factor, but
there is definitely more headroom to play with, since typical memory accesses can
take hundreds of processor cycles. Similar to on-chip caches, decompression lays 
on the critical path of the execution, and hence is the top priority in selecting
a proper compression algorithm. Prior attempts to use existing
software-based algorithms (e.g., Lempel-Ziv~\cite{lz}) were not successful~\cite{MXT}, because
even optimized versions of these algorithms for hardware had decompression latencies of 64 or more
cycles.

\subsection{On-Chip/Off-chip Buses}
Data compression is not only effective in providing higher capacity, it can also provide
higher effective bandwidth when applied to communication channels. We call
this effect \emph{bandwidth compression}.
For major memory communication channels (e.g., on-chip/off-chip buses), compression and decompression
are usually equally important, since both of them are directly added to the data transfer latency:
\emph{compression latency} (before sending the data), and \emph{decompression} latency 
(after the data is received). Hence, the challenge
is to properly balance both of these latencies without sacrificing the compression ratio.

It is possible to avoid some of these overheads, by storing and transferring the data in compressed form.
For example, if the main memory already stores compressed data, then there is no need to compress it again before
transferring it to the on-chip caches, etc. In a holistic approach, where compression is applied across many layers of the memory hierarchy 
(e.g., on-chip caches and main memory), it is possible that there is almost no overhead for bandwidth compression 
since both the source and the destination can store data in the same compressed form. 

\section{Quickly Locating Compressed Data}
While compression improves effective capacity and bandwidth, one challenge
is due to the fact that it generates data blocks in variable sizes.
It poses several challenges, and one of those challenges is the ability to
quickly locate the compressed data. In the uncompressed memory organization,
finding a certain cache line within a memory page is usually trivial: cache
line offset within a physical page is the same as the cache line offset within the virtual page.
Unfortunately, compression adds yet another layer of indirection, where cache
line offsets can vary significantly within a physical page, depending on compressed sizes of the
previous cache lines on the same page.

\textbf{For main memory}, this means that we either need to store the offsets of
all cache lines somewhere (either on-chip or in a different memory page) or
continuously compute those offsets (multiple additions of the previous cache
line sizes/offsets) from some metadata (which still needs to be stored
somewhere).  Both options can lead to (i) significant energy and latency
overheads and (ii) can significantly complicate the final design~\cite{MXT}.
It is important to mention that this challenge affects only main memory
compression because of the disparity in how the data is stored (e.g., 4KB page
granularity) and how it is accessed (e.g., 64B cache line granularity).  This
is usually not an issue for compressed cache organizations where tags and
actual cache blocks utilize simple mapping algorithms.  Similarly, it is not a
problem for transferring compressed data over on-chip/off-chip communication
channels, where data is usually transferred in small chunks (e.g., 16B flits in
on-chip interconnects). 

\section{Fragmentation}
Another challenge posed by the variable size blocks after compression is data
fragmentation.  \textbf{For on-chip caches}, the key issue is that after the
compressed block is stored in the data store, it has a fixed size, and then it
is immediately followed by another cache block (except for the last block).
The problem arises when this compressed cache line is updated with new data.
In that case, the cache line might not be compressed to the same size
as it was before, and hence there is not enough
space to simply store the new data for this cache block without moving data
around.  For a na\"{\i}ve compressed cache implementation, this could lead to
significant energy waste and design complexity when shuffling data around after cache writebacks.

\textbf{For main memory}, there can be two types of fragmentation: page level
and cache line level. Page level fragmentation happens due to the fact that it
is hard to support a completely flexible page size after compression, because
this would severely complicate the OS memory management process. Hence, in most
realistic designs (e.g., \cite{MMCompression}) only certain page sizes are
possible (e.g., 1KB, 2KB and 4KB).  This means that for every page that is not
compressed to exactly one of these sizes, its physical size would be rounded up to
the closest size that can fit this page. Cache line level fragmentation happens
due to the fact that many designs limit the number of compressed sizes for cache
lines within a particular page to reduce the amount of metadata to track per
cache line. Similar to page-level fragmentation, this means that many cache
lines could be padded to align with the smallest acceptable compressed block
size that fits them.

\section{Supporting Variable Size after Compression} 

The variable-sized nature of compression output causes significant challenges for
\textbf{on-chip/off-chip communication channels}.  For example, off-chip DRAM
buses are usually optimized to transfer one cache line (e.g., 64 bytes) at a time.
There is no easy mechanism (without changes to the existing DRAM) to
transfer smaller number of bytes faster. There are some exceptions with
GPU-oriented memories (e.g., GDDR5~\cite{gddr5}) where cache lines are
typically larger (128 bytes) and data buses are more narrow (32 bytes): hence
every cache line is transferred in four pieces, and data compression with
compression ratios up to 4$\times$ is possible without major changes to DRAM.  On-chip
interconnects usually transfer cache lines in several data chunks called flits.
In this case, compression ratio also limited by the granularity of the flits. 

\section{Data Changes after Compression}

Data compression inevitably changes the data itself, and,
unfortunately, sometimes these changes can lead to significant energy overhead.
There are several reasons for this. First, in every particular case,
it actually matters whether a 0 or 1 is transferred or stored. For example,
for the on-chip interconnect, that just transferred a 0 bit, 
transferring another 0 over the same pin that has just transferred a 0 is almost free in terms of energy, while transferring
1 would cost additional energy. Hence, higher number of switches on the interconnect wire (called
bit toggles) negatively affects energy efficiency of data communication.
Second, modern programming languages and compilers tend to store data in a regular
fashion such that data is usually nicely aligned at a 4/8-byte granularity.
This also nicely aligns with how the data is then transferred over communication
channels (e.g., 16-byte alignment for many modern on-chip networks). This means that
many similar bits are kept being transferred over the same pins, reducing the energy
cost of data transfers. Unfortunately, data compression frequently breaks this unspoken
assumption about ``nice'' data alignment, thereby significantly increasing the total number 
of bit toggles, and hence, increasing the energy of on-chip data transfers.

\section{Summary of Our Proposal}
In this dissertation, we aim to develop efficient solutions to overcome the
described challenges. 

To this end, we first propose a simple and fast yet
efficient compression algorithm that is suitable for on-chip cache compression (\textbf{Chapter 3}).
This algorithm solves one of the key challenges for cache compression:
achieving \emph{low decompression latency} (which is on the critical path of the
execution) while maintaining \emph{high compression ratio}.  Our algorithm is based on
the observation that many cache lines have data with a \emph{low dynamic range},
and hence can be represented efficiently using base-delta encoding. We
demonstrate the efficiency of the algorithm inspired by this observation
(called \emph{Base-Delta-Immediate Compression}) and the corresponding
compressed cache design.  

Second, we show that \emph{compressed block size} is
a new piece of information to be considered when making cache management decisions 
in a compressed (or even an uncompressed) cache.
Including this new piece of information  helps to
outperform state-of-the-art cache management mechanisms.
To this end, we introduce \emph{Compression-Aware
Management Policies} described in \textbf{Chapter 4}.  

Third, we propose a new design for main memory compression, called
\emph{Linearly Compressed Pages} (\textbf{Chapter 5}).  This mechanism solves a
key challenge in realizing data compression in main memory -- the disparity between
how the data is stored (i.e. page granularity), and how it is accessed (i.e.
cache line granularity).  

Fourth, we show that bandwidth
compression, both on-chip and off-chip, can be efficient in providing high
effective bandwidth increase in the context of modern GPUs. Importantly, 
we discover  that there is a new problem with bandwidth
compression that makes compression potentially energy inefficient -- number of \emph{bit toggles} (i.e. the number of
transitions between zeros and ones) 
increases significantly with compression, which leads to an increase in dynamic
energy. This problem was completely overlooked by the prior work on bandwidth
compression. We propose several potential solutions to this problem using our new
\emph{Energy Control} mechanisms (\textbf{Chapter 6}).

%%% Local Variables: 
%%% mode: latex
%%% TeX-master: "main"
%%% End: 

\chapter{Base-Delta-Immediate Compression}

\section{Introduction}
\label{bdi:sec:introduction}
\blfootnote{Originally published as ``Base-Delta-Immediate Compression: Practical Data
Compression for On-Chip Caches''in the 21st International Conference on Parallel
Architectures and Compilation Techniques, 2012~\cite{bdi}.}
To mitigate the latency and bandwidth limitations of accessing main
memory, modern microprocessors contain multi-level on-chip cache
hierarchies.  While caches have a number of design parameters and
there is a large body of work on using cache hierarchies more
effectively (e.g., \cite{iic,RRIP,dip,line-distillation,EAF,Seshadri1,Seznec1,mlp,Qureshi1,Johnson1,Johnson2,Tyson1}), one key
property of a cache that has a major impact on performance, die area,
and power consumption is its {\em capacity}.  The decision of how
large to make a given cache involves tradeoffs: while larger caches
often result in fewer cache misses, this potential benefit comes at
the cost of a longer access latency and increased area and power
consumption.

As we look toward the future with an increasing number of on-chip
cores, the issue of providing sufficient capacity in shared L2 and
L3 caches becomes increasingly challenging.  Simply scaling cache
capacities linearly with the number of cores may be a waste of
both chip area and power.  On the other hand, reducing the L2 and
L3 cache sizes may result in excessive off-chip cache misses,
which are especially costly in terms of latency and precious
off-chip bandwidth.

One way to potentially achieve the performance benefits of larger cache
capacity without suffering all disadvantages is to exploit {\em data
  compression}~\cite{fpc,register-caching,iic-comp,OldCompression,fvc,fvl}.  Data
compression has been successfully adopted in a number of different contexts
in modern computer systems~\cite{huffman,lz} as a way to conserve storage
capacity and/or data bandwidth (e.g., downloading compressed files over the
Internet~\cite{Networks} or compressing main memory~\cite{MXT}).  However,
it has not been adopted by modern commodity microprocessors as a way to
increase effective cache capacity.  Why not?

The ideal cache compression technique would be {\em fast}, {\em simple},
and {\em effective} in saving storage space.  Clearly, the resulting
compression ratio should be large enough to provide a significant upside,
and the hardware complexity of implementing the scheme should be low enough
that its area and power overheads do not offset its benefits.  Perhaps the
biggest stumbling block to the adoption of cache compression in commercial
microprocessors, however, is {\em decompression latency}.  Unlike cache
{\em compression}, which takes place in the background upon a cache fill
(after the critical word is supplied), cache {\em decompression} is on the
critical path of a {\em cache hit}, where minimizing latency is extremely
important for performance.  
In fact, because L1 cache hit times are of utmost 
importance, we only consider compression of the L2 caches and beyond
in this study (even though our algorithm could be applied to any cache).

Because the three goals of having {\em fast}, {\em simple}, and {\em
  effective} cache compression are at odds with each other (e.g., a very
simple scheme may yield too small a compression ratio, or a scheme with a very high
compression ratio may be too slow, etc.), the challenge is to find the right
balance between these goals.  Although several cache compression techniques
have been proposed in the past~\cite{fpc,c-pack,ZeroContent,iic-comp,fvc}, they
suffer from either a small compression ratio~\cite{ZeroContent,fvc},
high hardware complexity~\cite{iic-comp}, or large 
decompression latency~\cite{fpc,c-pack,iic-comp,fvc}.  
To achieve significant compression ratios while minimizing hardware complexity
and decompression latency, we propose a new cache compression technique 
called \textbf{Base-Delta-Immediate (\bdc)} compression.

\subsection{Our Approach: \bdc Compression}

The key observation behind \textbf{Base-Delta-Immediate~(\bdc)}
compression is that, for many cache lines, the data values stored
within the line have a {\em low dynamic range}: i.e., the relative
difference between values is small.  In such cases, the cache line can
be represented in a compact form using a common {\em base} value plus
an array of relative differences (``{\em deltas}''), whose combined
size is much smaller than the original cache line. (Hence the {\em
  ``base''} and {\em ``delta''} portions of our scheme's name).

We refer to the case with a single arbitrary base as {\em Base+Delta}
(\bd) compression, and this is at the heart of all of our designs.  To
increase the likelihood of being able to compress a cache line,
however, it is also possible to have {\em multiple bases}.  In fact,
our results show that for the workloads we studied, the best option is
to have {\em two bases}, where one base is always {\em zero}.  (The
deltas relative to zero can be thought of as small {\em immediate}
values, which explains the last word in the name of our \bdc
compression scheme.) Using these two base values (zero and something
else), our scheme can efficiently compress cache lines containing a
mixture of two separate dynamic ranges: one centered around an
arbitrary value chosen from the actual contents of the cache line
(e.g., pointer values), and one close to zero (e.g., small integer
values). Such mixtures from two dynamic ranges are commonly found
(e.g., in pointer-linked data structures), as we will discuss
later.

As demonstrated later in this chapter, \bdc compression offers the following
advantages: (i) a {\em high compression ratio} since it can exploit a
number of frequently-observed patterns in cache data (as shown using
examples from real applications and validated in our experiments); (ii)
{\em low decompression latency} since decompressing a cache line 
requires only a simple masked vector addition; and (iii) {\em relatively modest
  hardware overhead and implementation complexity}, since both the
compression and decompression algorithms involve only simple vector
addition, subtraction, and comparison operations.

\ignore{
This paper makes the following contributions:
\begin{itemize}

  \item We propose a new cache compression algorithm, Base-Delta-Immediate
    Compression (\bdc), which exploits the low dynamic range of
    values present in many cache lines to compress them to
    smaller sizes. Both the compression and decompression
    algorithms of \bdc have low latency and require only vector
    addition, subtraction and comparison operations.
  \item Based on the proposed \bdc compression algorithm, we
    introduce a new compressed cache design. This
    design achieves a high degree of compression at a lower decompression latency compared to two state-of-the-art
    cache compression techniques: Frequent Value
    Compression (FVC)~\cite{fvc} and Frequent Pattern Compression
    (FPC)~\cite{fpc}, which require complex and long-latency
    decompression pipelines~\cite{fpc-tr}.
  \item We evaluate the performance benefits of \bdc compared to
    a baseline system that does not employ compression, as well as
    against three state-of-the-art cache compression techniques~\cite{fpc,fvc,ZeroContent}.  
    We show that
    \bdc provides a better or comparable degree of compression for
    the majority of the applications we studied. It improves performance for
    both single-core (8.1\%) and multi-core workloads (9.5\%~/
    11.2\% for two- / four-cores). For many applications, compression with \bdc provides the performance
    benefit of doubling the uncompressed cache size of the baseline system.

\end{itemize}
}

\section{Background and Motivation}
\label{bdi:sec:background}

%% \begin{comment}
%% On-chip caches significantly reduce off-chip memory
%%   latencies by
%% \footnote{test}caching frequently used cache lines. When the working set of an
%% application fits into the on-chip cache storage, a majority of its
%% memory accesses are serviced by the cache leading to low latency
%% of memory access. On the other hand, if the working set is larger
%% than the cache size, most of the application's accesses go to
%% off-chip memory leading to much lower performance compared to
%% when the working set fits in the cache. As a result, increasing
%% the cache size will allow the processor to retain larger working
%% sets on-chip providing significant performance improvements.
%% \end{comment}
Data compression is a powerful technique for storing large amounts of
data in a smaller space. Applying data compression to an on-chip
cache can potentially allow the cache to store more cache lines in
compressed form than it could have if the cache lines were not
compressed. As a result, a compressed cache has the potential to
provide the benefits of a larger cache at the area and the power
of a smaller cache.
%% Data accessed by applications (i.e., in-memory or in-cache data) has a lot
%% of redundancy.  One of the primary reasons for this redundancy is the
%% existence of data with {\em low dynamic range} (LDR) -- a sequence of values where
%% the differences between the values in the sequence are small compared to
%% the size of the values themselves.
%% %% COMMENTED OUT BY VIVEK: Repeated in a later paragraph and does
%% %% not flow well here.
%% %% Low dynamic range data is commonly
%% %% found in media data where there is a low gradient in the pixel color
%% %% values~\cite{LDR}, or in a table of pointers where all the pointers point
%% %% to the same memory region.
%% Such a sequence of values  can
%% be efficiently encoded by storing a base value and an array of differences
%% between the base and all values in the sequence.

%% There are multiple patterns that can lead to a low dynamic range,
%% and, hence, result in high redundancy in data accessed by
%% applications.  We summarize the most common of such patterns
%% below.

Prior work~\cite{fpc,fvc,MMCompression} has observed that there is a significant amount of
redundancy in the data accessed by real-world applications. There
are multiple patterns that lead to such redundancy. We summarize
the most common of such patterns below.

\textbf{Zeros:} Zero is by far the most frequently seen value in
application data~\cite{VL,MMCompression,fvc}.  There are various
reasons for this. For example, zero is most commonly used to
initialize data, to represent NULL pointers or false boolean
values, and to represent sparse matrices (in dense form). 
In fact, a majority
of the compression schemes proposed for compressing memory data either
base their design fully around zeros~\cite{MMCompression,ZeroContent,ZeroValue,DynamicZero}, or treat zero as a special
case~\cite{fpc,vm-compression,fvl}.

\textbf{Repeated Values:} A large contiguous region of memory may
contain a single value repeated multiple
times~\cite{predictability}.  This pattern is widely present in
applications that use a common initial value for a large array, or
in multimedia applications where a large number of adjacent pixels
have the same color. Such a repeated value pattern can be easily
compressed to significantly reduce storage requirements.
Simplicity, frequent occurrence in memory, and high compression ratio make
repeated values an attractive target for a special consideration
in data compression~\cite{fpc}.

\textbf{Narrow Values:} A narrow value is a small value stored using a
large data type: e.g., a one-byte value stored as a four-byte
integer. Narrow values appear commonly in application data due to
over-provisioning or data alignment.  Programmers typically
provision the data types in various data structures for the worst case
even though a majority of the values may fit in a smaller data
type. For example, storing a table of counters requires the data type
to be provisioned to accommodate the maximum possible value for the
counters.  However, it can be the case that the maximum possible
counter value needs four bytes, while one byte might be enough to
store the majority of the counter values.  Optimizing such data
structures in software for the common case necessitates significant
overhead in code, thereby increasing program complexity and
programmer effort to ensure correctness.  Therefore, most programmers
over-provision data type sizes. As a result, narrow values present
themselves in many applications, and are exploited by different
compression techniques~\cite{fpc,vm-compression,narrow}.

\begin{table}[t]
\centering
    \begin{tabular}{|@{ }>{\scriptsize\bgroup}c<{\egroup}@{ }|>{\scriptsize\bgroup}c<{\egroup}@{ }|@{ }>{\scriptsize\bgroup}c<{\egroup}@{ }|
                    @{ }>{\scriptsize\bgroup}c<{\egroup}||>{\scriptsize\bgroup}c<{\egroup}@{ }|@{ }>{\scriptsize\bgroup}c<{\egroup}@{ }|
                    @{ }>{\scriptsize\bgroup}c<{\egroup}@{ }|@{ }>{\scriptsize\bgroup}c<{\egroup}@{ }|}
        %\toprule
         \hline
         \multirow{2}{*}{\rotatebox{30}{\textbf{}}}& \multicolumn{3}{c||}{\scriptsize \textbf{Characteristics}} & \multicolumn{4}{c|}{\scriptsize \textbf{Compressible data patterns}} \\
          \cline{2-8} 
          &   {Decomp. Lat.} & {Complex.} & {C. Ratio} & {Zeros} & {Rep. Val.}& {Narrow} & {LDR}\\
        %\cmidrule(rl){1-8}
         \hline
         ZCA~\cite{ZeroContent} & \textbf{Low} & \textbf{Low} & Low & \ding{52} & \ding{53} & \ding{53} & \ding{53} \\
         \hline
         %\cmidrule(rl){1-8}
         FVC~\cite{fvc} 
& High & High & Modest & \ding{52} & Partly & \ding{53} & \ding{53} \\
         \hline
         %\cmidrule(rl){1-8}
         FPC~\cite{fpc} 
& High & High & \textbf{High} & \ding{52} & \ding{52} & \ding{52} & \ding{53}  \\
         \hline
         %\cmidrule(rl){1-8}
         \bdc  & \textbf{Low} & Modest & \textbf{High} & \ding{52} & \ding{52} & \ding{52} & \ding{52} \\
         \hline
         %\bottomrule
    \end{tabular}%
 \caption{Qualitative comparison of \bdc with prior work. LDR: Low
   dynamic range. Bold font indicates desirable
     characteristics.}
  \label{tbl:comparison}%

\end{table}

\textbf{Other Patterns:} There are a few other common data
patterns that do not fall into any of the above three classes: a
table of pointers that point to different locations in the same
memory region, an image with low color gradient, etc. Such data
can also be compressed using simple techniques and has been
exploited by some prior proposals for main memory
compression~\cite{vm-compression} and image
compression~\cite{LDR}.

\begin{figure}[!htb]
\centering
\includegraphics[scale=0.55]{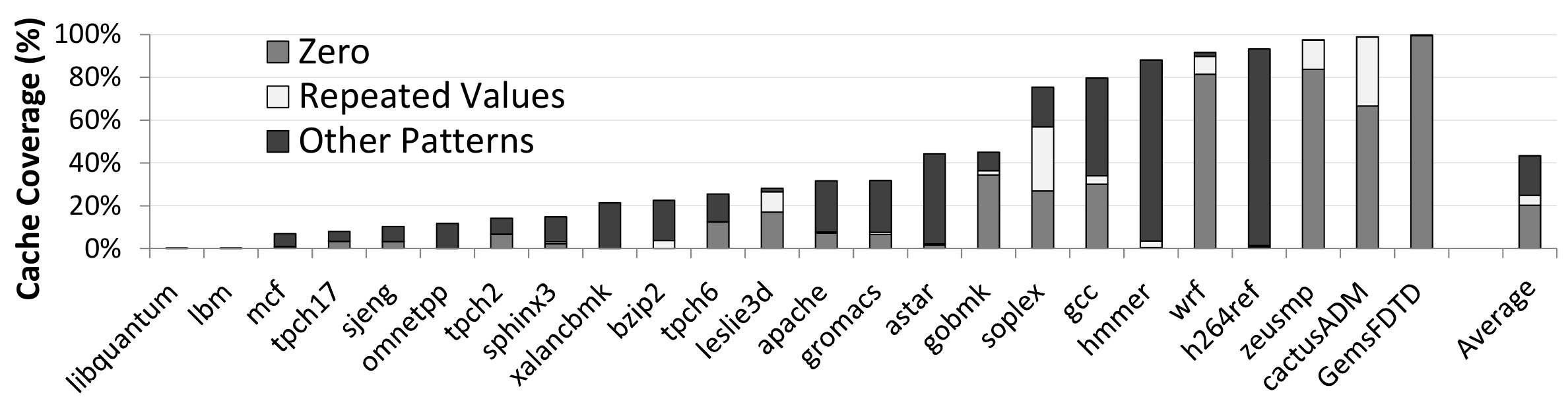}

 \caption{ Percentage of cache lines with different data patterns in a 2MB L2 cache.
          ``Other Patterns'' includes ``Narrow Values''.}
\label{fig:motivation2}

\end{figure}

In this work, we make two observations. First, we find that the above
described patterns are widely present in many applications (SPEC CPU
benchmark suites, and some server applications, e.g., Apache, TPC-H).
Figure~\ref{fig:motivation2} plots the percentage of cache lines that
can be compressed using different patterns.\footnote{The methodology
  used in this and other experiments is described in
  Section~\ref{sec:methodology}. We use a 2MB L2 cache unless
  otherwise stated.} As the figure shows, on average, 43\% of all
cache lines belonging to these applications can be compressed. This
shows that there is significant opportunity to exploit data
compression to improve on-chip cache performance.

Second, and more importantly, we observe that all the above commonly
occurring patterns fall under the general notion of \emph{low dynamic
  range} -- a set of values where the differences between the values
is much smaller than the values themselves. Unlike prior work, which
has attempted to exploit each of these special patterns individually
for cache compression~\cite{fpc,fvc} or main memory
compression~\cite{MMCompression,vm-compression}, our \textbf{goal} is
to exploit the general case of values with \emph{low dynamic range} to
build a simple yet effective compression technique.

\textbf{Summary comparison:}
Our resulting mechanism, base-delta-immediate (\bdc) compression,
strikes a sweet-spot in the tradeoff between decompression latency
(Decomp.~Lat.), hardware complexity of the implementation
(Complex.), and compression ratio (C. Ratio), as shown in
Table~\ref{tbl:comparison}.  The table qualitatively compares \bdc with three
state-of-the-art mechanisms: ZCA~\cite{ZeroContent}, which does zero-value
compression, Frequent Value Compression (FVC)~\cite{fvc},
and Frequent Pattern Compression (FPC)~\cite{fpc}.  (These mechanisms
are described in detail in Section~\ref{sec:comparison}.)
It also summarizes which data patterns (zeros, repeated
values, narrow values, and other low dynamic range patterns) 
are compressible with each mechanism. For modest complexity, \bdc 
is the only design to achieve both
low decompression latency and high compression ratio.

We now explain the design and rationale for our scheme in two parts.  In
Section~\ref{sec:bdc}, we start by discussing the core of our scheme, which
is \emph{Base+Delta~(\bd)} compression.  Building upon \bd, we then discuss
our full-blown \bdc compression scheme (with multiple bases) in
Section~\ref{sec:2-bdc}.

\section{Base + Delta Encoding:~Basic Idea}
\label{sec:bdc}
We propose a new cache compression mechanism, \emph{Base+Delta} (\bd)
compression, which unlike prior work~\cite{fpc,ZeroContent,fvc},
looks for compression opportunities at a cache line granularity --
i.e., \bd either compresses the entire cache line or stores the
entire cache line in uncompressed format. The key observation
behind \bd is that many cache lines contain data with low dynamic
range. As a result, the differences between the words within such
a cache line can be represented using fewer bytes than required to
represent the words themselves. We exploit this observation to
represent a cache line with low dynamic range using a common
\emph{base} and an array of \emph{deltas} (differences between
values within the cache line and the common base). Since the
\emph{deltas} require fewer bytes than the values themselves, the
combined size of the \emph{base} and the array of \emph{deltas}
can be much smaller than the size of the original uncompressed
cache line.

%% In contrast to prior cache compression approaches~\cite{fpc,
%%   ZeroContent, fvc}, our proposed mechanism,
%% \emph{Base+Delta~(\bd)} compression, looks for compression
%% opportunities at a cache line granularity -- i.e., either the
%% entire cache line is compressed or the entire cache line is stored
%% uncompressed. The key observation behind \bd is that many cache
%% lines contain data with low dynamic range -- the values in the
%% cache lines have a significant number of equal higher order bits.
%% As a result, differences between the words within the cache line
%% can be represented using fewer bytes than required to represent
%% the words themselves.  We exploit this observation to represent
%% words within such a cache line using an array of differences from
%% a common base value. The key idea is that if the words within a
%% cache line are sufficiently close in their values, then the
%% combined size of the common base and the array of differences will
%% be much smaller than the size of the original uncompressed cache
%% line.

The fact that some values can be represented in base+delta form
has been observed by others, and used for different purposes:
e.g. texture compression in GPUs~\cite{LDR} and also to save
bandwidth on CPU buses by transferring only deltas from a common
base~\cite{register-caching}.  To our knowledge, no previous work examined the
use of base+delta representation to improve on-chip cache
utilization in a general-purpose processor.

\begin{figure}[!h]
\centering
\includegraphics[scale=0.55]{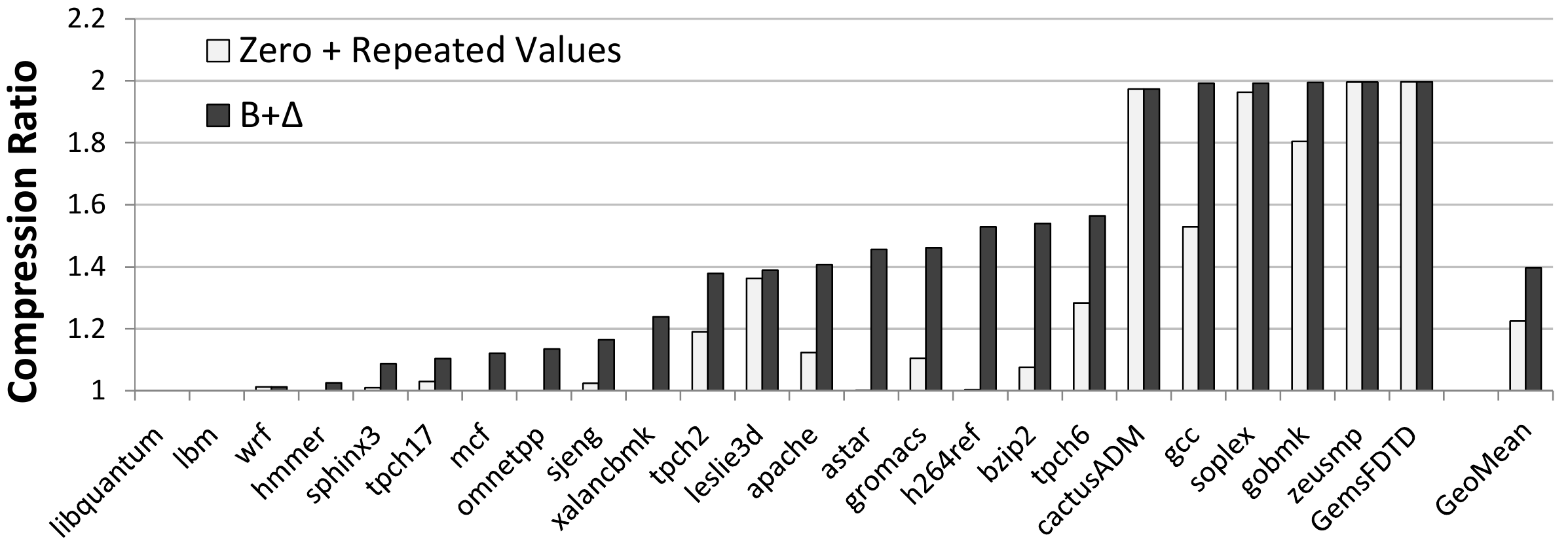}
 \caption{Effective compression ratio with different value patterns}
\label{fig:bdc-compressibility}
\end{figure}

To evaluate the applicability of the \bd compression technique for a
large number of applications, we conducted a study that compares the
effective compression ratio (i.e., effective cache size increase, see
Section~\ref{sec:methodology} for a full definition) of \bd against a
simple technique that compresses two common data patterns (zeros and
repeated values\footnote{Zero compression compresses an all-zero cache
line into a bit that just indicates that the cache line is
all-zero. Repeated value compression checks if a cache line has the
same 1/2/4/8 byte value repeated. If so, it compresses the cache line
to the corresponding value.}).  Figure~\ref{fig:bdc-compressibility}
shows the results of this study for a 2MB L2 cache with 64-byte cache
lines for applications in the SPEC CPU2006 benchmark suite, database
and web-server workloads (see Section~\ref{sec:methodology} for
methodology details).  We assume a design where a compression scheme
can store up to twice as many tags for compressed cache lines than the
number of cache lines stored in the uncompressed baseline cache
(Section~\ref{sec:design} describes a practical mechanism that
achieves this by using twice the number of tags).\footnote{This
assumption of twice as many tags as the baseline is true for all
compressed cache designs, except in Section~\ref{sec:res3}.} As the
figure shows, for a number of applications, \bd provides significantly
higher compression ratio (1.4X on average) than using the simple
compression technique. However, there are some benchmarks for
which \bd provides very little or no benefit
(e.g., \emph{libquantum}, \emph{lbm}, and \emph{mcf}). We will address
this problem with a new compression technique called \bdc in
Section~\ref{sec:2-bdc}.  We first provide examples from real
applications to show why \bd works.

% Before describing the
%\bdc technique, we provide examples from real applications to show
%why \bd works. We then describe the compression and decompression
%algorithms of \bd in detail, and lie the foundation for the
%\bdc compression algorithm.

%% We assume a design (see
%% Section~\ref{sec:design}) in which twice as many tags per set are
%% available as in conventional non-compressed cache, and compressed
%% cache line size is determined by the compression scheme, e.g., one
%% byte for zero compression.

%% The figure compares the potential of \bd compression against two
%% common data patterns \emph{combined} together: (1) all the differences are
%% zeros indicating that the entire cache line consists of the same
%% value (repeated values), (2) the entire cache line is filled with
%% zeros.  The figure indicates that there is a good potential for
%% \bd to be effective for a number of applications, but at the same
%% time there are applications, e.g., \emph{libquantum} and \emph{lbm}, that do not benefit
%% from compression using \bd. On average using \bd compression has the potential
%% to increase the effective cache size to 1.4X whereas compressing only zero
%% or repeated value patterns increases the size to 1.21X.
\subsection{Why Does \bd Work?}
\label{sec:examples}
\bd works because of: (1) regularity in the
way data is allocated in the memory (similar data values 
and types grouped together), and (2) low dynamic range of cache/memory data.  
The first reason is typically true due to the common usage of arrays to
represent large pieces of data in applications. The second reason is
usually caused either by the nature of computation, e.g., sparse
matrices or streaming applications; or by inefficiency
(over-provisioning) of data types used by many applications, e.g.,
4-byte integer type used to represent values that usually need
only 1 byte.  We have carefully examined different common data patterns
in applications that lead to \bd representation
 and summarize our observations in two
examples.

\begin{figure}[!h]
\centering
%\begin{minipage}[b]{0.5\linewidth}
  \includegraphics[scale=0.5]{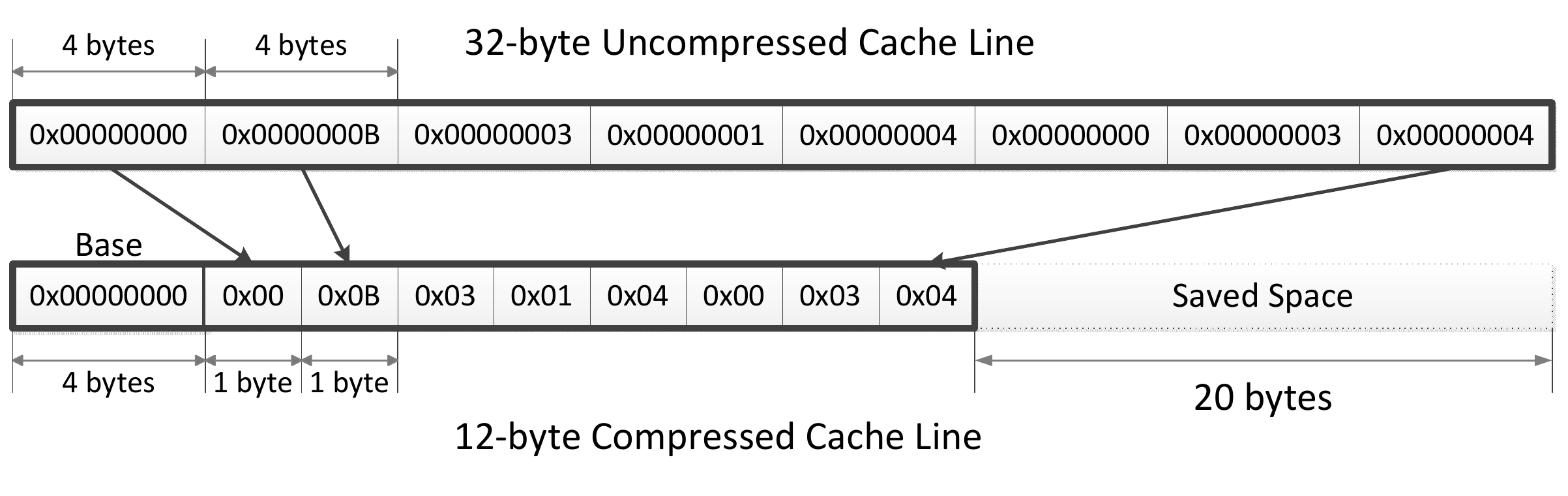}
  %\caption{Cache line from \emph{h264ref} compressed by \bd.}
  \caption{Cache line from \emph{h264ref} compressed with \bd}
  \label{fig:bdc-example}
%\end{minipage}
%\begin{minipage}[b]{0.5\linewidth}
  \includegraphics[scale=0.5]{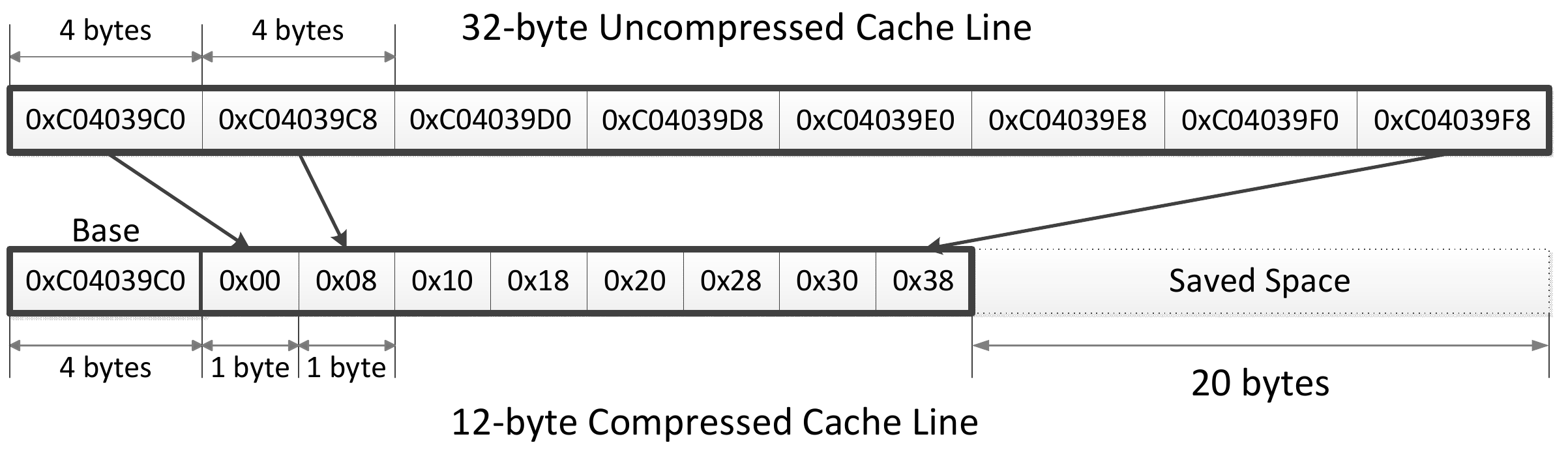}
  \caption{Cache line from \emph{perlbench} compressed with \bd}
  \label{fig:bdc-example2}

\end{figure}

Figures~\ref{fig:bdc-example} and \ref{fig:bdc-example2} show the compression of two
32-byte\footnote{We use 32-byte cache lines in our examples to save space.
64-byte cache lines were used in all evaluations (see Section~\ref{sec:methodology}).} 
cache lines from the applications \emph{h264ref} and
\emph{perlbench} using \bd. The first example from \emph{h264ref} shows
a cache line with a set of narrow values stored as 4-byte
integers. As Figure~\ref{fig:bdc-example} indicates, in this case,
the cache line can be represented using a single 4-byte base
value, $0$, and an array of eight 1-byte differences. As a result,
the entire cache line data can be represented using 12 bytes
instead of 32 bytes, saving 20 bytes of the
originally used space. Figure~\ref{fig:bdc-example2} shows a similar
phenomenon where nearby pointers are stored in the same cache line
for the \emph{perlbench} application. 

%Note that frequent value compression
%(FVC) can only compress the frequent values present within this
%cache line -- e.g., $0$ and $1$. All other values have to be
%represented as 4-byte integers. Frequent pattern compression

We now describe more precisely the compression and decompression
algorithms that lay at the heart of the \bd compression
mechanism.

\subsection{Compression Algorithm}

The \bd compression algorithm views a cache line as a set of
fixed-size values i.e., 8 8-byte, 16 4-byte, 
or 32 2-byte values for a 64-byte cache line. 
It then determines if the set of values can be
represented in a more compact form as a base value with a set of
differences from the base value. For analysis, let us assume that
the cache line size is $C$ bytes, the size of each value in the set is
$k$ bytes and the set of values to be compressed is $S = (v_1,
v_2, ..., v_n)$, where $n = \frac{C}{k}$. The goal of the
compression algorithm is to determine the value of the base,
$B^*$ and the size of values in the set, $k$, that provide
maximum compressibility. Once $B^*$ and $k$ are determined, the
output of the compression algorithm is $\{k, B^*, \Delta =
(\Delta_1, \Delta_2, ..., \Delta_n)\}$, where $\Delta_i = B^* -
v_i ~~\forall i \in \{1,..,n\}$.

\textbf{Observation 1:} The cache line is compressible \emph{only
  if} \\ ${\forall i}, \mathrm{max}(\mathrm{size}(\Delta_i)) < k$,
where $\mathrm{size}(\Delta_i)$ is the smallest number of bytes that is
needed to store $\Delta_i$.

%%% COMMENTED BY VIVEK
%This observation relates to whether the values within the cache
%line have low dynamic range.

In other words, for the cache line to be compressible, the number
of bytes required to represent the differences must be strictly
less than the number of bytes required to represent the values
themselves.

%%% COMMENTED BY VIVEK
% An implication of this observation is that if $k = 1$, then the
% size of each difference should be zero -- i.e., every byte
% within the cache line should be same.  We discuss this as a
% special case in Section~\ref{sec:special-cases}.

\textbf{Observation 2:} To determine the value of $B^*$, either the value of
$\mathrm{min}(S)$ or $\mathrm{max}(S)$ needs to be found.
  %the following equation: 
  %\begin{equation}
   % $ 
   % B^* = \frac{\mathrm{max}(S) + \mathrm{min}(S)}{2}.
   % \label{eqn:B*}
   % $
  %\end{equation}

The reasoning, where $\mathrm{max}(S)$/$\mathrm{min}(S)$ are the maximum and minimum values in the cache line, 
is based on the observation that
the values in the cache line are bounded by $\mathrm{min}(S)$ and
$\mathrm{max}(S)$.  And, hence, the optimum value for $B^*$ should
be between $\mathrm{min}(S)$ and $\mathrm{max}(S)$.  In fact, 
the optimum can be reached only for $\mathrm{min}(S)$, $\mathrm{max}(S)$, or 
exactly in between them. Any other value of $B^*$ can only increase the
number of bytes required to represent the differences.

 %% \footnote{Note that the optimum value
 %%   $B^*$ requires finding min and max values over the whole cache
 %%   line that may require complex hardware and introduce
 %%   additional latency to the \bdc compression algorithm.}.

%\begin{comment}
%  It is naturally
%  then to look for a simple approximation for $B^*$

%\item[]\textbf{Observation 2:} The value $B^*$ can be approximated by
%  picking one of the values from $S$ e.g., $v_1$
%\end{comment}

Given a cache line, the optimal version of the \bd compression algorithm needs to
determine two parameters: (1) $k$, the size of each value in $S$,
and (2) $B^*$, the optimum base value that gives the best possible
compression for the chosen value of $k$.

\textbf{Determining $k$.}  Note that the value of $k$ determines
how the cache line is viewed by the compression algorithm -- i.e.,
it defines the set of values that are used for
compression. Choosing a single value of $k$ for all cache lines
will significantly reduce the opportunity of compression. To
understand why this is the case, consider two cache lines, one
representing a table of 4-byte pointers pointing to some memory
region (similar to Figure~\ref{fig:bdc-example2}) and the other representing 
an array of narrow values
stored as 2-byte integers.  For the first cache line, the likely
best value of $k$ is $4$, as dividing the cache line into a set of
of values with a different $k$ might lead to an increase in dynamic range and reduce
the possibility of compression. Similarly, the likely
best value of $k$ for the second cache line is $2$.

Therefore, to increase the opportunity for compression by catering
to multiple patterns, our compression algorithm attempts to
compress a cache line using three different potential values of $k$
simultaneously: $2$, $4$, and $8$.
%\footnote{These values correspond to 
%the most common type sizes used in modern programming languages.}. 
The cache line is then
compressed using the value that provides the maximum compression
rate or not compressed at all.\footnote{ We restrict our search to these three values as almost all
basic data types supported by various programming languages have one of
these three sizes.}

%%% VIVEK: This might be the best place to talk about k=1

%As we mentioned before, $k = 1$ is a special case we discuss later in
%Section~\ref{sec:special-cases}.

\textbf{Determining $B^*$.}
For each possible value of $k$ $\in$ \{$2$, $4$, $8$\}, the cache line
is split into values of size $k$ and the best value for the base,
$B^*$ can be determined using Observation 2. However,
computing $B^*$ in this manner requires computing the maximum
or the minimum of the set of values, which adds logic complexity and
significantly increases the latency of compression.
%Although cache line compression is not on
%the critical path (as we will show later), this method requires additional logic to compute the maximum and minimum.

To avoid compression latency increase and reduce hardware
complexity, we decide to use the \emph{first} value from the set of values
as an approximation for the $B^*$.
For a compressible cache line with a low dynamic range, we find that
choosing the first value as the base instead of computing the
optimum base value reduces the average compression ratio only by 0.4\%.

\subsection{Decompression Algorithm}
To decompress a compressed cache line, the \bd decompression
algorithm needs to take the base value $B^*$ and an array of
differences $\Delta = {\Delta_1, \Delta_2, ..., \Delta_n}$, and generate the 
corresponding set of values $S = {(v_1, v_2, ..., v_n)}$. 
The value $v_i$ is simply given by $v_i = B^* + \Delta_i$. As a
result, the values in the cache line can be computed in
parallel using a SIMD-style vector adder. Consequently, the entire
cache line can be decompressed in 
the amount of time it takes to do an integer vector addition, using
a set of simple adders.

\section{\bdc Compression}
\label{sec:2-bdc}

\subsection{Why Could Multiple Bases Help?}
 \label{sec:examples.2}
 Although \bd proves to be generally applicable for many
 applications, it is clear that not every cache line can be
 represented in this form, and, as a result, some benchmarks do
 not have a high compression ratio, e.g., \emph{mcf}.  One common reason
 why this happens is that some of these applications can mix data
 of different types in the same cache line, e.g., structures of
 pointers and 1-byte integers. This suggests that if we apply \bd
 with multiple bases, we can improve compressibility for some of
 these applications.
 %%, and improve effective compression ratios for others.

\begin{figure}[ht!]
  \centering
  \includegraphics[scale=0.5]{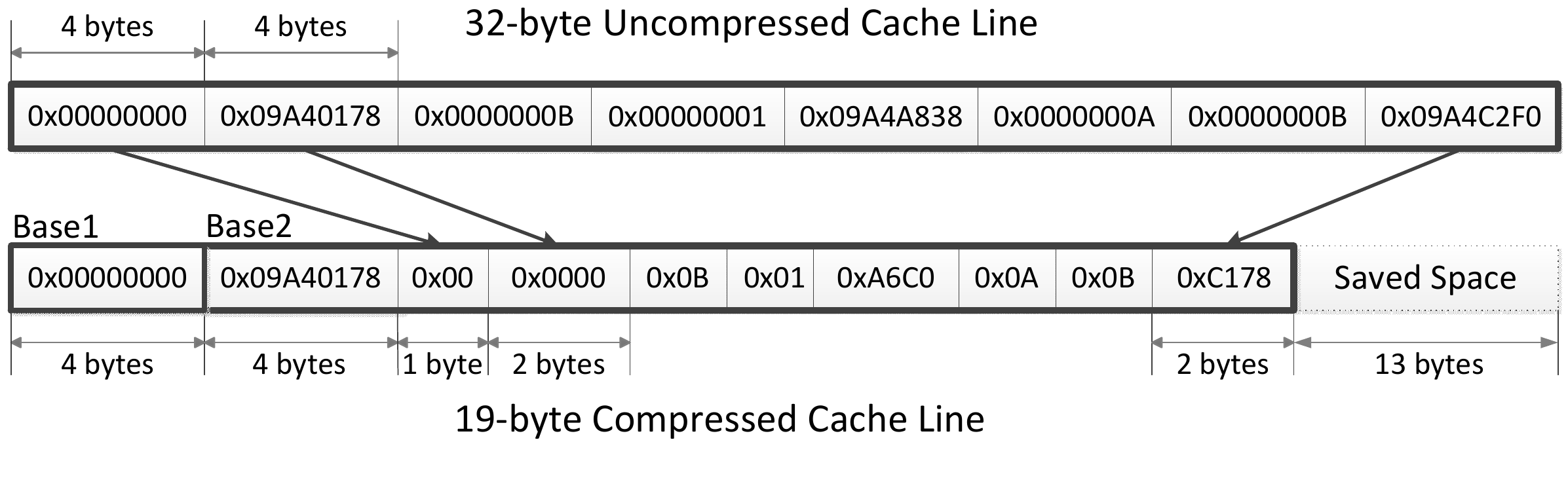}
  \caption{Cache line from \emph{mcf} compressed by \bd(two bases)}
  \label{fig:2-bases-example}
\end{figure}

Figure~\ref{fig:2-bases-example} shows a 32-byte cache line from
\emph{mcf} that is not compressible with a single base using \bd, because
there is no single base value that effectively compresses this
cache line. At the same time,
it is clear that if we use two bases, this cache line can
be easily compressed using a similar compression technique as in
the \bd algorithm with one base. As a result, the entire cache
line data can be represented using 19 bytes: 8 bytes for two bases
(\texttt{0x00000000} and \texttt{0x09A40178}), 5 bytes for five
1-byte deltas from the first base, and 6 bytes for three 2-byte deltas from the second base. 
This effectively saves 13 bytes of the 32-byte line.

As we can see, multiple bases can help compress more cache
lines, but, unfortunately, more bases can increase overhead (due to storage of the bases), 
and hence decrease effective compression ratio that can be achieved
with one base.
So, it is natural to ask \emph{how many bases are optimal for \bd
compression}?

\begin{figure}[ht!]
  \centering
  \includegraphics[scale=0.5]{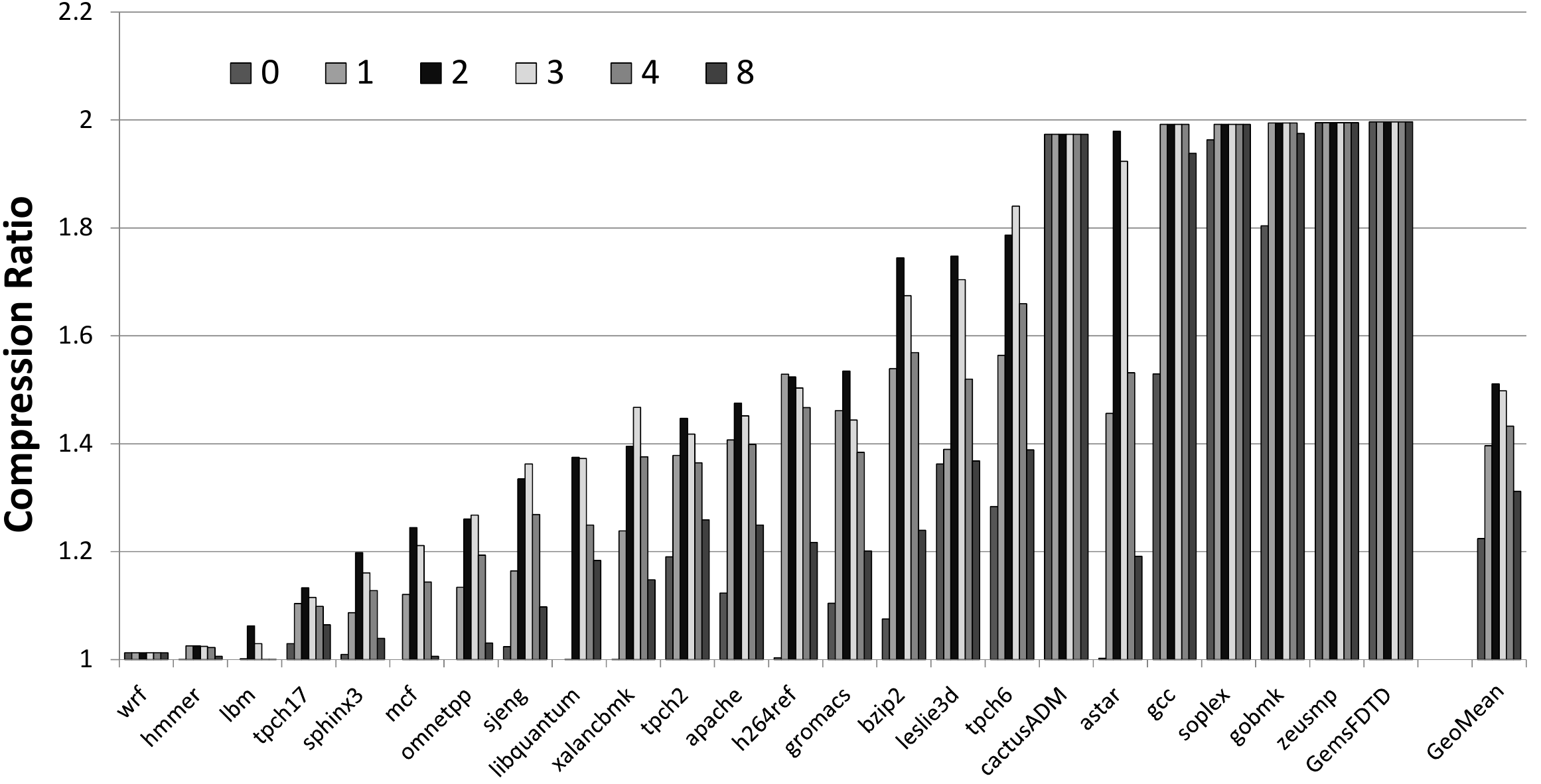}
 \caption{Effective compression ratio with different number of
 bases. ``0'' corresponds to zero and repeated value compression.}
\label{fig:multbases}
\end{figure}

In order to answer this question, we conduct an experiment where
we evaluate the effective compression ratio with different numbers
of bases (selected suboptimally using a greedy algorithm).
Figure~\ref{fig:multbases} shows the results of this
experiment. The ``0'' base bar corresponds to a mechanism that
compresses only simple patterns (zero and repeated values). These
patterns are simple to compress and common enough, so we can
handle them easily and efficiently without using \bd, e.g., a
cache line of only zeros compressed to just one byte for any
number of bases. We assume this optimization for all bars in
Figure~\ref{fig:multbases}.\footnote{If we do not assume this
optimization, compression with multiple bases will have very low
compression ratio for such common simple patterns.}

Results in Figure~\ref{fig:multbases} show that the empirically optimal number
of bases in terms of effective compression ratio is 2, with
some benchmarks having optimums also at one or three bases.  The
key conclusion is that \bd with two bases significantly
outperforms \bd with one base (compression ratio of 1.51 vs. 1.40 on average), suggesting that
it is worth considering for implementation. Note that having more
than two bases does not provide additional improvement in
compression ratio for these workloads, because the overhead of storing
more bases is higher than the benefit of compressing more cache lines.

Unfortunately, \bd with two bases has a serious
drawback: the necessity of finding a second base. 
The search for a second arbitrary base value (even a sub-optimal one) can add significant 
complexity to the compression hardware. This opens the question of
how to find two base values efficiently. We next propose a
mechanism that can get the benefit of compression with two bases
with minimal complexity.

\subsection{\bdc: Refining \bd with Two Bases and Minimal Complexity }
\label{sec:bdi}
Results from Section~\ref{sec:examples.2} suggest that the optimal (on average)
number of bases to use is two, but having an additional base has 
the significant shortcoming described above.  We observe that setting the second base to
zero gains most of the benefit of having an arbitrary second base value.
Why is this the case?

%%% VIVEK
%%% Percentage of the time one of the bases was zero.

Most of the time when data of different types are mixed in
the same cache line, the cause is an aggregate data type: e.g.,
a structure (\texttt{struct} in C). In many cases, this leads to the mixing of wide values with low dynamic
range (e.g., pointers) with narrow values (e.g., small integers).  
A first arbitrary base helps to compress wide values with low dynamic
range using base+delta encoding, while a second zero base is efficient enough
to compress narrow values separately from wide values.
%Programming languages like C/C++ constraint the alignment
%of a different size data. This results in padding or
%the addition of zero bits in front of smaller data types. And, hence,
%the second base set to zero is good enough for compressing these
%elements.  
Based on this observation, we refine the idea of \bd by
adding an additional implicit base that is always set to zero.  We
call this refinement \textbf{Base-Delta-Immediate} or
\textbf{\bdc} compression.

\begin{figure}
\centering
\includegraphics[scale=0.5]{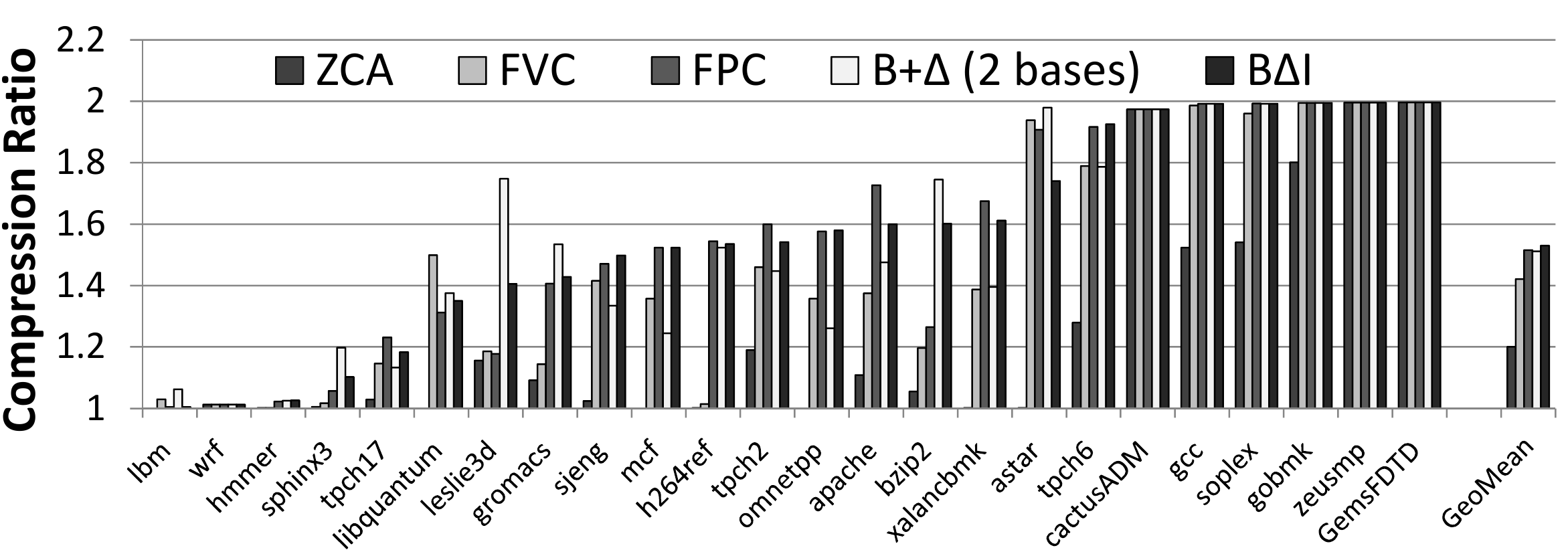}
 \caption{Compression ratio comparison of different algorithms: ZCA~\cite{ZeroContent}, 
          FVC~\cite{fvc}, FPC~\cite{fpc}, \bd (two arbitrary bases), and \bdc.
          Results are obtained on a cache with twice the tags to accommodate
          more cache lines in the same data space as an uncompressed cache.}
\label{fig:2-bdc-compressibility}
\end{figure}

There is a tradeoff involved in using \bdc instead of \bd with
two arbitrary bases. \bdc uses an implicit zero base as the second base, 
and, hence, it has less storage overhead, which means potentially higher
average compression ratio for cache lines that are compressible
with both techniques. \bd with two general bases uses more storage to store
an arbitrary second base value, but can compress more cache lines because the
base can be any value. As such, the compression ratio can potentially be better with
either mechanism, depending on the compressibility pattern of cache
lines. In order to evaluate this tradeoff, we compare in Figure~\ref{fig:2-bdc-compressibility}
the effective compression ratio of \bdc, \bd with two arbitrary bases,
and three prior approaches: ZCA~\cite{ZeroContent} (zero-based
compression), FVC~\cite{fvc}, and FPC~\cite{fpc}.\footnote{All
mechanisms are covered in detail in
Section~\ref{sec:comparison}. We provide a comparison of their
compression ratios here to give a demonstration of BDI's relative
effectiveness and to justify it as a viable compression mechanism.

}
%% The comparison results are
%% provided here to give a demonstration of intuition of \bdc's relative
%% compression ability and to justify it as a viable compression
%% mechanism.

 Although there are cases where \bd with two bases is better~---
 e.g., \emph{leslie3d} and \emph{bzip2}~--- on average, \bdc performs slightly
 better than \bd in terms of compression ratio (1.53
 vs. 1.51). We can also see that both mechanisms are better than
 the previously proposed FVC mechanism~\cite{fvc}, and competitive in terms of
 compression ratio with a more complex FPC compression
 mechanism. Taking into an account that \bd with two bases is
 also a more complex mechanism than \bdc, we conclude that our
 cache compression design should be based on the refined idea of
 \bdc.

 Now we will describe the design and operation of a cache that implements
 our \bdc compression algorithm.

\section{\bdc: Design and Operation}
\label{sec:design}

\subsection{Design}
\label{sec:design-design}
\begin{figure}[!htb]% 0.45\linewidth}

%\begin{minipage}[b]{0.5\linewidth}
\begin{center}
\includegraphics[scale=0.75]{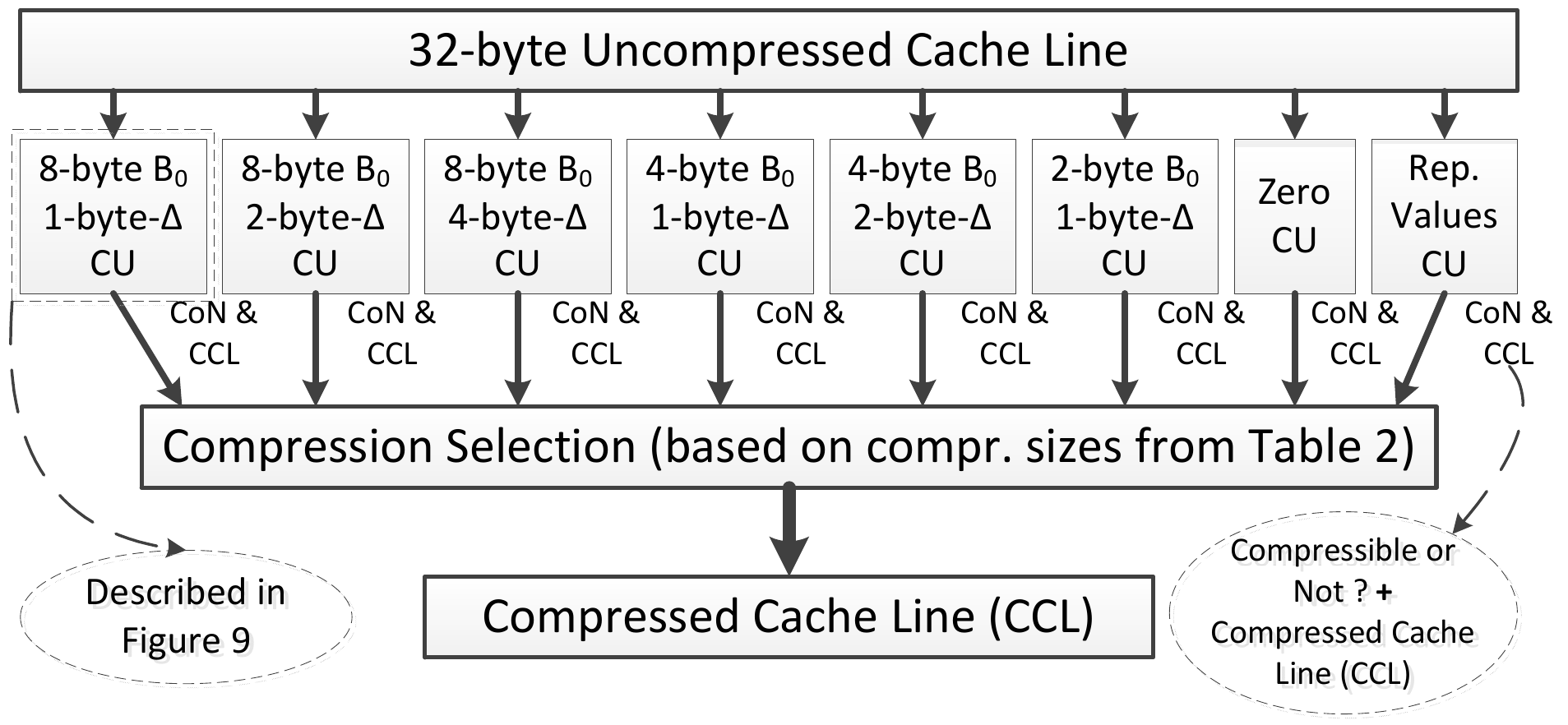}
 \caption{Compressor design. CU: Compressor unit.}
\label{fig:compression2}
\end{center}
\end{figure}

\textbf{Compression and Decompression}. We now describe the detailed design of the corresponding
compression and decompression logic.\footnote{For simplicity, we start with presenting the compression and decompression logic for \bd.
Compression for \bdc requires one more step, where elements are checked to be compressed with zero base; decompression logic only requires additional selector logic to
decide which base should be used in the addition. We describe the differences between \bdc and \bd designs later
in this section.} The compression logic consists of eight distinct compressor units: six
units for different base sizes (8, 4 and 2 bytes) and $\Delta$ sizes (4, 2 and 1 bytes), and two units for zero and repeated value compression (Figure~\ref{fig:compression2}).
Every compressor unit takes a cache line as an input, and outputs whether or not this cache line can be compressed with this unit. If it can be, the unit outputs  
the compressed cache line. The compressor selection logic is used to determine a set of compressor units that
can compress this cache line. If multiple compression options are available for the cache line (e.g., 8-byte base 1-byte $\Delta$ and 
zero compression), the selection logic chooses the one with the smallest compressed cache line size. 
Note that all potential compressed sizes are known statically and described in Table~\ref{tbl:ratios}.
All compressor units can operate in parallel. 

\begin{figure}[!htb]
\begin{center}%{l}{0.45\linewidth}
\centering
%\begin{minipage}[b]{0.5\linewidth}
\includegraphics[scale=0.8]{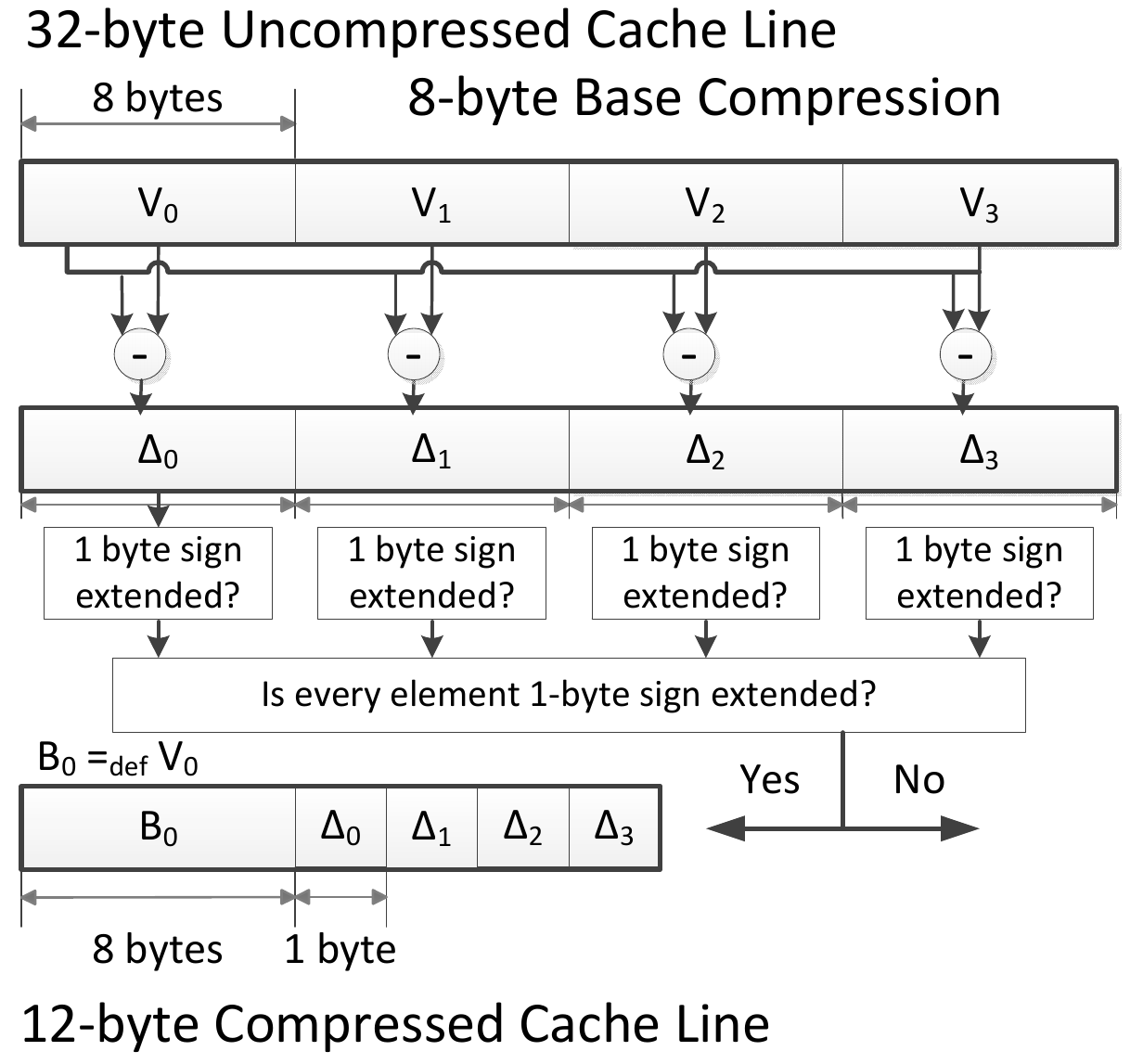}
 \caption{Compressor unit for 8-byte base, 1-byte $\Delta$}
\label{fig:compression}
\end{center}
\end{figure}

Figure~\ref{fig:compression} describes the organization of the 8-byte-base 1-byte-$\Delta$ compressor unit for a 32-byte cache line.
The compressor ``views'' this cache line as a set of four 8-byte elements ($V_0$, $V_1$, $V_2$, $V_3$), and in the first step,
computes the difference between the base element and all other elements. Recall that the base ($B_0$) is set to the first value ($V_0$), as
we describe in Section~\ref{sec:bdc}.
The resulting difference values ($ \Delta_0, \Delta_1, \Delta_2, \Delta_3$) are then checked 
to see whether their first 7 bytes are all zeros or ones (1-byte sign
extension check). If so, the resulting cache line can be stored as
the base value $B_0$ and the set of differences $\Delta_0, \Delta_1, \Delta_2, \Delta_3$, where each $\Delta_i$ requires only 1 byte. 
The compressed cache line size in this case is
12 bytes instead of the original 32 bytes. If the 1-byte sign extension check returns false (i.e., at least one
$\Delta_i$ cannot be represented using 1 byte), then the compressor unit cannot compress this cache line. 
The organization of all other compressor units is similar. 
This compression design can be potentially optimized, especially if hardware complexity is more critical than latency, e.g.,
all 8-byte-base value compression units can be united into one to avoid partial logic duplication.

\begin{table}[!htb]
 \begin{center}
 \begin{tabular}{|@{ }>{\scriptsize\bgroup}c<{\egroup}@{ }|@{ }>{\scriptsize\bgroup}c<{\egroup}@{ }|>{\scriptsize\bgroup}c<{\egroup}
                 |@{ }>{\scriptsize\bgroup}c<{\egroup}@{ }|>{\scriptsize\bgroup}c<{\egroup}||
                 @{ }>{\scriptsize\bgroup}c<{\egroup}@{ }|@{ }>{\scriptsize\bgroup}c<{\egroup}@{ }|@{ } >{\scriptsize\bgroup}c<{\egroup}@{ }|
                 >{\scriptsize\bgroup}c<{\egroup}|>{\scriptsize\bgroup}c<{\egroup}|}
 \hline
 {\textbf{Name}} & {\textbf{Base}}& {\textbf{$\Delta$}} & {\textbf{Size}} & \textbf{Enc.} &
 {\textbf{Name}} & {\textbf{Base}}& {\textbf{$\Delta$}} & {\textbf{Size}} &\textbf{Enc.} \\
  \hline
  \hline

 Zeros & 1 & 0 & 1/1 & 0000 &
 Rep.Values & 8 & 0 & 8/8 & 0001 \\\hline
 Base8-$\Delta$1 & 8 & 1 & 12/16 & 0010 &
 Base8-$\Delta$2 & 8 & 2 & 16/24 & 0011 \\\hline
 Base8-$\Delta$4 & 8 & 4 & 24/40 & 0100 &
 Base4-$\Delta$1 & 4 & 1 & 12/20 & 0101 \\\hline
 Base4-$\Delta$2 & 4 & 2 & 20/36 & 0110 &
 Base2-$\Delta$1 & 2 & 1 & 18/34 & 0111 \\\hline
 NoCompr. & N/A & N/A & 32/64 & 1111  \\\cline{1-5}
 \end{tabular}
 \end{center}
 \caption{\bdc encoding.  All sizes are in bytes. Compressed sizes (in bytes) are given for 32-/64-byte cache lines.}
 \label{tbl:ratios}
 \end{table}

%\end{minipage}
%\vspace{-1.0cm}
%\hspace{1.0cm}
%\begin{minipage}[b]{0.45\linewidth}

\begin{figure}[!htb]
\centering
\includegraphics[scale=0.8]{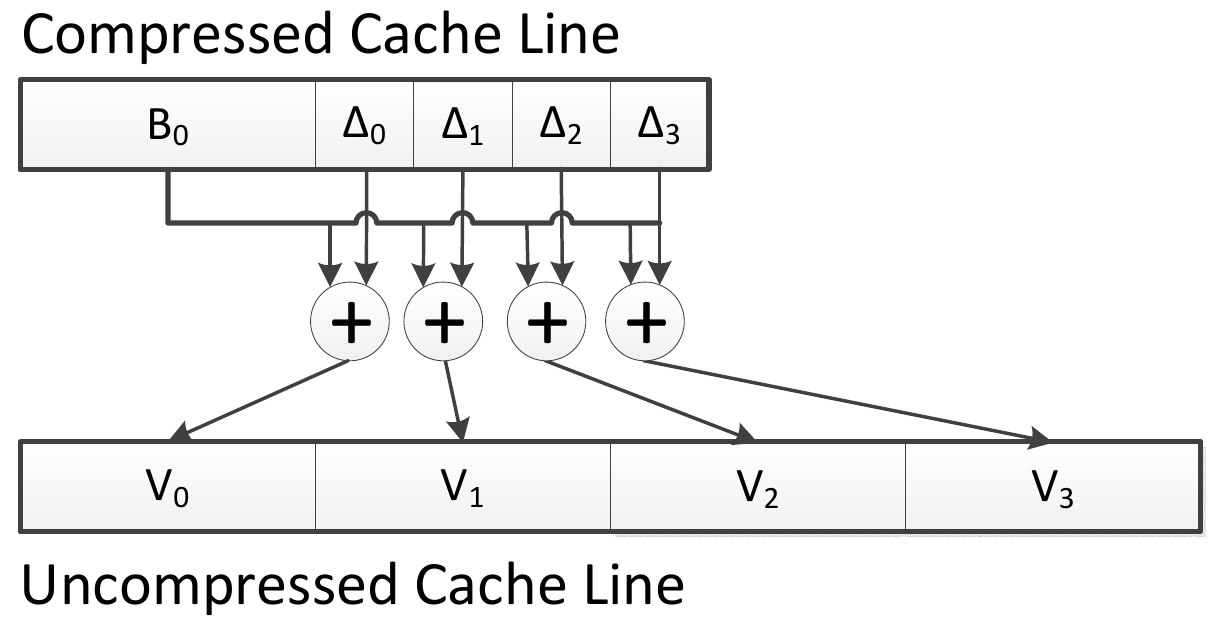}
\caption{Decompressor design}
\label{fig:decompression}
\end{figure}

Figure~\ref{fig:decompression} shows the latency-critical decompression logic. Its organization is simple: for a compressed
cache line that consists of a base value $B_0$ and a set of differences $\Delta_0, \Delta_1, \Delta_2,$ $\Delta_3$, only additions of
the base to the differences are performed to obtain the uncompressed cache line. Such decompression will take as long as the latency of an adder, 
and allows
the \bdc cache to perform decompression very quickly.

%compression ratio to have significant performance improvements for different applications.
%With these considerations in mind, we pick FVC and FPC
%cache compression designs as the most efficient and closest in spirit to
%our \bdc design.

\textbf{\bdc Cache Organization}. In order to obtain the benefits of compression, 
the conventional cache design requires certain changes.
Cache compression potentially allows more cache lines to be stored in the same data storage than a conventional uncompressed 
cache. But, in order to access these additional compressed cache lines, we need a way to address them.
%First, we need a way to represent more cache lines per set than the number of ways we have.
One way to achieve this is to have more tags~\cite{fpc}, e.g., twice as many,\footnote{We describe an implementation with the number of tags 
doubled and evaluate sensitivity to the number of tags in Section~\ref{sec:results}.} 
than the number we have in a conventional cache
of the same size and associativity. We can then use these additional tags as 
pointers to more data elements in the corresponding data storage.

\begin{figure}[hbt]
 \includegraphics[scale=0.7]{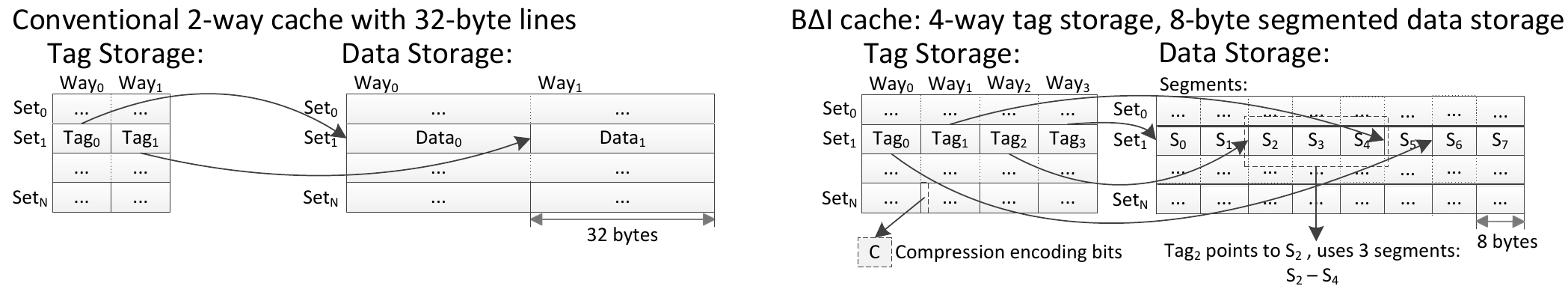}
 \caption{\bdc vs. conventional cache organization. Number of tags is doubled,
  compression encoding bits are added to every tag, data storage is the same in size, but
  partitioned into segments.}
\label{fig:2bdc}
\end{figure}

Figure~\ref{fig:2bdc} shows the required changes in the cache design. The conventional 2-way cache with 32-byte cache lines (shown on the top)
has a tag store with two tags per set, and a data store with two 32-byte cache lines per set. Every tag directly
maps to the corresponding piece of the data storage. In the \bdc design (at the bottom), we have twice as many
tags (four in this example), and every tag also has 4 additional bits to represent whether or not the line is compressed, and 
if it is, what compression type is used (see ``Encoding'' in Table~\ref{tbl:ratios}).
The data storage remains the same in size as before (2$\times$32 = 64 bytes), but it is separated into smaller fixed-size segments (e.g., 8 bytes in size in Figure~\ref{fig:2bdc}).
Every tag stores the starting segment (e.g., $Tag_2$ stores segment $S_2$) 
and the encoding for the cache block. By knowing the encoding we can easily know the number of segments used by the cache block. 

\ignore{
%%% REWRITTEN BELOW
\textbf{Storage Cost Analysis.}
This organization potentially allows storing twice as many cache lines
in the same data storage, because the number of tags in a set is
doubled. It requires modest increase in the tag store size (similar to
some other designs \cite{fpc-tr,iic,v-way}, see Table~\ref{tbl:cost}
for details) with only 1-2 cycle access latency increase depending on
the cache size (based on data from CACTI 5.3~\cite{cacti}).  According
to CACTI 5.3~\cite{cacti} the area increase with our tag organization
is only 2.3\% of the total area occupied by a 2MB 16-way L2, in
contrast to the 137\% times increase if we double both the size of the
data storage and associativity.
}

\textbf{Storage Cost Analysis.}
This cache organization potentially allows storing twice as many cache
lines in the same data storage, because the number of tags in a set is
doubled. As a result, it requires modest increase in the tag store
size (similar to some other designs \cite{fpc-tr,iic,v-way}. We
analyze the storage overhead in terms of raw additional bits in
Table~\ref{tbl:cost} for a baseline 16-way 2MB cache. We have also
used CACTI 5.3~\cite{cacti} to estimate the additional latency and
area cost of our proposed cache organization, using parameters for the
32nm technology node. Cache access latency increases by 1-2 cycles
(depending on cache size) for a 4GHz processor. On-chip cache area
increases by 2.3\%, but this increase is small compared to the 137\%
increase in area, which occurs if we double both the tag store and the
data store size (by doubling the associativity).\footnote{As we show
in Section~\ref{sec:results}, \bdc with our proposed cache organization
achieves performance that is within 1-2\% of a cache that has double
the tag and data store size.}

\begin{table}[h]
\centering
\begin{tabular}{|@{ }>{\scriptsize\bgroup}c<{\egroup}@{ }|
                 @{ }>{\scriptsize\bgroup}c<{\egroup}@{ }|
                 @{ }>{\scriptsize\bgroup}c<{\egroup}@{ }|}
  \hline
  { } & {\textbf{Baseline}} & {\textbf{\bdc}}\\
  \hline
  {Size of tag-store entry} & {21 bits} & {32 bits (+4--encoding, +7--segment pointer)} \\
  \hline
  {Size of data-store entry} & {512 bits} & {512 bits} \\
  \hline
  {Number of tag-store entries} & {32768} & {65536} \\
  \hline
  {Number of data-store entries} & {32768} & {32768} \\
   \hline
  {Tag-store size} & {84kB} & {256kB} \\
  \hline
  \hline
  {Total (data-store+tag-store) size} & {2132kB} & {2294kB} \\
  \hline
\end{tabular}
\caption{Storage cost analysis for 2MB 16-way L2 cache, assuming 64-byte cache lines, 8-byte segments, 
         and 36 bits for address space.}
\label{tbl:cost}
\end{table}

\textbf{Cache Eviction Policy.} In a compressed cache,
there are two cases under which multiple cache lines may need to be
evicted because evicting a single cache line (i.e., the LRU one in a
cache that uses the LRU replacement policy) may not create enough
space for the incoming or modified cache line.  First, when a new
cache line (compressed or uncompressed) is inserted into the
cache. Second, when a cache line already in the cache is modified such
that its new size is larger than its old size. In both cases, we
propose to use a slightly modified version of the LRU replacement
policy wherein the cache evicts multiple LRU cache lines to create
enough space for the incoming or modified cache line.\footnote{On
average, 5.2\% of all insertions or writebacks into the cache resulted
in the eviction of multiple cache lines in our workloads.}  such a
policy can increase the latency of eviction, it has negligible effect
on performance as evictions are off the critical path of
execution. Note that more effective replacement policies that take
into account compressed cache line sizes are possible -- e.g., a
policy that does not evict a zero cache line unless there is a need
for space in the tag store. We leave the study of such policies for
future work.

%% We also need a way to evict cache lines in this new cache
%% design. We propose to use a slightly modified version of the LRU
%% policy (although more effective replacement policies that can take
%% cache line sizes into account are possible).  When there is a need
%% to store a new cache line in the set and there are not enough
%% segments left to store it, we start evicting the least recently
%% used cache lines until we have enough space to add a new cache
%% line. This can result in multiple evictions per cache line
%% insertion. However, this is acceptable, since the eviction is not
%% on the critical path of execution.

%%Add info about other replacement policies attempts
\textbf{\bdc Design Specifics}.
So far, we described the common part in the designs of both \bd and \bdc. 
However, there are some specific differences between these two designs.

First, \bdc compression happens (off the critical path) in two steps (vs. only
one step for \bd).  
For a fixed $\Delta$ size, \emph{Step 1} attempts to compress all elements using an
implicit base of zero. \emph{Step 2} tries to compress those elements that
were not compressed in Step 1. The first uncompressible element of Step 1 
is chosen as the base for Step 2. The compression step stores
a bit mask, 1-bit per element indicating whether or not the
corresponding base is zero. Note that we keep the size of $\Delta$ (1, 2, or 4 bytes) 
the same for both bases.

Second, \bdc decompression is implemented as a masked addition of the base
(chosen in Step 2) to the array of differences. The elements to which 
the base is added depends on the bit-mask stored in the compression step.

%\pagebreak
\subsection{Operation}
We propose using our \bdc design at cache levels higher than L1
(e.g., L2 and L3). While it is possible to compress data in the L1
cache~\cite{fvc}, doing so will increase the critical path of
latency-sensitive L1 cache hits. This can result in significant
performance degradation for applications that do not benefit from
compression.

We now describe how a \bdc cache fits into a system with a 2-level
cache hierarchy (L1, L2 and main memory) with the L2 cache
compressed using \bdc~-- note that the only changes are to the L2
cache. We assume all caches use the writeback policy. There are
four scenarios related to the compressed L2 cache operation: 1) an
L2 cache hit, 2) an L2 cache miss, 3) a writeback from L1 to L2,
and 4) a writeback from L2 to memory.

First, on an L2 hit, the corresponding cache line is sent to the L1
cache. If the line is compressed, it is first decompressed before it
is sent to the L1 cache. Second, on an L2 miss, the corresponding
cache line is brought from memory and is sent to the L1 cache. In this
case, the line is also compressed and inserted into the L2
cache. Third, when a line is written back from L1 to L2, it is first
compressed. If an old copy of the line is already present in the L2
cache, the old (stale) copy is invalidated. The new compressed cache
line is then inserted into the L2 cache. Fourth, when a line is
written back from L2 cache to memory, it is decompressed before it is
sent to the memory controller. In both second and third scenarios,
potentially multiple cache lines might be evicted from the L2 cache
based on the cache eviction policy described in
Section~\ref{sec:design-design}.

\section{Related Work}
\label{sec:comparison}
Multiple previous works investigated the possibility of using
compression for on-chip caches~\cite{fvl,fpc,ZeroContent,ZeroValue,iic,c-pack} 
and/or memory~\cite{vm-compression,MXT,MMCompression}. All proposed designs
have different tradeoffs between compression ratio, 
decompression/compression latency and hardware complexity. 
%Depending on where the compression is proposed, e.g., the last level
%cache, L1 cache or memory, compression algorithms of different complexity
%were applied. 
The spectrum of proposed algorithms ranges
from general-purpose compression schemes e.g., the Lempel-Ziv algorithm~\cite{lz}, 
to specific pattern-based schemes, e.g., zero values~\cite{ZeroContent,ZeroValue} and
frequent values~\cite{fvc}.

The fundamental difference between \bdc and previous cache compression
mechanisms is that whereas prior techniques compress data at word
granularity -- i.e., each word within a cache line is compressed
separately, \bdc compresses data at cache-line granularity -- i.e.,
all the words within a cache line are compressed using the same
encoding or all the words within a cache line are stored uncompressed.
As a result, \bdc provides two major advantages. First, the
decompression of all words in the same cache line can be performed in
parallel (using a masked vector addition), since the starting point of
each word is known in the compressed cache line. In contrast,
compressing each word within a cache line separately, as in prior
works, typically serializes decompression as different words can be
compressed to different sizes, making the starting point of each word
in the compressed cache line dependent on the previous word. Second,
\bdc exploits correlation across words within a cache line, which can
lead to a better compression ratio -- e.g., when cache line consists
of an array of pointers. Prior works do not exploit this correlation
as they compress words individually.  As already summarized in Table
1, different prior works suffer from one or more of the following
shortcomings, which \bdc alleviates: 1) high decompression latency, 2)
low effective compression ratio, and 3) high hardware complexity. We
now describe the prior designs in more detail.

%% The fundamental difference between \bdc and previous cache compression
%% mechanisms is that instead of searching for patterns at a
%% word-granularity, which typically serializes decompression (because
%% consecutive words can be compressed to different sizes), \bdc aims to
%% build a compression/decompression mechanism for an entire cache line
%% that works in a \emph{parallel} manner for all values within the cache
%% line. This simplifies compression and especially decompression
%% logic, since only simple masked vector operations (e.g., addition) are
%% needed, and, at the same time, allows compression of cache lines that
%% are otherwise difficult to compress (e.g., an array of pointers). As
%% already summarized in Table 1, previous cache compression mechanisms
%% have one or more of the following shortcomings: (1) their
%% decompression latency is high mainly because of the sequential nature
%% of the pattern-based compression algorithms, (2)
%% they have low average effective compression ratio mostly because only
%% special patterns are covered, and, (3) they have relatively high
%% hardware complexity and/or overhead.

\subsection{Zero-based Designs}
Dusser et al.~\cite{ZeroContent} propose Zero-Content Augmented (ZCA) cache 
design where a conventional cache is augmented with a specialized cache
to represent zero cache lines. Decompression and compression latencies as well
as hardware complexity for the ZCA cache design are low. However, only applications
that operate on a large number of zero cache lines can benefit from this design.
In our experiments, only 6 out of 24 applications have enough
zero data to  benefit from ZCA (Figure~\ref{fig:2-bdc-compressibility}), leading
to relatively small performance improvements (as we show in Section~\ref{sec:results}).

Islam and Stenstr\"{o}m~\cite{ZeroValue} observe that 18\% of the dynamic loads actually
access zero data, and propose a cache design called Zero-Value Canceling
 where these loads can be serviced faster. Again, this can improve performance only
for applications with substantial amounts of zero data. Our proposal is more general
than these designs that are based only on zero values.

\subsection{Frequent Value Compression}

Zhang et al.~\cite{fvl} observe that a majority of values read or
written by memory operations come from a small set of 
frequently occurring values. 
%They refer to this phenomenon as frequent value
%locality. 
Based on this observation, they propose a compression
technique~\cite{fvc} that encodes frequent values present in cache lines with
fewer bits. They apply this technique to a direct-mapped L1 cache
wherein each entry in the cache can store either one uncompressed
line or two compressed lines. 

Frequent value compression (FVC) has three major drawbacks. First,
since FVC can only compress frequent values, it cannot exploit other
commonly found patterns, e.g., narrow values or stride patterns in
application data. As a result, it does not provide a high degree of
compression for most applications as shown in
Section~\ref{sec:results}.  Second, FVC compresses only the frequent
values, while other values stay uncompressed.  Decompression of such a
cache line requires sequential processing of every element (because
the beginning of the next element can be determined only after the
previous element is processed), significantly increasing the latency
of decompression, which is undesirable. Third, the proposed mechanism
requires profiling to identify the frequent values within an
application. Our quantitative results in Section~\ref{sec:results}
shows that \bdc outperforms FVC due to these reasons.
%This can either decrease the effectiveness of
%the mechanism if done statically or requires additional
%hardware to determine the frequent values if done
%dynamically. Identifying frequent values dynamically may require
%the cache to be flushed when new frequent values need to
%replace old ones.

\subsection{Pattern-Based Compression Techniques}

Alameldeen and Wood~\cite{fpc} propose frequent pattern
compression (FPC) that exploits the observation that a majority of
words fall under one of a few compressible patterns, e.g., if the upper 16
bits of a 32-bit word are all zeros or are all ones, all bytes
in a 4-byte word are the same. 
FPC defines a set of these patterns~\cite{fpc-tr} and then uses them to encode 
applicable words with fewer bits of data. For
compressing a cache line, FPC first divides the cache line into
32-bit words and checks if each word falls under one of seven
frequently occurring patterns. %It uses three bits to encode the%pattern. 
Each compressed cache line contains the pattern encoding
for all the words within the cache line followed by the additional
data required to decompress each word.

The same authors propose a compressed cache design~\cite{fpc} based on FPC
which allows the cache to store two times more compressed lines than
uncompressed lines, effectively doubling the cache size when all lines
are compressed. For this purpose, they maintain twice as many tag
entries as there are data entries.  Similar to frequent value
compression, frequent pattern compression also requires serial
decompression of the cache line, because every word can be compressed
or decompressed.  To mitigate the decompression latency of FPC, the
authors design a five-cycle decompression pipeline~\cite{fpc-tr}.  They
also propose an adaptive scheme which avoids compressing data if the
decompression latency nullifies the benefits of compression.

Chen et al.~\cite{c-pack} propose a pattern-based compression mechanism (called C-Pack) with several new features: 
(1) multiple cache lines can be compressed 
into one, (2) multiple words can be compressed in parallel; but parallel
decompression is not possible. 
Although the C-Pack design is more practical than FPC, it still has a 
high decompression latency (8 cycles due to serial decompression),
and its average compression ratio is lower than that of FPC.

\subsection{Follow-up Work}
Publication of this work~\cite{bdi} inspired several new proposals 
for hardware-oriented compression algorithms~\cite{sc2,hycomp,morc,kimbit},
and new compressed cache designs~\cite{dcc,scc,yacc}. 
Most of these works aim for higher compression ratios, but this happens at the cost
of much higher compression/decompression latency. This is why 
some of these works~\cite{morc,kimbit} are proposed in the context of modern
GPUs that are much more tolerant to increase in memory latency.

\section{Evaluation Methodology}
\label{sec:methodology}
We use an in-house, event-driven 32-bit x86 simulator whose front-end
is based on Simics~\cite{Simics}.  All configurations have either a
two- or three-level cache hierarchy, with private L1D caches.  Major
simulation parameters are provided in Table
\ref{tbl:simulation-parameters}.  All caches uniformly use a 64B cache
block size and LRU policy for replacement.  All cache latencies were
determined using CACTI~\cite{cacti} (assuming a 4GHz frequency), and
provided in Table~\ref{tbl:cache-latencies}.  We also checked that
these latencies match the existing last level cache implementations
from Intel and AMD, when properly scaled to the corresponding
frequency.\footnote{Intel Xeon X5570 (Nehalem) 2.993GHz, 8MB L3 - 35
  cycles~\cite{Nehalem}; AMD Opteron 2.8GHz, 1MB L2 - 13
  cycles~\cite{Opteron}.}  For evaluations, we use benchmarks from the
SPEC CPU2006 suite~\cite{SPEC}, three TPC-H queries~\cite{tpc}, and an
Apache web server (shown in Table~\ref{tbl:benchmarks}, whose detailed
description is in Section~\ref{sec:results}). All results are
collected by running a representative portion of the benchmarks for 1
billion instructions.

\begin{table}[ht]

 \centering
%\begin{minipage}[b]{0.7\linewidth}
%    \begin{tabular}{|>{\bgroup}l<{\egroup}|>{\bgroup}l<{\egroup}|}
\begin{tabular}{|>{\scriptsize\bgroup}l<{\egroup}|>{\scriptsize\bgroup}c<{\egroup}|}
         \hline
         Processor  &  1--4 cores, 4GHz, x86 in-order  \\
         \hline
         L1-D cache    &  32kB, 64B cache-line, 2-way, 1 cycle \\
         \hline
         L2 caches    &  0.5--16 MB, 64B cache-line, 16-way   \\
         \hline
         L3 caches    &  2--16 MB, 64B cache-line, 16-way  \\
         \hline
         Memory  & 300 cycle latency    \\
        \cline{1-2}
    \end{tabular}%
 \caption{Major parameters of the simulated system}
  \label{tbl:simulation-parameters}%
%\end{minipage}
%\begin{minipage}[b]{0.1\linewidth}
%\begin{tabular}{>{\scriptsize\bgroup}l<{\egroup}>{\scriptsize\bgroup}c<{\egroup}}
%  \toprule
%  Size & Latency  \\
%  \cmidrule(rl){1-2}
%  512kB & 15 \\ 1MB & 21 \\ 2MB & 27 \\
%  \cmidrule(rl){1-2}
%  4MB & 34 \\ 8MB & 41 \\ 16MB & 48\\
%  \bottomrule
%  \end{tabular}
  %\vspace{-0.2cm}
  %\caption{\small Cache hit latencies used in simulations (in cycles). \bdc caches have + 1-2 cycle latency on a hit/miss  and +1 cycle for decompression.}
  
%  \label{tbl:cache-latencies}
%\end{minipage}
\end{table}
%\begin{comment}

\begin{table}[ht]
% \begin{footnotesize}
\centering
  
  \begin{tabular}{|>{\scriptsize\bgroup}c<{\egroup}|>{\scriptsize\bgroup}c<{\egroup}|
                  >{\scriptsize\bgroup}c<{\egroup}|>{\scriptsize\bgroup}c<{\egroup}|
                  >{\scriptsize\bgroup}c<{\egroup}|>{\scriptsize\bgroup}c<{\egroup}| 
                }
  %\begin{tabular}{|>{\bgroup}l<{\egroup}|>{\bgroup}c<{\egroup}|
  %                |>{\bgroup}l<{\egroup}|>{\bgroup}c<{\egroup}|
  %                |>{\bgroup}l<{\egroup}|>{\bgroup}c<{\egroup}|}
  \hline%\toprule
  Size & Latency & Size & Latency & Size & Latency \\
  \hline%\cmidrule(rl){1-6}
  512kB & 15 & 1MB & 21 & 2MB & 27 \\
  \hline%\cmidrule(rl){1-6}
  4MB & 34 & 8MB & 41 & 16MB & 48\\
  \hline%\bottomrule
  \end{tabular}
  \caption{Cache hit latencies used in simulations (in cycles). \bdc caches have +1 cycle for 0.5--4MB (+2 cycle for others) on a hit/miss due
   to larger tag stores,  and +1 cycle for decompression.}
  \label{tbl:cache-latencies}
% \end{footnotesize}
\end{table}
%}
%\end{comment}

\textbf{{Metrics.}}  We measure performance of our benchmarks using
IPC (instruction per cycle), effective compression ratio (effective
cache size increase, e.g., 1.5 for 2MB cache means effective size of
3MB), and MPKI (misses per kilo instruction).  For multi-programmed
workloads we use the weighted speedup \cite{weightedspeedup,ws2} as
the performance metric: ($\sum_i \frac{IPC_i^{shared}} {{IPC}_i^{{alone}}}
$~).
%and instruction throughput ($\sum_i \textrm{IPC}_i$).
For bandwidth consumption we use BPKI (bytes transferred over bus per
thousand instructions~\cite{BPKI}).

Effective compression ratio for all mechanisms is computed without
meta-data overhead.  We add all meta-data to the tag storage, e.g.,
for \bdc, we add four bits to encode the compression scheme, and a bit
mask to differentiate between two bases. We include these in the tag
overhead, which was evaluated in Section~\ref{sec:design}.  Our
comparisons are fair, because we do not include this overhead in
compression ratios of previous works we compare to. In fact, the
meta-data overhead is higher for FPC (3 bits for each word).
  %If we add this overhead to the data instead of tags, then the
  %compression ratio will slightly decrease. If we take, as an example,
  %Base4-Delta2 from Table 2, its size will change from 36 bytes to
  %38.5 bytes (16 words * 1bit + 4 bits). Hence, the compression ratio
  %will decrease by 7%. This change won't affect zero and
  %repeated value cache lines. Note that the compression ratio of FPC
  %will decrease 19% with this overhead included (if we assume a
  %cache line with 2-byte narrow values).

We conducted a study to see applications' performance sensitivity to the increased L2 cache size (from
512kB to 16 MB). Our results show that there are benchmarks that are almost insensitive (IPC improvement less than 5\% with 32x increase in
cache size) to the size of the L2 cache: dealII, povray, calculix, gamess, namd, milc, and perlbench. This typically
means that their working sets mostly fit into the L1D cache, leaving almost no potential for any L2/L3/memory optimization.
Therefore, we do not present data for these applications, although we verified that our mechanism does not affect their performance.
%hile still running these applications in all our experiments,
%we exclude them from the data presented in Section~\ref{sec:results}.% to save space.

\textbf{{Parameters of Evaluated Schemes.}}
For FPC, we used a decompression latency of 5 cycles, and a segment size of 1 byte (as for \bdc)
to get the highest compression ratio as described in ~\cite{fpc-tr}. For FVC, we used
static profiling for 100k instructions to find the 7 most frequent values
as described in~\cite{fvc}, and a decompression latency of 5 cycles. For ZCA and \bdc, we used a decompression latency
of 1 cycle.

We also evaluated \bdc with higher decompression latencies (2-5
cycles).  \bdc continues to provide better performance, because for
most applications it provides a better overall compression ratio than
prior mechanisms. When decompression latency of \bdc increases from 1
to 5 cycles, performance degrades by 0.74\%.
% (same latency as reported for significantly more complex FPC
%decompression).

\textbf{{Internal Fragmentation.}}
In our simulations, we assumed that before every insertion, we
can shift segments properly to avoid fragmentation (implementable, but might be inefficient). 
We believe this is reasonable, because insertion happens off the critical path of
the execution. Previous work~\cite{fpc} adopted this assumption, and we treated all schemes equally 
in our evaluation. Several more recent works~\cite{dcc,scc,yacc} (after this work was published) looked
at more efficient ways of handling fragmentation.

%For evaluating fairness, we
%use the maximum slowdown metric.
%\vspace{-5mm}
%\begin{small}
%\begin{eqnarray*}
%  \textrm{Instruction Throughput} &=& \sum_i \textrm{IPC}_i\\
%  \textrm{Weighted Speedup} &=& \sum_i \frac{\textrm{IPC}_i^{\textrm{shared}}}
%         {\textrm{IPC}_i^{\textrm{alone}}}\\
  %\textrm{Maximum Slowdown} &=& \max_i \frac{\textrm{IPC}_i^{\textrm{alone}}}
  %       {\textrm{IPC}_i^{\textrm{shared}}}
%\end{eqnarray*}
%\end{small}
%\\\vspace{-20mm}
%Figure~\ref{fig:cache-size-sensitivity} shows the performance
%for the applications in
%SPEC CPU2006 benchmark suite, three TPC-H queries, and an Apache web server.
%The figure plots the \emph{instructions per cycle} for each application on a system
%using 1MB, 2MB, 4MB, 8MB, and 16MB of on-chip L2 cache compared to
%one using 512kB. As the figure indicates, for many
%applications, performance significantly improves by doubling the
%cache size.
% there are benchmarks that are almost insensitive to
%the size of the L2 cache: dealII, povray, calculix, gamess, namd, and milc. This typically
%means that their working sets fit into L1D cache, and, hence, can not get any benefit from the increased
%size of L2 cache. We exclude these benchmarks from a data presented in our experiments.

%\comment{
\begin{table}[!ht]
\centering
\begin{tabular}{
|>{\scriptsize\bgroup}c<{\egroup}|>{\scriptsize\bgroup}c<{\egroup}|>{\scriptsize\bgroup}c<{\egroup}|>{\scriptsize\bgroup}c<{\egroup}||
>{\scriptsize\bgroup}c<{\egroup}|>{\scriptsize\bgroup}c<{\egroup}|>{\scriptsize\bgroup}c<{\egroup}||
%>{\scriptsize\bgroup}c<{\egroup}|>{\scriptsize\bgroup}c<{\egroup}|>{\scriptsize\bgroup}c<{\egroup}||
>{\scriptsize\bgroup}c<{\egroup}|>{\scriptsize\bgroup}c<{\egroup}|>{\scriptsize\bgroup}c<{\egroup}|}

%|>{\scriptsize\bgroup}c<{\egroup}|>{\scriptsize\bgroup}c<{\egroup}|>{\scriptsize\bgroup}c<{\egroup}|}

\hline
  \textbf{Cat.} &
  \textbf{Name} & \textbf{Comp. Ratio} &  \textbf{Sens.} &
  \textbf{Name} & \textbf{Comp. Ratio} & \textbf{Sens.} &
%  \textbf{Name} & \textbf{Comp. Ratio} &  \textbf{Sens.} &
%  \textbf{Name} & \textbf{C. Ratio} &  \textbf{Sens.} &
  \textbf{Name} & \textbf{Comp. Ratio} & \textbf{Sens.} \\
 \hline
 \hline
  \multirow{3}{*}{\rotatebox{45}{LCLS}} 
  & gromacs & 1.43 / L  & L &
  hmmer & 1.03 / L  & L &
  lbm & 1.00 / L  & L \\ 
  \cline{2-10}
  
  & leslie3d & 1.41 / L  & L &
  sphinx & 1.10 / L  & L &
  tpch17 & 1.18 / L   & L \\ 
  \cline{2-10} 
 
& libquantum & 1.25 / L  & L &
   wrf & 1.01 / L   & L  \\ 
  \hline
  \hline

  \multirow{3}{*}{\rotatebox{45}{HCLS}} 
  & apache & 1.60 / H &  L &
  zeusmp & 1.99 / H   & L &
  gcc & 1.99 / H  &  L \\ 
  \cline{2-10}

  & gobmk & 1.99 / H  & L &
  sjeng & 1.50 / H    & L &
  tpch2 & 1.54 / H   & L \\
  \cline{2-10}
  
  & tpch6 & 1.93 / H  &  L &
  GemsFDTD & 1.99 / H & L &
  cactusADM & 1.97 / H & L \\ 
  \hline
  \hline

  \multirow{3}{*}{\rotatebox{45}{HCHS}} 
  & astar & 1.74 / H  &  H  &
  bzip2 & 1.60 / H  &  H &
  mcf & 1.52 / H   & H \\ 
  \cline{2-10}

  & omnetpp & 1.58 / H   & H &
  soplex & 1.99 / H   & H &
  h264ref & 1.52 / H   & H  \\
  \cline{2-10}
  
  & xalancbmk & 1.61 / H   & H &
  &  &  &
  &  &  \\
   \hline
\end{tabular}
  \caption{Benchmark characteristics and categories:
    \textbf{Comp. Ratio} (effective compression ratio for 2MB \bdc
    L2) and \textbf{Sens.} (cache size sensitivity). Sensitivity
    is the ratio of improvement in performance by going from 512kB
    to 2MB L2 (L - low ($\le$ 1.10) , H - high ($>$ 1.10)). For
    compression ratio: L - low ($\le$ 1.50), H - high ($>$
    1.50). \textbf{Cat.}  means category based on compression
    ratio and sensitivity.}
  \label{tbl:benchmarks}
\end{table}

\section{Results \& Analysis}
\label{sec:results}

\subsection{Single-core Results}
\label{sec:results-1-core}

\begin{figure}[h]
\centering  
%\begin{subfigure}[b]{0.23\textwidth}
\begin{minipage}[b]{0.45\linewidth}
    \includegraphics[width=0.9\textwidth]{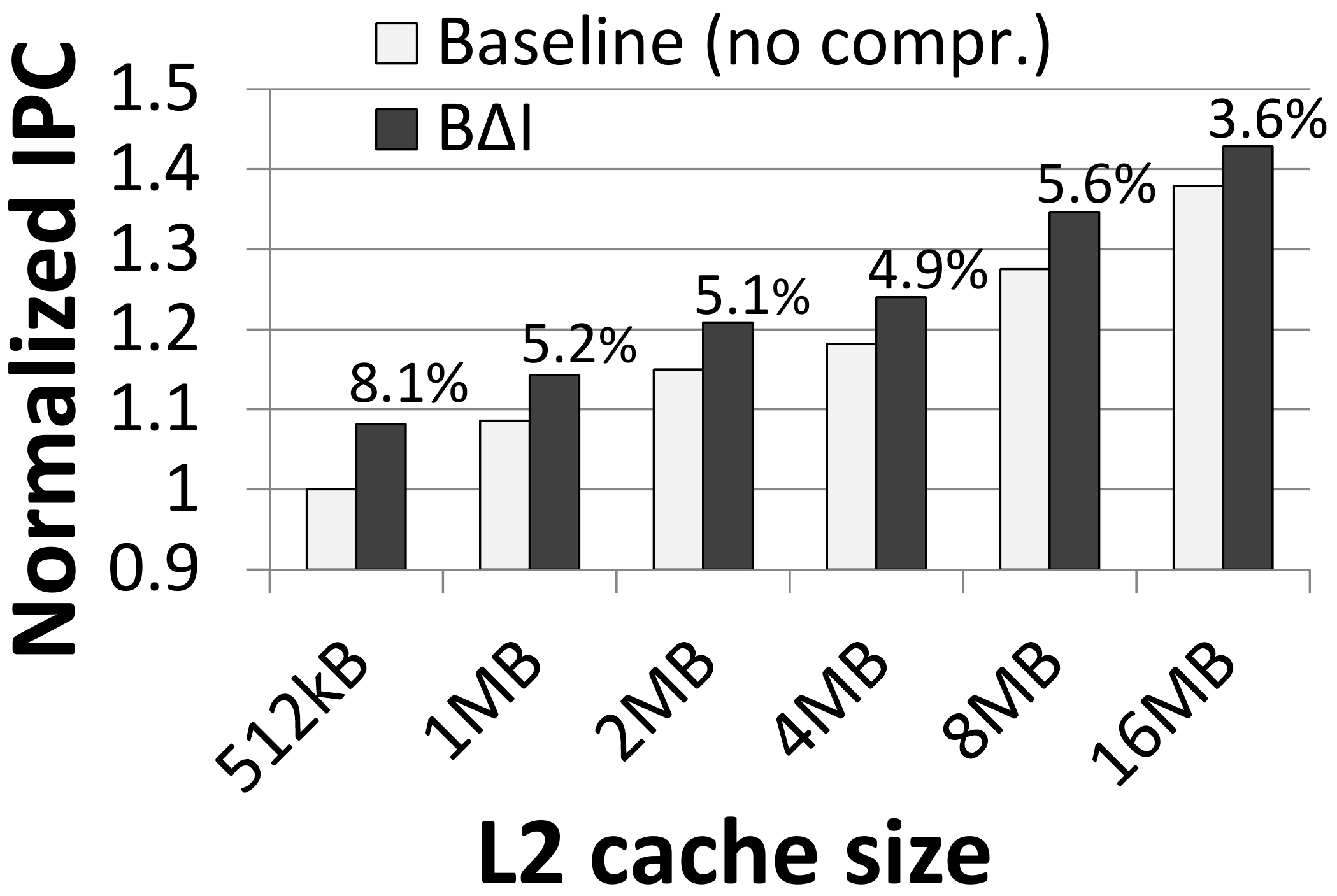}
    \caption{\small{(a) IPC}}
    \label{fig:L2RealGeoMean}
\end{minipage}
%  \end{subfigure}
  %\hspace{0.035cm}
 % \begin{subfigure}[b]{0.23\textwidth}
   \begin{minipage}[b]{0.45\linewidth}
    \includegraphics[width=0.9\textwidth]{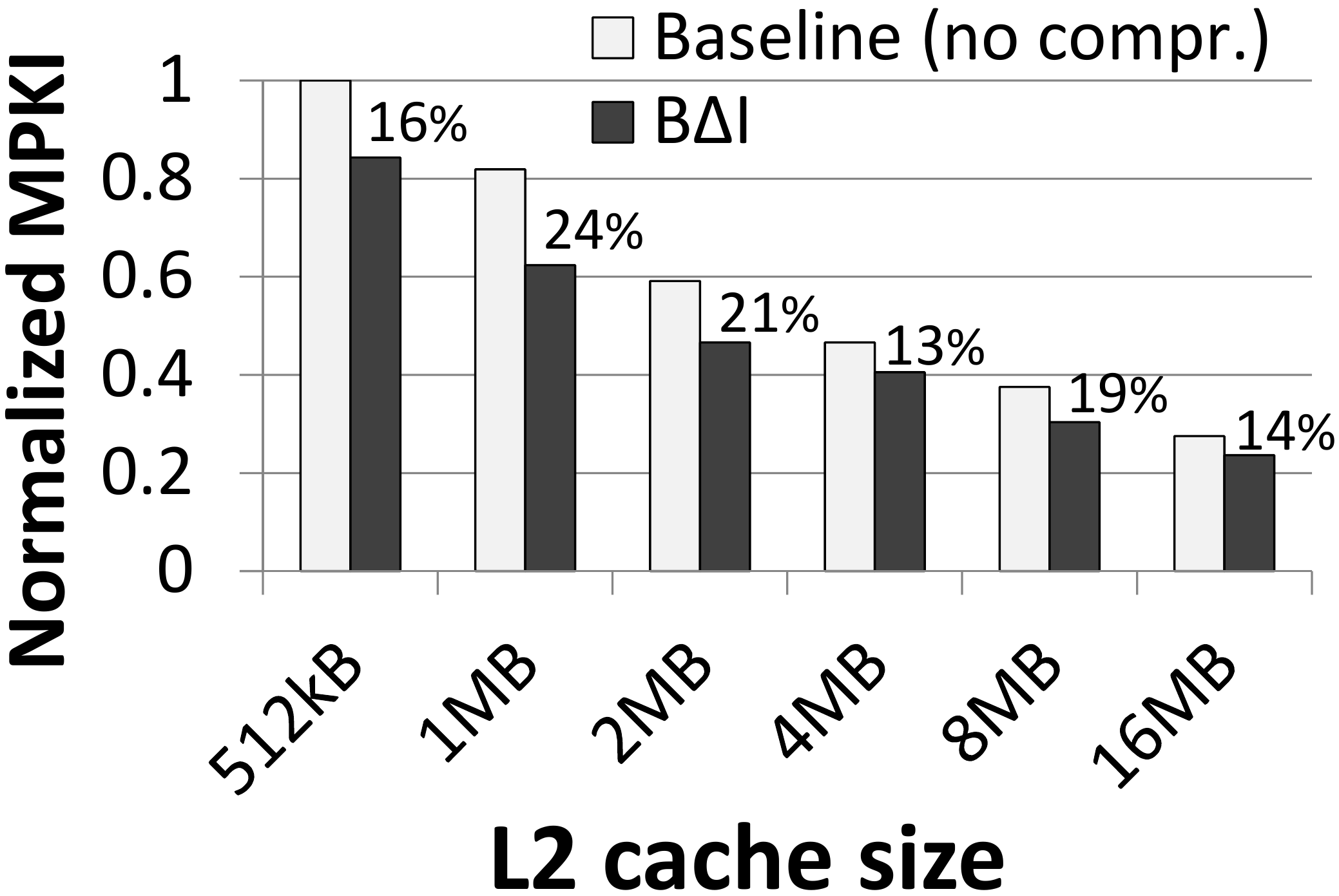}
    \caption{\small{(b) MPKI}}
    \label{fig:L2MPKI}
 \end{minipage}
 % \end{subfigure}
  \caption{Performance of \bdc with different cache sizes. Percentages
   show improvement over the baseline cache (same size).}
   \label{fig:L2RealAll}
  %\vspace{-0.2cm} 
\end{figure}

Figure~\ref{fig:L2RealAll}(a) shows the performance improvement of
our proposed \bdc design over the baseline cache design for various
cache sizes, normalized to the performance of a 512KB baseline
design. The results are averaged across all
benchmarks. Figure~\ref{fig:L2RealAll}(b) plots the corresponding results
for MPKI also normalized to a 512KB baseline design. Several
observations are in-order. First, the \bdc cache significantly
outperforms the baseline cache for all cache sizes. By storing cache
lines in compressed form, the \bdc cache is able to effectively store
more cache lines and thereby significantly reduce the cache miss rate
(as shown in Figure~\ref{fig:L2RealAll}(b)). Second, in most cases, \bdc
achieves the performance improvement of doubling the cache size. In
fact, the 2MB \bdc cache performs better than the 4MB baseline
cache. This is because, \bdc increases the effective cache size
\emph{without} significantly increasing the access latency of the data
storage. Third, the performance improvement of \bdc cache decreases
with increasing cache size. This is expected because, as cache size
increases, the working set of more benchmarks start fitting into the
cache. Therefore, storing the cache lines in compressed format has
increasingly less benefit. Based on our results, we conclude that \bdc
is an effective compression mechanism to significantly improve
single-core performance, and can provide the benefits of doubling the
cache size without incurring the area and latency penalties associated
with a cache of twice the size.

%% \begin{figure}[h]
 
%%  \subfloat[IPC]{\label{fig:L2RealGeoMean}
%%  \includegraphics[width=0.23\textwidth]{figures/L2RealGeoMean.pdf}}
%%  \hspace{0.035cm}
%%  \subfloat[MPKI]{\label{fig:L2MPKI}
%%  \includegraphics[width=0.23\textwidth]{figures/L2MPKI.pdf}}
 
%%  \caption{Performance of \bdc with different cache sizes. Percentages
%%    show improvement over the baseline cache (same size).}
%%  \vspace{-0.2cm} 
%% \end{figure}

\begin{comment}
The graph in Figure~\ref{fig:L2RealAllSizesSelective} represents
IPC for every benchmark for the cache size it was most sensitive
according to the sensitivity study described in
Section~\ref{sec:methodology}.  As we can see all benchmarks
presented have performance benefit from cache compression, but the
cache size at which the improvement is the highest varies across
different applications, and depends on the size of the working
set.

\begin{figure}[!h]
 \centering
 \includegraphics[scale=0.5]{chap3/figures/L2RealAllSizesSelective.pdf}
 
 \caption{IPC comparison for different cache sizes. Every
   benchmark presented for a cache size it is most sensitive to.}
\label{fig:L2RealAllSizesSelective}
\end{figure}
\end{comment}

\subsection{Multi-core Results}
\label{sec:mult}

When the working set of an application fits into the cache, the
application will not benefit significantly from compression even
though its data might have high redundancy. However, when such an
application is running concurrently with another cache-sensitive
application in a multi-core system, storing its cache lines in
compressed format will create additional cache space for storing the
data of the cache-sensitive application, potentially leading to
significant overall performance improvement.

\sloppypar{ To study this effect, we classify our benchmarks into four
  categories based on their compressibility using \bdc (low (LC) or
  high (HC)) and cache sensitivity (low (LS) or high
  (HS)). Table~\ref{tbl:benchmarks} shows the sensitivity and
  compressibility of different benchmarks along with the criteria used
  for classification. None of the benchmarks used in our evaluation
  fall into the low-compressibility high-sensitivity (LCHS)
  category. We generate six different categories of 2-core workloads
  (20 in each category) by randomly choosing benchmarks with different
  characteristics (LCLS, HCLS and HCHS).  }

\begin{figure}[ht!]
%\begin{minipage}[b]{0.5\linewidth}

\begin{center}
 \includegraphics[scale=0.5]{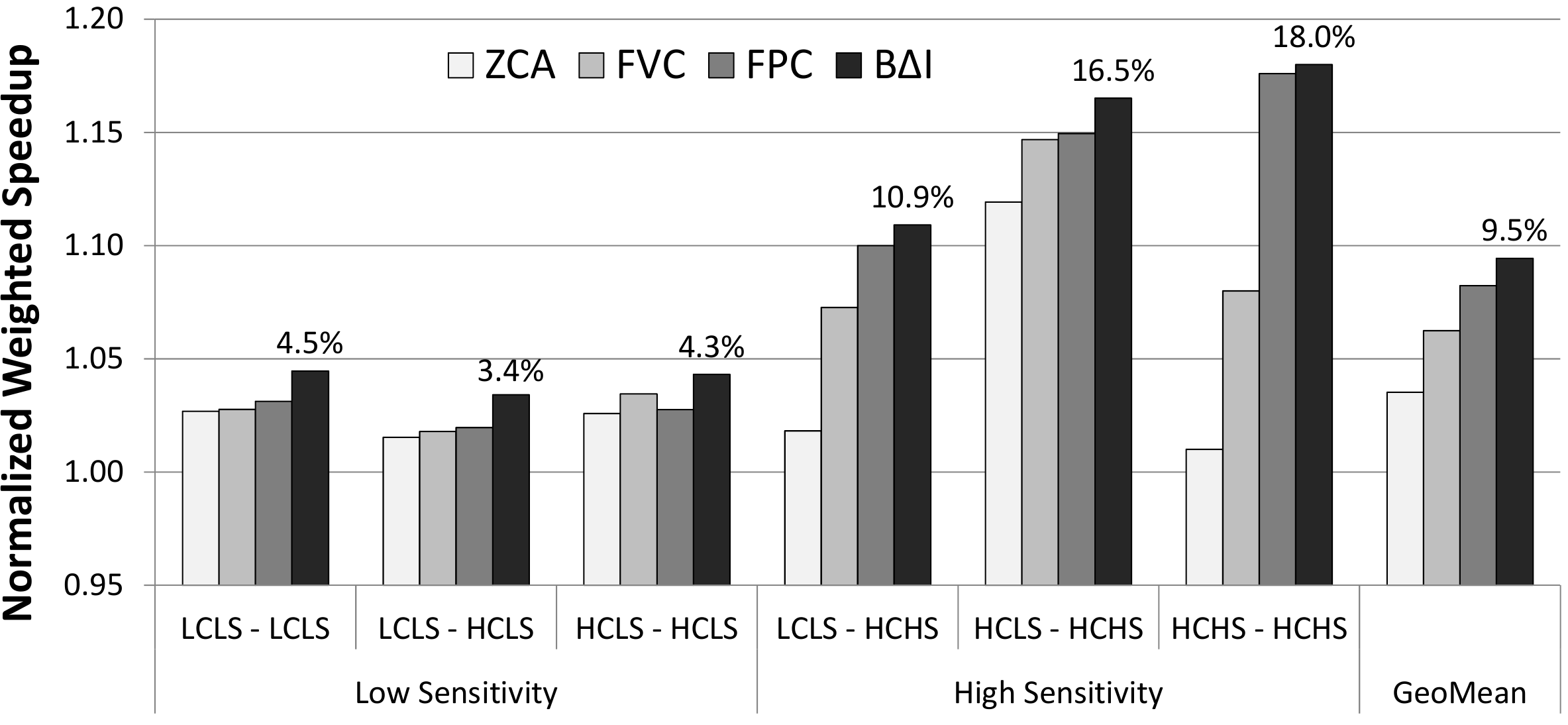}
\caption{Normalized weighted speedup for 2MB L2 cache, 2-cores. Percentages show improvement over the baseline uncompressed cache.}
\label{fig:l2ws2core2m}
  \end{center}
\end{figure}

Figure~\ref{fig:l2ws2core2m} shows the performance improvement
provided by four different compression schemes, namely, ZCA, FVC,
FPC, and \bdc, over a 2MB baseline cache design for different
workload categories. We draw three major conclusions.

First, \bdc outperforms all prior approaches for all workload
categories. Overall, \bdc improves system performance by 9.5\%
compared to the baseline cache design.

Second, as we mentioned in the beginning of this section, even though
an application with highly compressible data may not itself benefit
from compression (HCLS), it can enable opportunities for performance
improvement for the co-running application. This effect is clearly
visible in the figure. When at least one benchmark is sensitive to
cache space, the performance improvement of \bdc increases with
increasing compressibility of the co-running benchmark (as observed by
examining the bars labeled as High Sensitivity). \bdc provides the
highest improvement (18\%) when \emph{both} benchmarks in a workload
are highly compressible and highly sensitive to cache space
(HCHS-HCHS). As the figure shows, the performance improvement is not
as significant when neither benchmark is sensitive to cache space
irrespective of their compressibility (as observed by examining the
bars labeled Low Sensitivity).

Third, although FPC provides a degree of compression similar to \bdc for
most benchmarks (as we showed in Section~\ref{sec:bdi}, Figure~\ref{fig:2-bdc-compressibility})  
its performance improvement is lower than \bdc
for all workload categories. This is because FPC has a more
complex decompression algorithm with higher decompression latency
compared to \bdc. On the other hand, for high sensitivity
workloads, neither ZCA nor FVC is as competitive as FPC or \bdc in the
HCLS-HCHS category. This is because both ZCA and FVC have a
significantly lower degree of compression compared to
\bdc. However, a number of benchmarks in the HCLS category
(\emph{cactusADM}, \emph{gcc}, \emph{gobmk}, \emph{zeusmp}, and
\emph{GemsFDTD}) have high occurrences of zero in their
data. Therefore, ZCA and FVC are able to compress most of the
cache lines of these benchmarks, thereby creating additional space for the
co-running HCHS application.

\vspace{0.0cm}

We conducted a similar experiment with 100 4-core workloads with
different compressibility and sensitivity characteristics. We observed
trends similar to the 2-core results presented above. On average, \bdc
improves performance by 11.2\% for the 4-core workloads and it
outperforms all previous techniques. We conclude that \bdc, with its
high compressibility and low decompression latency, outperforms other
state-of-the-art compression techniques for both 2-core and 4-core
workloads, likely making it a more competitive candidate for adoption
in modern multi-core processors.

We summarize \bdc performance improvement against the baseline 2MB L2 cache (without compression) and 
other mechanisms in Table~\ref{tbl:summary}.
\begin{table}[ht]
\centering
    \begin{tabular}{|>{\scriptsize\bgroup}c<{\egroup}|>{\scriptsize\bgroup}c<{\egroup}|>{\scriptsize\bgroup}c<{\egroup}|>{\scriptsize\bgroup}c<{\egroup}|>{\scriptsize\bgroup}c<{\egroup}|}
          \hline
          \textbf{Cores}   &  \textbf{No Compression} & \textbf{ZCA} & \textbf{FVC} & \textbf{FPC} \\
         \hline
         1  &  5.1\% &   4.1\% & 2.1\% & 1.0\%  \\
         \hline
         2 & 9.5\% & 5.7\% &3.1\% & 1.2\% \\
         \hline
         4 & 11.2\%  & 5.6\% & 3.2\% & 1.3\% \\
         \hline
    \end{tabular}\vspace{-1mm}
 \caption{Average performance improvement of \bdc over other mechanisms: No Compression, ZCA, FVC, and FPC.} 
  \label{tbl:summary}%
\end{table}

\subsection{Effect on Cache Capacity}
\label{sec:res3}
\begin{figure}[!ht]
\centering
 \includegraphics[scale=0.6]{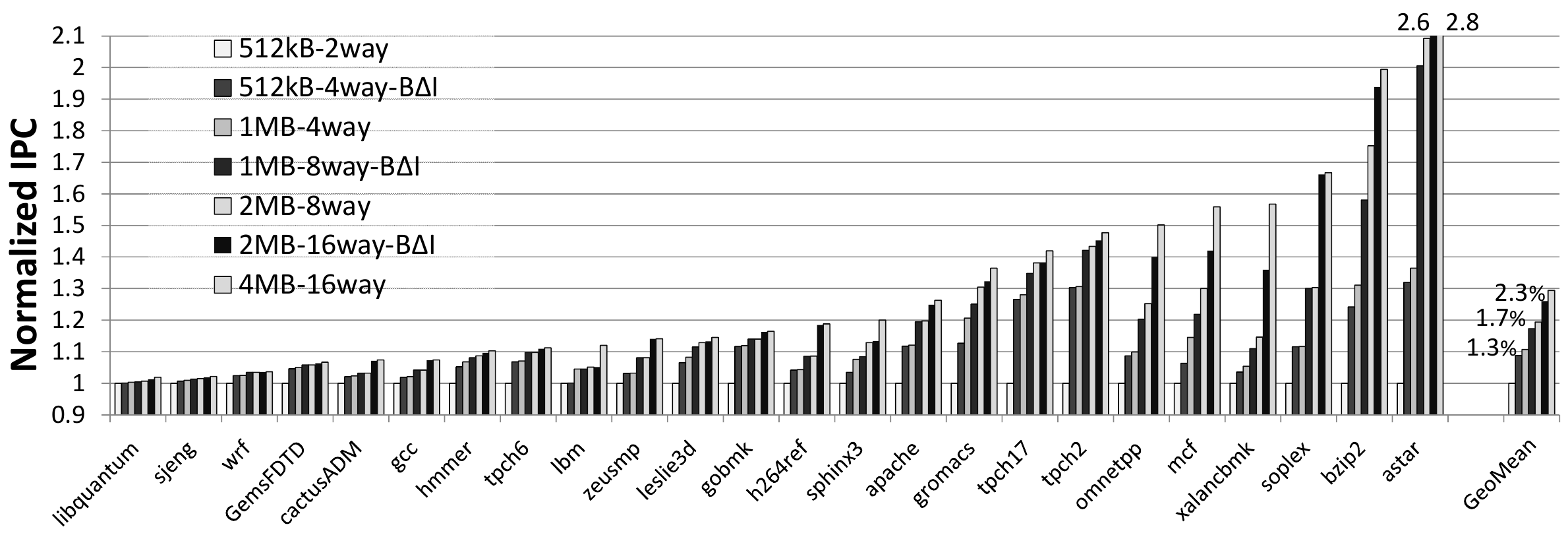}
 \caption{IPC comparison of \bdc against lower and upper
   limits in performance (from 512kB 2-way - 4MB 16-way L2
   cache). Percentages on the GeoMean bars show how close \bdc gets to the
           performance of the cache with twice the size (upper limit).}
\label{fig:L2PerfBoundaries}
\end{figure}

Our proposed \bdc cache design aims to provide the benefits of
increasing the cache size while not incurring the increased latency
of a larger data storage. To decouple the benefits of
compression using \bdc from the benefits of reduced latency
compared to a larger cache, we perform the following study. We
compare the performance of the baseline cache design and the \bdc
cache design by progressively doubling the cache size by doubling
the cache associativity. We fix the latency of accessing all
caches.

Figure~\ref{fig:L2PerfBoundaries} shows the results of this
experiment. With the same access latency for all caches, we expect
the performance of the \bdc cache (with twice the number of tags as the baseline)
to be strictly between the
baseline cache of the same size (lower limit) and the baseline cache of double
the size (upper limit, also reflected in our results). However, with its high
degree of compression, the \bdc cache's performance comes close to
the performance of the twice as-large baseline cache design for most
benchmarks (e.g., \emph{h264ref} and \emph{zeusmp}). On
average, the performance improvement due to the \bdc cache is
within 1.3\% -- 2.3\% of the improvement provided by a twice as-large baseline
cache. We conclude that our \bdc implementation (with twice the number of tags 
as the baseline) achieves performance improvement close to its upper bound potential
performance of a cache twice the size of the baseline.

\begin{figure}[htb]
\centering
 \includegraphics[scale=0.5]{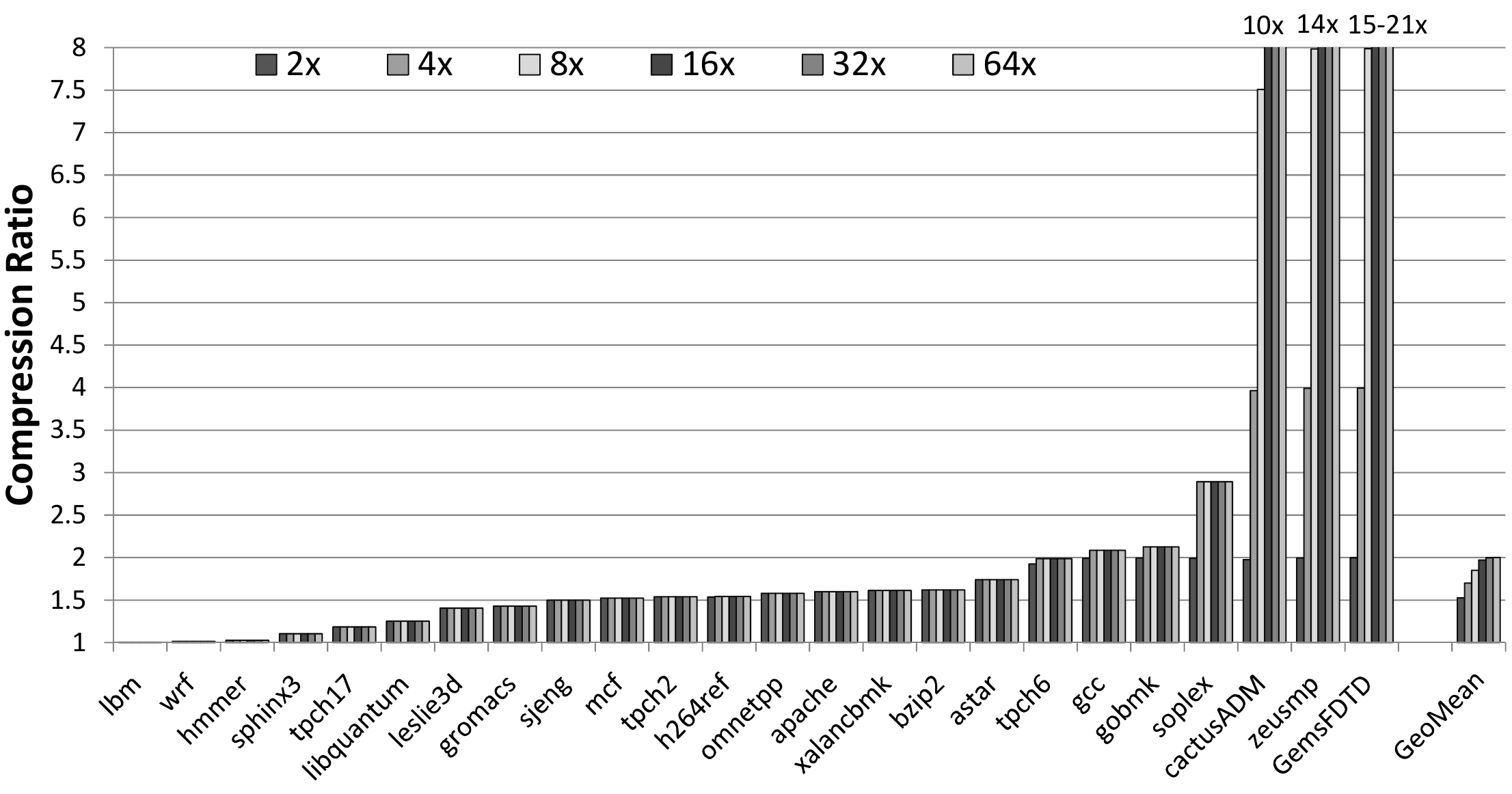}
 \caption{Effective compression ratio vs. number of tags}
\label{fig:L2MultTags}
\end{figure}

For an application with highly compressible data, the compression
ratio of the \bdc cache is limited by the number of additional
tags used in its design.  Figure~\ref{fig:L2MultTags} shows the
effect of varying the number of tags (from 2$\times$ to 64$\times$
the number of tags in the baseline cache) on compression ratio for
a 2MB cache. As the figure shows, for most benchmarks, except
\emph{soplex}, \emph{cactusADM}, \emph{zeusmp}, and
\emph{GemsFDTD}, having more than twice as many tags as the
baseline cache does not improve the compression ratio. The
improved compression ratio for the four benchmarks is primarily
due to the large number of zeros and repeated values present in
their data. At the same time, having more tags does not
benefit a majority of the benchmarks and also incurs higher
storage cost and access latency. Therefore, we conclude 
that these improvements likely do not justify the
use of more than 2X the tags in the \bdc cache design compared
to the baseline cache. 

%% As was shown in Figure~\ref{fig:2-bdc-compressibility}, \bdc
%% provides significant effective compression ratio, but this ratio
%% is limited by the number of additional tags in our design (we
%% showed results with doubled tag store). In order to understand how
%% this design decision affects compression ratio, we observe the
%% performance of \bdc as we vary the number of tags (from 2-64 times
%% the baseline) with a fixed L2 size (2MB).
%% Figure~\ref{fig:L2MultTags} shows that for most benchmarks having
%% more than twice as many tags does not improve effective
%% compression ratio, and, hence, performance stays the same. There
%% are four benchmarks (soplex, cactusADM, zeusmp, GemsFDTD) whose
%% effective compression ratio increases significantly (mostly
%% because of the significant numbers of zeros and repeated values).
%% However, these improvements likely do not justify the usage of
%% many more tags in the tag store, because this will lead to
%% increased L2 latency, as well as too more energy and area used by
%% the L2 cache.

\subsection{Effect on Bandwidth}
\begin{figure}[ht!]
  \centering
\includegraphics[scale=0.5]{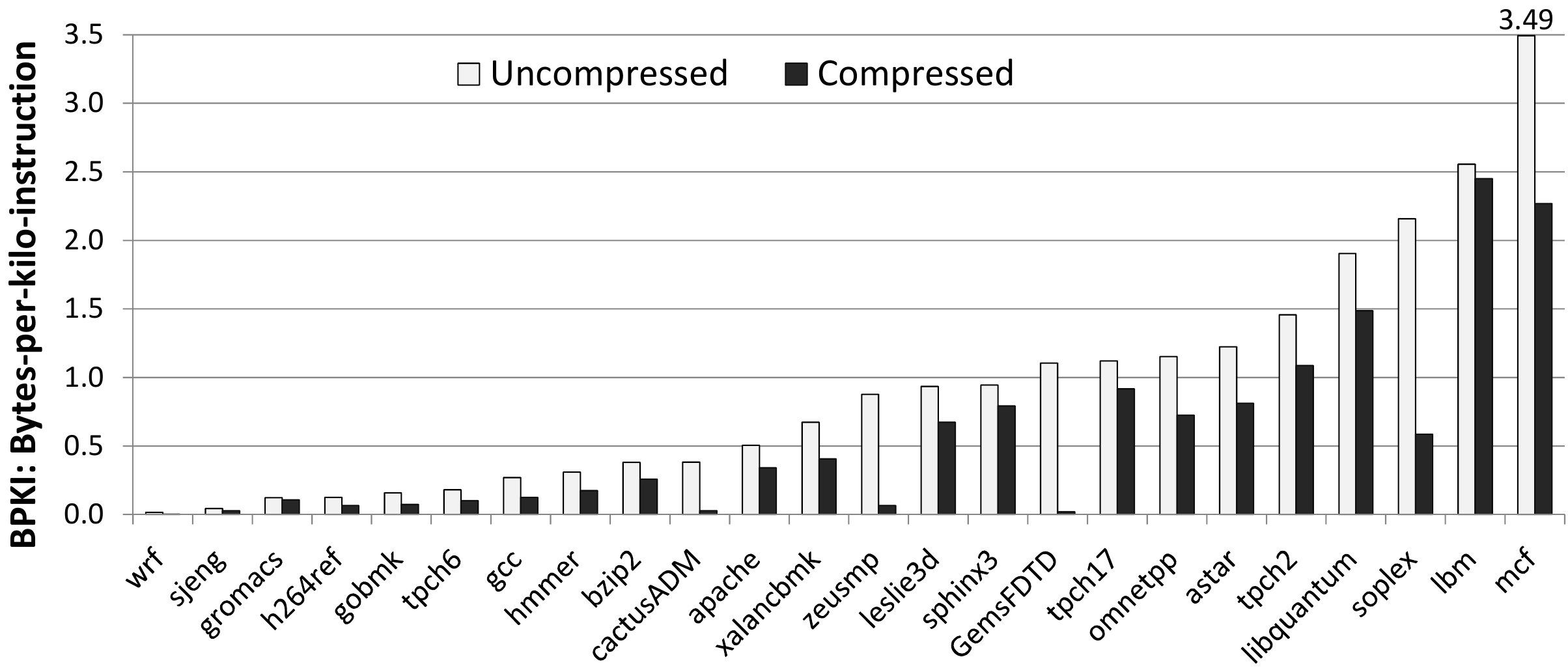}
 \caption{Effect of compression on bus bandwidth (in terms of BPKI)
   between L2 (256kB) and L3 (8MB)}
\label{fig:L3Bandwidth}
\end{figure}

In a system with a 3-level cache hierarchy, where both the L2 and
the L3 caches store cache lines in compressed format, there is an
opportunity to compress the traffic between the two caches. This
has two benefits: (1) it can lead to reduced latency of
communication between the two caches, and hence, improved system
performance, and (2) it can lower the dynamic power consumption of
the processor as it communicates less data between the two
caches~\cite{BandwidthCompression}. Figure~\ref{fig:L3Bandwidth}
shows the reduction in L2-L3 bandwidth (in terms of bytes per kilo instruction)
due to \bdc compression. We
observe that the potential bandwidth reduction with \bdc is as
high as 53X (for \emph{GemsFDTD}), and 2.31X on average. We
conclude that \bdc can not only increase the effective cache size,
but it can also significantly decrease the on-chip traffic.

%% A 3-level cache hierarchy has another potential benefit from
%% compression in addition to the enlarged cache capacity.  It is
%% possible to reduce the on-chip bandwidth between L2 and L3
%% caches. Since both caches store data in the compressed form, it is
%% natural to send data in the compressed form as well. This will
%% reduce the required bandwidth traffic between the caches and,
%% hence, will improve performance and, potentially, reduce power
%% consumption
%% \cite{BandwidthCompression}. Figure~\ref{fig:L3Bandwidth} shows
%% the reduction in required L2-L3 bandwidth due to data
%% compression. We observe that there is a potential in bandwidth
%% reduction with \bdc that can be as high as 53X reduction for
%% GemsFDTD, and with an average reduction of 2.31X. We conclude that
%% \bdc's benefit is not limited to providing the effective cache
%% size increase, but can also significantly decrease on-chip
%% traffic.

%\newpage
%\vspace{0.2cm}
\subsection{Detailed Comparison with Prior Work}
To compare the performance of \bdc against state-of-the-art cache
compression techniques, we conducted a set of studies and
evaluated IPC, MPKI, and effective compression ratio
(Figure~\ref{fig:2-bdc-compressibility}) for single core
workloads, and weighted speedup (Figure~\ref{fig:l2ws2core2m}) for
two- and four-core workloads.
%All IPCs are normalized to the IPC value of the baseline 2MB L2.

\begin{figure}[!htb]
\centering
\includegraphics[scale=0.5]{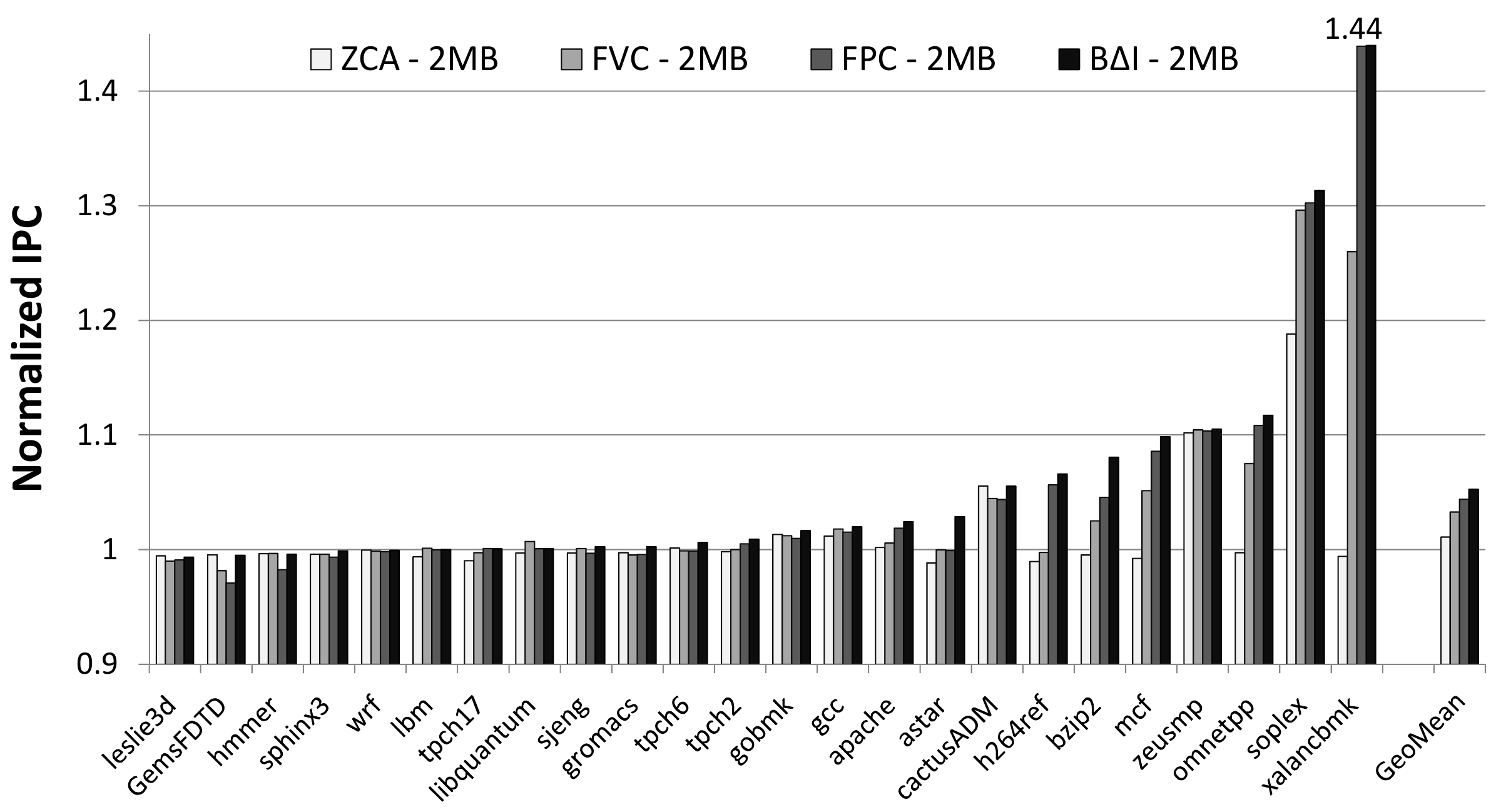}
\caption{Performance of \bdc vs. prior work for a 2MB L2
  cache}
\label{fig:L2IPCComparison2M}
\end{figure}

Figure~\ref{fig:L2IPCComparison2M} shows the improvement in IPC
using different compression mechanisms over a 2MB baseline
cache in a single-core system. As the figure shows, \bdc outperforms all prior approaches
for most of the benchmarks. For benchmarks that do not benefit from
compression (e.g, \emph{leslie3d}, \emph{GemsFDTD}, and
\emph{hmmer}), all compression schemes degrade performance
compared to the baseline. However, \bdc has the lowest performance degradation
 with its low 1-cycle decompression latency, and never degrades performance by more than
1\%. On the other hand, FVC and FPC degrade performance by as much
as 3.1\% due to their relatively high 5-cycle decompression latency. We also observe
that \bdc and FPC considerably reduce MPKI compared to ZCA and
FVC, especially for benchmarks with more complex data patterns
like \emph{h264ref}, \emph{bzip2}, \emph{xalancbmk}, \emph{hmmer},
and \emph{mcf} (not shown due to space limitations).

Based on our results, we conclude that \bdc, with its low
decompression latency and high degree of compression, provides the
best performance compared to all examined compression mechanisms.

\section{Summary}
\label{sec:conclusion}

In this chapter, we presented \bdc, a new and simple, yet efficient hardware cache
compression technique that provides high effective cache capacity
increase and system performance improvement compared to three
state-of-the-art cache compression techniques.  \bdc achieves these benefits by
exploiting the low dynamic range of in-cache data and representing cache
lines in the form of two base values (with one implicit base equal to
zero) and an array of differences from these base values.  We provide insights into why \bdc
compression is effective via examples of existing in-cache data
patterns from real programs.  \bdc's key advantage over previously
proposed cache compression mechanisms is its ability to have
low decompression latency (due to parallel decompression) 
while still having a high average
compression ratio.

%We describe the design and operation of a cache that can utilize data
%compressibility when it is available with small overhead. The proposed
%\bdc design does not require significant hardware costs and
%changes. 

We describe the design and operation of a cache that can utilize \bdc
compression with relatively modest hardware overhead.  Our extensive
evaluations across a variety of workloads and system configurations
show that \bdc compression in an L2 cache can improve system
performance for both single-core (8.1\%) and multi-core workloads
(9.5\%~/ 11.2\% for two/four cores), outperforming three
state-of-the-art cache compression mechanisms. In many workloads, the
performance benefit of using \bdc compression is close to the performance benefit of
doubling the L2/L3
cache size. In summary, we conclude that \bdc is an efficient and low-latency data
compression substrate for on-chip caches in both single- and
multi-core systems.

%%% Local Variables: 
%%% mode: latex
%%% TeX-master: "main"
%%% End: 

\chapter{Compression-Aware Cache Management}

%\section{Compression-Aware Management Policies}

\section{Introduction}
\label{sec:intro}
\blfootnote{Originally published as ``Exploiting Compressed Block Size as an
Indicator of Future Reuse'' in the 21st International Symposium on High Performance
Computer Architecture, 2015~\cite{camp}.}
Off-chip main memory latency and bandwidth are major performance bottlenecks in
modern systems.  Multiple levels of on-chip caches are used to hide the memory
latency and reduce off-chip memory bandwidth demand.  Efficient utilization of
cache space and consequently better performance is dependent upon the ability
of the cache replacement policy to identify and retain useful data. Replacement
policies, ranging from traditional (e.g., ~\cite{LRU,belady}) to
state-of-the-art (e.g.,~\cite{mlp,RRIP,EAF,lacs,rw-samira,dip}), work using a combination of
\textit{eviction} (identifies the block to be removed from the cache),
\textit{insertion} (manages the initial block priority), and \textit{promotion}
(changes the block priority over time) mechanisms.  In replacement policies
proposed for conventional cache organizations, these mechanisms usually work by
considering \emph{only} the locality of the cache blocks.
%In fact, Belady~\cite{belady} showed that it is possible to design an optimal
%replacement policy to maximize the cache hits by considering only the locality
%and reuse of the cache blocks.

A promising approach to improving effective cache capacity is to use cache compression (e.g.,~\cite{fvc,ecm,fpc,c-pack,iic-comp,bdi,dcc,sc2}). In compressed caches, data
compression algorithms, e.g., Frequent Pattern
Compression (FPC)~\cite{fpc-tr}, Base-Delta-Immediate
Compression (BDI)~\cite{bdi}, and Frequent Value Compression~\cite{fvc}, are used to achieve higher effective capacity (storing
more blocks of data) and to decrease off-chip bandwidth consumption
compared to traditional organizations without compression. This compression generates variable-size cache blocks, with larger
blocks consuming more cache space than smaller blocks.  However, most cache
management policies in these compressed cache designs do not use block size in cache management
decisions~\cite{fvc,fpc,c-pack,iic-comp,bdi,dcc,sc2}. 
Only one recent work---ECM~\cite{ecm}---uses the block size
information, but its effectiveness is limited by its coarse-grained 
(big vs.~small) view of block size.
The need to consider size along with temporal
locality is well known in the context of web
caches~\cite{elaarag1, elaarag2, size, lru-sp, luv}, but proposed
solutions rely on a recency list of \emph{all} objects in the
web cache~\cite{size} or consider frequency of object
accesses~\cite{lru-sp} and are usually prohibitively expensive to implement in
hardware for use with on-chip caches.

%ECM inserts ``big'' blocks with lower priority than
%``small'' blocks and evicts the biggest block in the eviction pool.

In this chapter, we propose a \textit{Compression-Aware Management Policy
(\carp{})} that takes into account compressed cache block size
along with temporal locality to improve the performance of compressed
caches.  Compared to prior work (ECM~\cite{ecm}), our policies first use a finer-grained
accounting for compressed block size and an optimization-based approach for eviction
decisions.  Second and more importantly,
we find that size is not only a measure of the cost of retaining a
given block in the cache, as previous works considered~\cite{ecm}, but it is sometimes 
also {\em an indicator of block reuse}.  
\carp\ contains two key components,
Minimal-Value Eviction (\mineviction{}) and Size-based Insertion
Policy (\insertionpolicy{}), which significantly improve the quality of replacement
decisions in compressed caches (see Section~\ref{camp:sec:results} for a comprehensive
analysis) at a modest hardware cost.

\textbf{Minimal-Value Eviction (\mineviction{}).}  \mineviction{} is
based on the observation that one should evict an uncompressed block
with good locality to make/retain room for a set of smaller
compressed blocks of the same total size, even if those
blocks individually have less locality, as long as the set of blocks collectively provides
more hits cumulatively. A special case of this is that when two blocks have
similar locality characteristics, it is preferable to evict the larger
cache block. \mineviction{} measures the \emph{value} of each block as a
combination of its locality properties and size. When an eviction is required (to
make space for a new block), \mineviction{} picks the block with the
least value as the victim.
%%%%%%%%%%%%%%%%%%%%%%%
\REM{
, with each cache block to indicate its relative importance. 
We make two key observations. 
First, it is possible to build a better replacement policy if the size is directly used in 
decision making process. The larger the size, the more space it occupies in the cache,
and hence its importance to the cache (to be useful) is less than that of a block of the a smaller size, but similar
priority. This observation leads to our first mechanism -- \emph{Minimal-Value Eviction (\mineviction)}.
The key idea behind \mineviction{} is that every block gets a \emph{value} assigned to it by a value function 
based on its expected reuse and size. \mineviction{} tries to maximize cache utilization, 
and if an eviction is needed (to create space for a new block), 
the first block to evict is the block with the currently \emph{minimal value}.}
%%%%%%%%%%%%%%%%%%%%%%%

\textbf{Size-based Insertion Policy (\insertionpolicy{}).}  SIP is
based on our new observation that the compressed size of a cache block
can sometimes be used as an indicator of its reuse characteristics. This is
because elements belonging to the same data structure and having the same access characteristics are sometimes (but not always)
compressed to the same size---e.g., in \emph{bzip2}~\cite{SPEC}, a
compressed block of 34 bytes (with BDI compression~\cite{bdi}) likely
belongs to one particular array with narrow values
(e.g., small values stored in large data types) as we show in Section~\ref{sec:size-reuse}---and these structures
more often than not have a specific pattern of access and/or reuse
distance.  

By dynamically inserting blocks of different sizes with either \emph{high priority}---e.g., in the most-recently-used 
position for the LRU policy (ensuring blocks stay in cache longer)---or \emph{low priority}---e.g., in the least-recently-used 
position for the LRU policy (ensuring blocks get evicted quickly unless reused shortly)---\insertionpolicy{} 
learns the reuse characteristics associated with various compressed block sizes and, if such an association exists, uses this information to maximize the hit ratio.

\begin{figure*}[!h]
  \centering
  \includegraphics[width=0.85\textwidth]{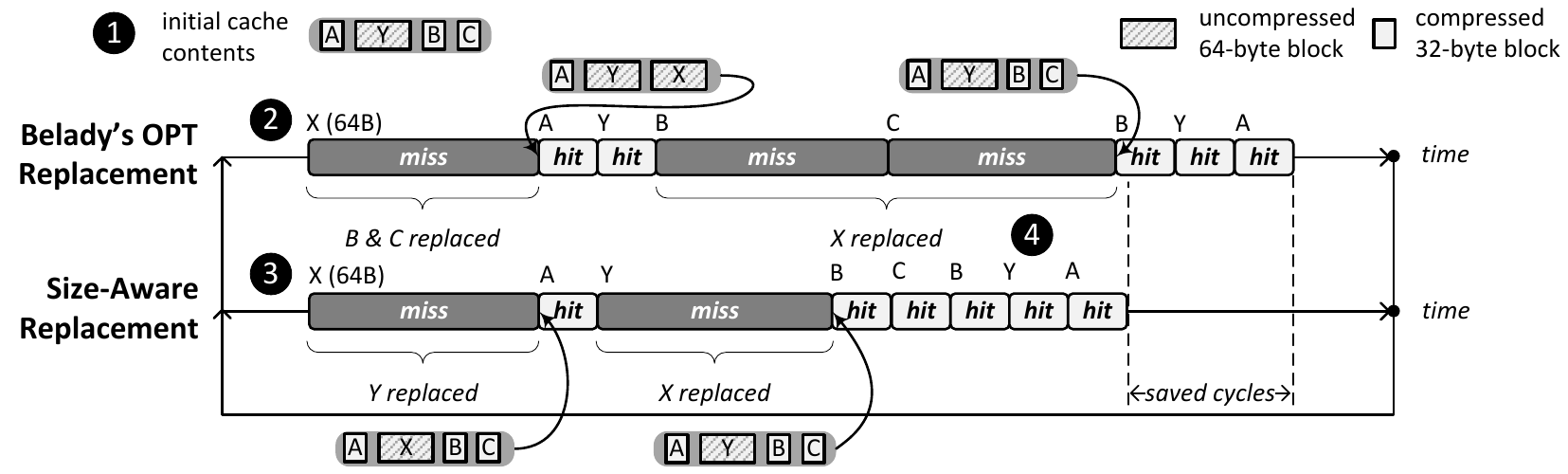}
  \caption{Example demonstrating downside of not including block size information in replacement decisions.}
%  \vspace{-0.2cm}
  \label{fig:belady}
\end{figure*}

As demonstrated later in this chapter, \carp{} (a combination of \mineviction{} and \insertionpolicy{}) works 
with both traditional compressed cache designs and compressed caches having decoupled tag and data stores
(e.g., V-Way Cache~\cite{v-way} and Indirect Index Cache~\cite{iic,iic-comp}). It is general enough to be used with 
different compression mechanisms and requires only modest hardware changes. Compared to prior work, \carp{} 
provides better performance, more efficient cache utilization, reduced off-chip bandwidth consumption, and an overall reduction in the memory 
subsystem energy requirements.

In summary, we make the following major contributions:
\begin{itemize}

\item We make the observation that the compressed size of a cache block can be indicative of its reuse. 
We use this observation to develop a new cache insertion policy for compressed caches, 
the Size-based Insertion Policy (\insertionpolicy{}), which uses the size of a compressed block as one of 
the metrics to predict its potential future reuse.

\item We introduce a new compressed cache replacement policy, Minimal-Value Eviction (\mineviction{}), 
      which assigns a value to each cache block based on both its size and its reuse and replaces the set of blocks with the least value.

\item We demonstrate that both policies 
%(which we call \carp{} when combined together)
 are generally applicable to different
      compressed cache designs (both with local and global replacement) and
      can be used with different compression algorithms (FPC~\cite{fpc} and BDI~\cite{bdi}).

\item We qualitatively and quantitatively compare \carp{} (\insertionpolicy{} + \mineviction{})
 to the conventional LRU policy and
      three state-of-the-art cache management policies:
      two size-oblivious policies (RRIP~\cite{RRIP} and a policy used in V-Way~\cite{v-way})
      and the recent ECM~\cite{ecm}.
      We observe that \carp{} (and its global variant G-\carp{})  
      can considerably (i) improve performance
      (by 4.9\%/9.0\%/10.2\% on average in single-/two-/four-core workload evaluations and up to 20.1\%),
     (ii) decrease off-chip bandwidth consumption (by 8.7\% in single-core), and (iii) decrease
      memory subsystem energy consumption (by 7.2\% in single-core) 
      on average for memory intensive workloads when compared with
      the best prior mechanism.

\end{itemize}

\section{Motivating Observations}
\label{sec:motivation}

%\begin{figure*}[!h]
%  \vspace{-0.4cm}
%  \centering
%  \includegraphics[width=0.90\textwidth]{figures/belady_example.pdf}
%  \caption{The downside of not including block size information in replacement decisions.}
%  \vspace{-0.2cm}
%  \label{fig:belady}
%\end{figure*}

Cache compression~\cite{fvc,ecm,fpc,c-pack,iic-comp,bdi,dcc,sc2} is a powerful mechanism that increases 
effective cache capacity and decreases off-chip bandwidth consumption.\footnote{
Data compression can be also effective in increasing the size of the main memory~\cite{MMCompression,lcp-tech,lcp-micro}
and reducing the off-chip memory bandwidth/energy consumption~\cite{lcp-micro,memzip}.} 
In this section, we show that cache compression adds an additional dimension
to cache management policy decisions -- \emph{the compressed block size} (or
simply \emph{the size}), which plays an important role in building more efficient management 
policies. We do this in three steps.

%Even with increased cache capacity, 
%efficient management of cached data is essential for maximizing performance. 
%In this section, it is first shown that 
%replacement decisions in compressed caches are improved by considering compressed cache block size along with temporal 
%locality to identify blocks to be evicted. Second, looking at cache compression reveals the challenges and opportunities involved in managing compressed caches.
%\REM{While
%these straightforward benefits of cache compression were studied in several
%recent works~\cite{bdi,c-pack,fpc}, very few works~\cite{ecm} look at the additional
%opportunities such as size-aware cache management policies created by cache compression.
%In this work, we show that cache compression adds an additional dimension
%to cache management policy decisions -- \emph{the compressed block size} (or
%simply \emph{the size}),
%that plays an important role in building more efficient management 
%policies. We do this in three steps. 
%}
%\textbf{Size matters.}
\subsection{Size Matters}
%\subsection{Impact of Cache Block Size on Replacement.} todo: add the footnote
%about MLP- unawareness
In compressed caches, one should design replacement policies that take into
account compressed cache block size along with locality to identify victim
blocks, because such policies can outperform existing policies that rely \emph{only}
on locality.  In fact, Belady's optimal algorithm~\cite{belady} that relies
only on locality (using perfect knowledge to evict the block that will be
accessed furthest in the future) is sub-optimal in the context of compressed
caches with variable-size cache blocks.  Figure~\ref{fig:belady} demonstrates
one possible example of such a scenario. In this figure, we assume that
cache blocks are one of two sizes: (i) uncompressed
64-byte blocks (blocks X and Y) and (ii) compressed 32-byte blocks (blocks A,
B, and C). We assume the cache capacity is 160 bytes. Initially (see \ding{202}), the 
cache contains four blocks: three compressed (A, B, C) and one
uncompressed (Y). Consider the sequence of memory requests X, A, Y, B,
C, B, Y, and A (see  \ding{203}). In this case, after a request for X, Belady's
algorithm (based on locality) evicts blocks B and C (to create 64 bytes of
free space) that will be accessed furthest into the future. Over the next four
accesses, this results in two misses (B and C) and two hits (A and Y). 

In contrast, a size-aware replacement policy can detect that it might be better
to retain a set of smaller compressed cache blocks that receive more hits
cumulatively than a single large (potentially uncompressed) cache block with
better locality.
%The intuition is that multiple smaller blocks cumulatively can provide more
%hits than a single large cache block.
For the access pattern discussed above, a size-aware replacement policy makes
the decision to retain B and C and evict Y to make space for X (see
\ding{204}). As a result, the cache experiences three hits (A, B, and C) and
only one miss (Y) and hence outperforms Belady's optimal
algorithm.\footnote{Later (see  \ding{205}), when there are three
requests to blocks B, Y, and A (all three hits), the final cache
state becomes the same as the initial one. Hence, this example can represent
steady state within a loop.} We conclude that using block size information in a
compressed cache can lead to better replacement decisions. 
%~\footnote{In this paper, we assume that data is uncompressed in main memory.
%Consequently, reducing miss rate results in a proportional reduction in memory
%bandwidth.  If, alternatively, compressed cache lines were stored in memory,
%64 bytes of compressed data would be fetched under both policies in this
%example.  However, compression-aware replacement policies could yield {\em
%both} fewer misses and fewer bytes fetched than compression-oblivious policies
%in other situations.} Note that if later (see  \ding{205}) there are three
%additional requests to blocks B, Y, and A (all three hits), the final cache
%state becomes the same as the initial one. Hence, this example can represent
%steady state within a loop.  

\subsection{Size Varies}
\begin{figure}[h!]
  \centering
  %\vspace{-0.2cm}
  \centering
  \includegraphics[width=0.65\textwidth]{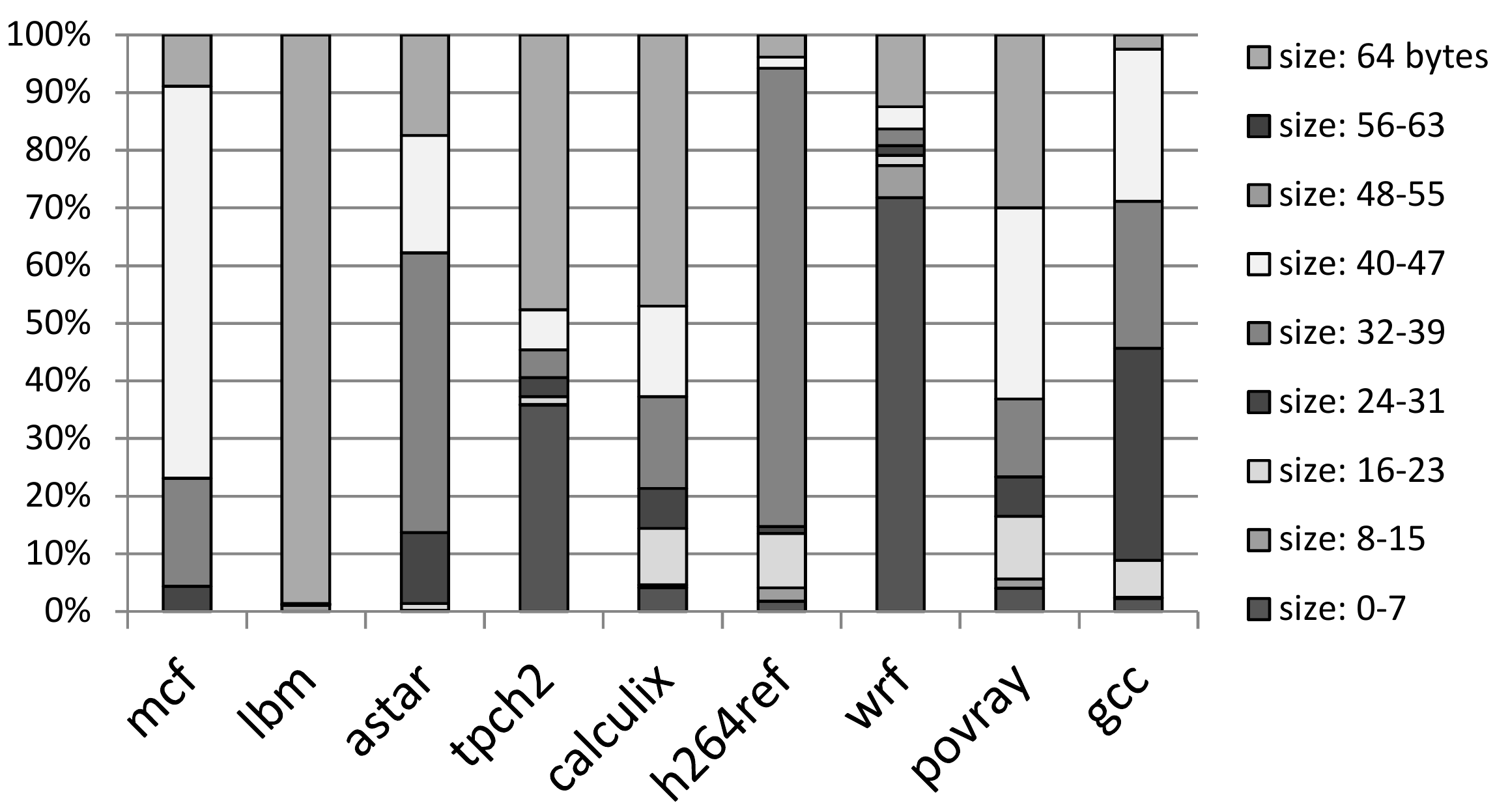}
\caption{Compressed block size distribution for representative applications with the BDI~\cite{bdi} compression algorithm.}
%\vspace{-0.2cm}
\label{fig:bdi}
\end{figure}

Figure~\ref{fig:bdi} shows the distribution of compressed cache block
sizes\footnote{Section~\ref{camp:sec:methodology} describes the details of our
evaluation methodology for this and other experiments.} for a set of
representative workloads given a 2MB cache employing the Base-Delta-Immediate
(BDI)~\cite{bdi} cache compression algorithm (our results with the
FPC~\cite{fpc} compression algorithm show similar trends).  Even though the
size of a compressed block is determined by the compression algorithm, under
both designs, \textbf{compressed cache block sizes can vary significantly},
both (i) within a single application (i.e., \emph{intra-application}) such as
in \emph{astar, povray}, and \emph{gcc} and (ii) between applications (i.e.,
\emph{inter-application}) such as between \emph{h264ref} and \emph{wrf}.

\REM{
In order to show that compressed block sizes significantly vary both within 
and between
multiple applications, we conducted an experiment\footnote{Section~\ref{camp:sec:methodology}
describes the details of our evaluation methodology for this and other experiments.} 
where we observed the cache block size distribution (collected from snapshots of the 2MB last-level-cache).
Figures~\ref{fig:bdi} and \ref{fig:fpc1} show the distributions of compressed block
sizes for a representative selection of applications from our workload pool.
In order to simplify both the data representation and analysis, we split
all possible sizes into 8-byte bins with one special bin for 64-byte (uncompressed)
blocks. 
We can draw two major conclusions from these figures.

First, the compressed block sizes vary significantly both 
(i) within an application (e.g., \emph{astar}), and
(ii) between the applications
(e.g., compare \emph{h264ref} and \emph{wrf}).
} 

Size variation within an application suggests that size-aware replacement policies could be effective for individual single-core workloads. 
Intra-application variation exists because applications have data that belong
to different common compressible patterns (e.g., zeros, repeated values, and narrow values~\cite{bdi})
and as a result end up with a mix of compressed cache block sizes.
In a system with multiple cores and shared caches, inter-application variation suggests that even 
if an application has a single dominant compressed cache block size (e.g., \emph{lbm, h264ref} and \emph{wrf}), 
running these applications together on different cores will result in the shared cache experiencing a mix of compressed cache block sizes. 
Hence, size-aware management of compressed caches can be even more important for efficient cache utilization in multi-core systems (as
we demonstrate quantitatively in Section~\ref{sec:multicore}).

\subsection{Size Can Indicate Reuse}
\label{sec:size-reuse}
We observe that elements belonging to the same data structure (within
an application) sometimes lead to cache blocks that compress to the
same size.  This observation provides a new opportunity: using
the compressed size of a cache block as an indicator of data reuse of the
block.

{\bf Intuition.} We first briefly provide intuition on why there can
be a relationship between compressed size and the reuse
characteristics of the cache block. As past work has shown, an
application's key data structures are typically accessed in a regular
fashion, with each data structure having an identifiable access
pattern~\cite{dataStructurePhase}. This regularity in accesses to a data
structure can lead to a dominant {\em reuse distance}~\cite{reuse} range for
the cache blocks belonging to the data structure.\footnote{Some prior works
(e.g.,~\cite{madcache,cachebasedonreusedist,singleUsage,DataCacheManagement})
captured this regularity by learning the relationship between the instruction
address and the reuse distance.}  
%Reuse
%distance is defined as the number of distinct cache block addresses
%accessed between two consecutive accesses to the same cache block
%address.
%The key shortcoming of such solutions is the addition of new hardware to learn the correlation (if any) 
%between PC and reuse distance. 
The same data structure can also have a dominant compressed cache
block size, i.e., a majority of the cache blocks containing the data
structure can be compressed to one or a few particular sizes (e.g., due
to narrow or sparse values stored in the elements of an array). For such
a data structure, the compressed cache block size can therefore be a
good indicator of the reuse behavior of the cache blocks. In fact,
different data structures can have different dominant compressed block
sizes and different dominant reuse distances; in such cases, the compressed
block size serves as a type of \emph{signature} indicating the reuse
pattern of a data structure's cache blocks.

{\bf Example to Support the Intuition.} To illustrate the connection
between compressed block size and reuse behavior of data structures
intuitively, Figure~\ref{fig:example} presents an example loosely based
on some of the data structures we observed in \emph{soplex}. 
%It shows why the compressed size can potentially be a
%good indicator of future reuse (as we will quantitatively demonstrate
%soon in Figure~\ref{fig:soplex}).\footnote{Note that our mechanisms
%  are applicable to a variety of applications
%  (Section~\ref{sec:results}) with very different data structures and
%  access patterns. The simplification in the code example is done for
%  clarity of demonstration.}
%
There are three data structures in this example: (i) array $A[N]$ of integer indexes that are smaller than
value $M$ (well-compressible with BDI~\cite{bdi} to 20-byte cache blocks), (ii) small
array $B[16]$ of floating point coefficients (incompressible, 64-byte cache blocks), and (iii) sparse matrix $C[M][N]$ with
the main data (very compressible zero values, many 1-byte cache blocks). These data structures not only have different compressed block sizes,
but also different reuse distances. Accesses to cache blocks for array $A$ occur only once every 
iteration of the outer loop (long reuse distance).
Accesses to cache blocks for array $B$ occur roughly every $16^{th}$ iteration of the inner loop
(short reuse distance).  Finally, the reuse distance of array $C$ is usually long, although it
is dependent on what indexes are currently stored in array $A[i]$.
Hence, this example shows that {\em compressed
  block size can indicate the reuse distance of a cache block}: 20-byte blocks (from $A$) usually
have long reuse distance, 64-byte blocks (from $B$) usually have short reuse 
distance, and 1-byte blocks (from $C$) usually have long reuse distance.
If a cache learns this relationship, it can prioritize 64-byte blocks over
20-byte and 1-byte blocks in its management policy.
As we show in Section~\ref{sec:sip}, our \insertionpolicy{} policy learns exactly this 
kind of relationship, leading to significant performance improvements for several applications 
(including \emph{soplex}), as shown in Section~\ref{sec:single-core}.\footnote{Note that our 
overall proposal also accounts for the size of the block, e.g., that a 64-byte block takes up more 
space in the cache than a 20-byte or 1-byte block, via the use of \mineviction{} policy (Section~\ref{sec:mve}).}

%This fine-grain relation between size and reuse can be exploited by a
%compression-aware cache management policy (see Section~\ref{sec:sip})
%to improve application's performance (see
%Section~\ref{sec:single-core} for \emph{soplex}).

%In this example, the compressed size of 20 bytes usually indicates a
%short reuse distance.  A coarse-grain approach with a single threshold
%is \emph{ineffective} for this case (a case that is quite common based
%on our experiments), because it cannot represent the \emph{fine-grain}
%correlation between the size and reuse (we have 20-byte blocks with
%short reuse distance and 1-/64-byte blocks with long reuse distance).
%This fine-grain relation between size and reuse can be exploited by a
%compression-aware cache management policy (see Section~\ref{sec:sip})
%to improve application's performance (see
%Section~\ref{sec:single-core} for \emph{soplex}).

\lstdefinestyle{customc}{
  belowcaptionskip=1\baselineskip,
  breaklines=true,
  frame=L,
  xleftmargin=\parindent,
  language=C,
  showstringspaces=false,
  basicstyle=\footnotesize\ttfamily,
  keywordstyle=\bfseries\color{green!40!black},
  commentstyle=\itshape\color{purple!40!black},
  identifierstyle=\color{blue},
  stringstyle=\color{orange},
}

\lstdefinestyle{customasm}{
  belowcaptionskip=1\baselineskip,
  frame=L,
  xleftmargin=\parindent,
  language=[x86masm]Assembler,
  basicstyle=\footnotesize\ttfamily,
  commentstyle=\itshape\color{purple!40!black},
}

\lstset{escapechar=@,style=customc}
%\lstset{language=C,numbers=left,stepnumber=1,numbersep=3pt}
\begin{figure}[t!]
%\vspace{-0.2cm}
%\hspace{-0.2cm}
\centering
\begin{lstlisting}
 int A[N];      // small indices: narrow values
 double B[16];  // FP coefficients: incompressible
 double C[M][N];// sparse matrix: many zero values
 for (int i=0; i<N; i++) {
    int tmp = A[i];
    for (int j=0; j<N; j++) {
      sum += B[(i+j)%16] * C[tmp][j];
    }
 } 
\end{lstlisting}
%\vspace{-0.2cm}
\caption{Code example: size and reuse distance relationship.}
\label{fig:example}
%\vspace{-0.2cm}
\end{figure}
%%% ONUR:
%In such cases, regular access patterns to such
%data structures can be linked to (and hence be indicated by) the
%compressed size of the cache blocks that store these data structures.
%. and hence predictable reuse 
%distances correlate with the compressed cache block size.
%Another key observation that we make is that a specific compressed block size
%within an application frequently correlates with a particular data structure,
%e.g., a certain array of elements stored in the compressed cache
%will likely have the same compressed block size
%for its data elements. This observation creates 
%an interesting opportunity. If an application's key data structures have
%certain dominant compressed sizes and these data structures
%also have some special properties when stored in the compressed cache, e.g.,
%certain average reuse distance,
%then compressed block size can be used as a predictor of these properties. 
{\bf Quantitative Evidence.} To verify the relationship between block
size and reuse, we have analyzed 23 memory-intensive applications' memory access traces
(applications described in Section~\ref{camp:sec:methodology}).  For every cache block
within an application, we computed the average distance (measured in
memory requests) between the time this block was inserted into the
compressed cache and the time when it was reused next. We then
accumulate this {\em reuse distance} information for all different
block sizes, where the size of a block is determined with the
BDI~\cite{bdi} compression algorithm. 
%We have verified a similar
%relationship exists with the FPC~\cite{fpc} algorithm.
%We then average the reuse distance across the compressed
%blocks of the same size. 

\ignore{
\begin{figure}[h!]
  %\vspace{-0.2cm}
  \centering
  \subfigure[bzip2 application]{\label{fig:reuse} \includegraphics[width=0.4\linewidth]{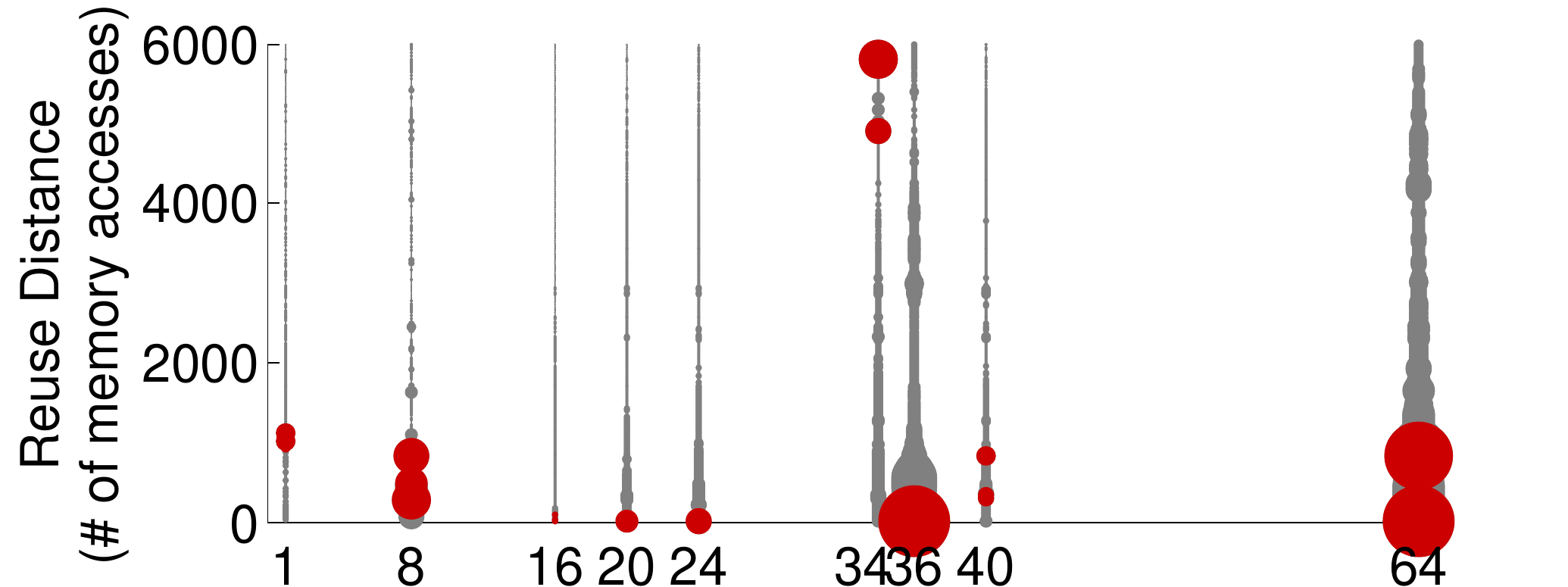}}
  %\caption{bzip2 application.}
  %}
  \subfigure[B]{\label{fig:leslie3d} \includegraphics[width=0.4\linewidth]{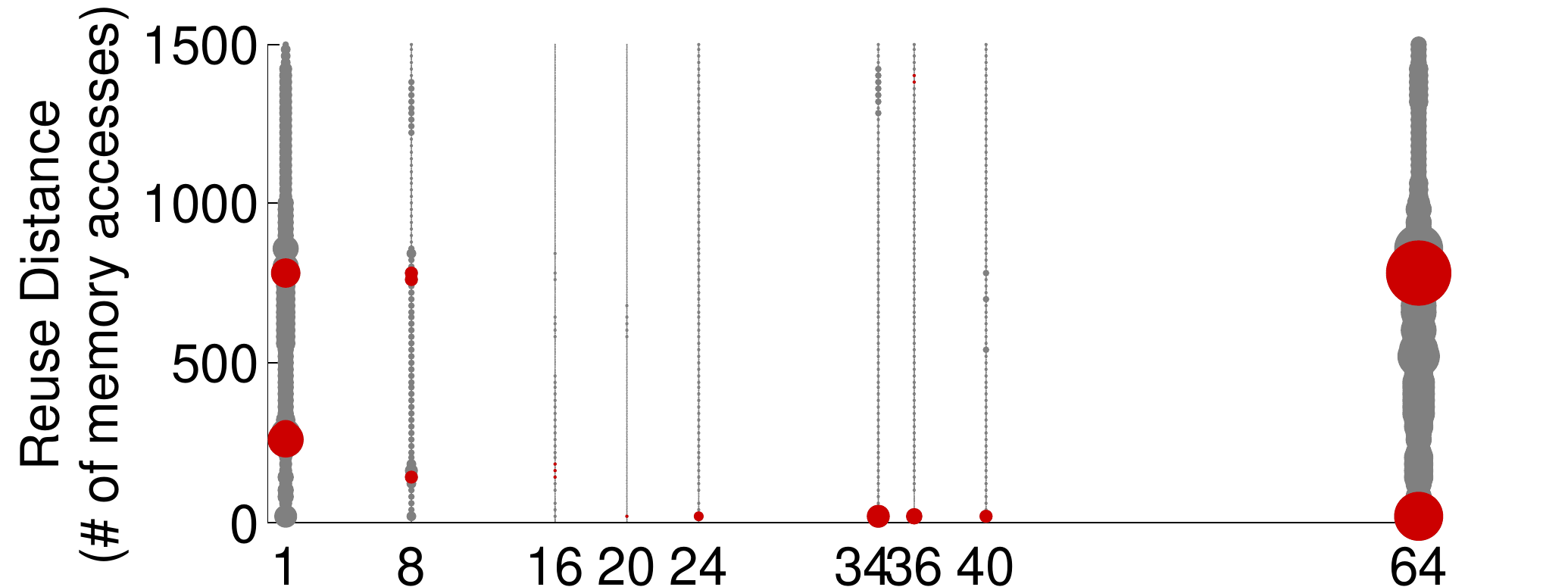}}
    
  %\begin{subfigure}[b]{0.49\linewidth}
  %\centering
  %\includegraphics[width=0.99\textwidth]{figures/leslie3d.pdf}
  %\caption*{\scriptsize{Size (bytes)}}
  %\caption{leslie3d application.}
  %\label{fig:leslie3d}
  %\end{subfigure}
  \caption{Plots demonstrate the relationship between the compressed block size and reuse distance.}
\end{figure}
}

\begin{figure}[tbh]
  \vspace{-0.1cm}
  \centering
  \subfigure[bzip2]{\label{fig:bzip2}\includegraphics[width=0.4\linewidth]{figures/bzip2.pdf}}
  \subfigure[sphinx3]{\label{fig:sphinx3}\includegraphics[width=0.4\linewidth]{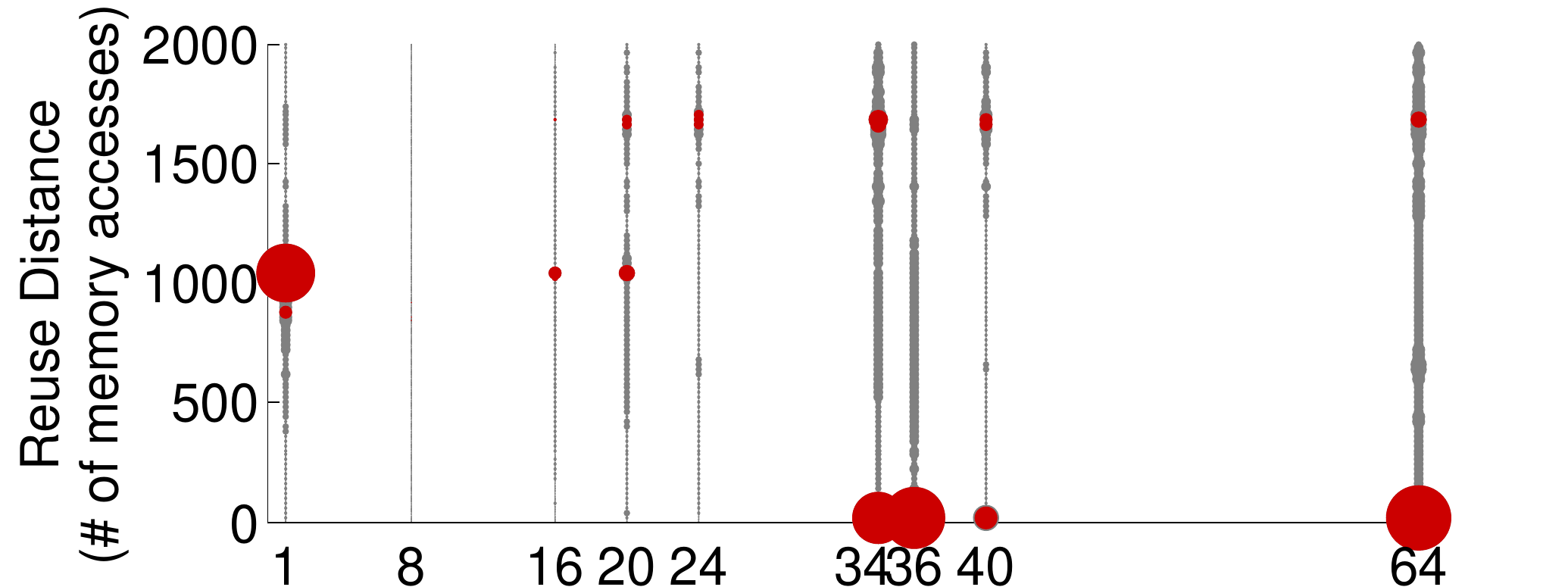}}
  \subfigure[soplex]{\label{fig:soplex}\includegraphics[width=0.4\linewidth]{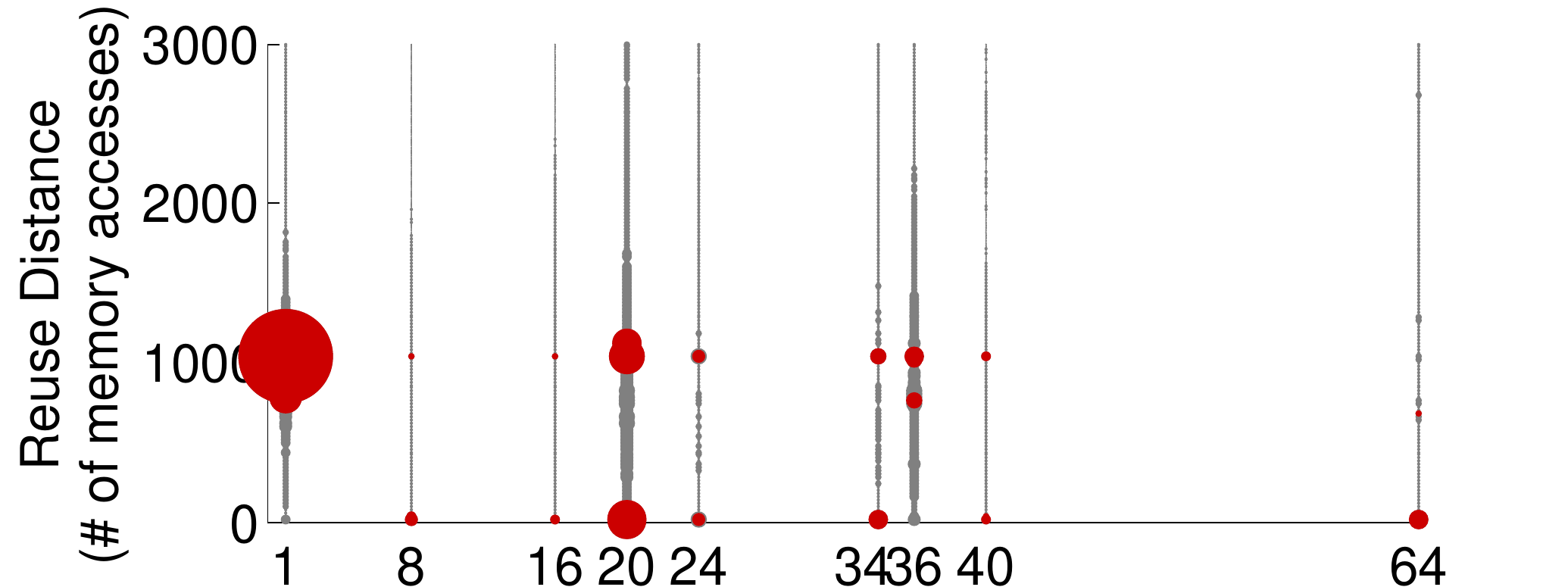}}
  \subfigure[tpch6]{\label{fig:tpch6}\includegraphics[width=0.4\linewidth]{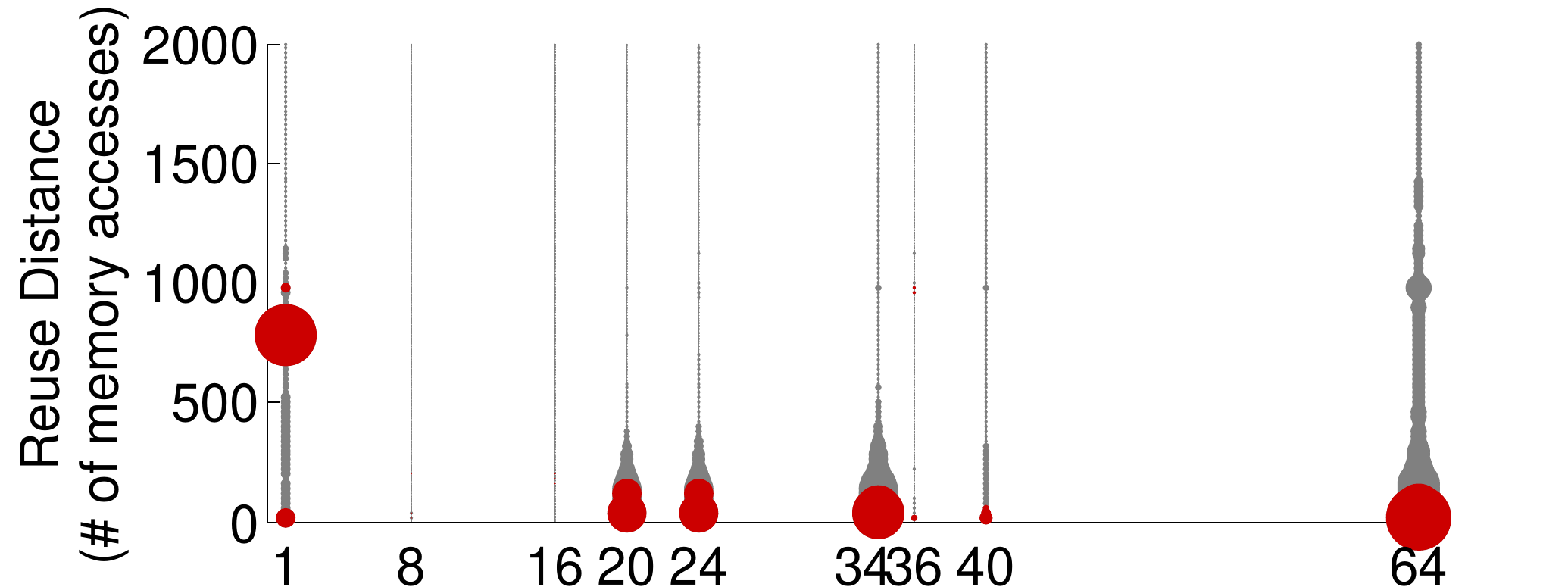}}
  %\subfigure[leslie3d]{\label{fig:leslie3d}\includegraphics[width=0.4\linewidth]{figures/leslie3d.pdf}}
  \subfigure[gcc]{\label{fig:gcc}\includegraphics[width=0.4\linewidth]{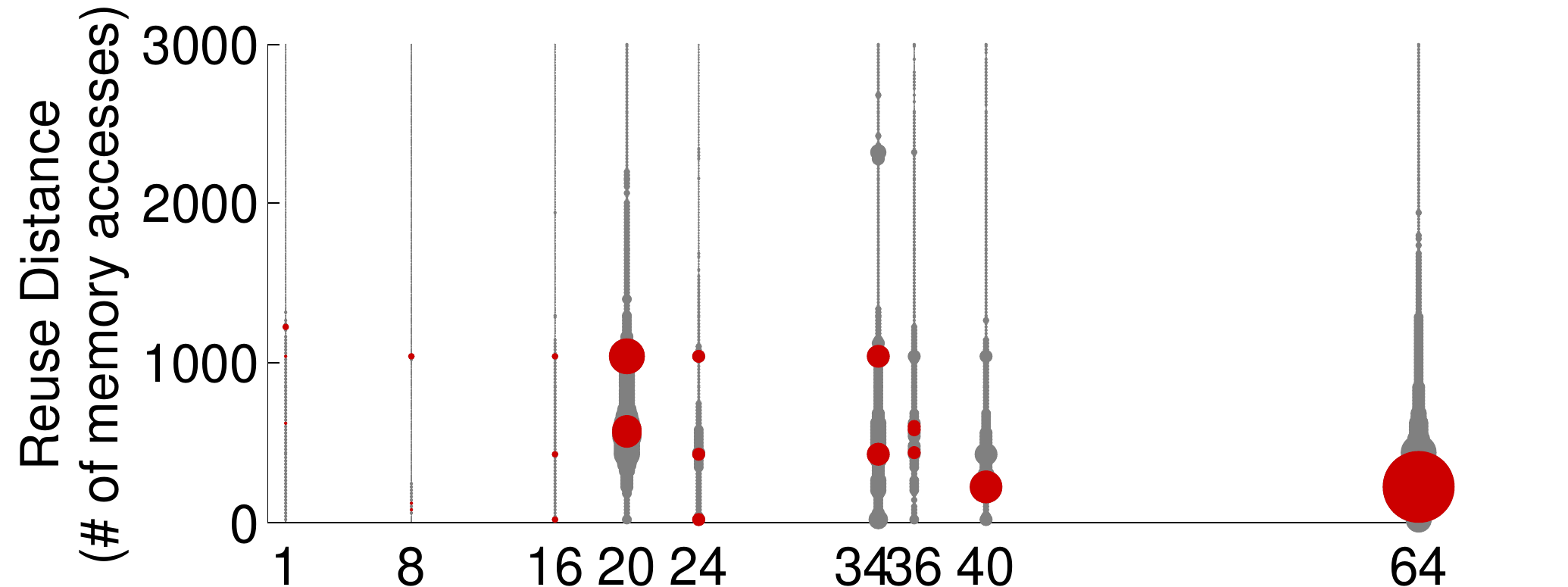}}
  %\subfigure[gobmk]{\label{fig:gobmk}\includegraphics[width=0.4\linewidth]{figures/gobmk.pdf}}
  \subfigure[mcf]{\label{fig:mcf}\includegraphics[width=0.4\linewidth]{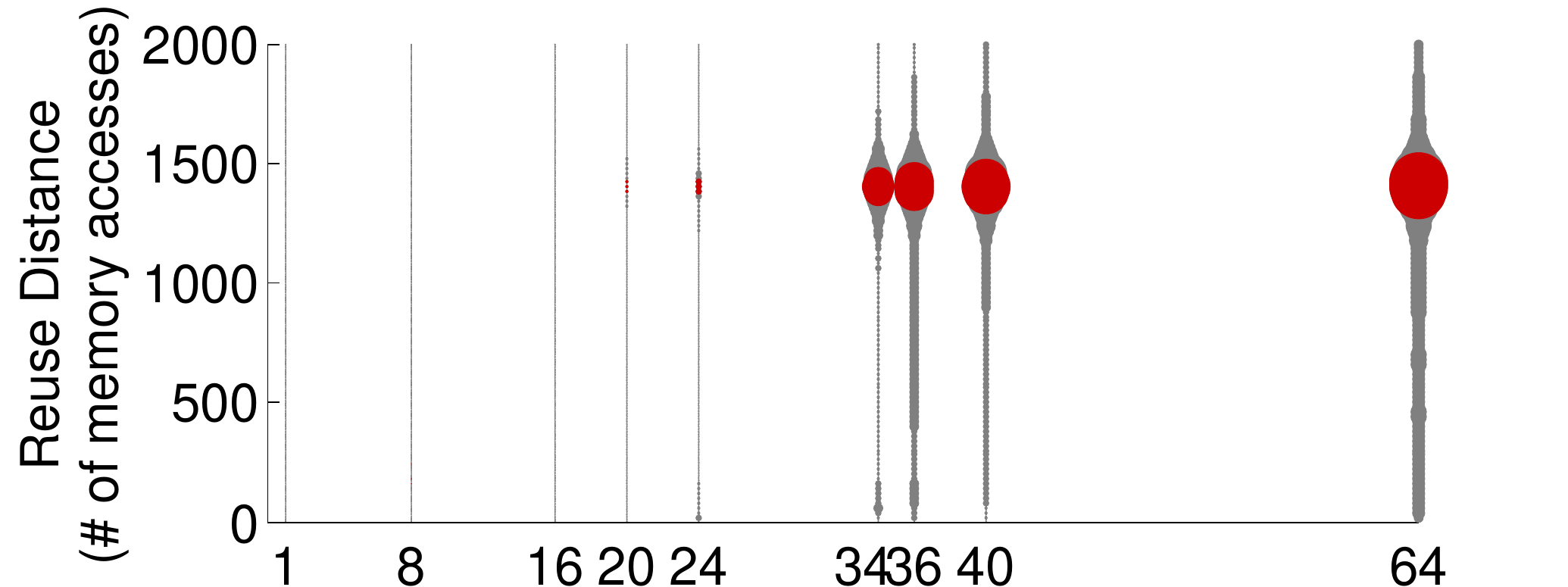}}
  \vspace{-0.2cm}
   \caption{Plots demonstrate the relationship between the compressed block size and reuse distance. Dark red circles correspond to the most frequent reuse distances for every size. The first five workloads ((a)--(e)) have some relation between size and reuse, while the last one (f) do not show that size is indicative of reuse.}
%   \vspace{-0.2cm}
\end{figure}

Figures~\ref{fig:bzip2}--\ref{fig:mcf} show the results of this
analysis for nine representative applications from our workload pool
(our methodology is described in Section~\ref{camp:sec:methodology}). In
five of these applications (\emph{bzip2}, \emph{sphinx3}, \emph{soplex},
\emph{tpch6}, \emph{gcc}), compressed
block size is an indicator of reuse distance (in other words, it can
be used to distinguish blocks with different reuse distances). In one
of the applications (\emph{mcf}), it is not.  Each
graph is a scatter plot that shows the reuse distance distribution
experienced by various compressed cache block sizes in these
applications.  There are nine possible compressed block sizes (based
on the description from the BDI work~\cite{bdi}).  The size of each
circle is proportional to the relative frequency of blocks of a
particular size that exhibit a specified reuse distance. The dark red
circles indicate the most frequent reuse distances (up to three) for
every size.

We make three major observations from these figures. 
First, there are many applications where block size is an indicator of reuse distance
(Figure~\ref{fig:bzip2}--\ref{fig:mcf}). For instance, in
\emph{bzip2} (Figure~\ref{fig:bzip2}), a large number of cache blocks
are 8, 36, or 64 (uncompressed) bytes and have a short reuse distance
of less than 1000. In contrast, a significant number of blocks are 34
bytes and have a large reuse distance of greater than 5000. This
indicates that the 34-byte blocks can be deprioritized by the cache
when running \emph{bzip2} to improve performance. Similarly, in
\emph{sphinx3}, \emph{tpch6}, and \emph{soplex} (Figures~\ref{fig:sphinx3}--\ref{fig:tpch6}), a significant number of
blocks are compressed to 1-byte with a long reuse distance of around 1000, 
whereas most of the blocks of other sizes have very short reuse distances of less than 100.
In general, we observe that data from 15 out of 23 of our evaluated applications show that block size is indicative of reuse~\cite{tr-camp}.
This suggests that a compressed block size can be used as an indicator
of future block reuse which in turn can be used to prioritize blocks of certain sizes (Section~\ref{sec:sip}),
improving application performance (e.g., see the effect on \emph{soplex} in Section~\ref{sec:single-core}).

Second, there are some applications where block size does not have a
relationship with reuse distance of the block (e.g., \emph{mcf}). 
For example, in \emph{mcf} (Figure~\ref{fig:mcf}), almost
all blocks, regardless of their size, have reuse distances around
1500. This means that block size is less effective as an indicator
of reuse for such applications (and the mechanism we
describe in Section~\ref{sec:sip} effectively avoids using block size
in cache management decisions for such applications).

Third, for applications where block size is indicative of reuse, there
is usually not a coarse-grained way to distinguish between block sizes
that are indicative of different reuse distances. In other words,
simply dividing the blocks into \emph{big} or \emph{small} blocks, as
done in ECM~\cite{ecm}, is not enough
to identify the different reuse behavior of blocks of
different sizes. The distinction between block sizes should be done at
a finer granularity. This is evident for \emph{bzip2}
(Figure~\ref{fig:bzip2}): while 8, 36, and 64-byte blocks have short
reuse distances, a significant fraction of the 34-byte blocks have
very long reuse distances (between 5000 and 6000). Hence, there is no single
block size threshold that would successfully {\em distinguish} blocks with high
reuse from those with low reuse. Data from other applications (e.g.,
\emph{soplex}, \emph{gcc}) similarly support this.

We briefly discuss why compressed size is sometimes not indicative of reuse behavior.  
First, data
stored in the data structure might be different, so multiple compressed sizes are
possible with the same reuse pattern (e.g., for \emph{mcf}). In this case,
blocks of different sizes are equally important for the cache. 
Second, blocks with the same size(s) can have multiple
different reuse patterns/distances (e.g., for \emph{milc} and \emph{gromacs}). 
In this case, size might not provide useful information to improve cache utilization, 
because blocks of the same size
can be of very different importance.

\section{\carp{}: Design and Implementation}
\label{sec:carp}

Our proposed Compression-Aware Management Policy (\carp{}) consists of two
components: Minimal-Value Eviction (\mineviction{}) and Size-based Insertion
Policy (\insertionpolicy{}). These mechanisms assume a compressed cache
structure where the compressed block size is available to the hardware making
the insertion and replacement decisions. Without the loss of generality, we assume 
that the tag-store contains double the
number of tags and is decoupled from the data-store to allow higher effective
capacity (as proposed in several prior works~\cite{fpc,bdi,c-pack}). We
also propose Global \carp{} (or G-\carp), an adaptation of \carp{} for a cache
with a global replacement policy.

In this section, we first provide the background information needed to
understand some of our mechanisms (Section~\ref{sec:background}).  Then, we
describe the design and implementation of each mechanism in depth
(Sections~\ref{sec:mve}-\ref{sec:gcarp}).  We detail the implementation of our
G-\carp{} mechanism assuming the structure proposed for the V-Way
cache~\cite{v-way}.
% (described in Section~\ref{sec:background}).  
None of the mechanisms require extensive hardware changes on top of the
baseline compressed cache designs (both local and global, see
Section~\ref{sec:complexity} for an overhead analysis).

\subsection{Background}
\label{sec:background}
Multiple size-oblivious cache management mechanisms (e.g.,~\cite{mlp,RRIP,EAF,lacs,rw-samira}) were proposed to improve
the performance of conventional on-chip caches (without compression).
Among them, we select RRIP~\cite{RRIP} as both
a comparison point in our evaluations and as a predictor of future re-reference in some
of our algorithms (see Section~\ref{sec:mve}). This selection is motivated both
by the simplicity of the algorithm and its state-of-the-art performance (as shown in \cite{RRIP}).

\textbf{RRIP.}
Re-Reference Interval Prediction (RRIP)~\cite{RRIP} uses an $M$-bit saturating counter per cache block
as a Re-Reference Prediction Value ($RRPV$) to predict the block's re-reference distance.
The key idea behind RRIP is to prioritize the blocks with lower predicted re-reference distance, as these blocks have higher
expectation of near-future reuse.
Blocks are inserted with a long re-reference interval prediction ($RRPV = 2^M-2$).
On a cache miss, the victim block is a block with a predicted distant re-reference interval ($RRPV = 2^M-1$).
If there is no such block, the $RRPV$ of all blocks is incremented by one and the process repeats until a victim is found.
On a cache hit, the $RRPV$ of a block is set to zero (near-immediate re-reference interval). Dynamic RRIP (DRRIP) uses set
dueling~\cite{mlp,dip} to select between the aforementioned policy (referred to as
SRRIP) and one that inserts blocks with a short re-reference interval
prediction with high probability and inserts blocks with a long re-reference
interval prediction with low probability.

\textbf{V-Way.}
The Variable-Way, or V-Way~\cite{v-way}, cache is a set-associative cache with a decoupled tag- and data-store.
The goal of V-Way is two-fold: providing flexible (variable) associativity together with a global replacement 
across the entire data store.
A defining characteristic is that there are more tag-entries than data-entries. 
Forward and backward pointers are maintained in the tag- and data-store to link the entries.
This design enables associativity to effectively vary on a per-set basis by
increasing the number of tag-store entries relative to data-store entries.
Another benefit is the implementation of a \emph{global replacement policy},
which is able to choose data-victims from anywhere in the data-store.  This is in
contrast to a traditional \emph{local replacement policy},
e.g.,~\cite{LRU,RRIP}, which considers data-store entries only within a single
set as possible victims.  The particular global replacement policy described in
\cite{v-way} (called Reuse Replacement) consists of a Reuse Counter Table (RCT) with a
counter for each data-store entry. Victim selection is done by starting at a
pointer (PTR) to an entry in the RCT and searching for the first counter equal
to zero, decrementing each counter while searching, and wrapping around if
necessary.  A block is inserted with an RCT counter equal to zero. On a hit,
the RCT counter for the block is incremented.  We use the V-Way design as a
foundation for all of our global mechanisms (described in
Section~\ref{sec:gcarp}).

\subsection{Minimal-Value Eviction (\mineviction)}
\label{sec:mve}

The key observation in our \mineviction{} policy is that evicting one or more important blocks of larger compressed size may be more beneficial 
than evicting several more compressible, less important blocks (see Section~\ref{sec:motivation}). 
The idea behind \mineviction{} is that each block has a value to the cache. This value is a function 
of two key parameters: (i) the likelihood of future re-reference and (ii) the compressed block size. 
For a given  $<$prediction of re-reference, compressed block size$>$ tuple, 
\mineviction{} associates \emph{a value with the block}. 
Intuitively, a block with higher likelihood of re-reference is more valuable 
than a block with lower likelihood of re-reference and is assigned a higher value. 
Similarly, a more compressible block is more valuable than a less compressible block 
because it takes up fewer segments in the data-store, potentially allowing for the caching of additional useful blocks. 
The block with the least value in the associativity set is chosen as the next victim for replacement---sometimes multiple blocks need to be evicted to make room for the newly inserted block.

In our implementation of \mineviction{},
the value $V_i$ of a cache block $i$ is computed as $V_i = p_i/s_i$, where $s_i$ is the compressed block size of block $i$ and
$p_i$ is a predictor of re-reference, such that a larger value of $p_i$ denotes block $i$ is more important and is predicted 
to be re-referenced sooner in the future. 
This function matches our intuition and is monotonically increasing with respect 
to the prediction of re-reference and monotonically decreasing with respect to the size. 
We have considered other functions with these properties (i.e., a weighted linear sum), but found the difference in performance to be negligible.

Our mechanism estimates $p_i$ using RRIP\footnote{Specifically, the version of RRIP that our mechanism uses is SRRIP. 
We experimented with DRRIP, but found it offered little performance improvement for our mechanisms compared to the additional complexity. 
All of our evaluations assume an RRPV width $M=3$. }~\cite{RRIP} as the predictor of future re-reference
due to its simple hardware implementation and state-of-the-art stand-alone performance.\footnote{Other alternatives considered (e.g., \cite{EAF}) provide only a binary value.}
As described in Section~\ref{sec:background}, 
RRIP maintains a re-reference prediction value (RRPV) for each cache block which predicts the re-reference distance. 
Since a larger RRPV denotes a longer predicted re-reference interval, we compute $p_i$ as $p_i=(RRPV_{MAX}+1-RRPV_i)$. 
Therefore, a block with a predicted short re-reference interval has more value than a comparable block with a predicted 
long re-reference interval. $p_i$ cannot be zero, because $V_i$ would lose dependence on $s_i$ and become size-oblivious.

Depending on the state of the cache, there are two primary conditions in which a victim block must be selected: 
(i) the data-store has space for the block to be inserted, but all tags are valid in the tag-directory, or 
(ii) the data-store does not have space for the block to be inserted 
(an invalid tag may or may not exist in the tag-directory). 
In the first case where the data-store is not at capacity, 
\mineviction{} relies solely on the predictor of re-reference or conventional replacement policy, such as RRIP.  
For the second case, the valid blocks within the set are compared based on $V_i$ and the set of blocks with 
the least value is evicted to accommodate the block requiring insertion. 

\mineviction{} likely remains off the critical path, but to simplify the microarchitecture, 
we eliminate division in the calculation of $V_i$ by bucketing block sizes such that $s_i$ is always a power of two, 
allowing a simple right-shift operation instead of floating point division. 
For the purposes of calculating $V_i$, $s_i=2$ for blocks of size 0B -- 7B, $s_i=4$ for blocks of size 8B -- 15B, 
$s_i=8$ for blocks of size 16B -- 31B, and so on. 
The most complex step, comparing blocks by value, can be achieved with a fixed multi-cycle 
parallel comparison.
%TODO: this may be too aggressive a simplification, our evaluation is based on bucketing into 8 buckets of size 8 and using a lookup table.

\subsection{Size-based Insertion Policy (\insertionpolicy)}
\label{sec:sip}

The key observation behind \insertionpolicy{} is that sometimes there is a relation between cache 
block reuse distance and compressed block size (as shown in Section~\ref{sec:size-reuse}). \insertionpolicy{} exploits this observation 
and inserts blocks of certain sizes with higher priority if doing so reduces the cache miss rate.  
%\insertionpolicy{} is effective because there is often a correlation between size and reuse distance either
%within an application or within an application phase. 
Altering the priority of blocks of certain sizes with short or long reuse distances 
helps to ensure that more important blocks stay in the cache.

At run-time, \insertionpolicy{} dynamically detects the set of sizes that, when inserted with higher priority, 
reduce the number of misses relative to a size-oblivious insertion policy. 
\insertionpolicy{} uses a simple mechanism based on dynamic set sampling~\cite{mlp} to make the 
prioritization decision for various compressed sizes.
It selects the best-performing policy among competing 
policies during a periodic training phase and applies that policy during steady state. 
The observation in dynamic set sampling is that sampling makes it possible to choose the better policy with only a 
relatively small number of sets selected from the Main Tag Directory (MTD) to have a corresponding set in an Auxiliary 
Tag Directory (ATD) participating in a tournament. Only the MTD is coupled with the data-store; the ATD is only 
for deciding which block size(s) should be inserted with high priority. Therefore, there are no performance degradations due to our sampling during training.

\begin{figure}[tb]
  \vspace{-0.3cm}
  %\centering
  %\begin{subfigure}[b]{0.45\linewidth}
  \centering
  \includegraphics[width=60mm]{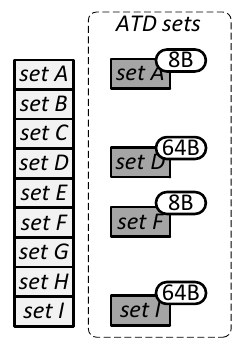}
  %\vspace{-0.2cm}
  \label{fig:sipImplementationAtd}
  %\end{subfigure}
  %\hspace{0.1cm}
  %\begin{subfigure}[b]{0.48\linewidth}
  %\centering
  \includegraphics[width=60mm]{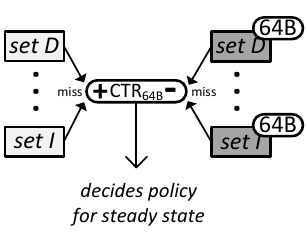}
  %\vspace{-0.2cm}
  \label{fig:sipImplementationCtr}
  %\end{subfigure}
  %\vspace{-0.1cm}
  \caption{Set selection during training and decision of best insertion policy based on difference in miss rate in MTD/ATD.}
\end{figure}

Let $m$ be the minimum number of sets that need to be sampled so that dynamic
set sampling can determine the best policy with high probability and $n$ be the
number of compressible block sizes possible with the compression scheme (e.g.,
8B, 16B, 20B, ..., 64B). In \insertionpolicy{}, the ATD contains $m\cdot{}n$
sets, $m$ for each of the $n$ sizes. As shown in
Figure~\ref{fig:sipImplementationAtd}, each set in the ATD is assigned one of
the $n$ sizes. The \emph{insertion policy} in these sets of the ATD differs from the
insertion policy in the MTD in that the assigned size is prioritized. For the
example in Figure~\ref{fig:sipImplementationAtd}, there are only two possible
block sizes. Sets A and F in the ATD \emph{prioritize} insertions of 8-byte blocks
(e.g., by increasing $p_i$).
Sets D and I prioritize the insertion of 64-byte blocks. Sets B, C, E, G, and H
are not sampled in the ATD. 
%Both tag directories implement the same replacement
%policy.

When a set in the MTD that has a corresponding set in the ATD receives a miss,
a counter ${CTR}_i$ is incremented, where \emph{i} is a size corresponding to
the prioritized size in the corresponding ATD set.  When an ATD set receives a
miss, it decrements ${CTR}_i$ for the size associated with the policy this set
is helping decide.  Figure~\ref{fig:sipImplementationCtr} shows the decision of
the output of ${CTR}_{64B}$.

For each of the possible compressed block sizes, a decision is made independently based on the result of the counter. 
If ${CTR}_i$ is negative, prioritizing blocks of size \emph{i} is negatively affecting miss rate 
(e.g., the insertion policy in the MTD resulted in fewer misses than the insertion policy in the ATD). 
Therefore,  \insertionpolicy{} does not prioritize blocks of size \emph{i}. 
Likewise, if ${CTR}_i$ is positive, prioritizing insertion of blocks of size \emph{i} is reducing the miss rate 
and  \insertionpolicy{} inserts size \emph{i} blocks with high priority for best performance. 
For $n$ different sizes, there are $2^n$ possible insertion schemes and any may be chosen by \insertionpolicy.

For simplicity and to reduce power consumption, the dynamic set sampling occurs during a periodic training phase\footnote{In our evaluations, we perform training for 10\% of the time. For example, for 100 million cycles every 1 billion cycles.}
at which time the insertion policy of the MTD is unaffected by \insertionpolicy{}. At the conclusion of the training phase,
a steady state is entered and the MTD adopts the chosen policies and prioritizes the insertion of blocks of sizes for which
$CTR$ was positive during training.

\insertionpolicy{} is general enough to be applicable to many replacement policies (e.g., LRU, RRIP, etc). 
In some cases (e.g., LRU), it is more effective to try inserting blocks with lower priority (e.g., LRU position) 
instead of higher priority as proposed above. We evaluate \insertionpolicy{} with RRIP where blocks by default
are inserted with a predicted long re-reference interval ($RRPV = 2^M-2$). Therefore, in the ATD sets, the appropriate sizes are prioritized 
and inserted with a predicted short re-reference interval ($RRPV=0$). For a 2MB cache with 2048 sets, we create 
an ATD with 32 sets for each of 8 possible block sizes. For simplicity, in our implementation we limit the number of sizes to eight by bucketing the sizes 
into eight size bins (i.e., bin one consists of sizes 0 -- 8B, bin two consists of sizes 9 -- 16B,\ldots, and bin eight consists of sizes 57 -- 64B).

\subsection{\carp{} for the V-Way Cache}
\label{sec:gcarp}

In addition to being an effective mechanism for the traditional compressed cache with 
a local replacement policy, 
the key ideas behind \carp{} are even more effective when applied to a cache with a decoupled tag- and data-store and a global
replacement policy, where the pool of potential candidates for replacement is much larger.
In this work, we apply these ideas to the V-Way cache~\cite{v-way} (described in Section~\ref{sec:background}) with
its decoupled tag- and data-store that increase the effectiveness of replacement algorithms.
To demonstrate this effectiveness, we propose Global \insertionpolicy{} (or G-\insertionpolicy) and Global \mineviction{} (or G-\mineviction). 
Together, we combine these into Global \carp{} (or G-\carp). 

%For a traditional cache structure, a local replacement policy considers only the blocks within a \emph{single set} for candidates to replace. 
%The V-Way cache~\cite{v-way} (described in Section~\ref{sec:background}),
%with its decoupled tag- and data-store, enables a global replacement decision where the pool of 
%potential candidates for replacement is much larger, increasing the effectiveness of caching.
%With some modest changes, V-Way cache can be extended to support data compression.

\begin{figure}[tb]
  \centering
  \includegraphics[width=80mm]{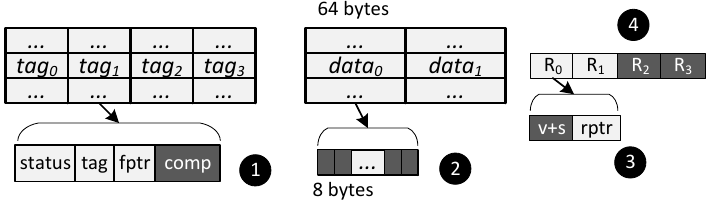}
  %\vspace{-0.4cm}
  \caption{V-Way + compression cache design.}
  \label{fig:vway+c}
  %\vspace{-0.2cm}
\end{figure}

\textbf{V-Way cache + compression.}
The V-Way cache~\cite{v-way} design can be enhanced with compression in four main steps (as shown
in Figure~\ref{fig:vway+c}).
First, the tag entries need to be extended with the encoding bits to represent a particular
compression scheme used for a cache block (e.g., 4 bits for BDI~\cite{bdi}, see \ding{202}). The number of tags
is already doubled in the V-Way cache. Second, the data store needs to be split into multiple
segments to get the benefit of compression (e.g., 8-byte segments, see \ding{203}). As in ~\cite{bdi}, every cache block
after compression consists of multiple adjacent segments.
Third, the reverse pointers ($R_{n}$) that are used to perform the replacement need to track not only 
the validity (v bit) but also the size of each block after compression (measured in the number of 8-byte segments, \ding{204}).
This simplifies the replacement policies, because there is no need to access the tags to find block sizes. 
Fourth, we double the number of reverse pointers per set, so that we can exploit the capacity benefits from compression (\ding{205}).

For a 2MB V-Way-based L2 cache with 64-byte cache blocks, the sizes of the \emph{fptr} and \emph{rptr}
pointers are 15 ($log_2{\frac{2MB}{64B}}$) and 16 ($log_2{\frac{2*2MB}{64B}}$) bits respectively. After compression is applied and assuming 8-byte segments, fptr would increase
by 3 bits to a total size of 18 bits.\footnote{Fptr and rptr pointers can be reduced in size (by 3 bits) by using regioning 
(as described later in Section~\ref{sec:gsip}).} A single \emph{validity} bit that was used in V-Way cache is now enhanced to 3
bits to represent 7 different sizes of the cache blocks after compression with BDI as well as the validity itself.

\textbf{G-\mineviction.}
\label{sec:gmve}
As in \mineviction, G-\mineviction{} uses a value function to calculate the value of blocks. 
The changes required are in (i) computing $p_i$ and (ii) selecting a pool of blocks from the large 
pool of replacement options to consider for one global replacement decision. To compute $p_i$, we propose 
using the reuse counters from the Reuse Replacement policy~\cite{v-way} as a predictor of future re-reference. 
As in the Reuse Replacement policy \cite{v-way} (see Section~\ref{sec:background}), each data-store entry has a counter. 
On insertion, a block's counter is set to zero. On a hit, the block's counter is incremented by one indicating its reuse.

For the second change, we implement global replacement by maintaining a pointer (PTR) to a reuse counter entry. 
Starting at the entry PTR points to, the reuse counters of 64 valid data entries are scanned, 
decrementing each non-zero counter by one (as in the Reuse Replacement policy). The 64 blocks are assigned a value, $V_i$, 
and the least-valued block(s) are evicted to accommodate the incoming block. 
64 blocks are chosen because it guarantees both an upper bound on latency and that evicting all 64 blocks 
(i.e., all highly compressed blocks) in the worst case will vacate enough data-store space for the incoming block.

A few applications (i.e., \emph{xalancbmk}~\cite{SPEC}) have a majority of
blocks of very similar sizes that primarily belong to two size bins of adjacent
sizes. When considering 64 such blocks, certain blocks in the smaller size bin
can essentially be ``stuck'' in the cache (i.e., there is only a very small
probability these blocks will be chosen as victim, because a block with the
same prediction of re-reference that belongs in the larger size bin is present
and will be chosen). This results from the microarchitectural simplifications
and approximate nature of the value function and can cause performance
degradations in a few cases. We address this shortcoming later in this section. 

\textbf{G-\insertionpolicy.}
\label{sec:gsip}
Dynamic set sampling (used by \insertionpolicy{}) motivates that only a select number of sets are required to be sampled to estimate the performance of competing policies \cite{mlp}. 
However, this assumption does not hold in a cache with global replacement, because evictions are not limited to the set in which 
a cache miss occurs and this interferes with sampling. For the V-Way cache, we propose instead a mechanism inspired by set dueling~\cite{dip} 
to select the optimal insertion policy.

To apply set dueling to G-\insertionpolicy,
we need to divide the data-store into $n$ (where $n$ is small; in our evaluations $n=8$) equal regions.
Instead of considering all blocks within the data-store, the replacement policy considers only the blocks within a particular region. 
This still allows considerably more replacement options than a traditional cache structure.
We observe that this division also simplifies the V-Way cache design  with negligible impact on performance.\footnote{G-\mineviction{} supports regions by simply maintaining one PTR per region.} 

\begin{figure}[tb]
  %\centering
  %\begin{subfigure}[b]{0.45\linewidth}
  \centering
  \includegraphics[width=60mm]{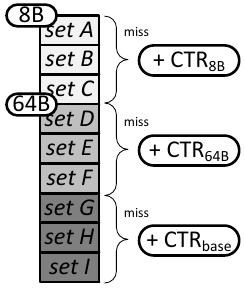}
  %\vspace{-0.2cm}
  %\vspace{-0.1cm}
  \label{fig:globalSipImplementationAtd}
  %\end{subfigure}
  %\hspace{0.1cm}
  %\begin{subfigure}[b]{0.45\linewidth}
  %\centering
  \includegraphics[width=60mm]{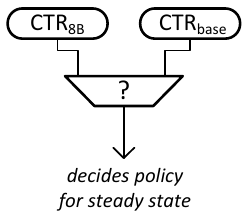}
  %\vspace{-0.2cm}
  \label{fig:globalSipImplementationCtr}
  %\end{subfigure}
  %\vspace{-0.3cm}
  \caption{Set selection during training and update of counters on misses to each region.}
\end{figure}

During a training phase, each region is assigned a compressed block size to prioritize on insertion. 
Figure~\ref{fig:globalSipImplementationAtd} shows this assignment for a simple cache with three regions and two block sizes, 8-byte and 64-byte. 
The third region is designated as a baseline (or control) region in which no blocks are inserted with higher priority. 
When a miss occurs within a region, the $CTR$ counter is incremented for that region. 
For example, in Figure~\ref{fig:globalSipImplementationAtd}, a miss to set A, B, or C increments ${CTR}_{8B}$.
 Likewise, a miss to set G, H, or I increments ${CTR}_{base}$ and so on. At the end of the training phase, 
the region $CTR$ counters are compared (see Figure~\ref{fig:globalSipImplementationCtr}). If ${CTR}_i < {CTR}_{base}$, blocks of size $i$ are inserted with higher priority in steady state in all regions. 
Therefore, G-\insertionpolicy{} detects at runtime the sizes that reduce the miss rate when inserted with higher priority than other blocks.

In our implementation, we have divided the data-store into eight
regions.\footnote{We conducted an experiment varying the number of regions (and
therefore the number of distinct size bins considered) from 4 to 64 and found having 8
regions performed best.} This number can be adjusted based on cache size.
Because one region is designated as the baseline region, we bin the possible
block sizes into seven bins and assign one range of sizes to each region.
During the training phase, sizes within this range are inserted with higher
priority. The training duration and frequency are as in \insertionpolicy{}.
Because training is short and infrequent, possible performance losses due to
set dueling are limited.

\textbf{G-\carp.} \label{sec:gcarp-implementation} G-\mineviction{} and
G-\insertionpolicy{} complement each other and can be easily integrated into
one comprehensive replacement policy referred to as G-\carp. We make one
improvement over the simple combination of these two orthogonal policies to
further improve performance in the few cases where G-\mineviction{} degrades
performance. During the training phase
of G-\insertionpolicy{}, we designate a region in which we insert blocks 
with simple Reuse Replacement instead of G-\mineviction{}. At the end of the
training phase, the $CTR$ for this region is compared with the control region
and if fewer misses were incurred, G-\mineviction{} is disabled 
in all regions at the steady state. In G-\mineviction{}-friendly applications, it remains enabled.

\subsection{Overhead and Complexity Analysis}
\label{sec:complexity}
%We believe the complexity and area overhead of our proposals are modest. 
Table~\ref{table:complexity} 
shows the storage cost of six cache designs: baseline uncompressed cache, BDI compressed cache with LRU, V-Way with and without compression, as well as \carp{} and G-\carp{}. 
On top of our reference 
cache with BDI and LRU (2384kB), \mineviction{} does not add any additional metadata and the dynamic set sampling in 
\insertionpolicy{} increases the cache size in bits by only 1.5\% (total \carp{} size: 2420kB). 
Adding BDI compression to V-Way cache with 2x tags and 8 regions increases cache size from 2458kB to 2556kB. 
G-\mineviction{}/G-\insertionpolicy{}/G-\carp{} do not add further metadata (with the exception of eight 16-bit counters for set-dueling 
in G-\insertionpolicy{}/G-\carp{}). In addition, none of the proposed mechanisms are on the critical path of the execution and
the logic is reasonably modest to implement (e.g., comparisons of CTRs). We conclude that the complexity and
storage overhead of \carp{} are modest.

\setlength{\tabcolsep}{.4em}
\begin{table}[h]\scriptsize
\centering
\vspace{-0.3cm}
\hspace{-0.0cm}
\begin{minipage}{\columnwidth}
  \hfill{}
  %\centering
 \begin{tabular}{|c|c|c|c|c|c|c|}
    %\toprule 
    \hline
    \textbf {} & \textbf{Base}  & \textbf{BDI} & \textbf{\carp{}} & \textbf{V-Way} & \textbf{V-Way+C} & \textbf{G-\carp{}} \\
    %\midrule
    \hline
    tag-entry(bits) & 21 & 35(\cite{bdi}) & 35 & 36~\footnote{+15 forward ptr; \textsuperscript{\textit{b}} +16 reverse ptr; 
\textsuperscript{\textit{c{}}}+1/8 set sampling in \textbf{\insertionpolicy{}}; \textsuperscript{\textit{d}}CTR's in \textbf{\insertionpolicy{}}; \textsuperscript{\textit{e}} +4 for comp. encoding; \textsuperscript{\textit{f}} +32 (2 reverse ptrs per data entry,
 13 bits each, and 2 extended validity bits, 3 bits each)} & 40~\textsuperscript{\textcolor{red}{\textit{e}}} & 40  \\
    %\midrule
    %\cmidrule(rl){1-7}
    \hline
    data-entry(bits) & 512 & 512 &  512 & 528~\textsuperscript{\textcolor{red}{\textit{b}}} & 544~\textsuperscript{\textcolor{red}{\textit{f}}} &  544 \\
    %\midrule
    %\cmidrule(rl){1-7}
    \hline
    \# tag entries & 32768 & 65536 & 73728~\textsuperscript{\textcolor{red}{\textit{c}}} &  65536 & 65536 & 65536 \\
    \hline
    \# data entries & 32768 & 32768 & 32768 & 32768 & 32768 & 32768 \\ 
    %\cmidrule(rl){1-7} %\hline
    %\# data-store entries & 32768 & 32768 & 32768 & 32768 & 32768 & 32768 & 32768 & 32768 & 32768 & 32768 \\
    %\cmidrule(rl){1-7} 
    \hline
    tag-store (kB) & 86 & 287 & 323 & 295 & 328 &  328 \\
    \hline
    data-store (kB) & 2097 & 2097 & 2097 & 2163 & 2228 &  2228 \\
    %\cmidrule(rl){1-7} 
    \hline
    other & 0 & 0 & 8*16~\textsuperscript{\textcolor{red}{\textit{d}}} & 0 & 0 & 8*16 \\
    %\cmidrule(rl){1-7} 
    \hline
    \hline
    \textbf{total (kB)} & 2183 & \textbf{2384} & \textbf{2420} & 2458 & \textbf{2556} & \textbf{2556} \\
    %\bottomrule
    \hline
  \end{tabular}
  \hfill{}
  \caption{Storage overhead of different mechanisms for a 2MB L2 cache. 
``V-Way+C'' means V-Way with compression.}
  %\vspace{-0.4cm}
  \label{table:complexity}
\end{minipage}
\vspace{-0.2cm}
\end{table}

\section{Qualitative Comparison with Prior Work}
\label{sec:related}

\subsection{Size-Aware Management in On-Chip Caches}
Baek et al. propose Effective Capacity Maximizer (ECM)~\cite{ecm}.  This
mechanism employs size-aware insertion and replacement policies for an on-chip
compressed cache.  Unlike size-oblivious DRRIP~\cite{RRIP} on which it is
built, ECM inserts big blocks with lower priority than small blocks.  The
threshold for what is considered a ``big'' block is determined dynamically at
runtime using an equation derived from heuristics and based on the current
effective capacity and physical memory usage. During replacement, the biggest
block in the eviction pool is selected as the victim.

ECM is the first size-aware policy employed for compressed on-chip caches. We
find that this approach has several shortcomings and underperforms relative to
our proposed mechanisms (as we show in Section~\ref{camp:sec:results}).  First, the
threshold scheme employed by ECM is coarse-grained and, especially in
multi-core workloads where a greater diversity of block sizes exists across
workloads, considering more sizes (as \carp{} does) yields  better performance.
Second, ECM's mechanism does not consider the relation between block reuse and
size, whereas \carp{} exploits the new observation that block size and reuse
can sometimes be related.  Third, due to ECM's complex threshold definition,  it
is unclear how to generalize ECM to a cache with global replacement, where size-aware
replacement policies demonstrate highest benefit (as shown in
Section~\ref{camp:sec:results}). In contrast, \carp{} is easily adapted to work with
such caches.

Recently, Sardashti and Wood propose the decoupled compressed cache (DCC)
design~\cite{dcc} that exploits both locality and decoupled sectored cache
design to avoid recompaction (and partially fragmentation) overhead in the
previous compressed cache designs. The DCC design is largely orthogonal to the
compression mechanisms proposed in this work and can be used in cojunction with
them.  

%TODO: describe all other problems: typos, tie to RRIP, ...

\subsection{Size-Aware Management in Web Caches} 
Prior works in web caches
have proposed many management strategies that consider object size, e.g.,
variable document size.  ElAarag and Romano~\cite{elaarag1, elaarag2} provide
one of the most comprehensive surveys.  While these proposed techniques serve
the same high-level purpose as a management policy for an on-chip cache (e.g., making
an informed decision on the optimal victim), they do so in a much different
environment.  Many proposed mechanisms rely on a recency list of \emph{all} objects in
the cache (e.g., \cite{size}) or consider frequency of object access (e.g.,
\cite{lru-sp}), which are prohibitively expensive techniques for an on-chip
cache.  In addition, these techniques do not consider a higher density of
information that comes with the smaller blocks after compression. This higher
density can lead to a higher importance of the smaller blocks for the cache,
which was mostly ignored in these prior mechanisms.
%In addition, many of these techniques try to maximize hit rate...instead of
%compressed objects with higher density of information (TODO).

Some prior works (e.g., \cite{luv, gd-size}) proposed function-based
replacement policies that calculate the value of an object much like our
proposed \mineviction{} policy. In particular, Bahn et al.~\cite{luv} proposed
a mechanism where the \textit{value} of a block is computed as the division of
re-reference probability and the relative cost of fetching by size.  Similar to
other function-based techniques, however, these inputs cannot efficiently be
computed or stored in hardware.  Our proposed technique does not suffer from
this problem and requires only simple metrics already built into on-chip
caches.

%An enhancement to this mechanism is Dynamic RRIP (DRRIP) that uses set
%dueling~\cite{mlp} to select the better performing of Static RRIP (described
%above) and Bimodal RRIP (BRRIP). 
%BRRIP is identical to SRRIP, 
%except it inserts blocks with a distant re-reference interval prediction with high probability 
%and inserts blocks with a long re-reference interval prediction with low probability.

\section{Methodology}
\label{camp:sec:methodology}

We use an in-house, event-driven 32-bit x86 simulator~\cite{MemSim} whose front-end is based
on Simics~\cite{Simics}.  All configurations have a two-level cache hierarchy,
with private L1 caches and a shared, inclusive L2 cache. Table~\ref{camp:tbl:simulation-parameters} 
provides major simulation parameters.  All caches
uniformly use a 64B cache block size.  All cache latencies were determined
using CACTI~\cite{cacti} (assuming a 4GHz frequency).  We also checked that
these latencies match the existing last-level cache implementations from Intel
and AMD, when properly scaled to the corresponding frequency.\footnote{Intel
Xeon X5570 (Nehalem) 2.993GHz, 8MB L3 - 35 cycles~\cite{Nehalem}; AMD Opteron
2.8GHz, 1MB L2 - 13 cycles~\cite{Opteron}.}  For single-core and multi-core
evaluations, we use benchmarks from the SPEC CPU2006 suite~\cite{SPEC}, two
TPC-H queries~\cite{tpc}, and an Apache web server.  All results are collected
by running a representative portion (based on PinPoints~\cite{pinpoints}) of
the benchmarks for 1 billion instructions. % (CPU) or to completion (GPU).  
We build our energy model based on McPAT~\cite{mcpat}, CACTI~\cite{cacti}, and
on RTL of BDI~\cite{bdi} synthesized with Synopsys Design Compiler 
with a 65nm library (to evaluate the energy of
compression/decompression with BDI).

\subsection{{Evaluation Metrics}}  We measure performance of our benchmarks using
IPC (instruction per cycle), effective compression ratio (effective
increase in L2 cache size without meta-data overhead, e.g., 1.5 for 2MB cache means effective size of
3MB), and MPKI (misses per kilo instruction).  For multi-programmed
workloads we use weighted speedup~\cite{weightedspeedup,ws2} as
the performance metric. %: ($\sum_i \frac{IPC_i^{shared}} {{IPC}_i^{{alone}}}
%$).
%Effective compression ratio for all mechanisms is computed without
%meta-data overhead~\cite{bdi}.  

\subsection{{Energy}} We measure the memory subsystem energy, which includes the static and dynamic energy
consumed by L1 and L2 caches, memory transfers, and DRAM, as well as
the energy of BDI's compressor/decompressor units. Energy results are normalized
to the energy of the baseline system with a 2MB compressed cache and an LRU
replacement policy. BDI was fully implemented in Verilog and synthesized to
create some of the energy results used in building our power model.
The area overhead of the compression and decompression logic is $0.014$ $mm^2$
(combined). Decompression power is 7.4 mW, and compression power is
20.59 mW on average.

Our results show that there are benchmarks that are almost insensitive (IPC
improvement is less than 5\% with 32x increase in cache size) to the size of the
L2 cache: dealII, povray, calculix, gamess, namd. This typically means that
their working sets mostly fit into the L1D cache, leaving almost no potential
for any L2/memory optimization.  Therefore, we do not present data in detail
for these applications, although we verified that our mechanism does not affect
their performance.

\begin{table}[t]
\vspace{-0.2cm}
 \centering
 \scriptsize{
%\begin{tabular}{|>{\scriptsize\bgroup}l<{\egroup}|>{\scriptsize\bgroup}c<{\egroup}|}
\begin{tabular}{|l|c|}
         \hline
         Processor  &  1--4 cores, 4GHz, x86 in-order  \\
         \hline
         L1-D cache    &  32KB, 64B cache-line, 2-way, 1 cycle, uncompressed \\
         \hline
         L2 caches    &  1--16 MB, 64B cache-line, 16-way, 15--48 cycles\\
         \hline
         Memory  & 300-cycle latency, 32 MSHRs    \\
        \cline{1-2}
    \end{tabular}%
}
 \caption{Major parameters of the simulated system.}
  \label{camp:tbl:simulation-parameters}%
 %\vspace{-0.4cm}
\end{table}

\subsection{{Parameters of Evaluated Schemes}}
For FPC (BDI), we used a decompression latency of 5 cycles~\cite{fpc-tr}
(1 cycle~\cite{bdi}), respectively.
We use a segment size of 1 byte  for both designs to get the highest compression ratio as described in~\cite{fpc-tr,bdi},
and an 8-byte segment size for V-Way-based designs. 
As in prior works~\cite{fpc,bdi}, we 
assume double the number of tags compared to the
conventional uncompressed cache (and hence the compression ratio cannot be larger than 2.0). 
%for both FPC and BDI designs.

\section{Results and Analysis}
\label{camp:sec:results}

\subsection{Single-core Results}
\subsubsection{Effect on Performance}
\label{sec:single-core}

\begin{figure}[htb]
  \centering
  \includegraphics[width=0.95\textwidth]{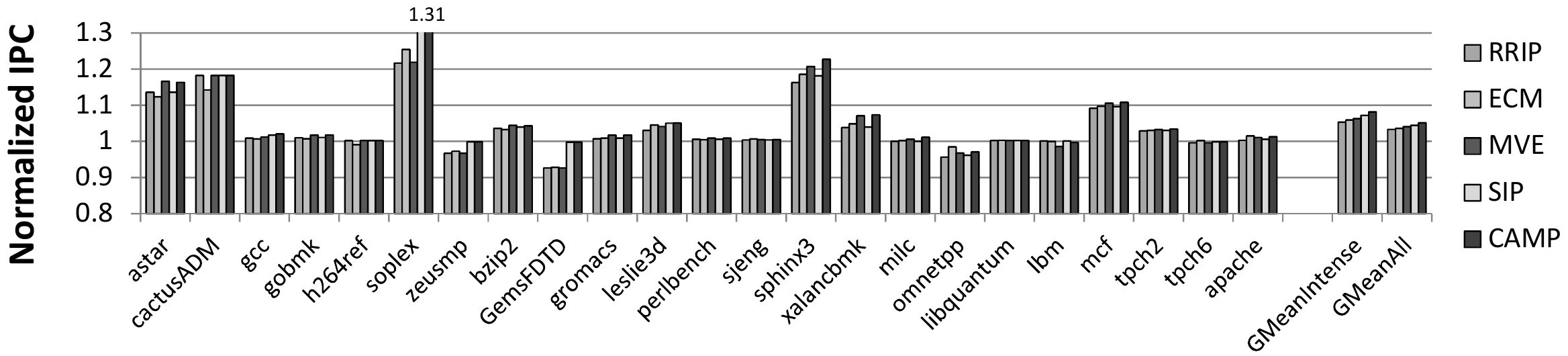}
%\vspace{-0.2cm}
  \caption{Performance of our local replacement policies vs. RRIP and ECM, normalized to LRU.}
  \label{fig:1-core}
%  \vspace{-0.1cm}
\end{figure}

\begin{figure*}[htb]
  \centering
  \includegraphics[width=0.95\textwidth]{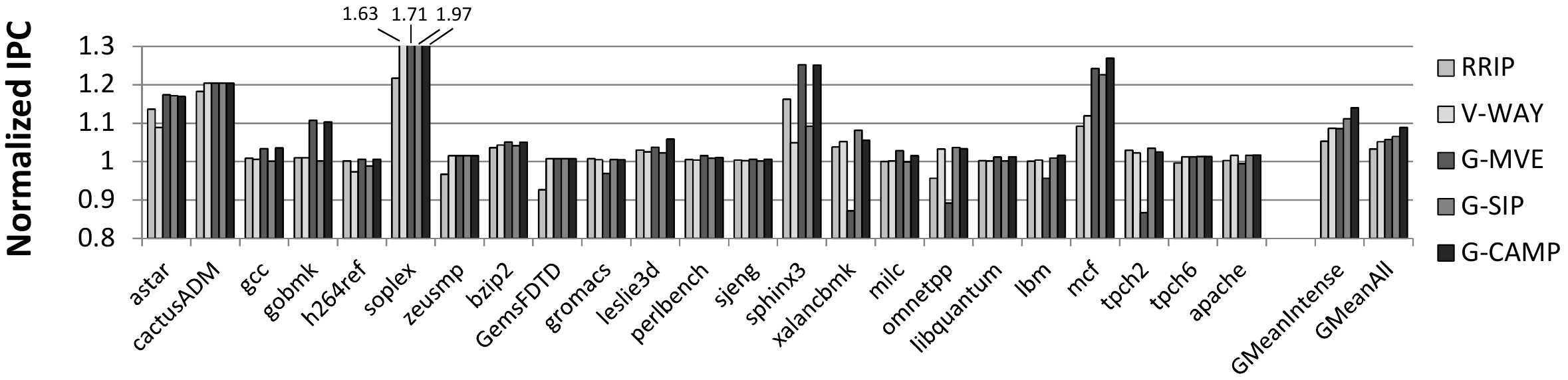}
% \vspace{-0.2cm} 
 \caption{Performance of our global replacement policies vs. RRIP and V-Way, normalized to LRU.}
  \label{fig:global-1-core}
\end{figure*}

Figures~\ref{fig:1-core} and~\ref{fig:global-1-core} show
the performance improvement of our proposed cache management policies
over the baseline design with a 2MB compressed\footnote{Unless
otherwise stated, we use 2MB BDI~\cite{bdi} compressed cache design.} L2 cache
and an LRU replacement policy. Figure~\ref{fig:1-core} compares
the performance of \carp's local version (and its components:
\mineviction{} and \insertionpolicy{}) over
(i) the conventional LRU policy~\cite{LRU}, (ii) the state-of-the-art
size-oblivious RRIP policy~\cite{RRIP}, and (iii) the recently proposed
ECM policy~\cite{ecm}. 
Figure~\ref{fig:global-1-core} provides the same comparison for 
G-\carp{} (with its components: G-\mineviction{} and G-\insertionpolicy{})
over (i) LRU, (ii) RRIP, and (iii) V-Way design ~\cite{v-way}.
Both figures are normalized to the performance of a BDI-cache with LRU replacement. 
Table~\ref{table:perf} summarizes our performance results.
Several observations are in order. 

\begin{table}[!ht]\small
 \centering
 \begin{tabular}{lccc}
    \toprule %hline
    \textbf{Mechanism}  & \textbf{LRU} & \textbf{RRIP} & \textbf{ECM} \\
    \midrule
    MVE & 6.3\%/-10.7\% & 0.9\%/-2.7\%  & 0.4\%/-3.0\%  \\
    %\midrule
    \cmidrule(rl){1-4}
    SIP & 7.1\%/-10.9\% & 1.8\%/-3.1\% & 1.3\%/-3.3\% \\
    %\midrule
    \cmidrule(rl){1-4} %\hline
    CAMP & \textbf{8.1\%/-13.3\%} & \textbf{2.7\%/-5.6\%} & \textbf{2.1\%/-5.9\%} \\
    \bottomrule
  \end{tabular}
  \begin{tabular}{lcccc}
    \toprule %hline
    \textbf{Mechanism}  & \textbf{LRU} & \textbf{RRIP} & \textbf{ECM} & \textbf{V-Way}\\
   \midrule
    G-MVE & 8.7\%/-15.3\% & 3.2\%/-7.8\%  & 2.7\%/-8.0\% & 0.1\%/-0.9\%\\
    %\midrule
    \cmidrule(rl){1-5}
    G-SIP  & 11.2\%/-17.5\% & 5.6\%/-10.2\% & 5.0\%/-10.4\% & 2.3\%/-3.3\%\\
    %\midrule
    \cmidrule(rl){1-5} %\hline
    G-CAMP & \textbf{14.0\%/-21.9\%} & \textbf{8.3\%/-15.1\%} & \textbf{7.7\%/-15.3\%} & \textbf{4.9\%/-8.7\%}\\
    \bottomrule
  \end{tabular}
  \caption{Performance (IPC) / Miss rate (MPKI) comparison between our cache management policies and prior works, 2MB L2 cache.
            All numbers are pairwise percentage improvements over the corresponding comparison points and averaged across fourteen memory-intensive applications.}
  \label{table:perf}
\end{table}

First, our G-\carp{} and \carp{} policies outperform all prior
designs: LRU (by 14.0\% and 8.1\%), RRIP (by 8.3\% and 2.7\%), and ECM (by 7.7\% and 2.1\%)
on average across fourteen memory-intensive applications (\emph{GMeanIntense}, with MPKI $>$ 5).
These performance improvements come from both components in our design, which
significantly decrease applications' miss rates (shown in Table~\ref{table:perf}).
For example, \mineviction{} and G-\mineviction{} are the primary sources of improvements
in \emph{astar}, \emph{sphinx3} and \emph{mcf}, while \insertionpolicy{}
is effective in \emph{soplex} and \emph{GemsFDTD}.  
Note that if we examine all applications, then
G-\carp{} outperforms LRU, RRIP and ECM by 8.9\%, 5.4\% and 5.1\% (on average).
%across these applications.

Second, our analysis reveals that the primary reasons why \carp{}/G-\carp{}
outperforms ECM are: (i) ECM's coarse-grain view of the size (only large vs.
small blocks are differentiated), (ii) ECM's difficulty in identifying the
right threshold for an application.  For example, in \emph{soplex}, ECM defines
every block that is smaller than or equal to 16 bytes as a small block and
prioritizes it (based on ECM's threshold formula). This partially helps to
improve performance for some important blocks of size 1 and 16, but our
\insertionpolicy{} mechanism additionally identifies that it is even more
important to prioritize blocks of size 20 (a significant fraction of such
blocks have short reuse distance as we show in Section~\ref{sec:size-reuse}).
This in turn leads to much better performance in \emph{soplex} by using \carp{}
(and G-\carp{}).

Third, in many applications, G-\mineviction{} significantly improves
performance (e.g., \emph{soplex} and \emph{sphinx3}), but there are some
noticeable exceptions (e.g., \emph{xalancbmk}). Section~\ref{sec:gmve}
describes the main reason for this problem.  Our final mechanism (G-\carp),
where we use set dueling~\cite{dip} to dynamically detect such situations and
disable G-\mineviction{} (for these cases only) avoids this problem. As a
result, our G-\carp{} policy gets the best of G-\mineviction{} when it is
effective and avoids degradations otherwise. %in other cases.

Fourth, global replacement policies (e.g., G-\carp) are more effective in
exploiting the opportunities provided by the compressed block size. G-\carp{}
not only outperforms local replacement policies (e.g., RRIP), but also global
designs like V-Way (by 3.6\% on average across all applications and by
\emph{4.9\%} across memory intensive applications).

We summarize the performance gains and the decrease in the cache miss rate
(MPKI) for all our policies in Table~\ref{table:perf}.  Based on our results,
we conclude that our proposed cache management policies (G-\carp{} and \carp{})
are not only effective in delivering performance on top of the existing cache
designs with LRU replacement policy, but also provide significant improvement
over state-of-the-art mechanisms.

\subsubsection{Sensitivity to the Cache Size} 
The performance benefits of our policies are significant across a variety of different
systems with different cache sizes. 
%uncomment in tech report
%\begin{comment}

\begin{figure}[h]
  \centering
  \includegraphics[width=0.95\textwidth]{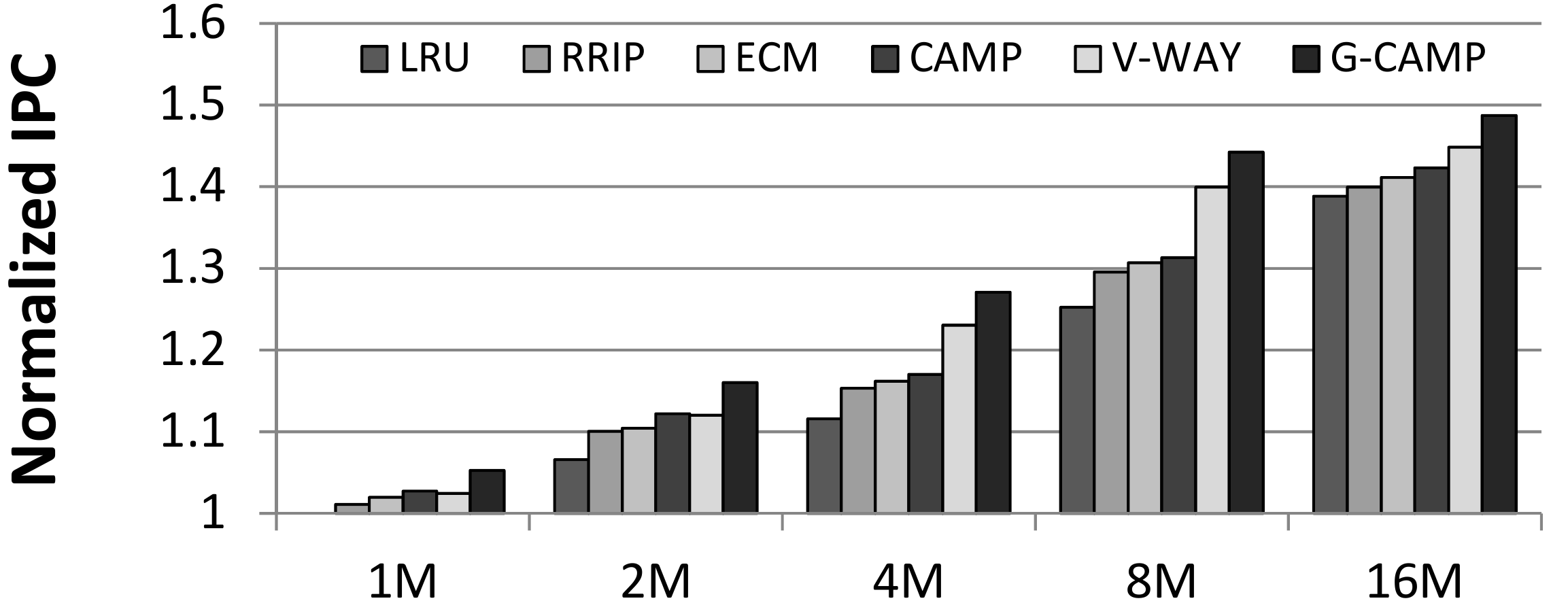}
  \caption{Performance with 1M -- 16MB L2 caches.}
  \label{fig:L2size}
  %\vspace{-0.2cm}
\end{figure}

Figure~\ref{fig:L2size} shows the performance of designs where
(i) L2 cache size varies from 1MB
to 16MB, and (ii) the replacement policies also vary: LRU, RRIP, ECM,
V-Way, \carp{}, and G-\carp{}.\footnote{All results are normalized to the performance of the 1MB compressed L2 cache with LRU
replacement policy. Cache access latency is modeled and adjusted appropriately for increasing cache size, using CACTI.}
Two observations are in order.

First, G-\carp{} outperforms all prior approaches for all corresponding cache
sizes. The performance improvement varies from 5.3\% for a 1MB L2 cache to as
much as 15.2\% for an 8MB L2 cache. \carp{} also outperforms all local replacement
designs (LRU and RRIP).
%\end{comment}
\begin{comment}
We analyze the performance of the cache designs where
(i) L2 cache size varies from 1MB
to 16MB, and (ii) the replacement policies vary as well: LRU, RRIP, ECM,
V-Way, \carp{}, and G-\carp{}.\footnote{We provide detailed results of this experiment in \cite{tr-camp}.}
We find that G-\carp{} outperforms all prior approaches for all corresponding cache sizes. The performance improvement varies from 5.3\% for 1MB L2 cache
to as much as 15.2\% for 8MB L2 cache. \carp{} also outperforms all local replacement designs (LRU and RRIP).
The effect of having size-aware cache management policies like G-\carp{} in many cases leads to performance that is better
than that of a twice as-large cache with conventional LRU replacement policy. We conclude that our management policies are efficient in achieving the performance
of higher capacity LLCs without making them physically larger.
\end{comment}

Second, the effect of having size-aware cache management policies like
G-\carp{}, in many cases, leads to performance that is better than that of a
twice-as-large cache with the conventional LRU policy (e.g, 4MB
G-\carp{} outperforms 8MB LRU).  In some cases (e.g., 8MB), G-\carp{}
performance is better than that of a twice-as-large cache with \emph{any other}
replacement policy.  We conclude that our management policies are efficient in
achieving the performance of higher-capacity last-level cache without making
the cache physically larger.  
%\end{comment}

\subsubsection{Effect on Energy}
By decreasing the number of transfers between LLC and DRAM, our
management policies also improve the energy consumption of the whole main
memory hierarchy. Figure~\ref{fig:energy} shows this effect on the memory
subsystem energy
%\footnote{The memory subsystem energy includes the static and dynamic energy
%consumed by L1 and L2 caches, memory transfers, and DRAM, as well as the
%energy of BDI's compressor/decompressor units. Results are normalized to the energy of the
%baseline system with 2MB compressed cache and LRU replacement policy. BDI was fully implemented in Verilog and synthesized to create some of the energy results used in building our power model.} 
for two of our mechanisms (\carp{} and G-\carp) and  three state-of-the-art mechanisms: (i) RRIP,
(ii) ECM, and (iii) V-Way. Two observations are in order.

\begin{figure}[h]
  \centering
  \includegraphics[width=0.95\textwidth]{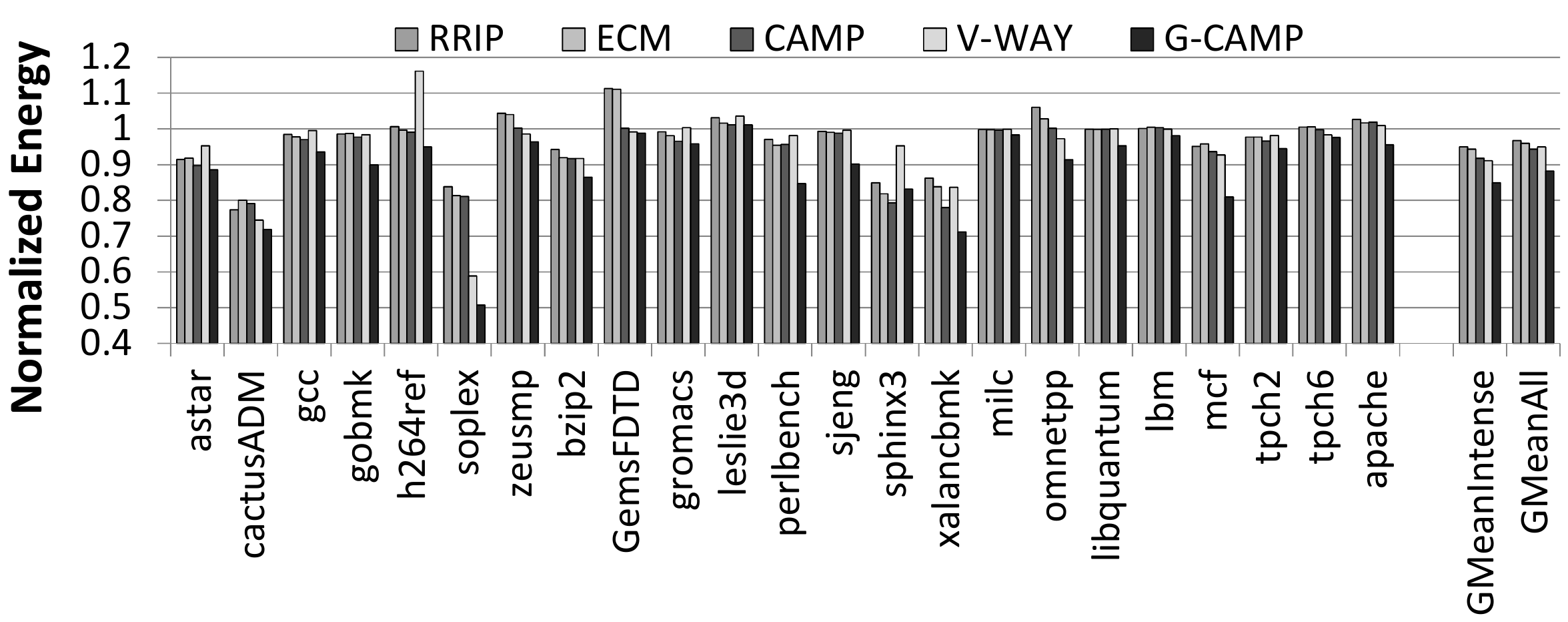}
 % \vspace{-0.6cm}
  \caption{Effect on memory subsystem energy.}
  \label{fig:energy}
\end{figure}

First, as expected, G-\carp{} is the most effective in decreasing energy consumption due
to the highest decrease in MPKI (described in Table~\ref{table:perf}). 
The total reduction in energy consumption is 15.1\% on average for memory-intensive workloads 
(11.8\% for all applications) relative to the baseline system and 7.2\% relative to the best prior mechanism.
We conclude that our cache management policies
are more effective in decreasing the energy consumption of the memory subsystem than previously-proposed mechanisms. 

Second, applications that benefit the most are usually the same applications that also have the highest
performance improvement and the highest decrease in off-chip traffic, e.g., \emph{soplex} and \emph{mcf}.
At the same time, there are a few exceptions, like \emph{perlbench}, that demonstrate
significant reduction in energy consumed by the memory subsystem, but do not show significant performance
improvement (as shown in Figures~\ref{fig:1-core} and~\ref{fig:global-1-core}). For these applications,
the main memory subsystem is usually not a performance bottleneck due to the relatively
small working set sizes that fit into the 2MB L2 cache and hence the relative improvements in the main memory subsystem
might not have noticeable effects on the overall system performance.

\subsubsection{Effect on Cache Capacity}
\begin{figure}[h]
  \centering
  \includegraphics[width=0.95\textwidth]{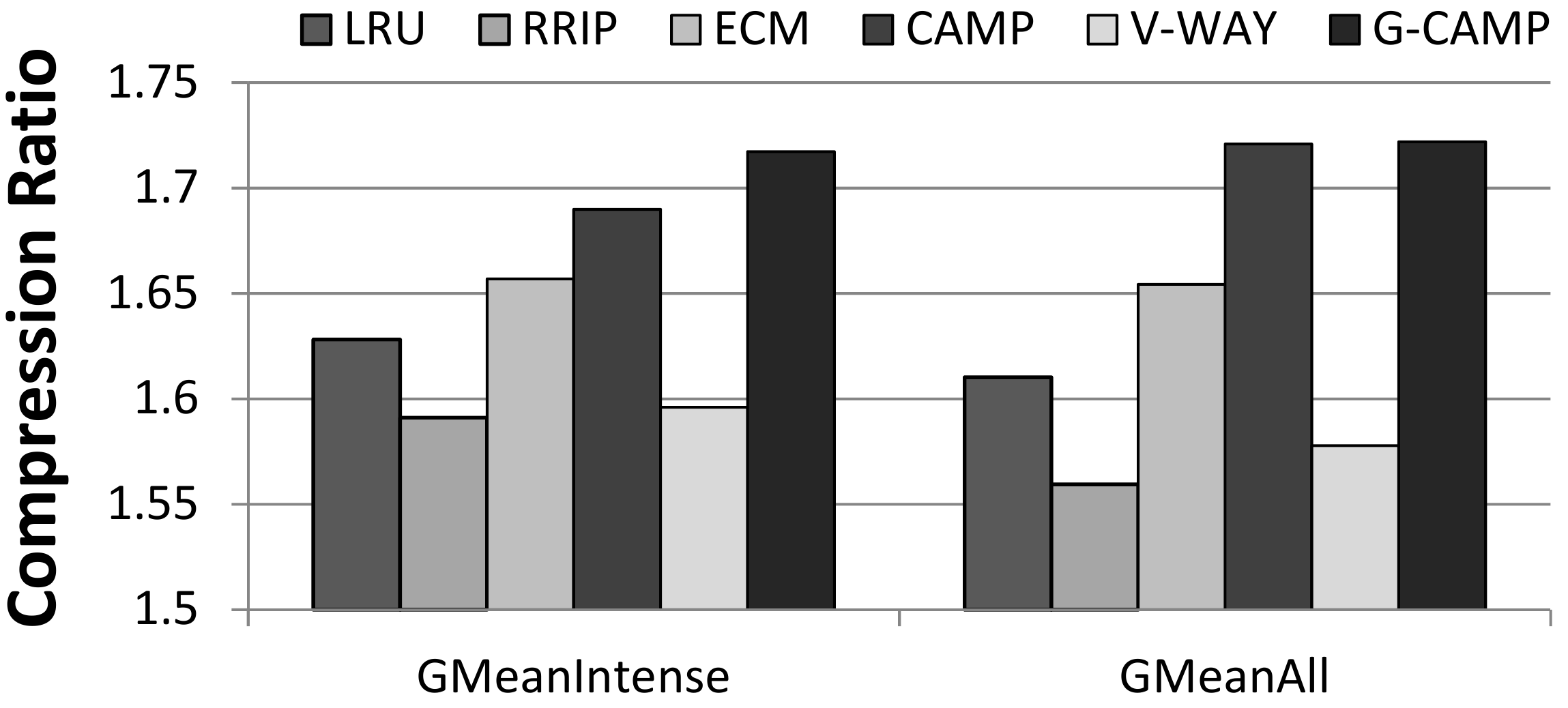}
  \caption{Effect on compression ratio with a 2MB L2 cache.}
  \label{fig:compratio}
\end{figure}

We expect that size-aware cache management policies increase the effective cache capacity
by increasing the effective compression ratio. 
Figure~\ref{fig:compratio} aims to verify this expectation by showing 
the average compression ratios for applications in our workload
pool (both the overall average and the average for memory-intensive applications).
We make two major observations.

First, as expected, our size-aware mechanisms (\carp{}/G-\carp{})
significantly improve effective compression ratio over corresponding size-oblivious mechanisms
(RRIP and V-Way) -- by 16.1\% and 14.5\% (on average across all applications).
The primary reason for this is that RRIP and V-Way are designed to be aggressive 
in prioritizing blocks with potentially higher reuse (better locality).
This aggressiveness leads to an even lower average compression ratio than that of the
baseline LRU design (but still higher performance shown in Section~\ref{sec:single-core}). 
Second, both \carp{} and G-\carp{} outperform ECM by 6.6\% and 6.7\% on average across
all applications for reasons explained in Section~\ref{sec:related}. 
We conclude that our policies achieve the highest effective cache ratio compression in the cache
compared to the other three state-of-the-art mechanisms. 

\subsubsection{Comparison with Uncompressed Cache}
Note that the overhead of using a compressed cache design is mostly due to the increased number of tags
(e.g, 7.6\% for BDI~\cite{bdi}).
If the same number of bits (or even a larger number, e.g., 10\%)
is spent on having a larger L2 cache (i.e., a 2.2MB \textit{uncompressed} L2 cache with RRIP
replacement), we find that the performance is 2.8\% lower than the performance of the
baseline system with 2MB \textit{compressed} L2 and LRU replacement, and 12.1\% lower than the performance
of the system with the 2MB L2 cache and G-\carp{} policy. We conclude that using a compressed cache with \carp{} 
provides a reasonable tradeoff in complexity for significantly higher performance.
\begin{comment}
\subsubsection{\insertionpolicy{} with Uncompressed Cache}
Our \insertionpolicy{} policy can be applied to a cache without a compressed data-store, but still with knowledge of a block's compressibility. 
Evaluating such a design interestingly isolates the effects of smarter replacement and increased cache capacity. 
The performance improvement of our mechanisms ultimately comes from increasing the utility of the cache 
which may mean increasing capacity, smarter replacement, or both. Our evaluations with G-\insertionpolicy{} without 
compression shows a 2.2\% performance improvement over an uncompressed V-Way cache, and a 1.3\% performance improvement over the state-of-the-art
PC-based mechanism~\cite{ship} without an overhead of storing a special hardware table. This performance comes from better 
management only, through the use of compressibility as an indicator of reuse. 
\end{comment}
\subsection{Multi-core Results}
\label{sec:multicore}
%include in tech report
\begin{comment}
When the cache blocks from the working set of an application 
are compressed to mostly the same
size, it is hard to expect that size-aware cache management policies would provide significant
benefit. However, when different applications are running together in the multi-core
system with a shared last-level-cache (LLC), there is a high chance that different 
applications will have different compressed sizes. As a result, 
we hypothesize that there is much more
room for improvement with size-aware management in multi-core systems.
\end{comment}

%To test this hypothesis, 
We classify our applications into two distinct
categories (\emph{homogeneous} and \emph{heterogeneous}) based on the distributions of the compressed sizes that they have.
A homogeneous application is expected to have very few different compressed sizes
for its data (when stored in the LLC). 
A heterogeneous application, on the other hand, has many different sizes. 
To formalize this classification, we first collect the access counts for different sizes for every application. 
Then, we mark the size with the highest access count as a ``peak'' and scale all 
other access counts with respect to this peak's access count. 
If a certain size within an application has over 10\% of the peak access count, it is also marked as a peak.
The total number of peaks is our measure of the application's heterogeneity with respect to block size. 
If the application's number of peaks exceeds two, 
we classify it as heterogeneous (or simply \emph{Hetero}).  Otherwise, 
the application is considered to be homogeneous (or simply \emph{Homo}).
This classification matches our intuition that applications that have only one or two common sizes (e.g.,
one size for uncompressed blocks and one size for most of the compressed blocks) should
be considered homogeneous. These two classes enable us to construct three different
2-core workload groups: (i) Homo-Homo, (ii) Homo-Hetero, and (iii) Hetero-Hetero. 
We generate 20 2-core workloads per group (60 total) by randomly selecting
applications from different categories.

\begin{figure}[h]
  \centering
  %\begin{subfigure}[b]{0.49\linewidth}
  % \centering
  \subfigure[Local replacement]{\label{fig:2-core} \includegraphics[width=65mm]{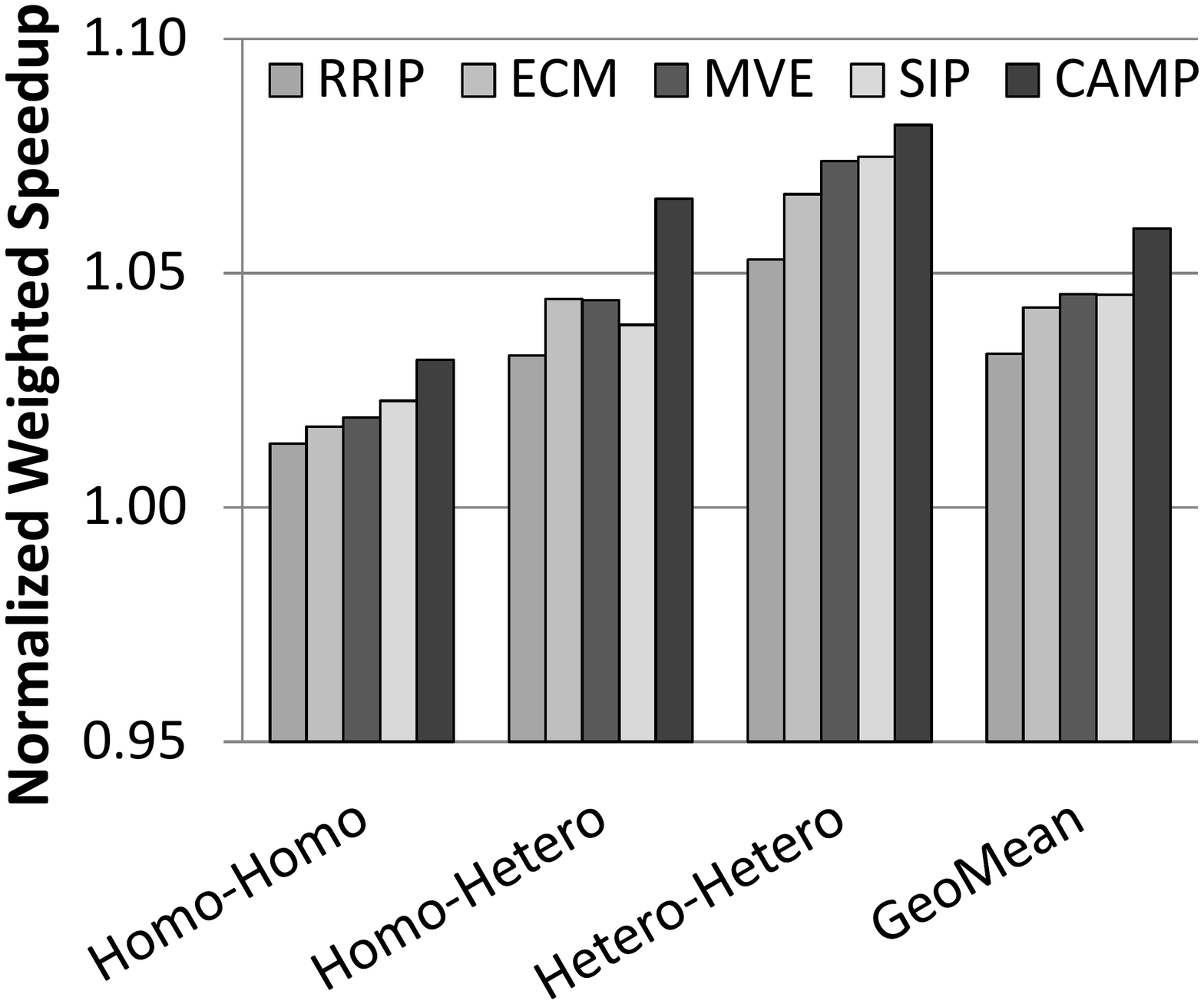}}
  %\end{subfigure}
  %\begin{subfigure}[b]{.49\linewidth}
  % \centering
  \subfigure[Glocal replacement]{\label{fig:global-2-core}\includegraphics[width=65mm]{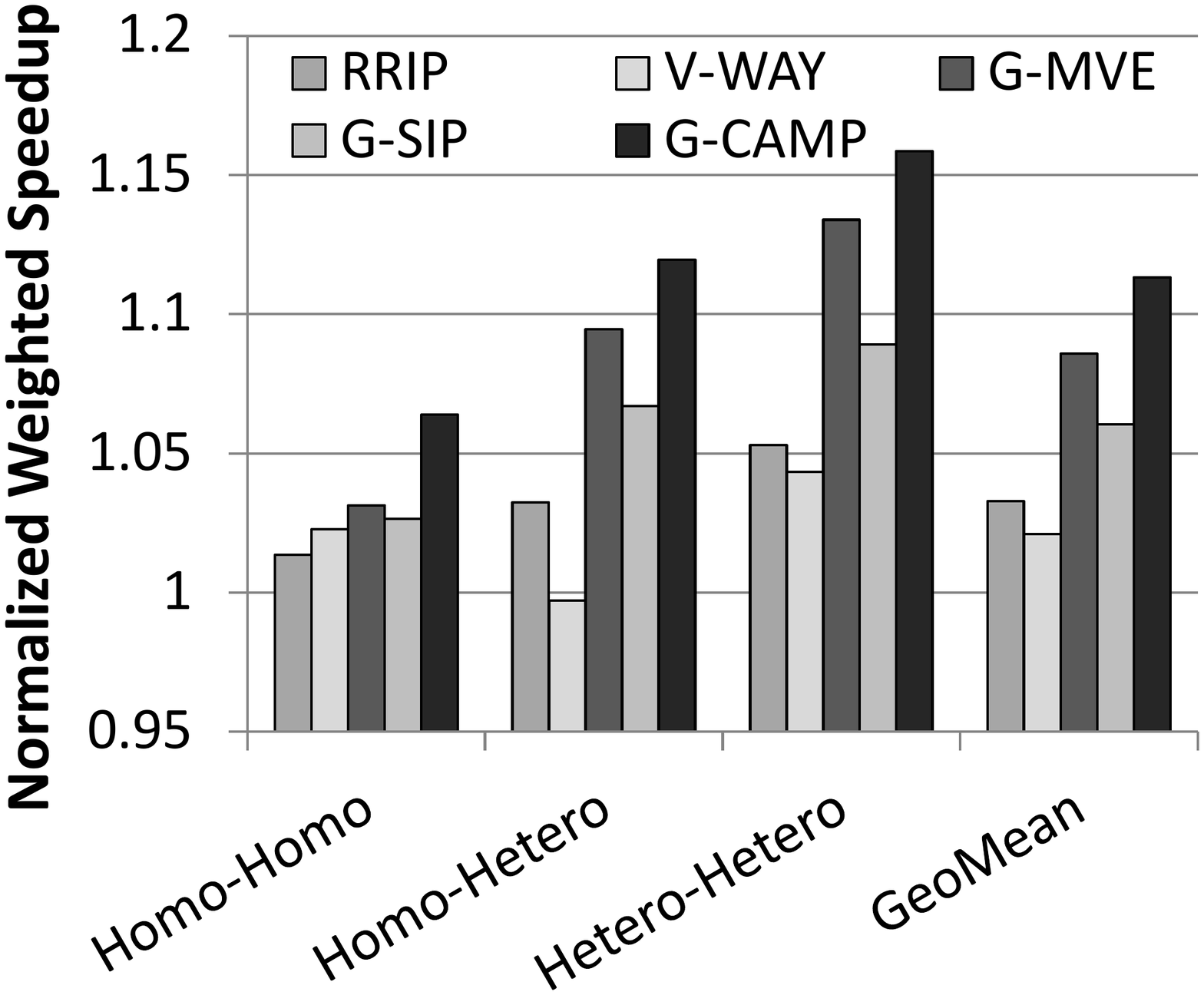}}
  \caption{Normalized weighted speedup, 2-cores with 2MB L2.}
\end{figure}

Figures~\ref{fig:2-core} and~\ref{fig:global-2-core} show the performance
improvement provided by all \carp{} designs as well as previously proposed designs:
(i) RRIP, (ii) ECM, and (iii) V-Way over a 2MB baseline compressed cache design with
LRU replacement. We draw three major conclusions. 

First, both G-\carp{} and \carp{} outperform all prior approaches in all categories.
Overall, G-\carp{} improves system performance by 11.3\%/7.8\%/6.8\% over
LRU/RRIP/ECM (\carp{} improves by 5.9\%/2.5\%/1.6\% over the same designs).
The effect on system fairness, i.e., maximum slowdown~\cite{TCM,ATLAS,reetu,fairness,f2} by our mechanisms is negligible.

Second, the more heterogeneity 
present, the higher the performance improvement with our size-aware 
management policies. This effect is clearly visible in both figures,
and especially for global replacement policies in Figure~\ref{fig:global-2-core}.
G-\carp{} achieves the highest improvement (15.9\% over LRU and 10.0\% over RRIP) when
both applications are heterogeneous, and hence there are more opportunities in size-aware
replacement.

Third, when comparing relative performance of \mineviction{} vs. \insertionpolicy{} from
Figure~\ref{fig:2-core} and the similar pair of G-\mineviction{} vs. G-\insertionpolicy{} from
Figure~\ref{fig:global-2-core}, we notice that in the first pair the relative performance
is almost the same, while in the second pair G-\mineviction{} is significantly better than G-\insertionpolicy{}.
The primary reason for this difference is that G-\mineviction{} can get more benefit from global cache replacement, because
it can easily exploit size variation between different sets. At the same time, G-\insertionpolicy{} 
gets its performance improvement from the relation between the size and corresponding data reuse, which
does not significantly change between local and global replacement.  

We conducted a similar experiment\footnote{We increased the LLC size to 4MB to provide the
same core to cache capacity ratio as with 2-cores.} with 30 4-core workloads and observe similar trends to the 2-core
results presented above. G-\carp{} outperforms the best prior mechanism by 8.8\%
on average across all workloads (by 10.2\% across memory-intensive workloads). 
%We conclude that our policies provide
%better performance than state-of-the-art mechanisms in a multi-core environment.

\subsection{Sensitivity to the Compression Algorithm}
So far, we have presented results only for caches that use BDI
compression~\cite{bdi}, but as described in Section~\ref{sec:motivation},
our proposed cache management policies are applicable to different compression
algorithms. We verify this by applying our mechanisms to a compressed cache
design based on the FPC~\cite{fpc} compression algorithm.  Compared to an
FPC-compressed cache with LRU replacement, \carp{} and G-\carp{} 
improve performance of memory-intensive applications by 7.8\% and 10.3\%
respectively. 
%Over the best prior mechanism ECM, \carp{} and G-\carp{} improve performance
%by 1.5\% and 3.9\% respectively. 
We conclude that our cache management policies are effective for different
compression designs where they deliver the highest overall performance when
compared to the state-of-the-art mechanisms. 

\subsection{\insertionpolicy{} with Uncompressed Cache}

Our \insertionpolicy{} policy can be applied to a cache {\em without}
a compressed data-store, while still using knowledge of a \emph{block's
  compressibility as an indicator of reuse}.  We evaluate such a
design to isolate the ``reuse prediction'' benefit of
\insertionpolicy{} independently of its benefits related to cache
compression. Our single-/two-core evaluations of G-\insertionpolicy{}
show a 2.2\%/3.1\% performance improvement over an uncompressed LRU cache design,
and a 1.3\%/1.2\% performance improvement over the
state-of-the-art PC-based cache management mechanism~\cite{ship} (evaluated as comparison 
to a state-of-the-art ``reuse predictor'').\footnote{ In
contrast to~\cite{ship}, \insertionpolicy{} does not require a special
hardware table and tracking of PC with cache blocks.}  We conclude that
using compressibility as an indicator of future reuse can improve the
performance of even uncompressed caches.

%\insertionpolicy{}'s performance improvement comes from better
%uncompressed cache management through the use of compressibility as an
%indicator of reuse.

%\subsubsection{Sensitivity to the Bin Size}

%\subsubsection{In-Order vs. Out-Of-Order}

\section{Summary}
\label{camp:sec:conclusion}

In this chapter, we presented Compression-Aware Management Policies (\carp)
-- a set of new and simple, yet efficient \emph{size-aware} replacement
policies for compressed on-chip caches. \carp{} improves system performance
and energy efficiency compared to three state-of-the-art
cache replacement mechanisms. Our policies are based on two key
observations. First, we show that
direct incorporation of the compressed cache block size into replacement
decisions can be a basis for a more efficient replacement policy.
Second, we find that the compressed block size can be used as an indicator
of a block's future reuse in some applications. Our extensive evaluations show
that \carp{}, applied to modern last-level-caches (LLC), improves performance
by 4.9\%/9.0\%/10.2\% (on average for memory-intensive workloads)
for single-core/two-/four-core workloads over the best
state-of-the-art replacement mechanisms we evaluated. We conclude that \carp{} is an efficient and low-complexity management policy for
compressed caches in both single- and multi-core systems.
We also hope that our observation that compressed block size indicates reuse behavior could be useful in other contexts.

%\input{chap4/database.tex}

%%% Local Variables: 
%%% mode: latex
%%% TeX-master: "main"
%%% End: 

\chapter{Main Memory Compression: Linearly Compressed Pages}

\section{Introduction}
\label{intro}
\blfootnote{Originally published as ``Linearly Compressed Pages: 
A Low Complexity, Low Latency Main Memory Compression Framework'' 
in the 46th International Symposium on Microarchitecture, 2013~\cite{lcp-micro}.}
Main memory, commonly implemented using DRAM technology, is a
critical resource in modern systems. To avoid the devastating performance
loss resulting from frequent page faults, main memory capacity must be
sufficiently provisioned to prevent the target workload's working set
from overflowing into the orders-of-magnitude-slower backing store (e.g.,
hard disk or flash).  

Unfortunately, the required minimum memory
capacity is expected to increase in the future due to two major
trends: (i) applications are generally becoming more data-intensive with
increasing working set sizes, and (ii) with more cores integrated onto the same
chip, more applications are running concurrently on the system, thereby
increasing the aggregate working set size.  Simply scaling up main memory
capacity at a commensurate rate is unattractive for two reasons: (i) DRAM
already constitutes a significant portion of the system's cost and power
budget~\cite{energy,Yixin1,MemoryScaling}, and (ii) for signal integrity reasons, today's high frequency
memory channels prevent many DRAM modules from being connected to the same
channel~\cite{signal}, effectively limiting the maximum amount of DRAM in a system unless
one resorts to expensive off-chip signaling buffers~\cite{BoB}.

If its potential could be realized in practice, {\em data compression}
would be a very attractive approach to effectively increase main memory
capacity without requiring significant increases in cost or power, because a
compressed piece of data can be stored in a smaller amount of physical
memory.  Further, such compression could be hidden from application (and most 
system\footnote{We assume that main memory compression is made visible to
  the memory management functions of the operating system (OS). 
%Unless the main memory is completely managed by hardware, it is difficult to make it transparent to the OS. 
  In Section~\ref{sec:background-prior-work}, we discuss the drawbacks of a
  design that makes main memory compression mostly transparent to the
  OS~\cite{MXT}.}) software by materializing the uncompressed data as it is brought into the processor cache.
Building upon the observation that there is significant redundancy
in in-memory data, previous work has proposed a variety of techniques for
compressing data in caches~\cite{fvc,fpc,fpc-tr,fvl,bdi,iic-comp,c-pack} and in main
memory~\cite{MXT,MMCompression,vm-compression,the-compression-cache,adaptive-compressed-caching}.

\subsection{Shortcomings of Prior Approaches}

A key stumbling block to making data compression practical is that
\emph{decompression} lies on the critical path of accessing any
compressed data. Sophisticated compression algorithms, such as
Lempel-Ziv and Huffman encoding~\cite{lz,huffman}, typically achieve
high compression ratios at the expense of large decompression
latencies that can significantly degrade performance. To counter this
problem, prior work~\cite{fvl,fpc-tr,bdi} on cache compression
proposed specialized compression algorithms that exploit regular
patterns present in in-memory data, and showed that such specialized
algorithms have reasonable compression ratios compared to more complex
algorithms while incurring much lower decompression latencies.

% (which is mostly
% a function of the compression algorithms supported by the framework).\footnote{Note
% that effective compression ratio (as well as decompression latency) 
% depends mostly on the compression algorithm (e.g., FPC~\cite{fpc} or BDI~\cite{bdi})
% employed by the main memory compression framework, and less significantly
% on the framework itself.}

%Although prior works have used such simpler algorithms to compress
%cache lines in on-chip caches~\cite{fvc,fpc,fpc-tr,bdi}, applying
%these techniques directly to main memory is difficult for three
%reasons.\footnote{Note that our goal, similarly to that of other works
%  on main memory compression~\cite{MMCompression, MXT}, is to have a
%  main memory compression mechanism that compresses data at the
%  granularity of last-level cache line, because main memory is
%  accessed at the cache line granularity. At the same time, we aim
%  to find a mechanism that is
%  compatible with compression employed in on-chip caches, minimizing
%  the number of compression/decompressions performed.}  

%\textbf{Challenges in Main Memory Compression.} 
%To satisfy these requirements, we need to overcome three challenges.
\sloppypar{
While promising, applying compression algorithms, sophisticated or
simpler, to compress data stored in main memory requires first
overcoming the following three challenges.  First, {\em main memory
  compression complicates memory management}, because the operating
system has to map fixed-size virtual pages to variable-size physical
pages.  Second, because modern processors employ on-chip caches with
tags derived from the physical address to avoid aliasing between
different cache lines (as physical addresses are unique, while virtual
addresses are not), {\em the cache tagging logic needs to be modified}
in light of memory compression to take the main memory address
computation off the critical path of latency-critical L1 cache
accesses.  Third, in contrast with normal virtual-to-physical address
translation, the physical page offset of a cache line is often
different from the corresponding virtual page offset, because compressed
physical cache lines are smaller than their corresponding virtual
cache lines.  In fact, the location of a compressed cache line in a
physical page in main memory depends upon the sizes of the compressed
cache lines that come before it in that same physical page.  As a
result, accessing a cache line within a compressed page in main memory
{\em requires an additional layer of address computation to compute
  the location of the cache line in main memory} (which we will call
the \emph{main memory address}).  This additional {\em main memory
  address computation} not only adds complexity and cost to the
system, but it can also increase the latency of accessing main memory
(e.g., it requires up to 22 integer addition operations in one prior
design for main memory compression~\cite{MMCompression}), which in
turn can degrade system performance.
}

While simple solutions exist for these first two challenges (as we
describe later in Section~\ref{lcp:sec:design}), prior attempts to
mitigate the performance degradation of the third challenge are either
costly or inefficient~\cite{MXT,MMCompression}.  One approach (IBM
MXT~\cite{MXT}) aims to reduce the number of main memory accesses, the
cause of long-latency main memory address computation, by adding a
large (32MB) uncompressed cache managed at the granularity at which
blocks are compressed (1KB). If locality is present in the program,
this approach can avoid the latency penalty of main memory address
computations to access compressed data. Unfortunately, its benefit
comes at a significant additional area and energy cost, and the
approach is ineffective for accesses that miss in the large cache. A
second approach~\cite{MMCompression} aims to hide the latency of main
memory address computation by speculatively computing the main memory address of
{\em every} last-level cache request in parallel with the cache access
(i.e., before it is known whether or not the request needs to access main 
memory). While this approach can effectively reduce the performance
impact of main memory address computation, it wastes a significant
amount of energy (as we show in Section~\ref{sec:results-energy}) because
many accesses to the last-level cache do not result in an access to main
memory. 

%In our evaluations across a variety of workloads, we see that
%XXX\% of last-level cache requests do not need to access main memory.

%%% ONUR: Above: Power or energy? Let's be consistent (and correct, of
%%% course)
%%% Also, can you fill in the XXX?

\begin{comment}
%% ONUR: Old version
One approach (IBM MXT~\cite{MXT}) adds a large (32MB) uncompressed
cache managed at the granularity at which blocks are compressed
(1KB). While this large cache can avoid some requests that need to
access main memory (if locality is present in the program) by reducing
the number of requests that suffer the latency penalty of the main
memory address computation (see Section~\ref{lcp:sec:background}),
unfortunately this benefit comes at the expense of the significant
additional area and power cost that are required for such a large
cache.  Another approach is to overlap a cache line's main memory
address computation with the last-level cache
access~\cite{MMCompression}. However, this latter approach wastes
power (as we show in Section~\ref{sec:results-energy}) because not all
accesses to the last-level cache result in an access to main memory.
\end{comment}

\subsection{Our Approach: Linearly Compressed Pages}

We aim to build a main memory compression framework that
neither incurs the latency penalty for memory accesses nor requires
power-inefficient hardware.  Our goals are: (i) having low
complexity and low latency (especially when performing
memory address computation for a cache line within a compressed page), 
(ii) being compatible with compression employed
in on-chip caches (thereby minimizing the number of compressions/decompressions
performed), and (iii) supporting compression algorithms with high compression
ratios.

To this end, we propose a new approach to
compress pages, which we call \emph{Linearly Compressed Pages} (LCP). The
key idea of LCP is to compress all of the cache lines within a given page to the
same size. Doing so simplifies the computation of the physical
address of the cache line, because the page offset is simply the product of
the index of the cache line and the compressed cache line size (i.e., 
it can be calculated
using a simple shift operation). Based on
this idea, a target compressed cache line size is determined for each page.
Cache lines that cannot be compressed to the target size for its page are called
\emph{exceptions}. All exceptions, along with the metadata required to
locate them, are stored separately in the same compressed page.
If a page requires more space in compressed form than in uncompressed form,
then this page is not compressed. The page table indicates the form in which
the page is stored.

The LCP framework can be used with any compression algorithm.
We adapt two previously proposed
compression algorithms (Frequent Pattern Compression (FPC)~\cite{fpc} and
Base-Delta-Immediate Compression (BDI)~\cite{bdi}) to fit the requirements of LCP, and
show that the resulting designs can significantly improve effective main memory
capacity on a wide variety of workloads.  

Note that, throughout this chapter, we assume that compressed cache
lines are decompressed before being placed in the processor caches.
LCP may be combined with compressed cache designs by storing
compressed lines in the higher-level caches (as in~\cite{fpc,bdi}), but
the techniques are largely orthogonal, and for clarity, we present an
LCP design where only main memory is compressed.\footnote{We show the
results from combining main memory and cache compression in our technical report~\cite{lcp-tech}.}

An additional, potential benefit of compressing data in main memory,
which has not been fully
explored by prior work on main memory compression, is {\em memory bandwidth
reduction}. When data are stored in compressed format in main memory, multiple
consecutive compressed cache lines can be retrieved at the cost of accessing a
single uncompressed cache line. Given the increasing demand on main memory
bandwidth, such a mechanism can significantly reduce the memory bandwidth
requirement of applications, especially those with high spatial locality. 
Prior works on bandwidth compression~\cite{LinkCompression,fvc-bus,GPUBandwidthCompression}
assumed efficient variable-length off-chip data transfers that are hard to achieve
with general-purpose DRAM (e.g., DDR3~\cite{micron-ddr3}).
We propose a mechanism that enables the memory controller to retrieve multiple
consecutive cache lines with a single access to DRAM, with negligible additional
cost. Evaluations show that our mechanism provides significant bandwidth
savings, leading to improved system performance.

In summary, we make the following contributions:
%\vspace{-0.3cm}
\begin{itemize}
\item We propose a new main memory compression framework---{\em Linearly
  Compressed Pages} (LCP)---that solves the problem of efficiently computing the
  physical address of a compressed cache line in main memory with much
  lower cost and complexity than prior proposals.  We also demonstrate
  that {\em any} compression algorithm can be adapted to fit the requirements of
  LCP.
%\vspace{0.2cm}
\item We evaluate our design with two state-of-the-art compression
  algorithms (FPC~\cite{fpc} and BDI~\cite{bdi}), and observe that it can
  significantly increase the effective main memory capacity (by 69\% on
  average).
\item We evaluate the benefits of transferring compressed cache lines over the
  bus between DRAM and the memory controller and observe that it can considerably 
  reduce memory bandwidth consumption (24\% on average), and improve overall
  performance by 6.1\%/13.9\%/10.7\% for single-/two-/four-core workloads,
  relative to a system without main memory compression. LCP also decreases 
the energy consumed by the main memory subsystem (9.5\% on average over the best prior
mechanism).
\end{itemize}

\section{Background on Main Memory Compression}
\label{lcp:sec:background}

Data compression is widely used in storage structures to
%to improve the efficiency of storage
%structures and communication channels. By reducing the amount of redundancy in
%data, compression increases 
increase the effective capacity and bandwidth without significantly
increasing the system cost and power consumption. One primary
downside of compression is that the compressed data must be decompressed
before it can be used. Therefore, for latency-critical applications, using
complex dictionary-based compression algorithms~\cite{lz}
% (such as Lempel-Ziv~\cite{lz} or Huffman encoding~\cite{Huffman})
significantly degrades performance due to their
high decompression latencies. Thus, prior work on compression of in-memory
data has proposed simpler algorithms with low decompression latencies and
reasonably high compression ratios, as discussed next.

\subsection{Compressing In-Memory Data}

Several studies~\cite{fvl,fpc-tr,bdi,fpc} have shown that in-memory data has
exploitable patterns that allow for simpler compression
techniques. Frequent value compression (FVC)~\cite{fvl} is based on the
observation that an 
application's working set is often dominated by a small set of values.
FVC exploits this observation by
encoding such frequently-occurring 4-byte values with fewer bits. Frequent pattern
compression (FPC)~\cite{fpc-tr} shows that a majority of words (4-byte elements)
in memory fall under a few frequently occurring patterns. FPC compresses
individual words within a cache line by encoding the frequently occurring
patterns with fewer bits. 
Base-Delta-Immediate (BDI) compression~\cite{bdi} observes that, in many cases, words
co-located in memory have small differences in their values.
BDI compression encodes a cache line as a
base-value and an array of differences that represent the deviation either from the
base-value or from zero (for small values) for each word.
%
%Due to the simplicity of these compression algorithms, their decompression
%latencies are much shorter than those of dictionary-based algorithms:~5 cycles
%for FVC/FPC~\cite{fpc-tr} and~1 cycle for BDI~\cite{bdi} in contrast to 64
%cycles for a variant of Lempel-Ziv~\cite{lz} used in IBM MXT~\cite{MXT}. These
These three low-latency compression algorithms have been proposed for on-chip
caches, but can be adapted for use as part of the main memory compression framework
proposed in this chapter. 
%Unfortunately, in order to apply such compression
%algorithms to main memory, the following challenges must be addressed.

\subsection{Challenges in Memory Compression}
\label{sec:background-challenges}

LCP leverages the fixed-size memory pages of modern systems for the
basic units of compression.  However, three challenges arise from the
fact that different pages (and cache lines within a page) compress to
different sizes depending on data compressibility.
  
%The key requirement in building an efficient hardware-based main memory 
%compression framework is low complexity and low latency. 
%There are three challenges that need to be addressed to satisfy
%this requirement. 

% Figures~\ref{fig:challenge1}, \ref{fig:challenge2}
% and \ref{fig:challenge3} pictorially show these three challenges.

\begin{figure}[h!]
  \centering
  \includegraphics[width=0.8\textwidth]{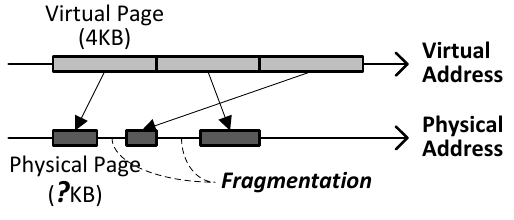}
  \caption{Main Memory Page Mapping Challenge}
  \label{fig:challenge2}
\end{figure}

\textbf{Challenge 1: Main Memory Page Mapping.} Irregular page sizes in main memory complicate the
memory management module of the operating system for two reasons (as shown in
Figure~\ref{fig:challenge2}). First, the operating system needs to allow
mappings between the fixed-size virtual pages presented to software and the variable-size physical
pages stored in main memory. Second, the operating system must implement mechanisms to efficiently
handle fragmentation in main memory.

\textbf{Challenge 2: Physical Address Tag Computation.} On-chip caches
(including L1 caches) typically employ tags derived from the physical address of the cache line
 to avoid aliasing, and
% -- i.e., two cache lines
%having the same virtual address but different physical address (homonyms) or
%vice versa (synonyms)~\cite{protection-lookaside-buffer, cache-memories}. 
in such systems, every cache access requires the physical address of
the corresponding cache line to be computed.  Hence, because the main
memory addresses of the compressed cache lines differ from the nominal
physical addresses of those lines, care must be taken that the
computation of cache line tag does not lengthen the critical path of
latency-critical L1 cache accesses.
% (Figure~\ref{fig:challenge3}).

%\begin{figure}[h!]
%  \vspace{-0.2cm}
%  \centering
%  \includegraphics[width=0.25\textwidth]{figures/challenge3.pdf}
%  \caption{Physically Tagged Caches Challenge}
%  \label{fig:challenge3}
%  \vspace{-0.3cm}
%\end{figure}

\begin{figure}[h!]
%  \vspace{-0.3cm}
  \centering
  \includegraphics[width=0.8\textwidth]{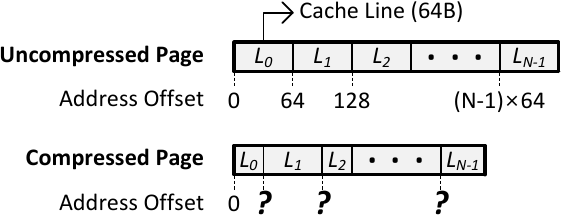}
  \caption{Cache Line Address Computation Challenge}
  \label{fig:challenge1}
%  \vspace{-0.0cm}
\end{figure}

\textbf{Challenge 3: Cache Line Address Computation.} 
%Unlike on-chip last-level caches, which are managed and
%accessed at the same granularity (e.g., a 64-byte cache line), main memory is
%managed and accessed at different granularities. Most modern systems employ
%virtual memory and manage main memory at a large page granularity (e.g., 4kB or
%8kB), but access data in main memory at a smaller cache line
%granularity. 
When main memory
is compressed, different cache lines within a page can be compressed to
different sizes. The main memory address of a cache line is therefore dependent
on the sizes of the compressed cache lines that come before it in the page.  As
a result, the processor (or the memory controller) must explicitly compute the
location of a cache line within a compressed main memory page before accessing
it (Figure~\ref{fig:challenge1}), e.g., as in~\cite{MMCompression}. This
computation not only increases complexity, but can also lengthen the critical
path of accessing the cache line from both the main memory and the physically
addressed cache. Note that systems that do \emph{not} employ main 
memory compression do not suffer from this problem because
the offset of a cache line within the physical page is the \emph{same}
as the offset of the cache line within the corresponding virtual page.

As will be seen shortly, while prior research efforts have considered
subsets of these challenges, this work is the first design that
provides a holistic solution to all three challenges, particularly Challenge 3,
with low latency and low (hardware and software) complexity.

\subsection{Prior Work on Memory Compression}
\label{sec:background-prior-work}
%Multiple previous works investigated the possibility of using compression for main
%memory~\cite{MXT,MMCompression,ccc,kaplan-thesis,adaptive-compressed-caching,the-compression-cache,GPUBandwidthCompression}.
%Among them, two in particular are the most closely related to the design

Of the many prior works on using compression for main
memory (e.g.,~\cite{MXT,MMCompression,vm-compression,kaplan-thesis,adaptive-compressed-caching,the-compression-cache,GPUBandwidthCompression}),
two in particular are the most closely related to the design
proposed in this chapter, because both of them are mostly hardware designs.
We describe these two designs along with their shortcomings.
% and defer
%the discussion of other techniques to Section~\ref{sec:relatedwork}.

%%% ONUR: Phrasing below is weird. IBM MXT is an entity that proposes
%%% a design? Or is IBM MXT a design itself. Rephrase.

%%% VIVEK: Addressed

Tremaine {\em et al.}~\cite{pinnacle} proposed a memory controller design, Pinnacle,
based on IBM's Memory Extension Technology (MXT)~\cite{MXT} that employed
Lempel-Ziv compression~\cite{lz} to manage main memory.  To address the three
challenges described above, Pinnacle employs two techniques.
First, Pinnacle internally uses a 32MB last level cache managed at a 1KB granularity, same as
the granularity at which blocks are compressed.  This cache reduces the number
of accesses to main memory by exploiting locality in access patterns, thereby
reducing the performance degradation due to the address computation (Challenge
3). However, there are several drawbacks to this technique: (i) such a large cache adds
significant area and energy costs to the memory controller, (ii) the approach requires the main
memory address computation logic to be present and used when an access misses in
the 32MB cache, and (iii) if caching is not effective (e.g., due to lack of locality or
larger-than-cache working set sizes), this approach cannot reduce the
performance degradation due to main memory address computation. Second, to avoid
complex changes to the operating system and on-chip cache-tagging logic,
Pinnacle introduces a \emph{real} address space between the virtual and physical
address spaces. The real address space is uncompressed and is twice the size of
the actual available physical memory. The operating system maps virtual pages to
same-size pages in the real address space, which addresses Challenge 1. On-chip
caches are tagged using the real address (instead of the physical address, which
is dependent on compressibility), which effectively solves Challenge 2. On a
miss in the 32MB cache, Pinnacle maps the corresponding real address to the
physical address of the compressed block in main memory, using a memory-resident
mapping-table managed by the memory controller. Following this, Pinnacle
retrieves the compressed block from main memory, performs decompression and
sends the data back to the processor. Clearly, the additional access to the
memory-resident mapping table on every cache miss significantly increases the
main memory access latency. In addition to this, Pinnacle's decompression
latency, which is on the critical path of a memory access, is 64 processor
cycles.

%%% ONUR: ``To make things even worse'' does not sound good. Rephrase.
%%% Vivek: Fixed

%%% ONUR: Look at the footnote I added above.
%%% Vivek: What kind of optimizations are you thinking of?

%%% ONUR: Look at how I organized shortcomings of the ``large cache
%%% approach''. Always better to be methodical like that.

Ekman and Stenstr\"{o}m~\cite{MMCompression} proposed a main memory compression
design to address the drawbacks of MXT. In their design, the operating system
maps the uncompressed virtual address space directly to a compressed physical
address space. To compress pages, they use a variant of the Frequent Pattern
Compression technique~\cite{fpc,fpc-tr}, which has a much smaller decompression latency
(5 cycles) than the Lempel-Ziv compression in Pinnacle (64
cycles). To avoid the long latency of a cache line's main memory
address computation (Challenge 3), their design overlaps this computation with
the last-level (L2) cache access. For this purpose, their design extends the
page table entries to store the compressed sizes of all the lines within the
page. This information is loaded into a hardware structure called the
\emph{Block Size Table} (BST). On an L1 cache miss, the BST is accessed in
parallel with the L2 cache to compute the exact main memory address of the
corresponding cache line. While the proposed mechanism reduces the latency
penalty of accessing compressed blocks by overlapping main memory address
computation with L2 cache access, the main memory address computation is
performed on {\em every} L2 cache access (as opposed to only on L2 cache misses 
in LCP). This leads to significant wasted work
and additional power consumption. 
Even though BST has the same number of entries as the translation lookaside
buffer (TLB), its size is at least twice that of the TLB~\cite{MMCompression}.
%Moreover, the size of the BST is at least
%twice the size of the translation look-aside buffer (TLB) (number of entries is
%the same).  
This adds to the complexity and power consumption of the
system significantly. To address Challenge 1, the operating system uses multiple
pools of fixed-size physical pages. This reduces the complexity of managing
physical pages at a fine granularity. Ekman and Stenstrom~\cite{MMCompression} 
do not address Challenge 2.

%%% ONUR: Please discuss zero based designs at least in related work... And,
%%% definitelyt evaluate them.

In summary, prior work on hardware-based main memory compression mitigate the
performance degradation due to the main memory address computation problem
(Challenge 3) by either adding large hardware structures that consume
significant area and power~\cite{MXT} or by using techniques that require
energy-inefficient hardware and lead to wasted energy~\cite{MMCompression}.  

\section{Linearly Compressed Pages}
\label{sec:basic}
In this section, we provide the basic idea and a brief overview of our
proposal, Linearly Compressed Pages (LCP), which overcomes the
aforementioned shortcomings of prior proposals.  Further details will
follow in Section~\ref{lcp:sec:design}.
\vspace{-0.1cm}
\subsection{LCP: Basic Idea}
\label{sec:basic-lcp}

The main shortcoming of prior approaches to main memory compression is
that different cache lines within a physical page can be compressed to
different sizes based on the compression scheme. As a result, the
location of a compressed cache line within a physical page depends on
the sizes of all the compressed cache lines before it in the same
page. This requires the memory controller to explicitly perform this
complex calculation (or cache the mapping in a large, energy-inefficient
structure) in order to access the line.

To address this shortcoming, we propose a new approach to compressing
pages, called the \emph{Linearly Compressed Page} (LCP). The key idea
of LCP is to \emph{use a fixed size for compressed cache lines within
a given page} (alleviating the complex and long-latency main memory
address calculation problem that arises due to variable-size cache
lines), and yet still enable a page to be compressed even if not all
cache lines within the page can be compressed to that fixed size
(enabling high compression ratios).

%%% ONUR: See above rewrite of the key idea, and below fixes to go
%%% along with the rewrite.

Because all the cache lines within a given page are compressed to the same
size, the location of a compressed cache line within the page is
simply the product of the index of the cache line within the page and
the size of the compressed cache line---essentially a linear scaling
using the index of the cache line (hence the name \emph{Linearly
  Compressed Page}). LCP greatly simplifies the task of computing a
cache line's main memory address. For example, if all cache lines
within a page are compressed to $16$ bytes, the byte offset of the
third cache line (index within the page is 2) from the start of the
physical page is $16 \times 2 = 32$, if the line is compressed. This
computation can be implemented as a simple shift operation.

%%gpekhime: we should move this to Section 4, this not really
%% a basic idea
\begin{comment}
There are two key design choices made in LCP to improve compression
ratio in the presence of fixed-size compressed cache lines. First, the
target size for the compressed cache lines can be different for
different pages, depending on the algorithm used for compression and
the data stored in the pages. Our LCP-based framework identifies this
target size for a page when the page is compressed for the first time
(or recompressed), as we will describe in
Section~\ref{sec:design-algos}. Second, not all cache lines
within a page can be compressed to a specific fixed size. Also, a
cache line which is originally compressed to the target size may later
become incompressible due to a write. One approach to handle such
cases is to store the entire page in uncompressed format even if a
single line cannot be compressed into the fixed size. However, this
inflexible approach can lead to significant reduction in the benefits
from compression and may also lead to frequent
compression/decompression of entire pages. To avoid these problems,
LCP stores such incompressible cache lines of a page separately from
the compressed cache lines (but still within the page), along with the
metadata required to locate them.
\end{comment}
%%% ONUR: See the tweaks I made above to strengthen.

\begin{figure}[tb]
%  \vspace{-0.2cm}
  \centering
  \includegraphics[width=0.8\textwidth]{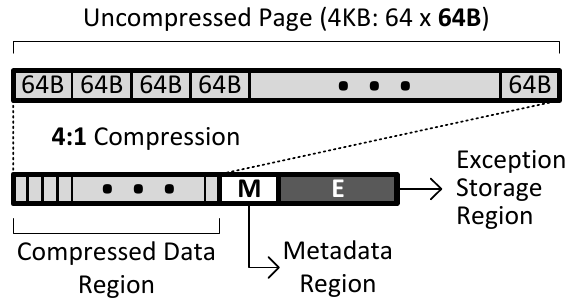}
  \caption{Organization of a Linearly Compressed Page}
  %\vspace{-0.1cm}
  \label{fig:page-organization}
\end{figure}

Figure~\ref{fig:page-organization} shows the organization of an example Linearly
Compressed Page, based on the ideas described above. In this example, we assume
that a virtual page is 4KB, an uncompressed cache line is 64B, and
the target compressed cache line size is 16B. 

As shown in the figure, the LCP
contains three distinct regions.
The first region, \emph{the compressed data region}, contains a 16-byte slot for
each cache line in the virtual page. If a cache line is compressible, the
corresponding slot stores the compressed version of the cache line. However, if
the cache line is not compressible, the corresponding slot is assumed to contain
invalid data. In our design, we refer to such an incompressible cache line as an
``exception''. The second region, \emph{metadata}, contains all the necessary
information to identify and locate the exceptions of a page. We provide more details
on what exactly is stored in the metadata region in
Section~\ref{sec:design-lcp-organization}. The third region, \emph{the exception
  storage}, is the place where all the exceptions of the LCP are stored in their
uncompressed form. Our LCP design allows the exception storage to contain unused
space. In other words, not all entries in the exception storage may store valid
exceptions. As we will describe in Section~\ref{lcp:sec:design}, this enables the
memory controller to use the unused space for storing future exceptions, and also
simplifies the operating system page management mechanism.

Next, we will provide a brief overview of the main
memory compression framework we build using LCP.

%%% ONUR: Metadata explanation above needs more. Tell exactly what is
%%% stored there (for each cache line: whether the line is an
%%% exception, and the start address of the exception's data within
%%% the page, ... what else?).

%%% Vivek: provided a forward pointer to the design section

%%% ONUR: In general, flow is OK, but:
%%% 1) you need to tighten the English
%%% 2) strive for a better organization. Are you missing something; a key design decision that needs to be conveyed to the reader quickly?
%%%    it would be good to write down the key issues to be handled and see if you cover all of them

%\vspace{-0.2cm}
\subsection{LCP Operation}
\begin{figure}[h!]
  \centering
  \includegraphics[width=0.9\linewidth]{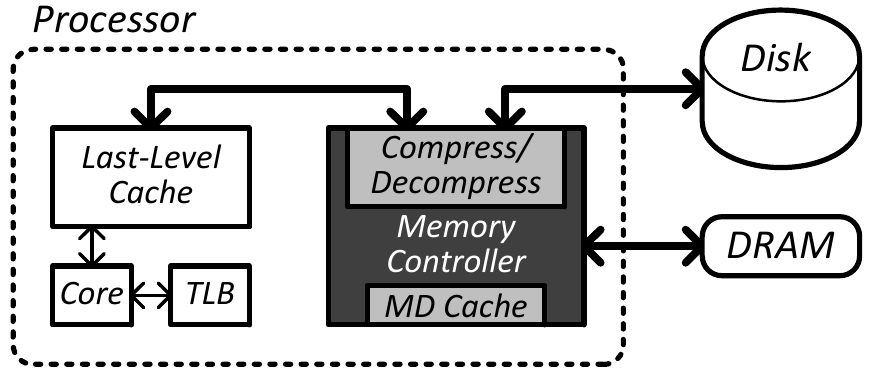}
  \caption{Memory request flow}
  \label{fig:request-flow}
\end{figure}

%\vspace{-0.1cm}
\label{sec:basic-mcf-overview}
%% gpekhime: Staring a major reorganizations here
%% Replacing mechanican description with the flow of the request
%% 
Our LCP-based main memory compression framework consists of components
that handle three key issues: (i) page compression, % and recompression, 
(ii) cache line reads from main memory, and (iii) cache
line writebacks into main memory. 
 Figure~\ref{fig:request-flow} shows
 the high-level design and operation.
% of the LCP-based framework.\footnote{
% Section~\ref{sec:design} presents a detailed description of our 
% design and its operation.}

\begin{comment}
Our LCP-based main memory compression framework consists of components
that handle three key issues: 1) page compression/recompression, 2)
handling a cache line read from main memory, and 3) handling a cache
line writeback into main memory. We briefly provide an overview of
each of these components below. Section~\ref{lcp:sec:design} presents a 
detailed description of our design.
\end{comment}

%\begin{comment}
%\end{comment}

%\vspace{-0.05cm}
%\subsubsection{Component 1: Page (Re)Compression}

\textbf{Page Compression.} 
When a page is accessed for the first time from disk, the operating system (with
the help of the memory controller) first determines whether the page is compressible
using the compression algorithm employed by the framework (described in
Section~\ref{sec:design-algos}). If the page is compressible, the OS allocates a
physical page of appropriate size and stores the compressed page (LCP) in the
corresponding location. It also updates the relevant portions of the
corresponding page table mapping to indicate (i) whether the page is compressed, 
and if so, (ii) the compression scheme used to compress the page (details in 
Section~\ref{sec:design-page-table-extension}).

\label{sec:basic-controller-read-operation}

%%% ONUR: How does the memory controller know the request is to an LCP
%%% vs to a normal page? Clarify this quickly. TLB stores whether or
%%% not a page is compressed, right? 

%The second component of our framework is related to the operation of
%\textbf{Handling Cache Line Read.} 

\textbf{Cache Line Read.}
When the memory controller receives a \emph{read request} for a cache line
within an LCP, it must find and decompress the data. 
Multiple design solutions are possible to perform this task efficiently. 
A na\"{i}ve way of reading a cache line from an LCP
would require at least two accesses to the corresponding page in main
memory. First, the memory controller accesses the \emph{metadata} in the LCP
to determine whether the cache line is stored in the compressed format. 
Second, based on the result, the controller either (i) accesses the
cache line from the \emph{compressed data
region} and decompresses it, or (ii) accesses it uncompressed from 
the \emph{exception storage}. 

To avoid two accesses to main memory, we propose two optimizations that
enable the controller to retrieve the cache line with the latency of
just \emph{one} main memory access in the common case. 
First, we add a
small \emph{metadata (MD) cache} to the memory controller that caches the
metadata of the recently accessed LCPs---the controller
avoids the first main memory access to the metadata in cases when the
metadata is present in the MD cache. Second, in cases when the metadata
is not present in the metadata cache, the controller speculatively
assumes that the cache line is stored in the compressed format and
\emph{first} accesses the data corresponding to the cache line from
the compressed data region. The controller then \emph{overlaps} the
latency of the cache line decompression with the access to the
metadata of the LCP. In the common case, when the speculation is
correct (i.e., the cache line is actually stored in the compressed format),
this approach significantly reduces the latency of serving the read
request. In the case of a misspeculation (uncommon case), the memory
controller issues another request to retrieve the cache line from the
exception storage.

%%% ONUR: above is OK, but can be made more clear by being more
%%% methodical. I still do not have a good feel of the whole
%%% system. Maybe you need to provide a big picture including the TLB,
%%% page table, metadata cache, compressor, decompressor, ..., and any
%%% other important components and how they fit together. A picture is
%%% easier to follow the text with than no picture.

%%% ONUR: Can you prove to the reader that you can fit all metadata
%%% into 64 bytes? Without knowing what is in the metadata it is hard
%%% to believe you and this bugs me as a reader.
%%%
%%% Vivek: 64 bytes in the figure will go away.

%\vspace{-0.05cm}
%\subsubsection{Component 3: Handling Cache Line Writeback}
\label{sec:basic-controller-writeback-operation}

%The third component of our framework deals with the operation of the
\textbf{Cache Line Writeback.}
If the memory controller receives a request for a cache line
\emph{writeback}, it then attempts to compress
the cache line using the compression scheme associated with the
corresponding LCP. Depending on the original state of the cache line
(compressible or incompressible), there are four different
possibilities: the cache line (1) was compressed and stays
compressed, (2) was uncompressed and stays uncompressed, (3) was
uncompressed but becomes compressed, and (4) was compressed but
becomes uncompressed. In the first two cases, the memory controller
simply overwrites the old data with the new data at the same location
associated with the cache line. %, as determined by the metadata stored in
%the page. 
In case 3, the memory controller frees the exception storage slot
for the cache line and writes the compressible data in the compressed
data region of the LCP. (Section~\ref{sec:design-lcp-organization}
provides more details on how the exception storage is managed.) In
case 4, the memory controller checks whether there is enough space in the  
exception storage region to store the uncompressed cache line. 
If so, it stores the cache
line in an available slot in the region. If there are no free exception
storage slots in the exception storage region of the page, the memory controller
traps to the operating system, which migrates the page to a new
location (which can also involve page recompression). 
In both cases 3 and 4, the memory controller appropriately
modifies the LCP metadata associated with the cache line's page.

Note that in the case of an LLC writeback to main memory (and
assuming that TLB information is not available at the LLC), the cache tag entry
is augmented with the same bits that are used to augment page table entries. 
Cache compression mechanisms, e.g., FPC~\cite{fpc} and BDI~\cite{bdi},
already have the corresponding bits for encoding, so that the tag
size overhead is minimal when main memory compression is used together
with cache compression.

\section{Detailed Design}
\label{lcp:sec:design}
%As described in Section~\ref{sec:background}, there are three main
%challenges that have to be addressed to implement a hardware-based
%compression framework in main memory: (i) providing a low-complexity and
%low-latency mechanism to compute the main memory address of a compressed
%cache line, (ii) supporting variable-size physical pages in the
%operating system, and (iii) modifying the cache tagging logic to take a
%cache line's physical address computation off the critical path of the
%cache access. In the previous section, we provided a high-level
%overview of the Linearly Compressed Page (LCP) organization and the
%operation of our main memory compression framework. 
In this section,
we provide a detailed description of LCP, along with the changes to
the memory controller, operating system and on-chip cache tagging
logic. In the process, we explain how our proposed design addresses
each of the three challenges (Section~\ref{sec:background-challenges}).

\newcommand{\vps}{\mathcal{V}}
\newcommand{\pps}{\mathcal{P}}
\newcommand{\cls}{\mathcal{C}}
\newcommand{\ccls}{\mathcal{C}^*}
\newcommand{\clc}{n}
\newcommand{\nex}{n_{ex}}
\newcommand{\navail}{n_{avail}}
\newcommand{\msize}{\mathcal{M}}

\newcommand{\cbit}{{\small{\texttt{c-bit}}}\xspace}
\newcommand{\ctype}{{\small{\texttt{c-type}}}\xspace}
\newcommand{\csize}{{\small{\texttt{c-size}}}\xspace}
\newcommand{\cbase}{{\small{\texttt{c-base}}}\xspace}
\newcommand{\pbase}{{\small{\texttt{p-base}}}\xspace}
\newcommand{\minsize}{{\small{\texttt{m-size}}}\xspace}

\newcommand{\ebit}{{\small{\texttt{e-bit}}}\xspace}
\newcommand{\zbit}{{\small{\texttt{z-bit}}}\xspace}
\newcommand{\vbit}{{\small{\texttt{v-bit}}}\xspace}
\newcommand{\eindex}{{\small{\texttt{e-index}}}\xspace}

\vspace{0.2cm}
\subsection{Page Table Entry Extension}
\label{sec:design-page-table-extension}

\sloppypar

\begin{figure}[h!]
  \vspace{-0.0cm}
  \centering
  \includegraphics[width=0.9\linewidth]{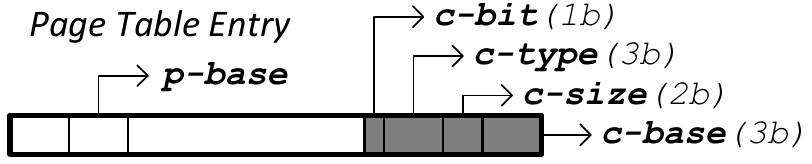}
  \caption{Page table entry extension.}
  \label{fig:pte-extension}
  \vspace{-0.0cm}
\end{figure}

To keep track of virtual pages that are stored in compressed format in
main memory, the page table entries need to be extended to store
information related to compression (Figure~\ref{fig:pte-extension}). 
In addition to the information
already maintained in the page table entries (such as the base address
for a corresponding physical page, \pbase), each virtual page in the
system is associated with the following pieces of metadata: (i) \cbit, a bit
that indicates if the page is mapped to a compressed physical page
(LCP), (ii) \ctype, a field that indicates the compression scheme used to
compress the page, (iii) \csize, a field that indicates the size of the LCP,
and (iv) \cbase, a \pbase extension that enables LCPs to
start at an address not aligned with the virtual page size.
The number of bits required to store \ctype, \csize and \cbase
depends on the exact implementation of the framework. 
In the implementation we evaluate, we assume 3 bits for \ctype (allowing 8 possible different compression encodings),
2 bits for \csize (4 possible page sizes: 512B, 1KB, 2KB, 4KB),
and 3 bits for \cbase (at most eight 512B compressed pages 
can fit into a 4KB uncompressed slot).
Note that existing systems usually have
enough unused bits (up to 15 bits in Intel x86-64 systems~\cite{ia64}) 
in their PTE entries that can be used by LCP without increasing
the PTE size. 

When a virtual page is compressed (the \cbit is set), all the
compressible cache lines within the page are compressed to the same
size, say $\ccls$. The value of $\ccls$ is uniquely determined by the
compression scheme used to compress the page, i.e., the \ctype
(Section~\ref{sec:design-algos} discusses determining the \ctype for a page).
We next describe the LCP organization in more detail.
%will defer the discussion on how to determine the \ctype for a page to
%Section~\ref{sec:design-algos}. Now, assuming the \ctype (and hence, $\ccls$) is
%already known for a page, we describe the corresponding LCP
%organization in more detail.

\subsection{LCP Organization}
\label{sec:design-lcp-organization}

We will discuss each of an LCP's three regions in turn.
We begin by defining the following symbols:
$\vps$ is the virtual page size of the system (e.g., 4KB);
$\cls$ is the uncompressed cache line size (e.g., 64B);\footnote{
Large pages (e.g., 4MB or 1GB) can be supported with LCP through
minor modifications that include scaling the corresponding
sizes of the metadata and compressed data regions.
The exception area metadata keeps the exception index for every cache line on a compressed page.
This metadata can be partitioned into multiple 64-byte cache lines that can be handled similar to 4KB pages.
The exact ``metadata partition'' can be easily identified based on the cache line index within a page.
}
$\clc = \frac{\vps}{\cls}$ is the number of cache lines per virtual page (e.g., 64);
and $\msize$ is the size of LCP's metadata region.
In addition, on a per-page basis, we define
$\pps$ to be the compressed physical page size;
$\ccls$ to be the compressed cache line size; and
$\navail$ to be the number of slots available for exceptions.

\vspace{-0.05cm}
\subsubsection{Compressed Data Region}
\begin{figure}[t]
  \centering
  \includegraphics[width=0.9\linewidth]{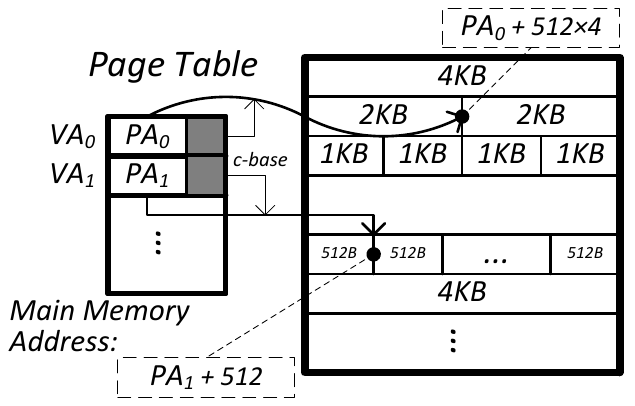}
  \caption{Physical memory layout with the LCP framework.}
  \label{fig:macroview}
\end{figure}

The compressed data region is a contiguous array of $\clc$ slots each of size
$\ccls$. Each one of the $\clc$ cache lines in the virtual page is mapped to
one of the slots, irrespective of whether the cache line is compressible or
not. Therefore, the size of the compressed data region is $\clc\ccls$. This
organization simplifies the computation required to determine the main memory
address for the compressed slot corresponding to a cache line. More
specifically, the address of the compressed slot for the $i^{th}$ cache line
can be computed as $\pbase + \minsize*\cbase + (i-1)\ccls$, where the first two terms correspond
to the start of the LCP ($\minsize$ equals to the minimum page size, 512B in
our implementation) and the third indicates the offset within the LCP of the 
$i^{th}$ compressed slot (see Figure~\ref{fig:macroview}).
Thus, computing the main memory address
of a compressed cache line requires one multiplication (can be implemented as a shift) 
and two additions
%where \cbase~\footnote{We store \cbase value as three bits, but 
%in address calcluations 
%we use the value that is padded with 9 zeros at the end (to reflect the number
%of 512B it represents).} is the physical page base address and $\ccls$ is the
%compressed cache line size. Computing the main memory address
%of a cache line requires one multiplication and one addition,
independent of $i$ (fixed latency). This computation requires a lower latency and
simpler hardware than prior approaches (e.g., up to 22 additions 
in the design proposed in \cite{MMCompression}), thereby
efficiently addressing Challenge 3 (cache line address
computation). 
% Our approach also has fixed latency to compute the main
% memory address as address computation operations performed 
% independently of the cache line's location in the page.

%\vspace{-0.05cm}
\subsubsection{Metadata Region}
\begin{figure}[h]
%  \vspace{-0.3cm}
  %\vspace{-0.2cm}
  \centering
  \includegraphics[width=0.8\linewidth]{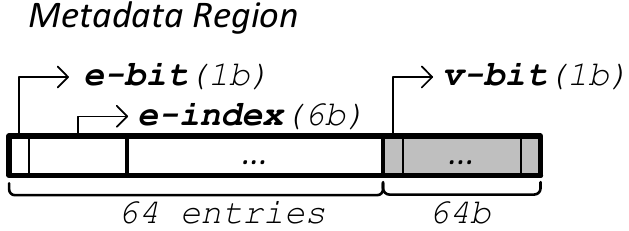}
  \caption{Metadata region, when $\clc=64$.}
  \label{fig:md-region}
  %\vspace{-0.1cm}
\end{figure}

The metadata region of an LCP contains two parts (Figure~\ref{fig:md-region}).
The first part stores two pieces of information 
for each cache line in the virtual
page: (i) a bit indicating whether the cache line is incompressible, i.e.,
whether the cache line is an \emph{exception}, \ebit, and (ii) the index
of the cache line in the exception storage,
\eindex. 
If the \ebit is set for a cache line, then the
corresponding cache line is stored uncompressed in
location \eindex in the exception storage. 
The second part of the metadata region is a \emph{valid} bit (\vbit) vector to track 
the state of the slots in
the exception storage. If a \vbit is set, it indicates that the
corresponding slot in the exception storage is used by some
uncompressed cache line within the page.

The size of the first part depends on the size of \eindex, which in turn depends
on the number of exceptions allowed per page. Because the number of exceptions
cannot exceed the number of cache lines in the page ($\clc$), we will need at
most $1 + \lceil \log_2 \clc \rceil$ bits for each cache line in the first part of
the metadata. For the same reason, we will need at most $\clc$ bits in the bit
vector in the second part of the metadata. Therefore, the size of the metadata
region is given by 
%begin{equation}
$ \msize = \clc(1 + \lceil \log_2 \clc \rceil) + \clc ~\textrm{bits}$.
%end{equation}
Since $\clc$ is fixed for the entire system, the size of the metadata region
($\msize$) is the same for all compressed pages (64B in our implementation).
%%% ONUR: Implicit in the explanation is that you kjnow the starting point of the exception storage.

%\vspace{-0.05cm}
\subsubsection{Exception Storage Region}
The third region, the exception storage, is the place where all incompressible
cache lines of the page are stored. If a cache line is present in the
location \eindex in the exception storage, its main memory address
can be computed as: 
$\pbase + \minsize*\cbase + \clc\ccls + \msize + \eindex \cls $.
%location \eindex in the exception storage, its main memory address can be computed
%as:
%$\cbase + \clc\ccls +
%  \msize + \eindex \cls $.
The number of slots available in the
exception storage ($\navail$) is dictated by the size of the compressed physical
page allocated by the operating system for the corresponding LCP. The
following equation expresses the relation between the physical page size
($\pps$), the compressed cache line size ($\ccls$) that is determined by
\ctype, and the number of available slots in the exception storage ($\navail$):
\vspace{-0.4cm}\\
\begin{equation}
  \navail = \lfloor(\pps - (n\ccls + \msize))/\cls\rfloor
  \label{eqn:avail-exceptions}
\vspace{-0.05cm}
\end{equation}

\noindent
As mentioned before, the metadata region contains a bit vector that is used to
manage the exception storage. When the memory controller assigns an exception
slot to an incompressible cache line, it sets the corresponding bit in the bit
vector to indicate that the slot is no longer free. If the cache line later
becomes compressible and no longer requires the exception slot, the memory
controller resets the corresponding bit in the bit vector. In the next section,
we describe the operating system memory management policy that determines the
physical page size ($\pps$) allocated for an LCP, and hence, the number of
available exception slots ($\navail$).

%%% ONUR: Above is OK, but can be tightened...

\subsection{Operating System Memory Management}
The first challenge related to main memory compression is to provide operating
system support for managing variable-size compressed physical pages --
i.e., LCPs. Depending on the compression scheme employed by the framework,
different LCPs may be of different sizes. Allowing LCPs of arbitrary sizes would
require the OS to keep track of main memory at a very fine granularity. It could
also lead to fragmentation across the entire main memory at a
fine granularity. As a result, the OS would need to maintain large amounts of
metadata to maintain the locations of individual pages and the free space, which
would also lead to increased complexity.

To avoid this problem, our mechanism allows the OS to manage main memory using a
fixed number of pre-determined physical page sizes -- e.g., 512B, 1KB, 2KB,
4KB (a similar approach was proposed in~\cite{berger-thesis}
to address the memory allocation problem).
For each one of the chosen sizes, the OS maintains a pool of allocated
pages and a pool of free pages. When a page is compressed for the first time or
recompressed due to overflow (described in Section~\ref{sec:design-handling-overflows}), 
the OS chooses the
smallest available physical page size that fits the compressed page. For
example, if a page is compressed to 768B, then the OS allocates a physical page
of size 1KB. 
For a page with a given size, the available number of
exceptions for the page, $\navail$, can be determined using
Equation~\ref{eqn:avail-exceptions}.

\subsection{Changes to the Cache Tagging Logic}

As mentioned in Section~\ref{sec:background-challenges}, modern
systems employ physically-tagged caches to avoid aliasing
problems. However, in a system that employs main memory
compression, using the physical (main memory) address to tag cache
lines puts the main memory address computation on the critical path of
L1 cache access (Challenge 2). To address this challenge, we modify the
cache tagging logic to use the tuple $<$physical page base address,
cache line index within the page$>$ for tagging cache lines. This
tuple maps to a unique cache line in the system, and hence avoids aliasing
problems without requiring the exact main memory
address to be computed. The additional 
index bits are stored within the cache line tag.

\subsection{Changes to the Memory Controller}

In addition to the changes to the memory controller operation
described in Section~\ref{sec:basic-mcf-overview}, our LCP-based
framework requires two hardware structures to be added to
the memory controller: (i) a small metadata cache to accelerate main
memory lookups in LCP, and (ii) compression/decompression hardware to
perform the compression and decompression of cache lines.

%\vspace{-0.05cm}
\subsubsection{Metadata Cache}
\label{sec:design-metadata-cache}

As described in Section~\ref{sec:basic-controller-read-operation},
a small metadata cache in the memory controller enables our approach, 
in the common case, to retrieve a compressed cache block in a single
main memory access. This cache stores the metadata
region of recently accessed LCPs so that the metadata for subsequent
accesses to such recently-accessed LCPs can be retrieved directly from
the cache. In our study, we find that a small 512-entry metadata cache 
(32KB\footnote{We evaluated the
sensitivity of performance to MD cache size and find that 32KB is the
smallest size that enables our design to avoid most of the performance loss
due to additional metadata accesses.})
can service 88\% of the metadata accesses on average across all our
workloads. Some applications have lower hit rate,
especially \emph{sjeng} and \emph{astar}~\cite{SPEC}. An analysis of 
%the source code of 
these applications reveals that their memory accesses exhibit very low locality.
As a result, we also observed a low TLB hit rate for these applications.
Because TLB misses are costlier than MD
cache misses (the former requires multiple memory accesses), the low MD
cache hit rate does not lead to significant performance degradation
for these applications.

We expect the MD cache power to be much lower than the power consumed
by other on-chip structures (e.g., L1 caches), because the MD cache is accessed 
much less frequently (hits in any on-chip cache
do not lead to an access to the MD cache). 

%\vspace{-0.05cm}
\subsubsection{Compression/Decompression Hardware}

Depending on the compression scheme employed with our LCP-based
framework, the memory controller should be equipped with the hardware
necessary to compress and decompress cache lines using the
corresponding scheme. Although our framework does not impose any
restrictions on the nature of the compression algorithm, it is
desirable to have compression schemes that have low complexity and
decompression latency -- e.g., Frequent Pattern Compression
(FPC)~\cite{fpc} and Base-Delta-Immediate Compression
(BDI)~\cite{bdi}. In Section~\ref{sec:design-algos}, we provide more
details on how to adapt any compression algorithm to fit the
requirements of LCP and also the specific changes we made to FPC and
BDI as case studies of compression algorithms that we adapted to the
LCP framework.

%%% ONUR: Tighten. I know I added more for clarity.

%%% ONUR: Tighten. I removed the following redundancy, which I believe was
%%% discussed previously:
%When a cache line switches from the incompressible state to the compressible
%state, the memory controller simply needs to write the new data in the slot
%reserved for the cache line in the compressed data region, and free the old
%slot from the exception storage.

\subsection{Handling Page Overflows}
\label{sec:design-handling-overflows}
As described in Section~\ref{sec:basic-controller-writeback-operation}, when a
cache line is written back to main memory, the cache line may
switch from being compressible to being incompressible. When this
happens, the memory controller should explicitly find a slot in the exception
storage for the uncompressed cache line. However, it is possible that all
the slots in the exception storage are already used by other exceptions in the
LCP. We call this scenario a \emph{page overflow}.  A page overflow increases
the size of the LCP and leads to one of two scenarios: (i)~the LCP still requires a
physical page size that is smaller than the uncompressed virtual page size
(type-1 page overflow), and (ii)~the LCP now requires a physical page size that is
larger than the uncompressed virtual page size (type-2 page overflow).

Type-1 page overflow simply requires the operating system to migrate the LCP to
a physical page of larger size (without recompression). The OS first allocates a new page and copies the
data from the old location to the new location. It then modifies the mapping for
the virtual page to point to the new location. While in transition, the page is
locked, so any memory request to this page is delayed. In our evaluations, we
stall the application for 20,000 cycles\footnote{
To fetch a 4KB page, we need to access 64 cache lines (64 bytes each). In the worst
case, this will lead to 64 accesses to main memory, most of which are likely to
be DRAM row-buffer hits. Since a row-buffer hit takes 7.5ns, the total time to fetch
the page is 495ns. On the other hand, the latency penalty of two
context-switches (into the OS and out of the OS) is around 4us~\cite{ContextSwitch}. 
%, assuming each context-switch takes about 2us~\cite{ContextSwitch}. 
Overall, a type-1 overflow takes around 4.5us.
For a 4.4Ghz or slower processor, this is less than 20,000 cycles. 
}
when a type-1 overflow occurs; we also find
that (on average) type-1 overflows happen less than once
per two million instructions.
We vary this latency between 10,000--100,000 cycles and 
observe that the benefits of our framework (e.g., bandwidth compression)
far outweigh the overhead due to type-1 overflows. 
%Note that
%the exact latency for type-1 overflow is hard to calculate without changing the
%operating system. 

%Since the data in the old page is
%still valid, this operation may not have to stall the thread for read requests
%to the page. 

%However, if there are no additional cores to run the OS handler or
%if a write request is generated for the page, the application may have to stall
%till the migration operation is complete. In any case, in our evaluations, we
%stall the application for 20000 cycles when a type-1 overflow occurs.

In a type-2 page overflow, the size of the LCP exceeds the uncompressed virtual
page size. Therefore, the OS attempts to recompress the page, possibly using a
different encoding (\ctype). % that fits well with the new data of the page. 
Depending on whether the page is compressible or not, the OS allocates a
new physical page to fit the LCP or the uncompressed page, and migrates the data
to the new location. The OS also appropriately modifies the \cbit, \ctype and
the \cbase in the corresponding page table entry. Clearly, a type-2 overflow
requires more work from the OS than a type-1 overflow. However,
we expect page overflows of type-2 to occur rarely. In fact, we
never observed a type-2 overflow in our evaluations.

\subsubsection{Avoiding Recursive Page Faults}
There are two types of pages that require special consideration:
(i) pages that keep internal OS data structures, e.g., pages 
containing information required to handle page faults,
and (ii) shared data pages that have more than one page table entry (PTE) mapping
to the same physical page. Compressing pages of the first type
can potentially lead to recursive page fault handling.
The problem can be avoided if the OS sets a special 
\emph{do not compress} bit, e.g.,
as a part of the page compression encoding, so that the memory controller
does not compress these pages. The second type of pages (shared pages) 
require consistency across multiple page table entries,
such that when one PTE's compression information changes, 
the second entry is updated as well. There are two possible solutions 
to this problem. First, as with the first type of pages, 
these pages can be marked as \emph{do not compress}.
Second,
the OS could maintain consistency of the shared PTEs by performing 
multiple synchronous PTE updates (with accompanying TLB shootdowns). 
While the second solution can potentially 
lead to better average compressibility, the first solution (used in
our implementation) is simpler and requires minimal changes inside the OS.

Another situation that can potentially lead to a recursive fault is 
the eviction of dirty cache lines from the LLC to DRAM due to some page overflow
handling that leads to another overflow. In order to solve this problem,
we assume that the memory controller has a small dedicated portion of the
main memory that is used as a scratchpad to store cache lines needed
to perform page overflow handling. Dirty cache lines that are evicted
from LLC to DRAM due to OS overflow handling are stored in this buffer space.
The OS is responsible to minimize the memory footprint of the overflow handler.
Note that this situation is expected to be very rare in practice, because
even a single overflow is infrequent.

\subsubsection{Handling Special Cases}
There are several types of scenarios that require special attention: (i) rapid
changes in compressibility (e.g., highly compressed page overwritten with
non-compressible data), (ii) multiple back-to-back page overflows.
The first scenario leads to the increase in the number of page overflows
that are costly and time-consuming. This situation is common when the page
is initialized with some values (frequently zero values), and then after
some period of time multiple updates (e.g., writebacks) bring completely 
different data into this page. For zero pages the solution is simply not
storing them at all - only one bit in TLB buffer, until there are not
enough writebacks happen to these page to estimate its compressibility.
For other pages, especially the ones that are allocated (e.g., through malloc),
but never been updated, we also delay compression until there is not enough
evidence that this page can be successfully compressed. These simple optimizations
allow us to avoid major sources of the page overflows. 

The second scenario, while possible in practice, was extremely rare in our experiments.
Nevertheless, one possible solution we consider to this problem, is to detect 
the situations like this, and when the number of back to back page overflows exceeds
certain threshold, start to decompress this applications' data in the background to avoid
further overflows.

\subsection{Compression Algorithms}
\label{sec:design-algos}

Our LCP-based main memory compression framework can be employed with
any compression algorithm. In this section, we describe how to adapt a
generic compression algorithm to fit the requirements of the LCP
framework. Subsequently, we describe how to adapt the two compression algorithms
used in our evaluation.

\newcommand{\cf}{f_c}
\newcommand{\df}{f_d}

%\vspace{-0.05cm}
\subsubsection{Adapting a Compression Algorithm to Fit LCP}
\label{sec:generic-compression}
Every compression scheme is associated with a compression function, $\cf$, and
a decompression function, $\df$.
To compress a virtual page into the corresponding LCP using the compression
scheme, the memory controller carries out three steps. In the first step, the
controller compresses every cache line in the page using $\cf$ and feeds the
sizes of each compressed cache line to the second step. In the second step, the
controller computes the total compressed page size (compressed data + metadata
+ exceptions, using the formulas from Section~\ref{sec:design-lcp-organization}) 
for each of a fixed set of target compressed cache line sizes and selects a target 
compressed cache line size $\ccls$ that minimizes the overall LCP size. In the
third and final step, the memory controller classifies any cache line whose
compressed size is less than or equal to the target size as compressible and all
other cache lines as incompressible (exceptions). The memory controller uses
this classification to generate the corresponding LCP based on the organization
described in Section~\ref{sec:basic-lcp}. 

To decompress a compressed cache line of the page, the memory
controller reads the fixed-target-sized compressed data and feeds it
to the hardware implementation of function $\df$. 
%Note that the memory controller already knows the size of the uncompressed cache line.
%($\cls$) is already known to the memory controller.

%\vspace{-0.05cm}
\subsubsection{FPC and BDI Compression Algorithms}
\label{sec:design-prev-algos}
Although any compression algorithm can be employed with our framework
using the approach described above, it is desirable to use compression
algorithms that have low complexity hardware implementation and low
decompression latency, so that the overall complexity and latency of the design are
minimized. For this reason, we adapt to fit our LCP framework two state-of-the-art compression
algorithms that achieve such design points in the 
context of compressing in-cache data: (i) Frequent Pattern
Compression~\cite{fpc}, and (ii) Base-Delta-Immediate
Compression~\cite{bdi}.

%%% ONUR: specifically designed to compress in-memory data --> More
%%% specifically in-cache data, right?

Frequent Pattern Compression (FPC) is based on the observation that a
majority of the words accessed by applications fall under a small set
of frequently occurring patterns~\cite{fpc-tr}. FPC compresses each
cache line one word at a time. Therefore, the final compressed size of
a cache line is dependent on the individual words within the cache
line. To minimize the time to perform the compression search procedure described
in Section~\ref{sec:generic-compression}, we limit the search to
four different target cache line sizes: 16B, 21B, 32B and 44B (similar
to the fixed sizes used in \cite{MMCompression}). 
%We call this variant of FPC as \emph{FPC-Fixed}.

Base-Delta-Immediate (BDI) Compression is
based on the observation that in most cases, words co-located in memory have
small differences in their values, a property referred to as \emph{low dynamic
  range}~\cite{bdi}. BDI encodes cache lines with such low dynamic range using a
base value and an array of differences ($\Delta$s) of words within the cache
line from either the base value or from zero. The size of the final compressed cache line
depends only on the size of the base and the size of the $\Delta$s. To employ BDI within
our framework, the memory controller attempts to compress a page with different
versions of the Base-Delta encoding as described by Pekhimenko {\em et al.}~\cite{bdi} and
then chooses the combination that minimizes the final compressed page size
(according to the search procedure in Section~\ref{sec:generic-compression}).

\begin{comment}
For our GPU workloads, multiple values are typically packed into a word -- e.g.,
three components of a color. As a result, in a number of cases, we observed
cache lines for which the most significant byte of the values within the cache
line are different while the remaining bytes are fixed. The original BDI
algorithm will not be able to compress such cache lines as the differences
between the words will be large. However, if words of such cache lines are
shifted cyclically (by one byte), they can then be compressed using BDI. We call
this modification to BDI as BDI-rotate and evaluate it in
Section~\ref{sec:gpu-bandwidth}.
\end{comment}

%%% ONUR: the memory controller attempts to compress a page with
%%% different sizes for the base??? UNCLEAR. 

%VIVEK: I don't understand the extension to BDI.

%%% ONUR: The explanation of what changes you exactly amde to
%%% ``adapt'' the FPC and BDI algorithms to the LCP framework is not
%%% clear. What are these changes?

\section{LCP Optimizations}
In this section, we describe two simple optimizations to our proposed LCP-based
framework: (i) memory bandwidth reduction via compressed cache lines, 
and (ii) exploiting zero pages and cache lines for higher bandwidth utilization. 

\subsection{Enabling Memory Bandwidth Reduction}
\label{sec:opt-bandwidth}
One potential benefit of main memory compression that has not been examined in
detail by prior work on memory compression is bandwidth
reduction.\footnote{Prior
  work~\cite{register-caching,fvc-bus,LinkCompression,GPUBandwidthCompression} looked at the
  possibility of using compression for bandwidth reduction between the memory
  controller and DRAM. While significant reduction in bandwidth consumption is 
  reported, prior work achieve this reduction either at the cost of increased
  memory access latency~\cite{register-caching,fvc-bus,LinkCompression}, 
  as they have to both compress and decompress a cache
  line for every request, or based on a specialized 
  main memory design~\cite{GPUBandwidthCompression}, e.g., GDDR3~\cite{gddr3}.}  
When cache lines are stored in compressed format in
main memory, multiple consecutive compressed cache lines can be retrieved at the
cost of retrieving a single uncompressed cache line. For example, when cache
lines of a page are compressed to 1/4 their original size, four
compressed cache lines can be retrieved at the cost of a single uncompressed cache line
access. This can significantly reduce the bandwidth requirements of
applications, especially those with good spatial locality. We propose two
mechanisms that exploit this idea.

In the first mechanism, when the memory controller needs to access a cache line
in the compressed data region of LCP, it obtains the data from multiple
consecutive compressed slots (which add up to the size of an uncompressed cache
line). However, some of the cache lines that are retrieved in this manner may not be
\emph{valid}. To determine if an additionally-fetched cache line is valid or
not, the memory controller consults the metadata corresponding to the LCP. If a
cache line is not valid, then the corresponding data is not
decompressed. Otherwise, the cache line is decompressed and then stored in the
cache.

The second mechanism is an improvement over the first mechanism, where the
memory controller additionally predicts if the additionally-fetched cache lines
are \emph{useful} for the application. For this purpose, the memory controller
uses hints from a multi-stride prefetcher~\cite{stride-prefetching}. In this
mechanism, if the stride prefetcher suggests that an additionally-fetched cache
line is part of a useful stream, then the memory controller stores that cache
line in the cache. This approach has the potential to prevent cache lines that
are not useful from polluting the
cache. Section~\ref{sec:results-prefetching-hints} shows the effect of this
approach on both performance and bandwidth consumption.

Note that prior work~\cite{register-caching,fvc-bus,LinkCompression,GPUBandwidthCompression}
assumed that when a cache line is compressed, only the compressed amount
of data can be transferred over the DRAM bus, thereby freeing the bus for the future
accesses. Unfortunately, modern DRAM chips are optimized for full cache block
accesses~\cite{variable-reads}, so they would need to be modified to support such
smaller granularity transfers. Our proposal does not require modifications to DRAM
itself or the use of specialized DRAM such as GDDR3~\cite{gddr3}.

%%vspace{0.6cm}
\subsection{Zero Pages and Zero Cache Lines}
\label{sec:opt-zeros}
Prior work~\cite{ZeroContent,fvc,fpc,MMCompression,bdi} observed that in-memory
data contains a significant number of zeros at two granularities: all-zero
pages and all-zero cache lines. Because this pattern is quite common, we propose two
changes to the LCP framework to more efficiently compress such occurrences of
zeros. First, one value of the page compression encoding (e.g., \ctype of 0) is
reserved to indicate that the entire page is zero. When accessing data from a
page with \ctype$=0$, the processor can avoid any LLC or DRAM
access by simply zeroing out the allocated cache line in the L1-cache. Second,
to compress all-zero cache lines more efficiently, we can add another bit per cache line
to the first part of the LCP metadata. This bit, which we call the \zbit, indicates
if the corresponding cache line is zero. Using this approach, the memory
controller does not require any main memory access to retrieve a cache line with
the \zbit set (assuming a metadata cache hit).

\section{Methodology}
\label{lcp:sec:methodology}

\begin{table}[t]\footnotesize
\vspace{-0.05cm}
 \centering
%\begin{minipage}[b]{0.7\linewidth}
    \begin{tabular}{ll}
         \toprule %\hline
         CPU Processor  &  1--4 cores, 4GHz, x86 in-order  \\
         \cmidrule(rl){1-2} %\hline
         CPU L1-D cache    &  32KB, 64B cache-line, 2-way, 1 cycle \\
         \cmidrule(rl){1-2} %\hline
         CPU L2 cache    &  2 MB, 64B cache-line, 16-way,  20 cycles  \\
         \cmidrule(rl){1-2} %\hline
         Main memory & 2 GB, 4 Banks, 8 KB row buffers, \\ 
         & 1 memory channel, DDR3-1066~\cite{micron-ddr3} \\
         \cmidrule(rl){1-2} %\hline
         LCP Design & Type-1 Overflow Penalty: 20,000 cycles \\
          \bottomrule%\hline
          %\hline
         %GPU Processor  &  1--16 cores, 8 warps per core, 1.4GHz  \\
         %\hline
         %GPU L1-D cache    &  16kB, 256B cache-line, 2-way, 1 cycle \\
         %\hline
         %GPU L2 cache    &  64 kB, 256B cache-line, 4-way,  5 cycles  \\
         %\hline
  
    \end{tabular}%
%\vspace{-0.1cm}
 \caption{\small Major Parameters of the Simulated Systems.}
  \label{lcp:tbl:simulation-parameters}%
\vspace{-0.2cm}
\end{table}

Our evaluations use an in-house, event-driven 32-bit x86 simulator whose front-end is based on Simics~\cite{Simics}.
%, and multi2sim~\cite{multi2sim} AMD Evergreen ISA~\cite{amd-gpu} simulator for GPU evaluations.
All configurations have private L1 caches and shared L2 caches.
Major simulation parameters are provided in Table \ref{lcp:tbl:simulation-parameters}.
We use benchmarks from the SPEC CPU2006 suite~\cite{SPEC}, four
TPC-H/TPC-C queries~\cite{tpc}, and an Apache web server. 
%For GPU evaluations, we use
%a set of AMD OpenCL benchmarks. 
All results are
collected by running a representative portion (based on PinPoints~\cite{pinpoints}) of the benchmarks for 1 billion
instructions. % (CPU) or to completion (GPU).
We build our energy model based on McPat~\cite{mcpat}, CACTI~\cite{cacti}, C-Pack~\cite{c-pack},
and the Synopsys Design
Compiler with 65nm library (to evaluate the energy of compression/decompression with BDI and 
address calculation in~\cite{MMCompression}). 

\textbf{{Metrics.}}
We measure the performance of our benchmarks
using IPC (instruction per cycle) and effective compression ratio (effective DRAM size increase,  
e.g., a compression ratio of 1.5 for 2GB DRAM means that the compression scheme
% employed with this DRAM
achieves the size benefits of a 3GB DRAM). %, and L2 MPKI (misses per kilo instruction).
For multi-programmed workloads we use the weighted speedup~\cite{weightedspeedup} performance metric:
($\sum_i \frac{IPC_i^{shared}} {{IPC}_i^{{alone}}}$).
%and instruction throughput ($\sum_i \textrm{IPC}_i$).
For bandwidth consumption we use BPKI (bytes transferred over bus per thousand instructions~\cite{BPKI}).

\textbf{{Parameters of the Evaluated Schemes.}}  As reported in the respective
previous works, we used a decompression latency of 5 cycles for
FPC and 1 cycle for BDI. 

%We did
%sweep a type-1 overflow latency from 10,000 to 100,000 cycles and found that
%the impact on performance was negligible.~\footnote{Type-1 overflows
%are very rare (see Section~\ref{sec:results}). 
%Prior work on main memory compression~\cite{MMCompression}
%also used 10,000 to 100,000 cycle range for such overflows.}
 
%For BDI-rotate, we
%assumed a 2-cycle latency due to the additional rotation step.

\section{Results}
\label{lcp:sec:results}

In our experiments for both single-core and multi-core systems, we 
compare five different designs that employ different main memory compression
strategies (frameworks) and different compression algorithms:
(i) \textit{Baseline} system with no compression,
(ii) robust main memory compression (\textit{RMC-FPC})~\cite{MMCompression},
(iii) and (iv) LCP framework with both FPC and BDI compression algorithms (\textit{LCP-FPC} 
and \textit{LCP-BDI}), and (v) \textit{MXT}~\cite{MXT}. 
Note that it is fundamentally possible
to build a RMC-BDI design as well, but we found that it leads to either low energy efficiency (due to
an increase in the BST metadata table entry size~\cite{MMCompression} with many more encodings in BDI) or
low compression ratio (when the number of encodings is artificially decreased). Hence, for brevity,
we exclude this potential design from our experiments. 

In addition, we evaluate two hypothetical designs: Zero Page
Compression (\textit{ZPC}) and Lempel-Ziv (\textit{LZ})\footnote{Our
  implementation of LZ performs compression at 4KB page-granularity
  and serves as an idealized upper bound for the in-memory compression
  ratio. In contrast, MXT employs Lempel-Ziv at 1KB granularity.} to
show some practical upper bounds on main memory compression.
Table~\ref{table:schemes} summarizes all the designs.

\begin{table}[t]\footnotesize
\vspace{-0.05cm}
  \centering
  \begin{tabular}{lll}
    \toprule %hline
    \textbf{Name} & \textbf{Framework} & \textbf{Compression Algorithm}\\
    \midrule%\cmidrule(rl){1-3} %\midrule %\hline
    \textit{Baseline} & None & None\\
    \cmidrule(rl){1-3}
    \textit{RMC-FPC} & RMC~\cite{MMCompression} & FPC~\cite{fpc}\\
    \cmidrule(rl){1-3} %\hline
    \textit{LCP-FPC} & LCP & FPC~\cite{fpc}\\
    \cmidrule(rl){1-3} %\hline
    \textit{LCP-BDI} &  LCP & BDI~\cite{bdi}\\
    \cmidrule(rl){1-3} %\hline
    \textit{MXT} & MXT~\cite{MXT} & Lempel-Ziv~\cite{lz}\\
    %\cmidrule(rl){1-3}
    \midrule
    \midrule
    %\cmidrule(rl){1-3}
    \textit{ZPC} & None & Zero Page Compression\\
    \cmidrule(rl){1-3}
    \textit{LZ} & None & Lempel-Ziv~\cite{lz}\\
    \bottomrule
  \end{tabular}
  \caption{List of evaluated designs.}
  \vspace{-0.2cm}
  \label{table:schemes}
\end{table}

\subsection{Effect on DRAM Capacity}

\begin{figure}[htb]
  \centering
  \includegraphics[width=0.95\textwidth]{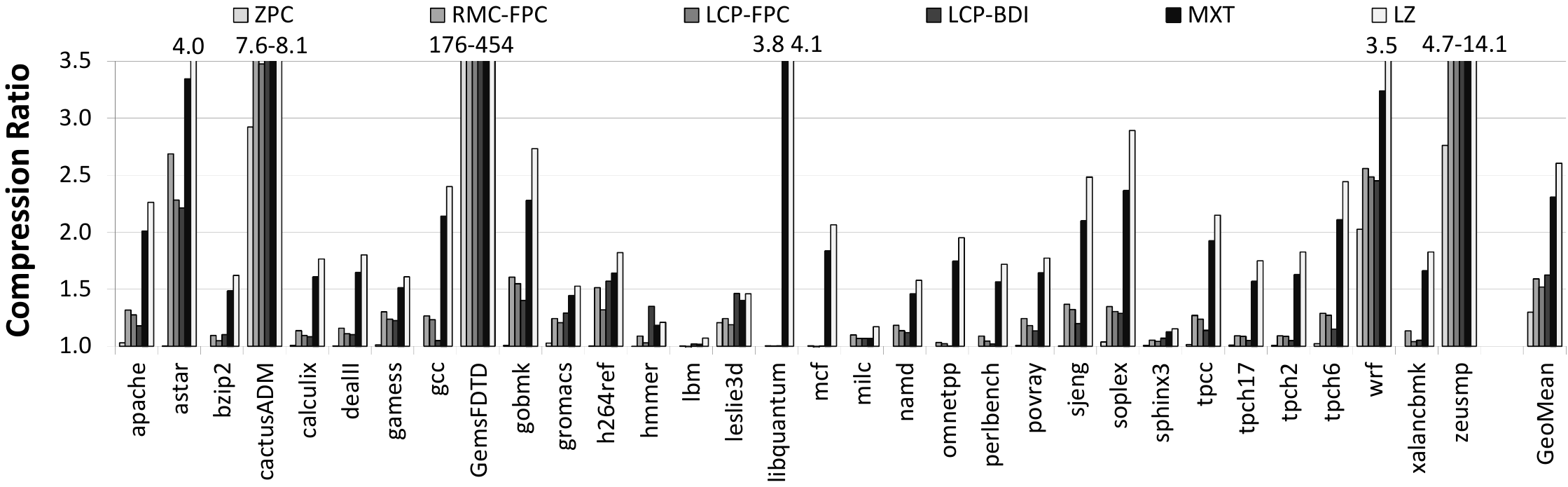}
  \caption{Main memory compression ratio.}
  \label{fig:capacity}
\end{figure}

Figure~\ref{fig:capacity} compares the compression ratio of all the designs
described in Table~\ref{table:schemes}.
We draw two major conclusions. First, as expected, MXT,
which employs the complex LZ algorithm, has the highest average compression
ratio (2.30) of all practical designs and performs closely to our idealized LZ
implementation (2.60). At the same time, LCP-BDI
provides a reasonably high compression ratio (1.62 on average),
outperforming RMC-FPC (1.59), and LCP-FPC (1.52).
(Note that LCP could be used with both BDI and FPC algorithms together, and the average compression
ratio in this case is as high as 1.69.)

Second, while the average compression ratio of
ZPC is relatively low (1.29), it greatly improves the
effective memory capacity for a number of applications (e.g., {\em GemsFDTD}, 
{\em zeusmp},
and {\em cactus\-ADM}). This justifies our design decision of handling zero pages
at the TLB-entry level. We conclude that our LCP framework
achieves the goal of high compression ratio.
    
%Figure~\ref{fig:capacity} shows the results of this comparison.  First, as
%expected, MXT has the highest average compression ratio (2.30) of all practical
%designs and is close to the LZSS compression ratio (2.60). At the same time, LCP
%with both compression techniques provides reasonable compression ratio (up to
%1.69 with BDI+FPC-fixed), outperforming previously proposed DRAM compression
%design~\cite{MMCompression} based on FPC (1.59). 

%Second, noticable number of applications, e.g., GemsFDFD, zeusmp, and CactusADM,
%have significant number of zero pages (the average is 1.29), which justifies our
%design decision of handling zeros specifically at TBL-entry level.         

\subsubsection{Distribution of Compressed Pages}
\begin{figure}[htb]
  %\vspace{0.1cm}
  \includegraphics[width=0.95\textwidth]{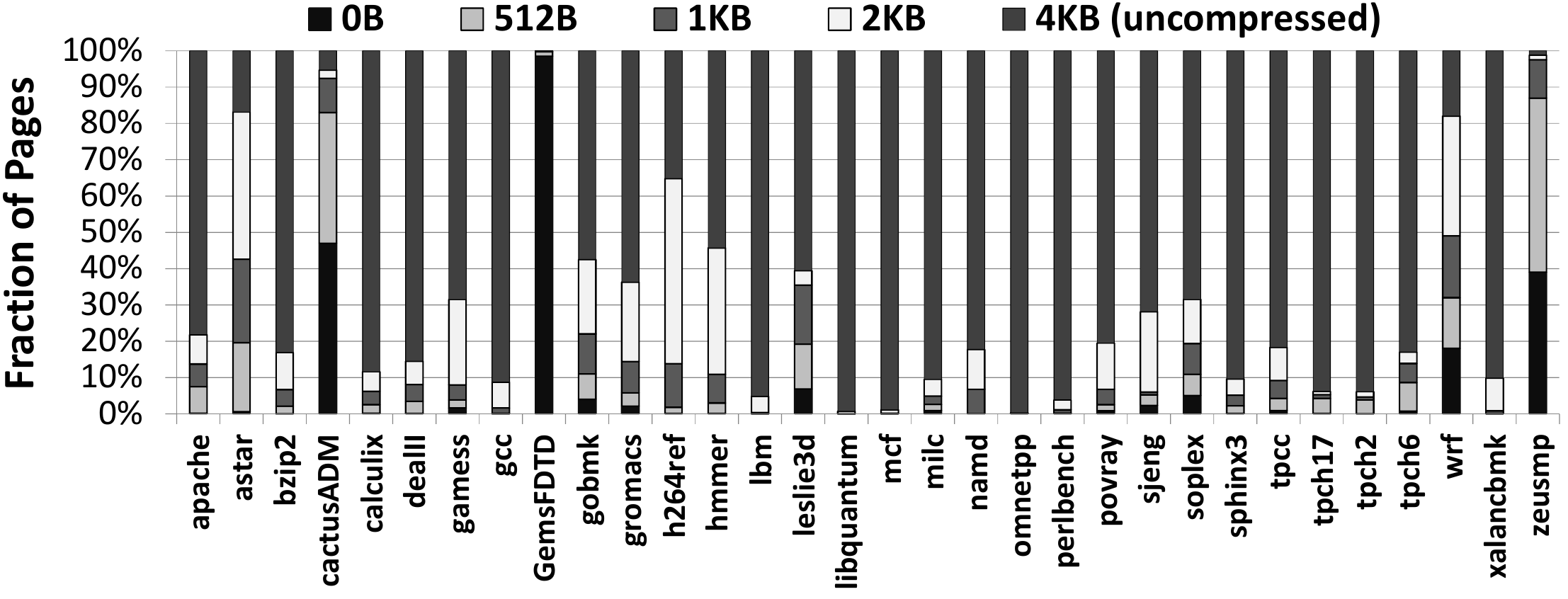}
  \caption{Compressed page size distribution with LCP-BDI.}
  \label{fig:distribution}
\end{figure}

The primary reason why applications have different compression ratios is the 
redundancy difference in their data. This leads to the situation where every application
has its own distribution of compressed pages with different sizes (0B, 512B, 1KB, 2KB, 4KB).
Figure~\ref{fig:distribution} shows these distributions for the applications in our study when using the LCP-BDI design.
As we can see, the percentage of memory pages of every size in fact significantly varies between the applications,
leading to different compression ratios (shown in Figure~\ref{fig:capacity}). For example, 
\emph{cactusADM} has a high compression ratio due to many 0B and 512B pages (there is a significant number of zero cache lines in its data), 
while \emph{astar} and \emph{h264ref} get most of their compression with 2KB pages due to cache lines with low dynamic range~\cite{bdi}.

\subsubsection{Compression Ratio over Time}
\begin{figure}[htb]
  \includegraphics[width=0.95\textwidth]{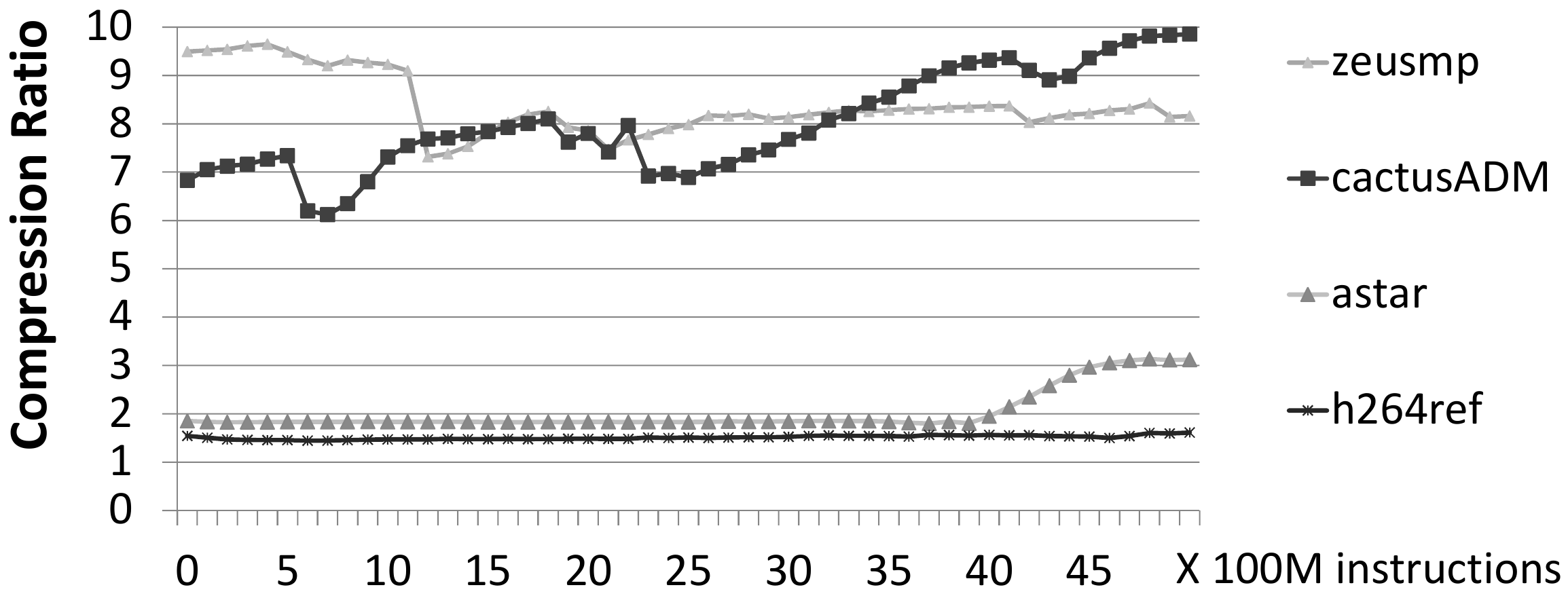}
  \caption{Compression ratio over time with LCP-BDI.}
  \label{lcp:fig:compression}
%  \vspace{-0.1cm}
\end{figure}

To estimate the efficiency of LCP-based compression over time, we conduct
an experiment where we measure the compression ratios of our applications
every 100 million instructions (for a total period of 5 billion instructions).
 The key observation we make is that the compression ratio for most of the applications
is stable over time (the difference between the highest and the lowest ratio is within 10\%). 
Figure~\ref{lcp:fig:compression}
shows all notable outliers from this observation: \emph{astar}, 
\emph{cactusADM}, \emph{h264ref}, and \emph{zeusmp}. 
Even for these applications, the compression ratio stays relatively constant for a long period of time, although
there are some noticeable fluctuations in compression ratio (e.g., for \emph{astar} 
at around 4 billion instructions, for \emph{cactusADM} at around 500M instructions). We attribute this behavior 
to a phase change within an application that sometimes leads to changes in the applications' data. Fortunately, these cases are infrequent and do not
have a noticeable effect on the application's performance (as we describe in Section~\ref{sec:results-perf}). 
We conclude that the capacity benefits provided by the LCP-based frameworks are usually stable over long
periods of time.

%\vspace{0.4cm}
\subsection{Effect on Performance}
\label{sec:results-perf}
Main memory compression can improve performance in two major ways: 
(i) reduced memory bandwidth requirements, which can enable less contention 
on the main memory bus, an
increasingly important bottleneck in systems, and
(ii) reduced memory footprint, which can reduce long-latency disk accesses. 
% In our system performance
%evaluations, we do not take into account the former benefit as we do not model
%disk accesses (i.e., we assume that the uncompressed working set fits entirely
%in memory). However, 
We evaluate the performance improvement due to memory
bandwidth reduction (including our optimizations for compressing zero values
described in Section~\ref{sec:opt-zeros}) in Sections~\ref{lcp:sec:single-core} 
and~\ref{sec:multi-core}.
% Evaluations using our LCP-based frameworks
% show that the performance gains due to the bandwidth reduction more than
% compensate for the slight increase in memory access latency due to decompression.
We also evaluate the decrease in page faults in Section~\ref{sec:page-faults}.

\subsubsection{Single-Core Results}
\label{lcp:sec:single-core}
Figure~\ref{fig:IPC} shows the performance of single-core workloads using 
three key evaluated designs (RMC-FPC, LCP-FPC, and LCP-BDI) normalized to 
the \textit{Baseline}. 
% We draw two major conclusions from the figure.
Compared against an uncompressed system (\textit{Baseline}), the LCP-based designs
(LCP-BDI and LCP-FPC) improve performance by 6.1\%/5.2\% and also 
outperform RMC-FPC.\footnote{Note that
in order to provide a fair comparison, we enhanced the RMC-FPC approach with 
the same optimizations we did for LCP, e.g., bandwidth compression.
The original RMC-FPC design reported an average degradation in performance~\cite{MMCompression}.}
We
conclude that our LCP framework is effective in improving performance by
compressing main memory.

\begin{figure}[htb]
  \centering
  \includegraphics[width=0.95\textwidth]{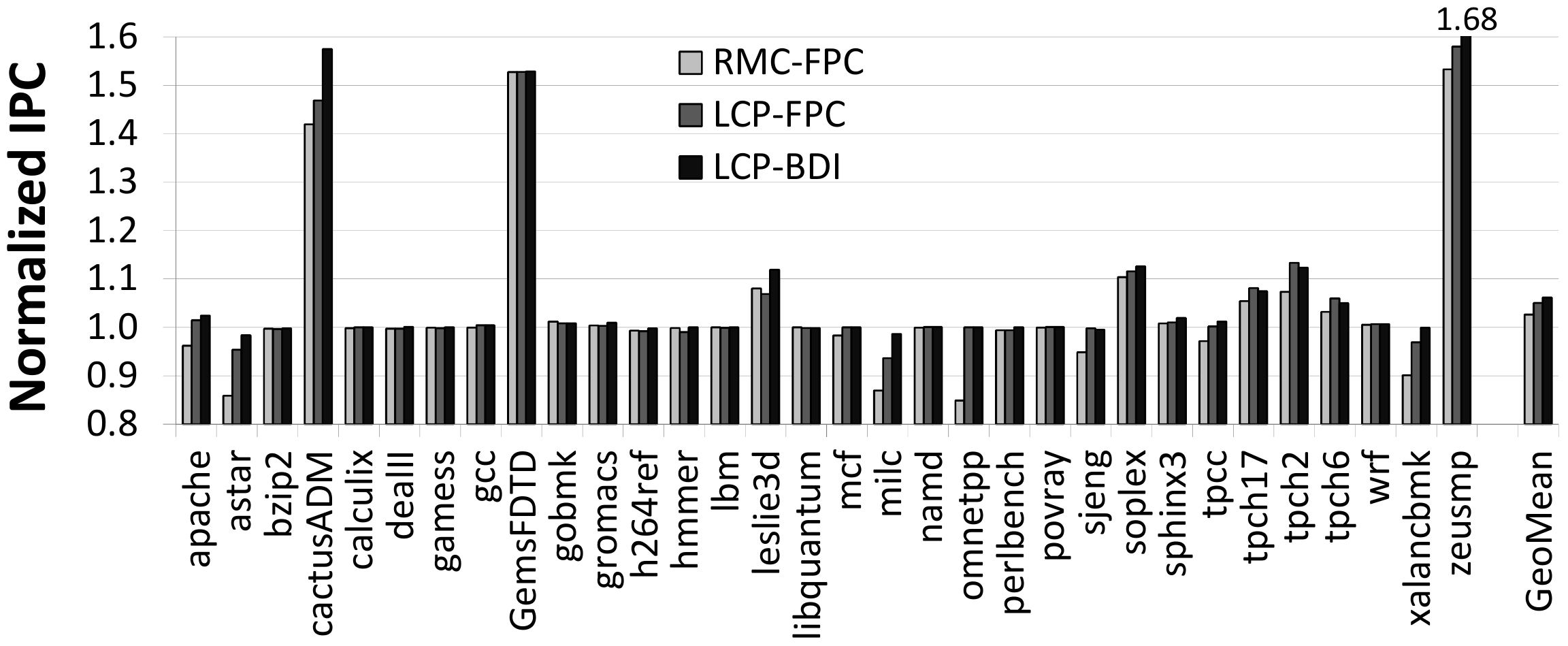}
  \caption{Performance comparison (IPC) of different compressed designs for the single-core system.}
  \label{fig:IPC}
\end{figure}

%Note that the (FPC, FPC) scheme has not been proposed before, but we evaluate it
%for completeness. 

%Figure~\ref{fig:IPC} shows the performance improvement of LCP-based designs
%(NoCompr-BDI, BDI-BDI, BDI-BDI+FPC-fixed) over the baseline design without
%compression, as well as comparison with two prior cache compression designs
%(BDI-NoCompr and FPC-Nocompr in our taxonomy, see
%Section~\ref{sec:design-taxonomy}), and two\footnote{We do not provide
%performance numbers for MXT, because its design requires 32MB L3 that operates
%on large 1kB blocks. Making MXT's last level cache comparable to other designs
%produces the unfairly low performance (DRAM accesses very expensive in MXT).  We
%believe that setting other designs with 1kB cache blocks also gives unfair
%advantage to MXT.} compressed DRAM designs (NoCompr-FPC, FPC-FPC). Note that
%FPC-FPC design never has been proposed before, but we evaluate it for
%completeness.

Note that LCP-FPC outperforms RMC-FPC (on average) despite
having a slightly lower compression ratio. This is mostly 
due to the lower overhead when accessing metadata information 
(RMC-FPC needs two memory accesses to \emph{different} main memory pages
in the case of a BST table miss, while LCP-based framework 
performs two accesses to the same main memory page that can be pipelined). 
This is especially noticeable in several applications, e.g., \emph{astar},
\emph{milc}, and \emph{xalancbmk} that have low metadata table (BST) hit rates (LCP can
also degrade performance for these applications).
We conclude that our LCP framework is more effective in improving
performance than RMC~\cite{MMCompression}.

\subsubsection{Multi-Core Results}
\label{sec:multi-core}
When the system has a single core, the memory bandwidth pressure may
not be large enough to take full advantage of the bandwidth benefits of main
memory compression.
% (explained in Section~\ref{sec:opt-bandwidth}).
However, in a multi-core system where multiple applications are
running concurrently, savings in bandwidth (reduced number of memory
bus transfers) may significantly increase the overall system
performance.

%When only a single application is running in the system, the bandwidth pressure
%may not be significant enough to get the benefit from bandwidth compression.
%However, if multiple applications are running concurrently, sending multiple
%requests to DRAM at the same time, the decrease in number of transfers over the
%bus between LLC and DRAM becomes more critical to the overall system
%performance.

\begin{figure}[htb]
  %\centering
  \centering
  \includegraphics[width=0.6\textwidth]{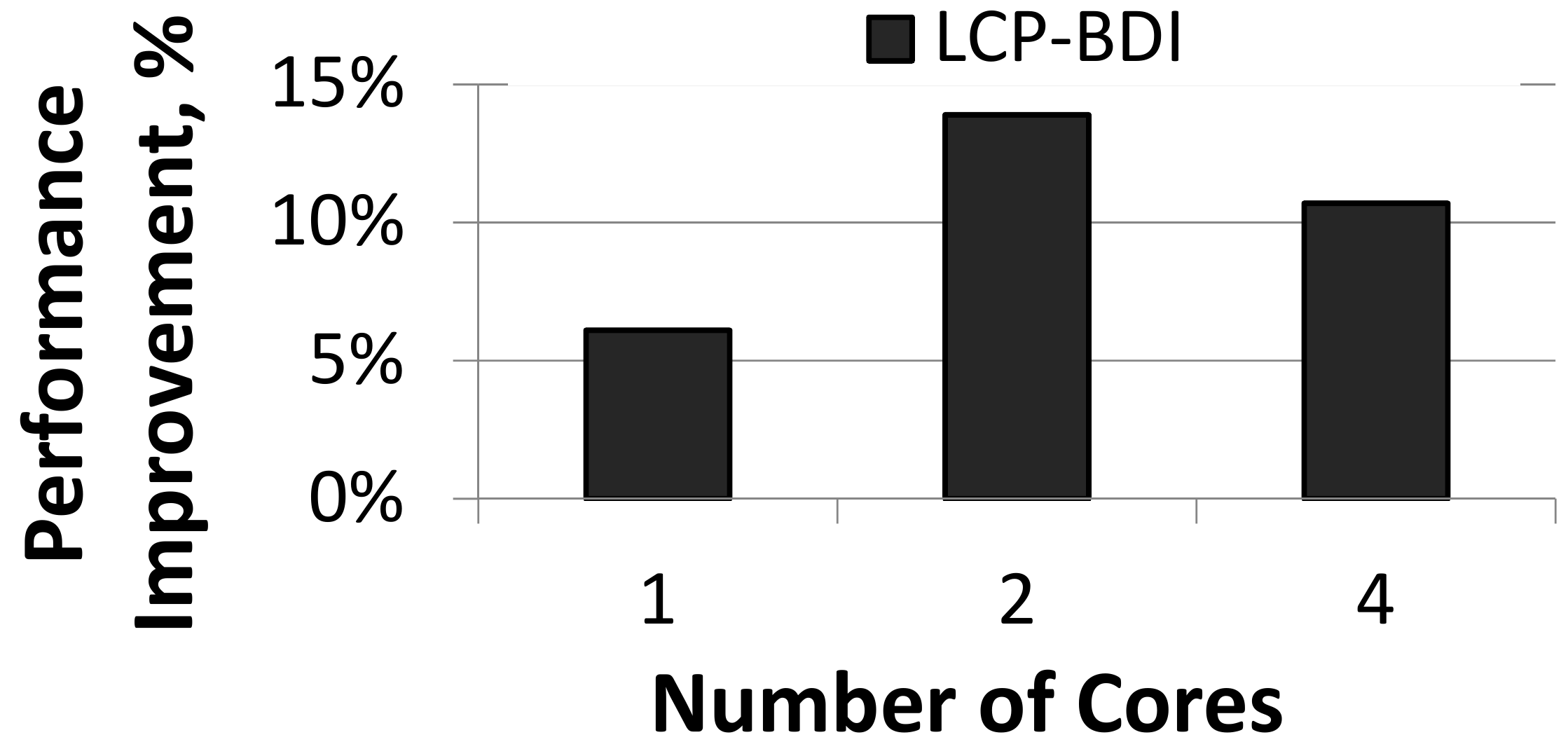}
  \caption{Average performance improvement (weighted speedup).}
  \label{fig:ws}
\end{figure}

To study this effect, we conducted experiments using 100 randomly
generated multiprogrammed mixes of applications (for both 2-core and
4-core workloads). Our results show that the bandwidth benefits of memory 
compression are
indeed more pronounced for multi-core workloads. Using our LCP-based
design, LCP-BDI, the average performance
improvement (normalized to the performance of the \textit{Baseline}
  system without compression) is 13.9\% for 2-core workloads and
10.7\% for 4-core workloads. We summarize our multi-core performance
results in Figure~\ref{fig:ws}.

We also vary the last-level cache size (1MB -- 16MB) for both single 
core and multi-core systems across all evaluated workloads. We find 
that LCP-based designs outperform the \emph{Baseline} across all evaluated 
systems (average performance improvement for single-core 
varies from 5.1\% to 13.4\%), 
even when the L2 cache size of the system is as large as 16MB.

 \begin{figure}[htb]
  \centering
  \includegraphics[width=0.6\textwidth]{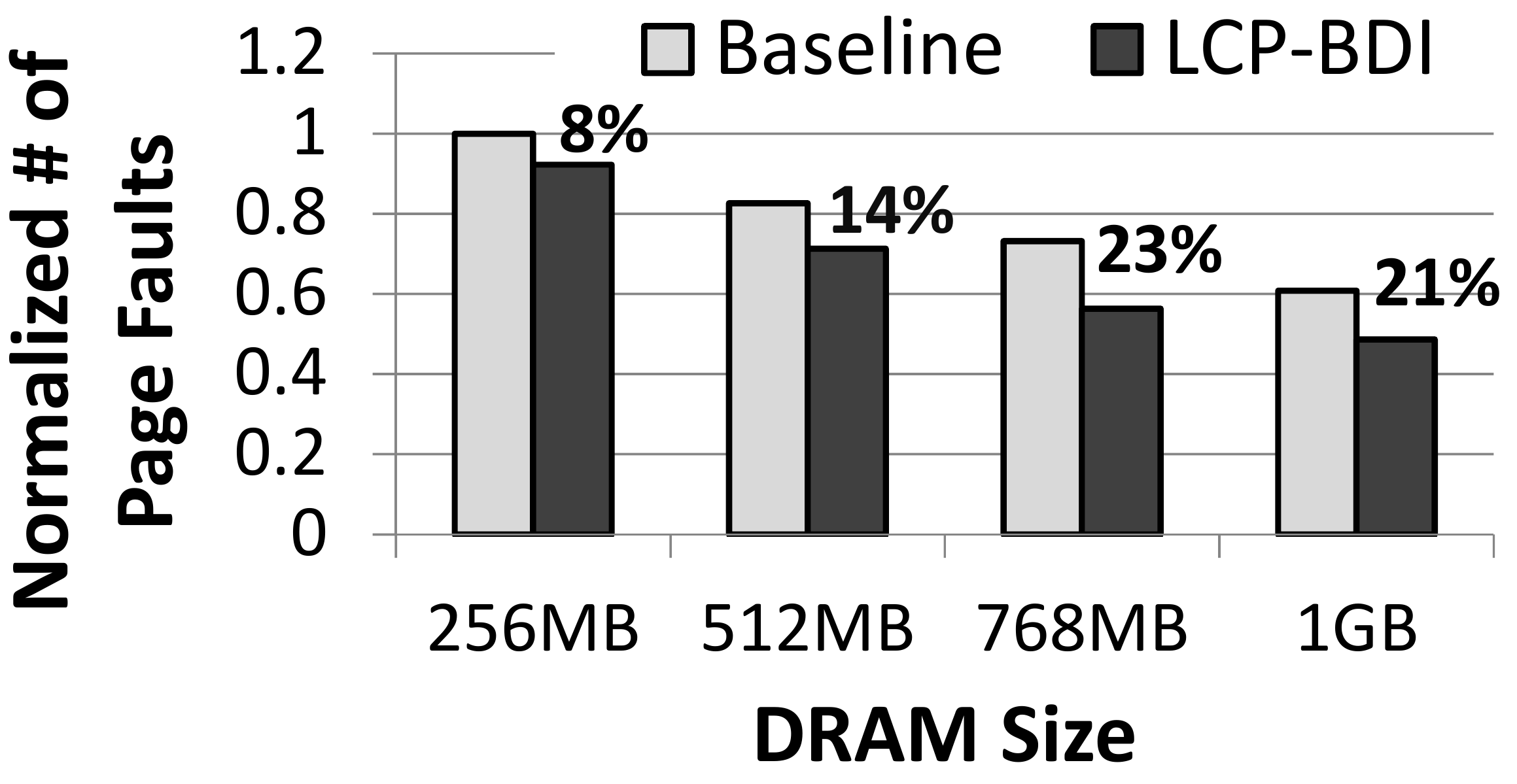}
  \caption{Number of page faults (normalized to \textit{Baseline} with 256MB).}
  \label{fig:pf}
%\caption{Performance (with 2 GB DRAM) and number of page faults (varying DRAM
% size) using LCP-BDI.}
% \vspace{0.1cm}
\end{figure}

\subsubsection{Effect on the Number of Page Faults}
\label{sec:page-faults}
Modern systems are usually designed such that concurrently-running 
applications have enough main memory to avoid most of the potential 
capacity page faults. At the same time, if the applications' total working set size
exceeds the main memory capacity, the 
increased number of page faults can significantly affect performance.
To study the effect of the LCP-based framework (LCP-BDI) on the number of page faults, 
we evaluate twenty randomly generated 
16-core multiprogrammed mixes of applications from our benchmark
set. We also vary the main memory capacity from 256MB to 1GB (larger memories usually
lead to almost no page faults for these workload simulations). 
Our results (Figure~\ref{fig:pf}) show that the LCP-based framework
(LCP-BDI) can decrease the number of page faults by 21\% on average (for 1GB DRAM)
when compared with the \textit{Baseline} design with no compression. We conclude
that the LCP-based framework can significantly decrease the number of page
faults, and hence improve system performance beyond the benefits it provides due
to reduced bandwidth.

\subsection{Effect on Bus Bandwidth and Memory Subsystem Energy}
\label{sec:results-bandwidth}
\label{sec:results-energy}

When DRAM pages are compressed, the traffic between the LLC and DRAM can be
reduced. This can have two positive effects: (i) reduction in the
average latency of memory accesses, which can lead to improvement in the overall
system performance, and (ii) decrease in the bus energy consumption due to the
decrease in the number of transfers. 

%In a system, where DRAM pages are stored in the compressed form, there is an
%opportunity to compress the traffic between the LLC and DRAM. This can have
%multiple positive effects: 1) reduction in average latency of memory accesses,
%which can lead to improvement in the overall system performance, 2) decrease in
%the bus power consumption due to the decrease in the number of transfers. 

Figure~\ref{fig:bandwidth} shows the reduction in main memory
bandwidth between LLC and DRAM (in terms of bytes per kilo-instruction,
normalized to the \textit{Baseline}
system with no compression) using different
compression designs. 
The key observation we make from this figure is that 
there is a strong correlation between bandwidth compression
and performance improvement (Figure~\ref{fig:IPC}). Applications that
show a significant reduction in bandwidth consumption (e.g., \emph{GemsFDTD},
\emph{cactusADM}, \emph{soplex}, \emph{zeusmp}, \emph{leslie3d}, 
and the four \emph{tpc} queries) also see large performance
improvements. There are some noticeable exceptions to this
observation, e.g., \emph{h264ref}, \emph{wrf} and \emph{bzip2}. Although the memory bus
traffic is compressible in these applications, main memory bandwidth
is not the bottleneck for their performance.

\begin{figure}[htb]
  \centering
  \includegraphics[width=0.95\textwidth]{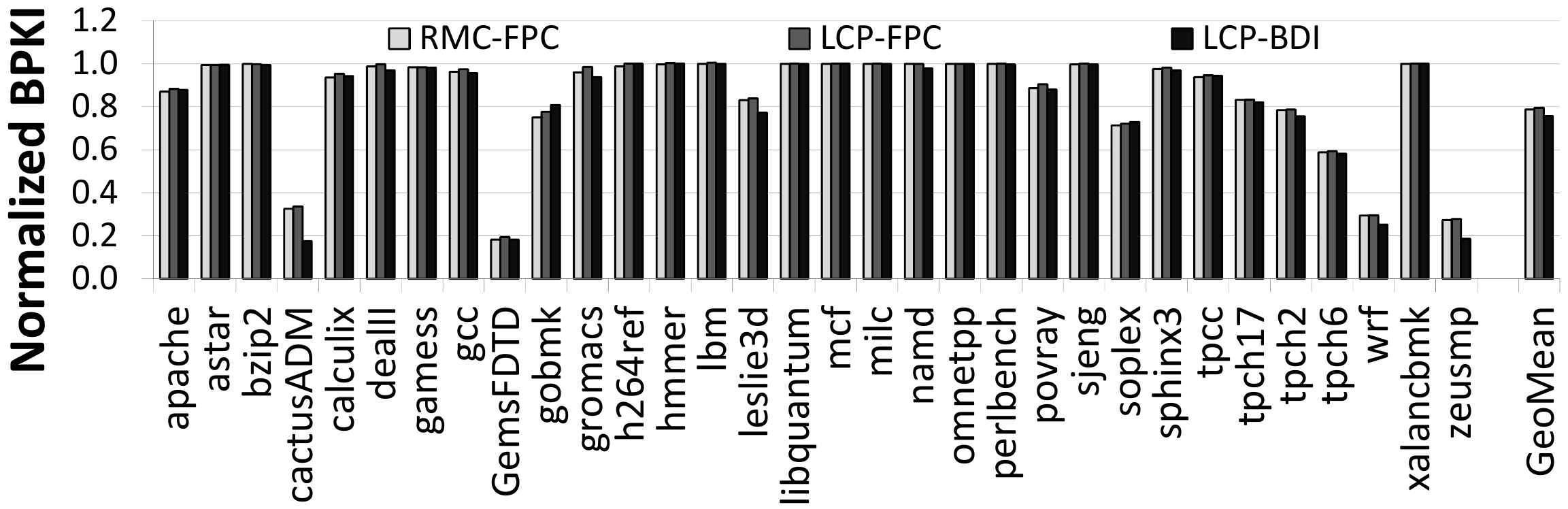}
  \caption{Effect of different main memory compression schemes on memory bandwidth.}
  \label{fig:bandwidth}
\end{figure}

\begin{figure}[htb]
  \centering
  \includegraphics[width=0.95\textwidth]{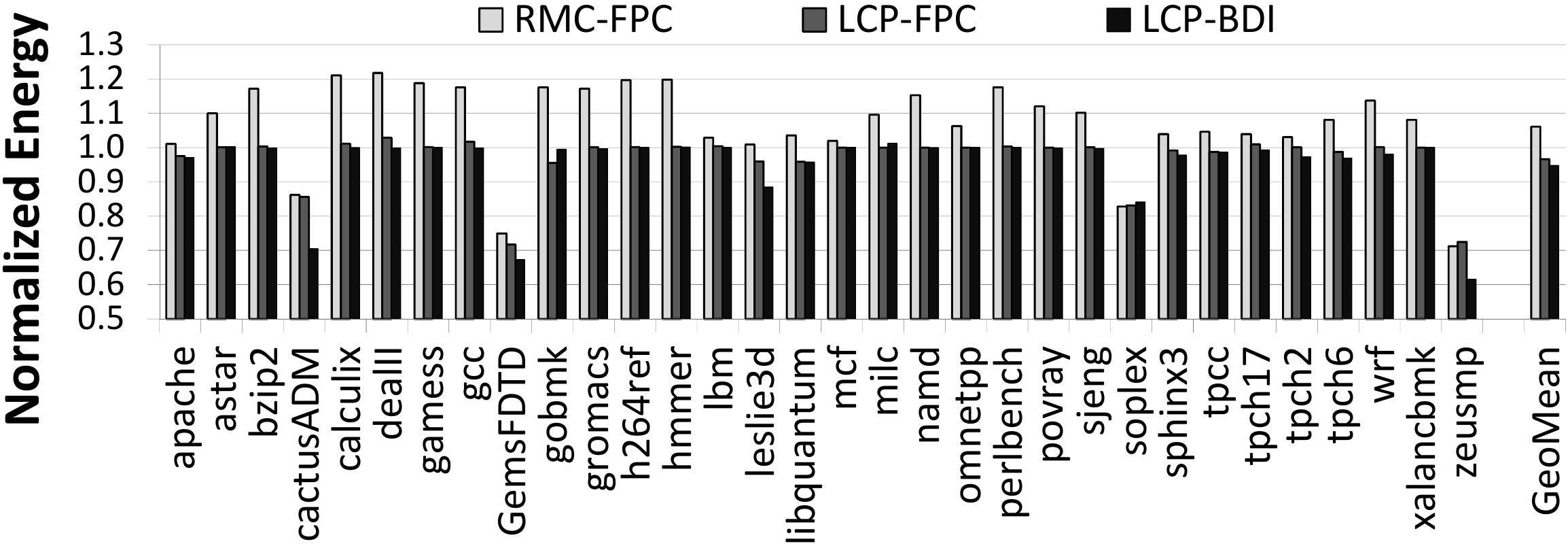}
  \caption{Effect of different main memory compression schemes on memory subsystem energy.}
  \label{lcp:fig:energy}
\end{figure}

Figure~\ref{lcp:fig:energy} shows the reduction in memory subsystem
 energy of three systems that employ main memory compression---RMC-FPC,
LCP-FPC, and LCP-BDI---normalized to the energy of \textit{Baseline}.
The memory subsystem energy includes the static and
dynamic energy consumed by caches, TLBs, memory transfers, and DRAM,
plus the energy of additional components due to main memory compression: 
BST~\cite{MMCompression}, MD cache,
address calculation, compressor/decompressor units.
Two observations are in order.

First, our LCP-based designs (LCP-BDI and LCP-FPC) improve the memory subsystem energy by 5.2\% / 3.4\% on 
average over the \textit{Baseline}
design with no compression, and by 11.3\% / 9.5\%  over the state-of-the-art design (RMC-FPC)
based on~\cite{MMCompression}. This is especially noticeable for bandwidth-limited
applications, e.g., \emph{zeusmp} and \emph{cactusADM}. We conclude that our framework for main memory
compression enables significant energy savings, mostly due to the decrease
in bandwidth consumption.

Second, RMC-FPC consumes significantly more energy than \textit{Baseline}
(6.1\% more energy on average, as high as 21.7\% for \emph{dealII}). The primary
reason for this energy consumption increase is the physical address calculation that RMC-FPC
speculatively performs on \emph{every} L1 cache miss (to avoid increasing the memory latency
due to complex address calculations). 
The second reason is the frequent (every L1 miss) accesses
to the BST table (described in Section~\ref{lcp:sec:background}) that holds the 
address calculation information. 

Note that other factors, e.g., compression/decompression energy overheads or 
different compression ratios, are not the reasons for this energy consumption increase. 
LCP-FPC uses the same compression algorithm as RMC-FPC (and even has a slightly lower
compression ratio), but does not increase energy consumption---in fact, LCP-FPC
improves the energy consumption due to its decrease in consumed bandwidth.
We conclude that our LCP-based framework is a more energy-efficient main memory compression
framework than previously proposed designs such as RMC-FPC.

\subsection{Analysis of LCP Parameters}
\subsubsection{Analysis of Page Overflows}

\begin{figure}[htb]
  \centering
  \includegraphics[width=0.95\textwidth]{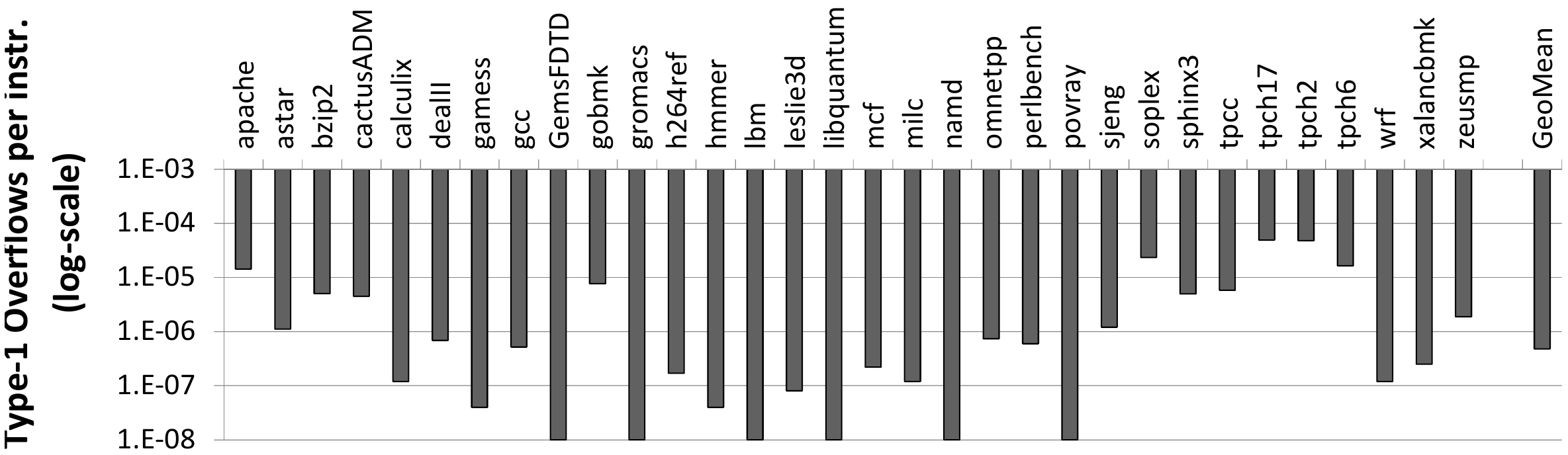}
  \caption{Type-1 page overflows for different applications.}
  \label{fig:overflows}
\end{figure}

As described in Section~\ref{sec:design-handling-overflows}, page
overflows can stall an application for a considerable duration. As we
mentioned in that section, we did not encounter any type-2 overflows
(the more severe type) in our simulations. Figure~\ref{fig:overflows}
shows the number of type-1 overflows per instruction. The y-axis uses
a log-scale as the number of overflows per instruction is very
small. As the figure shows, on average, less than one type-1 overflow
occurs every one million instructions. Although such overflows are
more frequent for some applications (e.g., \emph{soplex} and the three 
\emph{tpch} queries), our
evaluations show that this does not degrade performance in spite of
adding a 20,000 cycle penalty for each type-1 page
overflow.\footnote{We varied the type-1 overflow latency from 10,000 to
  100,000 cycles and found that the impact on performance was
  negligible as we varied the latency. Prior work on main memory
  compression~\cite{MMCompression} also used 10,000 to 100,000 cycle
  range for such overflows.}  In fact, these applications gain
significant performance from our LCP design. The main reason for this
is that the performance benefits of bandwidth reduction far outweigh
the performance degradation due to type-1 overflows. We conclude that
page overflows do not prevent the proposed LCP framework from
providing good overall performance.

%% Page size of overflow (increase over the current compressed page size limit) is
%% the most time consuming event that can be triggered in LCP framework.  As
%% described in Section~\ref{sec:methodology}, more common inner
%% overflows\footnote{As demonstrated in \cite{MMCompression} and in our own
%%   experiments, overflows over the uncompressed page size are orders of magnitude
%%   less frequent than inner overflows, so we united both of them in our results
%%   presentation for simplicity.} can take up to 10000 cycles, and this page is
%% locked during the required recompression. So, it is critical to show that such
%% events are rare in applications' execution.

%% Figure~\ref{fig:overflows} shows the number of overflows per committed
%% instruction, and since the numbers are relatively small, we put them in the
%% log-scale (the shorter the bar - the more frequent are page size overflows). As
%% we can see, the average number of overflows per instruction is less than
%% $10^{-6}$, which means a single overflow happens less than once per one million
%% instruction.  At the same time, there are few applications, e.g., soplex, that
%% have relatively high number of page size overflows, but our evaluations show
%% that it does significantly decrease performance of these few applications,
%% mostly because requests to other pages can still be serviced in parallel. We
%% conclude that page size overflows will typically not prevent LCP-based designs
%% from achieving good overall performance.

%\vspace{-0.05cm}
\subsubsection{Number of Exceptions}

\begin{figure}[htb]
  \centering
  \includegraphics[width=0.95\textwidth]{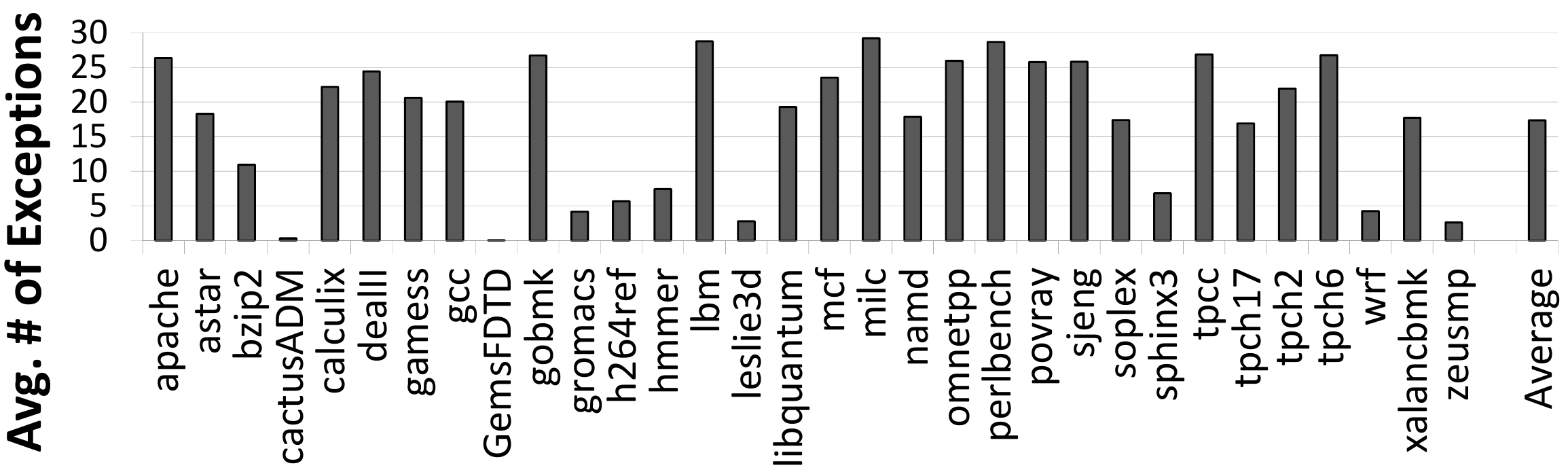}
  \caption{Average number of exceptions per compressed page for different applications.}
  \label{fig:exceptions}
%  \vspace{0.3cm}
\end{figure}

The number of exceptions (uncompressed cache lines) in the LCP framework is critical for two
reasons. First, it determines the size of the physical page required
to store the LCP. The higher the number of exceptions, the larger the
required physical page size. Second, it can affect an application's
performance as exceptions require three main memory accesses on an MD
cache miss (Section~\ref{sec:basic-mcf-overview}). We studied the
average number of exceptions (across all compressed pages) for each
application. Figure~\ref{fig:exceptions} shows the results of these
studies.

The number of exceptions varies from as low as 0.02/page for \emph{GemsFDTD}
to as high as 29.2/page in \emph{milc} (17.3/page on average). The average
number of exceptions has a visible impact on the compression ratio of
applications (Figure~\ref{fig:capacity}). An application with a high
compression ratio also has relatively few exceptions per page. Note
that we do not restrict the number of exceptions in an LCP. As long as
an LCP fits into a physical page not larger than the uncompressed page
size (i.e., 4KB in our system), it will be stored in compressed
form irrespective of how large the number of exceptions is. This is
why applications like \emph{milc} have a large number of exceptions per
page. We note that better performance is potentially achievable by
either statically or dynamically limiting the number of exceptions per
page---a complete evaluation of the design space is a part of our
future work.

%% One of the critical parameters in the LCP page organization is the total number
%% of elements in the list of exceptions. This list defines a flexible part in the
%% compressed page organization, and for the best compressibility we would like to
%% keep it as small as possible. So, it is interesting to conduct an experiment
%% where the average number of exceptions per page is evaluated, and, hence,
%% indirectly see the effectiveness of our compression framework with the selected
%% compression algorithms.

%% Figure~\ref{fig:exceptions} presents the results of this experiment. As the
%% figure shows, the number of exceptions can vary significantly from as low as
%% 0.02 exceptions per page in GemsFDFD to as high as 29.2 in milc (average is
%% 17.3). These results reflects the effective compression ratios we observed in
%% Figure~\ref{fig:capacity}, typically application with high compression ratio has
%% relatively low number of exclusions.  One of the reasons why the number of
%% exceptions can be so high for some applications is that we do not restrict the
%% number of exceptions artificially in any way, so if the compressed page fits
%% into some permitted compressed page size, e.g., 512B or 4kB, then it would be
%% stored in the compressed form.

%\end{comment}
%\vspace{0.2cm}
\subsection{Comparison to Stride Prefetching}
\label{sec:results-prefetching-hints}

\begin{figure}[thb]
%  \centering
%  \includegraphics[width=0.49\textwidth]{figures/PrefetchIPC.pdf}
  \includegraphics[width=0.9\textwidth]{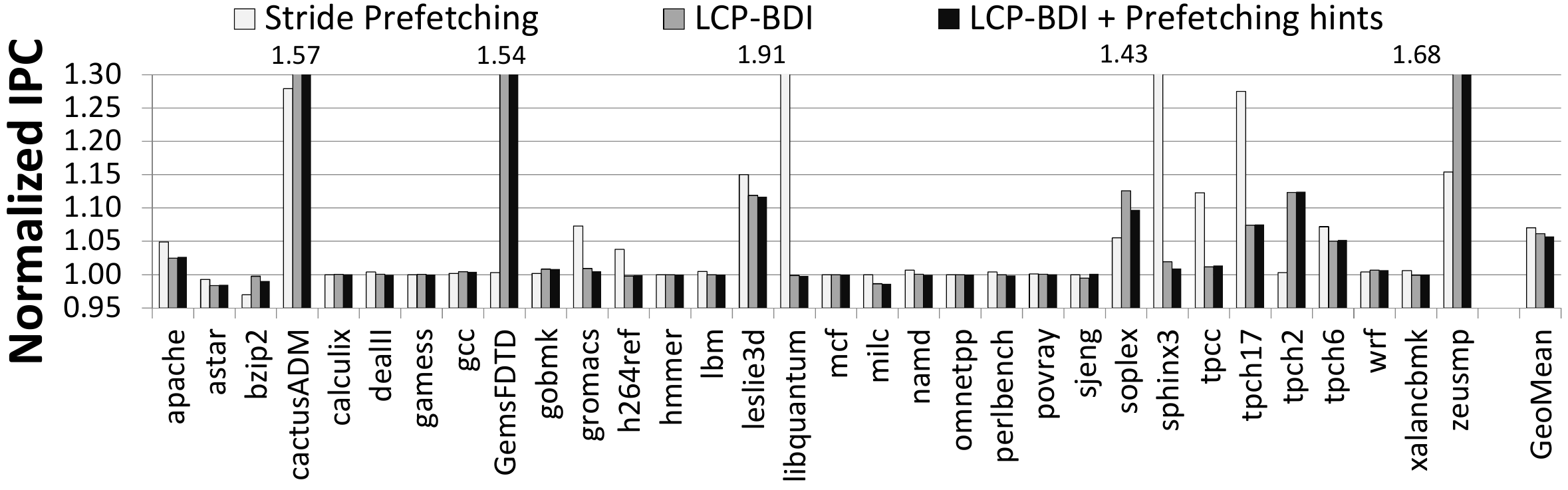}
  \caption{Performance comparison with stride prefetching, and using prefetcher
    hints with the LCP-framework.}
  \label{fig:pref-ipc}
\end{figure}

\begin{figure}[htb]
%  \centering
%\hspace{-0.1in}
%  \includegraphics[width=0.49\textwidth]{figures/PrefetchBandwidth.pdf}
  \includegraphics[width=0.9\textwidth]{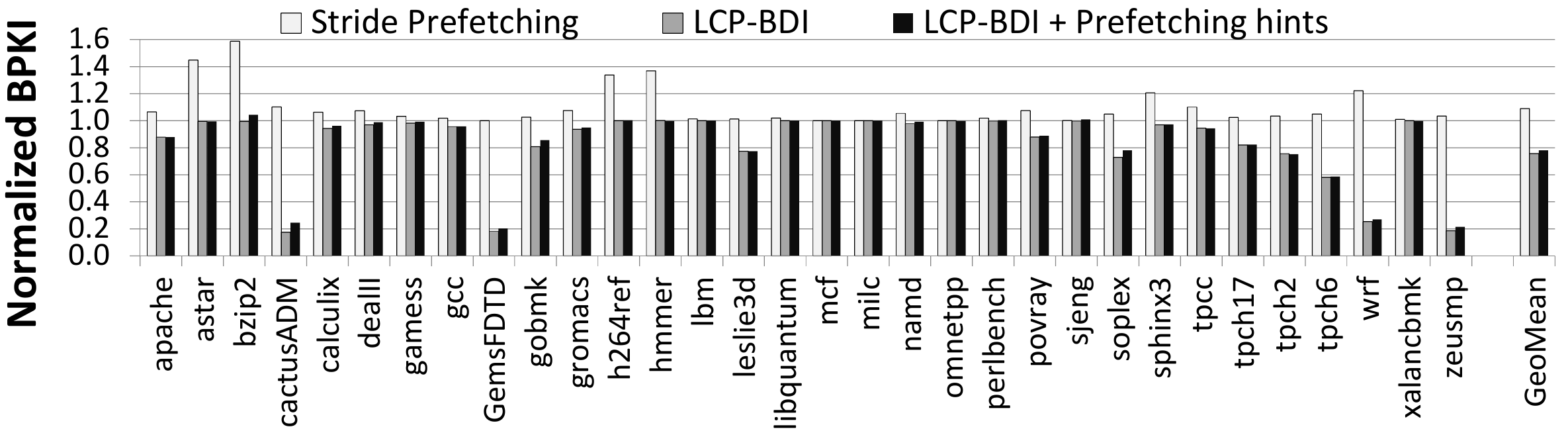}
  \caption{Bandwidth comparison with stride prefetching.}
  \label{fig:pref-bandwidth}
%\vspace{-0.1cm}
\end{figure}

Our LCP-based framework improves performance due to its ability to
transfer multiple compressed cache lines using a single memory
request. Because this benefit resembles that of prefetching cache lines
into the LLC, we compare our LCP-based design to a
system that employs a stride
prefetcher implemented as described in \cite{stride-prefetching}. Figures~\ref{fig:pref-ipc} and
\ref{fig:pref-bandwidth} compare the performance and bandwidth
consumption of three systems: (i)~one that employs stride prefetching,
(ii)~one that employs LCP-BDI, and (iii)~one that employs LCP-BDI along with hints
from a prefetcher to avoid cache pollution due to bandwidth compression
(Section~\ref{sec:opt-bandwidth}). Two conclusions are in order.

%%% ONUR: What isthe configuration of the prefetcher?

%%% ONUR: There is a lot of potential here... You can do very
%%% interesting research and write a great paper. All resonable
%%% systems employ prefetching, so exploring this interaction is
%%% critical.

%%% ONUR: I see Alaa's paper on prefetching + compression is not cited
%%% here. Any particular reason? Citing widely is important. It is a
%%% relevant paper. If you are stingy about citing, you will run into
%%% problems.

First, our LCP-based designs (second and third bars) are competitive 
with the more general stride prefetcher for all but a few applications (e.g.,
\emph{libquantum}). The primary reason is that a stride prefetcher
can sometimes increase the memory bandwidth consumption of an
application due to inaccurate prefetch requests. On the other hand,
LCP obtains the benefits of prefetching without increasing (in fact,
while significantly reducing) memory bandwidth consumption.

Second, the effect of using prefetcher hints to avoid cache pollution
is not significant. The reason for this is that our systems employ a
large, highly-associative LLC (2MB 16-way) which is less susceptible
to cache pollution. Evicting the LRU lines from such a cache has
little effect on performance, but we did observe the benefits of this
mechanism on multi-core systems with shared caches (up to 5\% performance improvement
for some two-core workload mixes---not shown).

\section{Summary}
\label{lcp:sec:conclusion}

%With increasing number of cores integrated on a single chip and increasing
%memory footprint of individual applications, there is an increasing demand for
%main memory capacity. 
% in future systems. 
Data compression is a promising
technique to increase the effective main memory capacity without significantly
%increasing the cost and power consumption of the system. 
increasing cost and power consumption.  As we described in this chapter,
the primary challenge in incorporating compression in main memory is
to devise a mechanism that can efficiently compute the main memory
address of a cache line without significantly adding complexity, cost,
or latency. Prior approaches to addressing this challenge are either
relatively costly or energy inefficient.

We proposed a new main memory compression framework, called \emph{Linearly
  Compressed Pages} (LCP), to address this problem. The two key ideas of LCP are to use a fixed
size for compressed cache lines within a page (which simplifies main
memory address computation) and to enable a page to be compressed even
if some cache lines within the page are incompressible (which enables
high compression ratios). We showed that any compression algorithm can be
adapted to fit the requirements of our LCP-based framework.

We evaluated the LCP-based framework using two state-of-the-art
compression algorithms (Frequent Pattern Compression and
Base-Delta-Immediate Compression) and showed that it can significantly
increase effective memory capacity (by 69\%) and reduce page fault rate (by 23\%). 
We showed that storing
compressed data in main memory can also enable the memory controller
to reduce memory bandwidth consumption (by
24\%), leading to significant performance and energy
improvements on a wide variety of single-core and multi-core systems
with different cache sizes. 
%We illustrate that our framework
%integrates well with compression in last-level caches. 
Based on our
results, we conclude that the proposed LCP-based framework provides an
effective approach for designing low-complexity and low-latency 
compressed main memory.

\chapter{Toggle-Aware Bandwidth Compression}

\section{Introduction}
\label{sec:introduction}
\blfootnote{Originally published as ``Toggle-Aware Bandwidth Compression for GPUs'' 
in the 22nd International Symposium on High Performance
Computer Architecture, 2016~\cite{toggles-hpca}, and as
``Toggle-Aware Compression for GPUs'' in Computer Architecture Letters, 2015~\cite{toggles-cal}.}

%%% ONUR: I removed references to the three walls. These are obvious
%%% and these papers are actually not the first to introduce them,
%%% really. Better to save our real estate space to more relevant
%%% references, e.g., in compression.

Modern data-intensive computing forces system designers to deliver
good system performance under multiple constraints: shrinking power
and energy envelopes ({\em power wall}), increasing memory latency
({\em memory latency wall}), and scarce and expensive bandwidth
resources ({\em bandwidth wall}). While many different techniques have
been proposed to address these issues, these techniques often offer a
trade-off that improves one constraint at the cost of
another. Ideally, system architects would like to improve one or more
of these system parameters, e.g., on-chip and off-chip\footnote{Communication
channel between the last-level cache and main memory.} bandwidth
consumption, while simultaneously avoiding negative effects on other
key parameters, such as overall system cost, energy, and latency
characteristics. One potential way of addressing multiple constraints
is to employ dedicated hardware-based \emph{data compression}
mechanisms (e.g.,~\cite{fvc,fpc,c-pack,bdi,sc2}) across different data
links in the system. Compression exploits the high data redundancy
observed in many modern applications~\cite{bdi,dcc,sc2,caba} and can
be used to improve both capacity (e.g., of caches, DRAM, non-volatile
memories~\cite{fvc,fpc,c-pack,bdi,sc2,lcp-micro,memzip,camp,caba,buri}) and
bandwidth utilization (e.g., of on-chip and off-chip interconnects
~\cite{reetu,CompressionPrefetching,LinkCompression,GPUBandwidthCompression,lcp-micro,memzip,caba}). Several
recent works focus on bandwidth compression to decrease memory traffic
by transmitting data in a compressed form in both
CPUs~\cite{lcp-micro,LinkCompression,CompressionPrefetching} and
GPUs~\cite{GPUBandwidthCompression,lcp-micro,caba}, which results in
better system performance and energy consumption. Bandwidth
compression proves to be particularly effective in GPUs because they
are often bottlenecked by memory
bandwidth~\cite{veynu,osp-isca13,OWL,sched,caba,SchedPIM,TOM,sch6}.  GPU applications
also exhibit high degrees of data
redundancy~\cite{GPUBandwidthCompression,lcp-micro,caba}, leading to
good compression ratios.
%\vspace{-0.2cm}
%\subsection{Data Compression can be Energy-inefficient} 
While data compression can dramatically reduce the number of bit
symbols that must be transmitted across a link, compression also
carries two well-known overheads: (1) latency, energy, and area
overhead of the compression/decompression hardware~\cite{fpc,bdi}; and
(2) the complexity and cost to support variable data
sizes~\cite{iic-comp,dcc,lcp-micro,memzip}. Prior work has addressed
solutions to both of these problems. For example, Base-Delta-Immediate
compression~\cite{bdi} provides a low-latency, low-energy
hardware-based compression algorithm. Decoupled and Skewed Compressed
Caches~\cite{dcc,skewedCompressedCache} provide a mechanism to efficiently manage data
recompaction and fragmentation in compressed caches.

\subsection{Compression \& Communication Energy} 
In this chapter, we make a new observation that there is yet another
important problem with data compression that must be addressed to
implement energy-efficient communication: transferring data
in compressed form (as opposed to uncompressed form) leads to a
significant increase in the number of {\em bit toggles}, i.e., the
number of wires that switch from 0 to 1 or 1 to 0\@. An increase in
bit toggle count causes higher switching
activities~\cite{nuca,tlc,desc} of wires, leading to higher dynamic
energy being consumed by on-chip or off-chip interconnects.  The bit toggle
count increases for two reasons. 
First, the compressed data has a higher per-bit entropy because the
same amount of information is now stored in fewer bits (the
``randomness'' of a single bit grows).
Second, the variable-size nature of compressed data, which can negatively
affect the word/flit data alignment in hardware.  Thus, in contrast to
the common wisdom that bandwidth compression saves energy (when it is
effective), our key observation reveals a new trade-off: energy savings
obtained by reducing bandwidth versus energy loss due to higher
switching energy during compressed data transfers. This observation
and the corresponding trade-off are the key contributions of this work.

%impedance. 
%At the same time, bandwidth
%compression, when applicable, may improve overall system energy through better
%performance (by providing a higher effective bandwidth). 
%Hence, in contrast to
%the common wisdom about bandwidth
%compression~\cite{GPUBandwidthCompression,memzip}, our observation offers an interesting
%trade-off: higher available bandwidth versus potentially higher dynamic energy
%of the data transfers. 

To understand (1) how applicable general-purpose data compression is
for real GPU applications, and (2) the severity of the problem, we use
six compression algorithms to analyze 221 discrete and mobile graphics
application traces from a major GPU vendor and 21 open-source,
general-purpose GPU applications. Our analysis shows that although
off-chip bandwidth compression achieves a significant compression
ratio (e.g., more than 47\% average effective bandwidth increase with
C-Pack~\cite{c-pack} across mobile GPU applications), it also
greatly increases the bit toggle count (e.g., 2.2$\times$ average
corresponding increase). This effect can significantly
increase the energy dissipated in the on-chip/off-chip interconnects,
which constitute a significant portion of the memory subsystem
energy.

% SK: this footnote doesn't seem necessary. If it is important enough,
% put it in the main text, not a footnote.
% 
% \footnote{For example, up to 80\% energy of the LLC caches is H-tree
%   capacitance interconnects~\cite{desc}.}
\vspace{-0.2cm}
\subsection{Toggle-Aware Compression} 
In this work, we develop two new techniques that make bandwidth compression for
on-chip/off-chip buses more energy-efficient by limiting the overall increase
in compression-related bit toggles. \emph{Energy Control (EC)} decides whether
to send data in compressed or uncompressed form, based on a model that accounts
for the compression ratio, the increase in bit toggles, and current bandwidth
utilization. The key insight is that this decision can be made in a
fine-grained manner (e.g., for every cache line), using a simple model to
approximate the commonly-used $Energy \times Delay$ and $Energy \times Delay^2$
metrics. In this model, $Energy$ is directly proportional to the bit toggle
count; $Delay$ is inversely proportional to the compression ratio and directly
proportional to the bandwidth utilization. Our second technique, \emph{Metadata
Consolidation (MC)}, reduces the negative effects of scattering the metadata
across a compressed cache line, which happens with many existing compression
algorithms~\cite{fpc,c-pack}. Instead, MC consolidates compression-related
metadata in a contiguous fashion.

Our toggle-aware compression mechanisms are generic and applicable to
different compression algorithms (e.g., Frequent Pattern Compression
(FPC)~\cite{fpc} and Base-Delta-Immediate (BDI)
compression~\cite{bdi}), different communication channels (on-chip/off-chip buses), 
and different architectures (e.g., GPUs,
CPUs, and hardware accelerators).  We demonstrate that these
mechanisms are mostly orthogonal to different data encoding schemes
also used to minimize the bit toggle count (e.g., Data Bus
Inversion~\cite{dbi}), and hence can be used together with them to
enhance the energy efficiency of interconnects.

Our extensive evaluation shows that our proposed mechanisms
can significantly reduce the negative effect of bit toggling
increase (in some cases the 2.2$\times$ increase
in bit toggle count is completely eliminated), while preserving most of the benefits
of data compression when it is useful -- hence the reduction in performance benefits from compression
is usually within 1\%. 
This efficient trade-off leads to the reduction 
in (i) the DRAM energy that is as high as 28.1\% for some applications (8.3\% average reduction),
and (ii) the total system energy (at most 8.9\%, 2.1\% on average).
Moreover, we can dramatically reduce the energy cost to support data compression over the on-chip interconnect.
For example, our toggle-aware compression mechanisms 
can reduce the original 2.1$\times$ increase in consumed energy with C-Pack compression algorithm to much more acceptable 1.1$\times$ increase.

\begin{comment}

In summary, we make the following contributions: 
\begin{itemize} 
\item
We make a new observation that hardware-based bandwidth compression applied to
on-chip/off-chip communication interfaces poses a new challenge for system
designers: a potentially significant increase in the bit toggle count as a
result of data compression. Without proper care, this increase can lead to
significant energy overheads when transferring compressed data that was not
accounted for in prior works.

\item We propose a set of new mechanisms to address this new challenge: Energy
Control, and Metadata Consolidation.

\item We provide a detailed analysis and evaluation of a large
  spectrum of GPU applications that justify both the usefulness of
  data compression for bandwidth compression in many real
  applications, as well as the existence of the bit toggle problem for
  bandwidth compression. Our proposed solutions can deliver most of
  the benefits of bandwidth compression with only minor increase in
  energy consumption, in contrast to 2.2$\times$ growth in the energy
  consumption with the baseline compressed design.

\end{itemize}
\end{comment}

\section{Background}
\label{toggles:sec:background}
%\subsection{Baseline GPU Architecture}

%\input{sections/2_1_gpu_background}

%\textbf{Data Compression.}
Data compression is a powerful mechanism that exploits the existing redundancy
in the applications' data to relax capacity and bandwidth requirements for many
modern systems.  Hardware-based data compression was explored in the context of
on-chip caches~\cite{fvc,fpc,c-pack,bdi,dcc,sc2} and main
memory~\cite{MXT,LinkCompression,MMCompression,lcp-micro,memzip}, but mostly
for CPU-oriented applications.  Several prior
works~\cite{LinkCompression,lcp-micro,GPUBandwidthCompression,memzip,caba,Reetu1} looked at
the specifics of memory bandwidth compression, where it is very critical to
decide where and when to perform compression and decompression.  

While these
works looked at energy/power benefits of bandwidth compression, the overhead of
compression was limited to the overhead of compression/decompression logic and
the overhead of the newly proposed mechanisms/designs. To the best of our
knowledge, this is the first work that looks at energy implications of
compression on the data transferred over on-chip/off-chip buses.
%\textbf{Energy-Efficient Data Communication.}
Depending on the type of the communication channel the data bits transferred have different effect on the energy
spent on communication. We summarize this effect for three major communication channel types.

\textbf{On-chip Interconnect.} For the full-swing on-chip interconnects, one of the dominant factors that defines the
energy cost of a single data transfer (commonly called a flit) is the activity
factor{\textemdash}the number of \emph{bit toggles} on the wires (communication
channel switchings from 0 to 1 or from 1 to 0).  The bit toggle count for a
particular flit depends on both the current flit's data and on the data that
was just sent over the same wires.  Several prior
works~\cite{dbi,desc,zhang-1998,nuca,tlc} looked at more
energy-efficient data communication in the context of on-chip
interconnects~\cite{desc} where the number of bit toggles can be reduced. The
key difference between our work and these prior works is that we aim to address
the specific effect of increase (sometimes a dramatic increase, see
Section~\ref{toggles:sec:motivation}) in bit toggle count due to data compression.  Our
proposed mechanisms (described in Section~\ref{sec:idea}) are mostly orthogonal
to these prior mechanisms and can be used in parallel with them to achieve even
larger energy savings in data transfers.  

\textbf{DRAM bus.} In the case of DRAM (e.g., GDDR5~\cite{jedec-gddr5}), the energy attributed to the
actual data transfer is usually less than the background and activate energy, but still
significant (16\% on average based on our estimation with the Micron power calculator~\cite{micron-power}).
%
%not the dominant factor anymore as in the case of
%on-chip interconnect\footnote{Background and activate power is usually higher
%than I/O power based on our estimation with the Micron power
%calculator~\cite{micron-power}.}, but still takes a significant portion of the
%DRAM energy (16\% on average based on our estimation). 
The second major distinction between on-chip
and off-chip buses, is the definition of bit toggles. In case of DRAM, bit
toggles are defined as the number of zero bits. 
%This is because DRAM
%transmission lines have high level termination. 
Reducing the number of signal
lines driving a low level (zero bit) results in reduced power dissipation in
the termination resistors and output drivers~\cite{jedec-gddr5}. To reduce the number
of zero bits, techniques like DBI (data-bus-inversion) are usually used. For example,
DBI is the part of the standard for GDDR5~\cite{jedec-gddr5} and DDR4~\cite{DDR4}. 
As we will show later in Section~\ref{toggles:sec:motivation}, these techniques are not
effective enough to handle the significant increase in bit toggles due to data compression.

%\new{
\textbf{PCIe and SATA.} For SATA and PCIe, data is transmitted in a serial fashion at much
higher frequencies than typical parallel bus interfaces. 
%Bit toggles within
%these high speed bus interfaces have different implications than that for on-chip
%or off-chit buses. First, since data
%is transmitted in a serial fashion, data alignment at larger byte sizes no
%longer plays a significant role in determining toggle rate. 
Under these conditions, bit toggles 
impose different design considerations and implications. Data is
transmitted across these buses without an accompanying clock signal which means
that the transmitted bits need to be synchronized with a clock signal by the
receiver. This \emph{clock recovery} requires \emph{frequent} bit toggles to
prevent loss in information. In addition, it is desirable that the \emph{running
disparity}{\textemdash}which is the difference in the number of one and zero
bits transmitted{\textemdash}be minimized. This condition is referred to as
\emph{DC balance} and prevents distortion in the signal. Data is typically
scrambled using encodings like the 8b/10b encoding~\cite{8b10b} to balance the
number of ones and zeros while ensuring frequent transitions. These
encodings have high overhead in terms of the amount of additional data
transmitted but obscure any difference in bit transitions with compressed or
uncompressed data. 
As the result, we do not expect further compression or toggle-rate
reduction techniques to apply well to interfaces like SATA and PCIe\@.

\textbf{Summary.} With on-chip interconnect, \emph{any bit transitions}
increase the energy expended during data transfers. In the case of DRAM, energy
spent during data transfers increases with an increase in \emph{zero} bits.
Data compression exacerbates the energy expenditure in both these channels. For
PCIe and SATA, data is scrambled before transmission and this obscures any
impact of data compression and hence, our proposed mechanisms are not
applicable to these channels.
%\input{sections/other_channels}

%In the context of GPUs, however, little work~\cite{GPUBandwidthCompression,lcp-tech} 
%has been done so far 

%to understand (i) how compressible the data is for general-purpose
%GPU applications, and (ii) what characteristics of the GPU architecture
%and applications running on it can be exploited for efficient
%data compression designs. In this work we aim to fulfill this gap by exploring
%a new and more flexible way to integrate data compression in the GPU design.

\section{Motivation and Analysis}
\label{toggles:sec:motivation}
%Various compression algorithms would achieve different compression ratios and
%as a result have different potential in providing higher effective bandwidth.

In this work, we examine the use of six compression algorithms for
bandwidth compression in GPU applications, taking into account bit
toggles: (i) \emph{FPC} (Frequent Pattern Compression)~\cite{fpc};
(ii) \emph{BDI} (Base-Delta-Immediate Compression)~\cite{bdi}; (iii)
\emph{BDI+FPC} (combined FPC and BDI)~\cite{lcp-micro}; (iv)
\emph{LZSS} (Lempel-Ziv compression)~\cite{lz,MXT}; (v)
\emph{Fibonacci} (a graphics-specific compression
algorithm)~\cite{fibonacci}; and (vi) \emph{C-Pack}~\cite{c-pack}. 
All of these compression algorithms explore different forms of redundancy
in memory data. For example, FPC and C-Pack algorithms look for different
static patterns in data (e.g., high order bits are zeros or the word consists
of repeated bytes). At the same time, C-Pack allows partial matching
with some locally defined dictionary entries that usually gives it better coverage than FPC\@.
In contrast, the BDI algorithm is based on the observation that the whole cache line of data
can be commonly represented as a set of one or two bases and the deltas from these bases.
This allows compression of some cache lines much more efficiently than FPC and even C-Pack,
but potentially leads to lower coverage. For completeness of our compression algorithms analysis,
we also examine the well-known software-based mechanism called LZSS, and the recently proposed graphics-oriented
Fibonacci algorithm. 

\begin{figure}[t!]
\centering
%\begin{subfigure}[b]{0.24\textwidth}
%\includegraphics[width=\textwidth]{figures/CompRatioGP.pdf}
%\caption{Discrete GPU applications.}
%\label{fig:cr-gpgpu}
%\end{subfigure}
%\begin{subfigure}[b]{0.24\textwidth}
%\includegraphics[width=\textwidth]{figures/CompRatioMobile.pdf}
%\caption{Mobile GPU applications.}
%\label{fig:cr-mobile}
%\end{subfigure}
\includegraphics[width=0.9\textwidth]{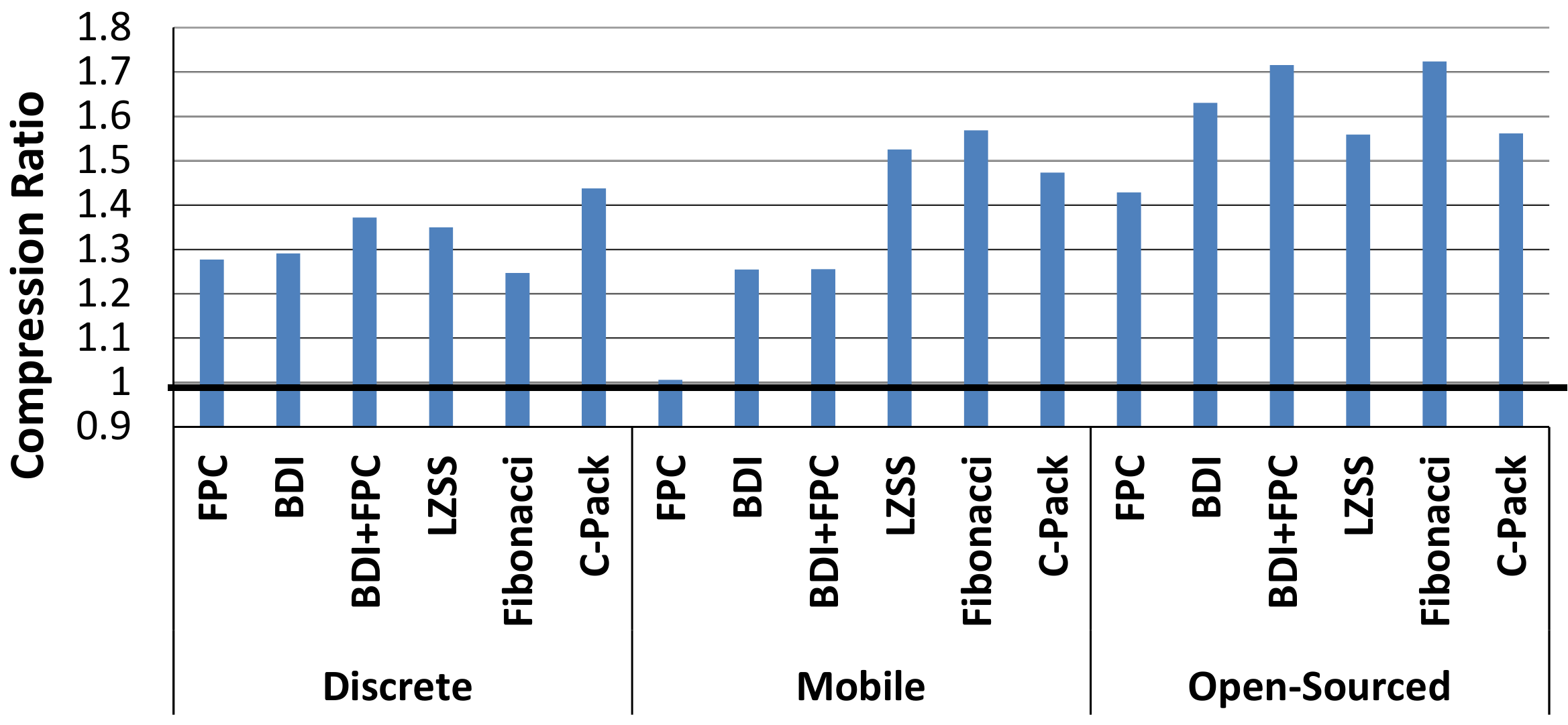}
\caption{Effective bandwidth compression ratios for various GPU
  applications and compression algorithms (higher bars are better).}
\label{fig:cr-all}
\end{figure}

To ensure our conclusions are practically applicable, we analyze
both the real GPU applications (both \emph{discrete} and \emph{mobile} ones)
with actual data sets provided by a major GPU
vendor and \emph{open-sourced} GPU computing applications~\cite{sdk,rodinia,mars,lonestar}.
The primary difference is that discrete applications have more single and double precision floating point
data, mobile applications have more integers, and open-source applications are in between. 
Figure~\ref{fig:cr-all} shows the potential of these six compression
algorithms in terms of effective bandwidth increase, averaged across
all applications.  These results exclude simple data patterns (e.g.,
zero cache lines) that are already handled by modern GPUs efficiently,
and assume practical boundaries on bandwidth compression ratios (e.g.,
for on-chip interconnect, the highest possible compression ratio is 4.0, 
because the minimum flit size is 32 bytes while the uncompressed packet size is 128 bytes).
 
First, for the 167 discrete GPU applications (left side of
Figure~\ref{fig:cr-all}), all algorithms provide substantial increase
in available bandwidth (25\%--44\% on average for different
compression algorithms). It is especially interesting that simple
compression algorithms are very competitive with the more complex
GPU-oriented \emph{Fibonacci} algorithm and the software-based
Lempel-Ziv algorithm~\cite{lz}. Second, for the 54 mobile GPU
applications (middle part of Figure~\ref{fig:cr-all}), bandwidth
benefits are even more pronounced (reaching up to 57\% on average with
the Fibonacci algorithm). Third, for the 21 open-sourced GPU computing applications the bandwidth benefits
are the highest (as high as 72\% on average with the Fibonacci and BDI+FPC algorithms). 
Overall, we conclude that existing
compression algorithms (including simple, general-purpose ones) can be
effective in providing high on-chip/off-chip bandwidth compression for
 GPU compute applications.

Unfortunately, the benefits of compression come with additional
costs. Two overheads of compression are well-known: (i) additional
data processing due to compression/decompression, and (ii) hardware
changes due to transfer variable-length cache lines.
% (e.g.,
%for NoCs, variable numbers of flits can be sent instead of the fixed number in
%the baseline). 
While these two problems are significant, multiple compression
algorithms~\cite{fpc,fvc,bdi,ZeroContent} have been proposed to minimize the
overheads of data compression/decompression. Several
designs~\cite{memzip,GPUBandwidthCompression,lcp-micro,caba} 
integrate bandwidth compression into existing memory
hierarchies.  In this work, we identify a new challenge with data
compression that needs to be addressed: the increase in the total
number of bit toggles as a result of compression.

\begin{figure}[t!] \centering
%\begin{subfigure}[b]{0.24\textwidth}
%\includegraphics[width=\textwidth]{figures/TogglesBaseGP.pdf}
%\caption{Discrete GPU applications.}
%\label{fig:tr-gpgpu}
%\end{subfigure}
%\begin{subfigure}[b]{0.24\textwidth}
%\includegraphics[width=\textwidth]{figures/TogglesBaseMobile.pdf}
%\caption{Mobile GPU applications.}
%\label{fig:tr-mobile}
%\end{subfigure}
\includegraphics[width=0.9\textwidth]{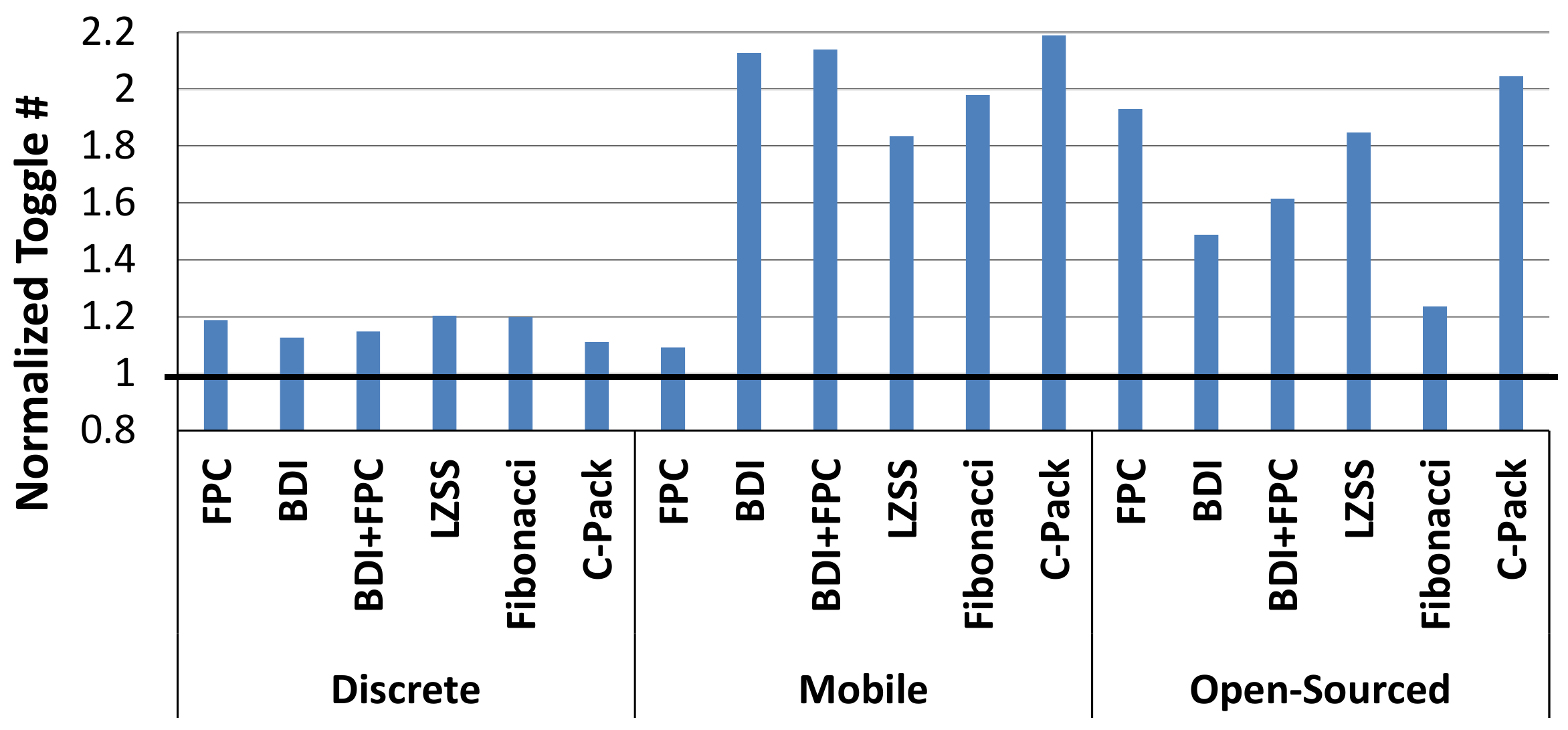}
\caption{Bit toggle count increase due to compression.}
\label{fig:tr-all}
\end{figure}

On-chip data communication energy is directly proportional to the
number of bit toggles on the communication
channel~\cite{nuca,tlc,desc}, due to the charging and discharging of
the channel wire capacitance with each toggle. Data compression may
increase or decrease the bit toggle count on the communication channel
for any given data. As a result, energy consumed for moving this data
can change. Figure~\ref{fig:tr-all} shows the increase in bit toggle
count for all GPU applications in our workload pool with the six
compression algorithms over a baseline that employs
zero line compression (as this is already efficiently done in modern GPUs).  
The total number of bit toggles is computed such that it
already includes the positive effects of compression (i.e., the
decrease in the total number of bits sent due to compression).
%We make two observations from this figure.

We make two observations.  First, all compression algorithms
consistently increase the bit toggle count.  The effect is significant
yet smaller (12\%--20\% increase) in discrete applications, mostly
because they include floating-point data, which already has high toggle
rates (31\% on average across discrete applications) and is less
amenable to compression. This increase in bit toggle count happens even though we transfer less data due to
compression. If this effect would be only due to the higher density of information per bit, 
we would expect the increase in the bit toggle rate (the relative percentage 
of bit toggles per data transfer), but not in the bit toggle count (the total number of bit toggles).

Second, the increase in bit toggle count is
more dramatic for mobile and open-sourced applications (right two-thirds of
Figure~\ref{fig:tr-all}), exceeding 2$\times$ in four cases.\footnote{The FPC algorithm is not as effective in compressing mobile
application data in our pool, and hence does not greatly affect bit
toggle count.} For all types of applications, the increase in bit
toggle count can lead to significant increase in the dynamic energy
consumption of the communication channels.

\begin{figure}[h!]
\centering
\includegraphics[width=0.9\textwidth]{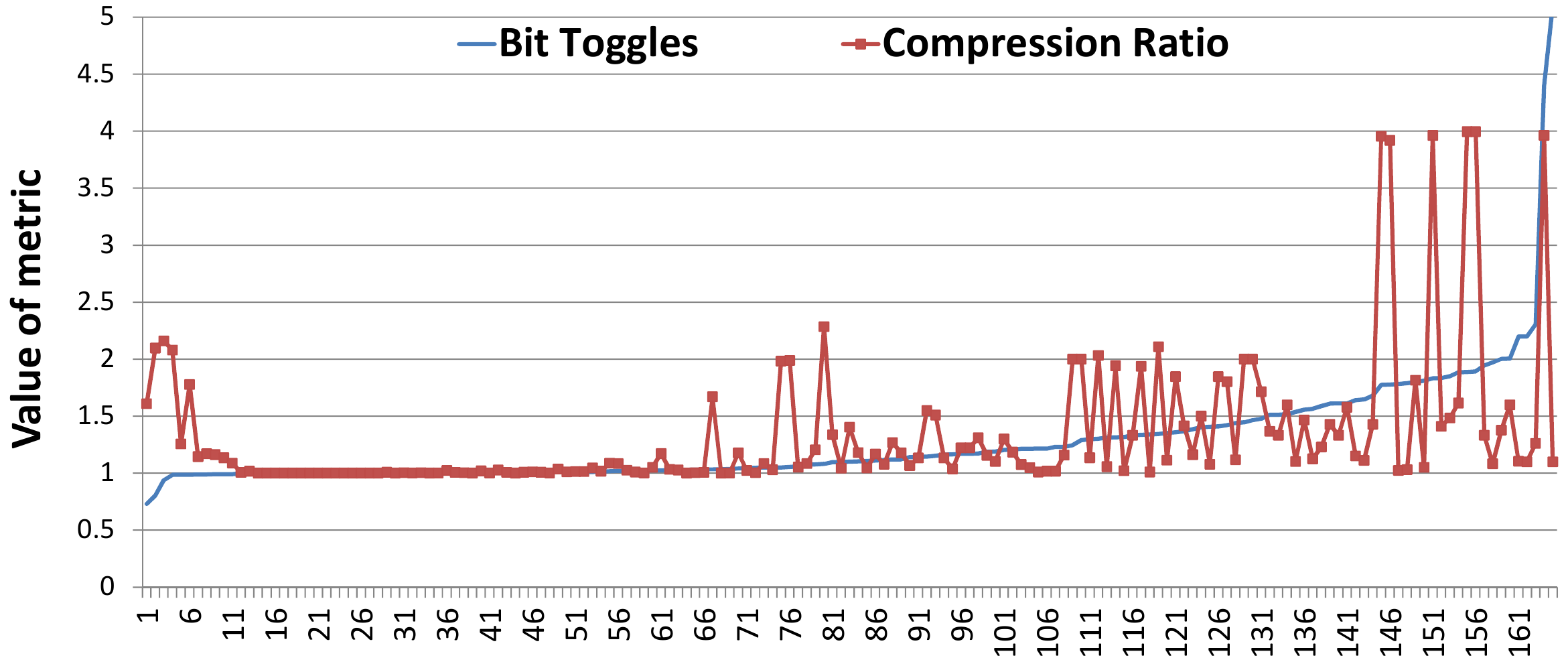}
\caption{Normalized bit toggle count vs. compression ratio (with the
  FPC algorithm) for each of the discrete GPU applications.}
\label{fig:cr-tr}
\end{figure}

We study the relationship between the achieved compression ratio and
the resultant increase in bit toggle count. Figure~\ref{fig:cr-tr}
shows the compression ratio and the normalized bit toggle count of
each discrete GPU application after compression with the FPC
algorithm.\footnote{We observe
similarly-shaped curves for other compression algorithms.} 
Clearly, there is a positive correlation between the
compression ratio and the increase in bit toggle count, although it is not
a simple direct correlation---higher compression ratio does not necessarily
means higher increase in bit toggle count. To make things worse,
the behaviour might change within an application due to phase and data
patterns changes. 

We draw two major conclusions from this study.
First, it strongly suggests that successful compression may lead to increased
dynamic energy dissipation by on-chip/off-chip communication channels
due to increased toggle counts. Second, these results show that any efficient solution
for this problem should probably be dynamic in its nature to adopt for 
data pattern changes during applications execution.

\begin{figure}[ht!]
\centering
\includegraphics[width=0.9\textwidth]{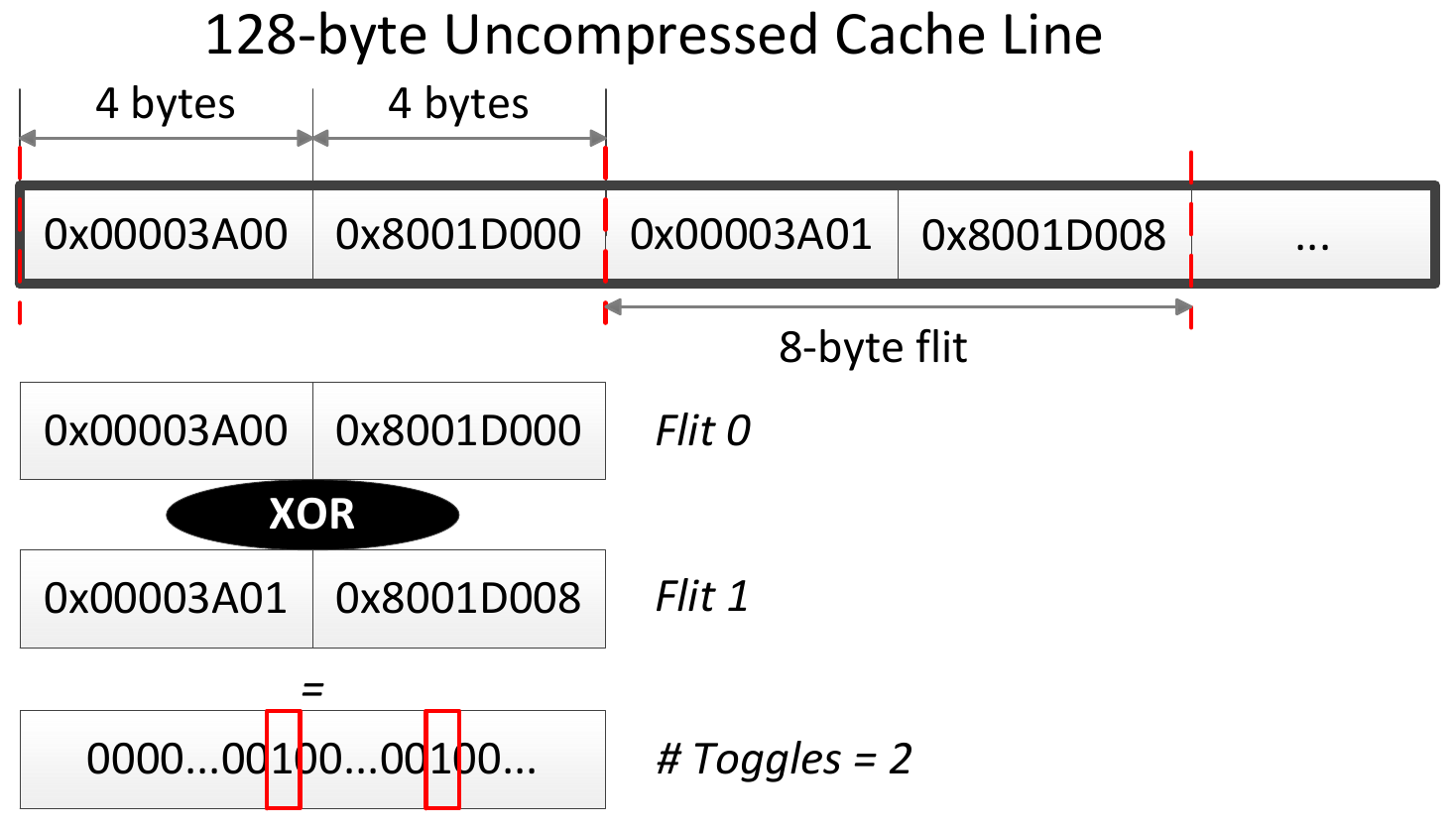}
\caption{Bit toggles without compression.}
\label{fig:nocomp-example}
\end{figure}

\begin{figure}[ht!]
\centering
\includegraphics[width=0.9\textwidth]{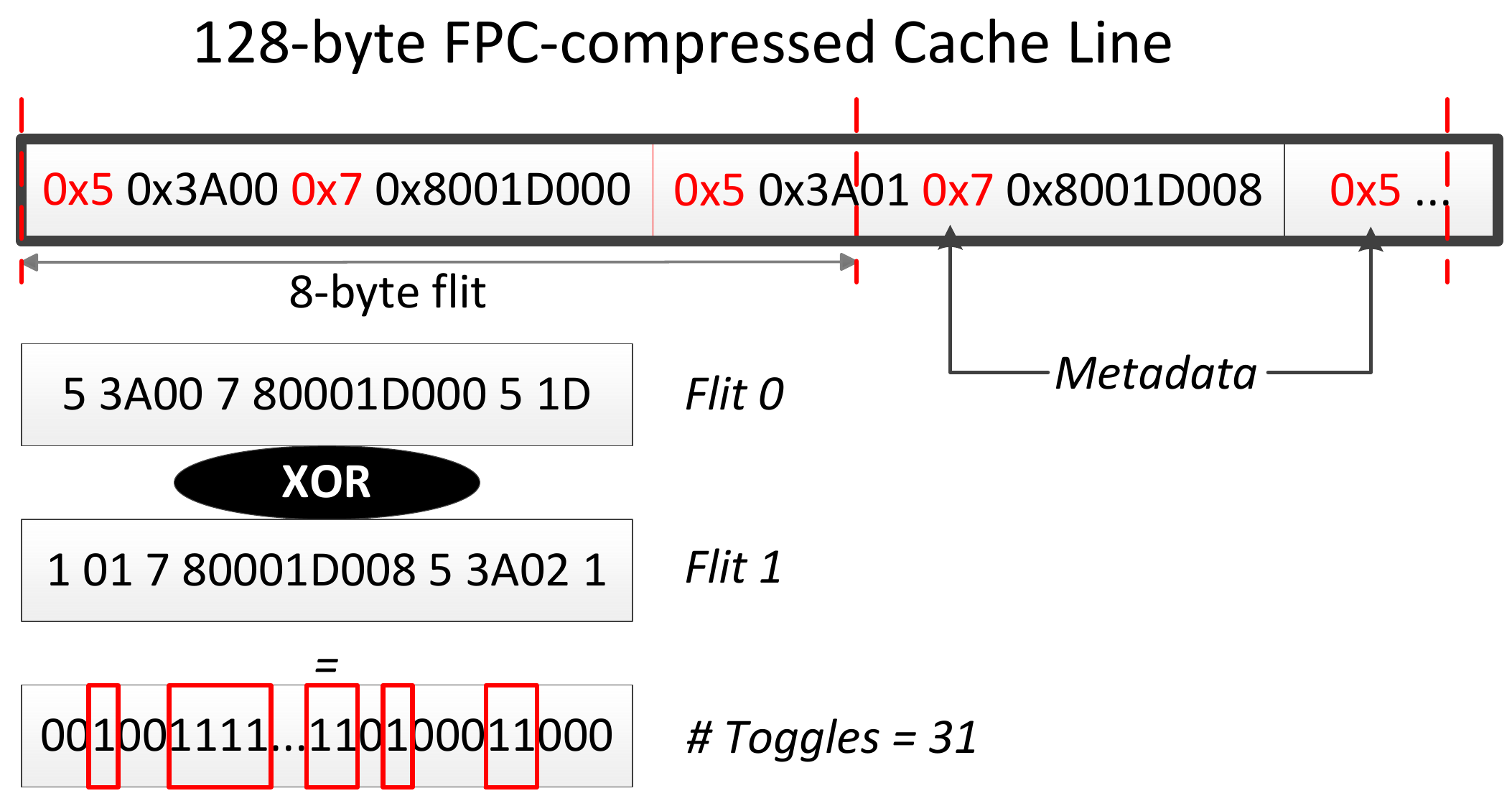}
\caption{Bit toggles after compression.}
\label{fig:fpc-example}
\end{figure}

To understand the toggle increase phenomenon, we examined several example cache lines
where bit toggle count increases significantly after compression.
Figures~\ref{fig:nocomp-example} and \ref{fig:fpc-example} show one of
these cache lines with and without compression (FPC), assuming 8-byte flits.

Without compression, the example cache line in
Figure~\ref{fig:nocomp-example}, which consists of 8-byte data
elements (4-byte indices and 4-byte pointers) has a very low number of
toggles (2 toggles per 8-byte flit). This low number of bit toggles is
due to the favourable alignment of the uncompressed data with the
boundaries of flits (i.e., transfer granularity in the on-chip
interconnect). With compression, the toggle count of the same cache
line increases significantly, as shown in Figure~\ref{fig:fpc-example}
(e.g., 31 toggles for a pair of 8-byte flits in this example). This
increase is due to two major reasons. First, because compression
removes zero bits from narrow values, the resulting higher per-bit
entropy leads to higher ``randomness'' in data and, thus, a larger
toggle count.  Second, compression negatively affects the alignment of
data both at the byte granularity (narrow values replaced with shorter
2-byte versions) and bit granularity (due to the 3-bit metadata
storage; e.g., $0\text{x}5$ is the encoding metadata used to
indicate narrow values for the FPC algorithm).

\section{Toggle-aware Compression}
\label{sec:idea}
\subsection{Energy vs. Performance Trade-off}
Data compression can reduce energy consumption and improve performance
by reducing communication bandwidth demands. At the same time, data
compression can potentially lead to significantly higher energy
consumption due to increased bit toggle count. To properly evaluate
this trade-off, we examine commonly-used metrics like $Energy \times
Delay $ and $Energy \times Delay^2$~\cite{ed1}.  We estimate these
metrics with a simple model, which helps to make
compression-related performance/energy trade-offs.
We define the $Energy$ of a single data transfer to be proportional to
the bit toggle count associated with it. Similarly, $Delay$ is defined
to be inversely proportional to performance, which we assume is
proportional to bandwidth reduction (i.e., compression ratio) and bandwidth utilization.  
The intuition behind this heuristic is that compression ratio reflects on how much
additional bandwidth we can get, while bandwidth utilization shows
how useful this additional bandwidth is in improving performance.
Based on the observations above, we develop two techniques to enable
{\em toggle-aware compression} to reduce the negative
effects of increased bit toggle count.
%A direct application of these metrics for a full application
%is impractical in our case, because it will lead to a single decision (send data compressed or 
%uncompressed) made for the whole application. To avoid this problem, in this work,
%we employ heuristics that are inspired by these metrics. 
%Instead of applying these metrics for a full application, we use them for individual
%cache lines.
\subsection{Energy Control (EC)}
\label{sec:ec}

\begin{figure}[h!]
\centering
\includegraphics[width=0.9\textwidth]{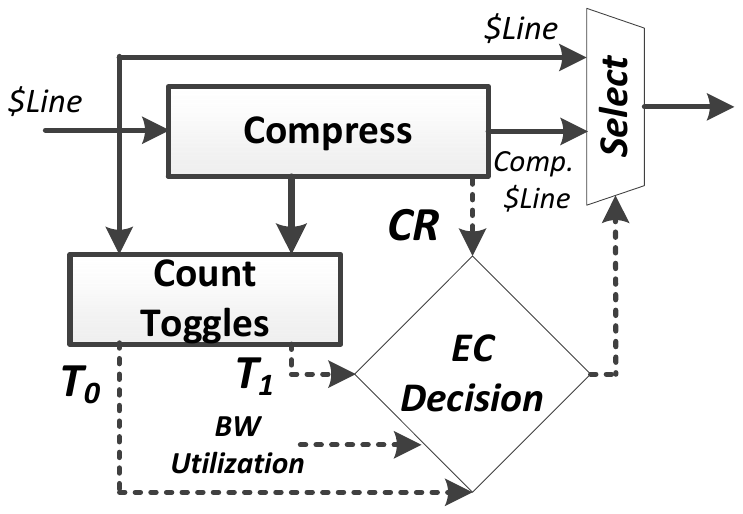}
\caption{Energy Control decision mechanism.}
\label{fig:ec-detailed}
\end{figure}

We propose a generic \emph{Energy Control} (EC) mechanism that can be applied
along with any current (or future) compression algorithm.\footnote{In this
work, we assume that only memory bandwidth is compressed, while on-chip caches
and main memory still store data in uncompressed form.} It aims to achieve high
compression ratio while minimizing the bit toggle count.  As shown in
Figure~\ref{fig:ec-detailed}, the Energy Control mechanism uses a generic
decision function that considers (i) the bit toggle count for transmitting the
original data ($T_{0}$), (ii) the bit toggle count for transmitting the data in
compressed form ($T_{1}$), (iii) compression ratio ($CR$), (iv) current
bandwidth utilization ($BU$), and possibly other metrics of interest that can be
gathered and analyzed dynamically to decide whether to transmit the data
compressed or uncompressed.  Using this approach, it is possible to achieve a
desirable trade-off between overall bandwidth reduction and increase/decrease in
communication energy. The decision function that compares the compression ratio
($A$) and toggle ratio ($B$) can be linear ($ A \times B > 1$, based on $Energy
\times Delay $) or quadratic ($ A \times B^{2} > 1$, based on $Energy \times
Delay^2$).\footnote{We also find a specific coefficient in the relative weight between $Energy$
and $Delay$ empirically.} 
Specifically, when the bandwidth utilization ($BU$) is very high
(e.g., $BU > 50\%$), we incorporate it into our decision function by
multiplying the compression ratio with $ \frac{1}{1 - BU} $, hence allocating more weight to the compression ratio. 
Since the data patterns during application execution could change drastically,
we expect our mechanism to be applied dynamically (either per cache line or a per
region of execution) rather than statically for the whole application execution.

%\evgeny{I think we need to rewrite this paragraph, energy control should rely on more generic metrics, that take into account MC for example, Delay as we defined it is not good enough, we need to generalize DELAY or to rewrite it}
%\gena{Remove the next couple of lines.}The decision function can also use as inputs other metrics,
%including application bandwidth requirements, power limitations,
%available voltage/frequency scaling options, and other system power
%management opportunities. We leave detailed explorations to future
%work.

%\gena{Lets add Energy-Aware reordering as well? As a mitigation technique? since it a full paper now - up to you.}
%response: not enough space and data to do it now.

%\subsection{Energy-Aware Reordering}
%In addition to EC, we develop a new mechanism for toggle energy reduction that
%we call Energy-Aware Reordering (EAR). Even though communication over the
%channels is usually performed in global units of communication (packets), those
%are usually broken into smaller chunks (flits). The key idea behind
%Energy-Aware Reordering is that a data transmission can potentially cost less
%in terms of energy if the flits (or packets) are reordered in such a way that
%the number of togles is lower than before. EAR aggregates several flits within
%a packet (local-reordering), or several outstanding packets
%(global-reordering), and then reorders them in a such way that the overall
%tnumber of toggles associated with the transmission of this data is minimized.
%EAR can be used with or without Energy Control, and even without compression
%(although we found that implementation of EAR is more efficient when the number
%of flits is relatively small - a common case after compression).

\subsection{Metadata Consolidation} 
\label{sec:mc}
Traditional energy-oblivious compression algorithms are not optimized
to minimize the bit toggle count. Most of these algorithms~\cite{c-pack,fpc,fibonacci}
have distributed metadata to efficiently track the redundancy in data, e.g.,
several bits per word to represent the current pattern used for encoding.
These metadata bits can significantly increase the bit toggle count as they
shift the potentially good alignment between different words within a cache line (Section~\ref{toggles:sec:motivation}).
It is possible to enhance these compression algorithms (e.g., FPC and C-Pack) such
that the increase in bit toggle count would be less after compression is applied. 
Metadata Consolidation (MC) is a new technique
that aims to achieve this. 
%Recall that one major reason for increased
%bit toggle count was the misalignment of compressed data with flit
%width due to alignment issues caused by compression-related metadata
%(Section 2). 
The key idea of MC is to
consolidate compression-related metadata into a {\em single contiguous
  metadata block} instead of storing (or, scattering) such metadata in
a fine-grained fashion, e.g., on a per-word basis. We can locate this
single metadata block either before or after the actual compressed
data (this can increase decompression latency since the decompressor
needs to know the metadata). The major benefit of MC is that it
eliminates misalignment at the bit granularity. In cases where a cache
line has a majority of similar patterns, a significant portion of the
toggle count increase can be avoided.

\begin{figure}[h!]
\centering
\includegraphics[width=0.9\textwidth]{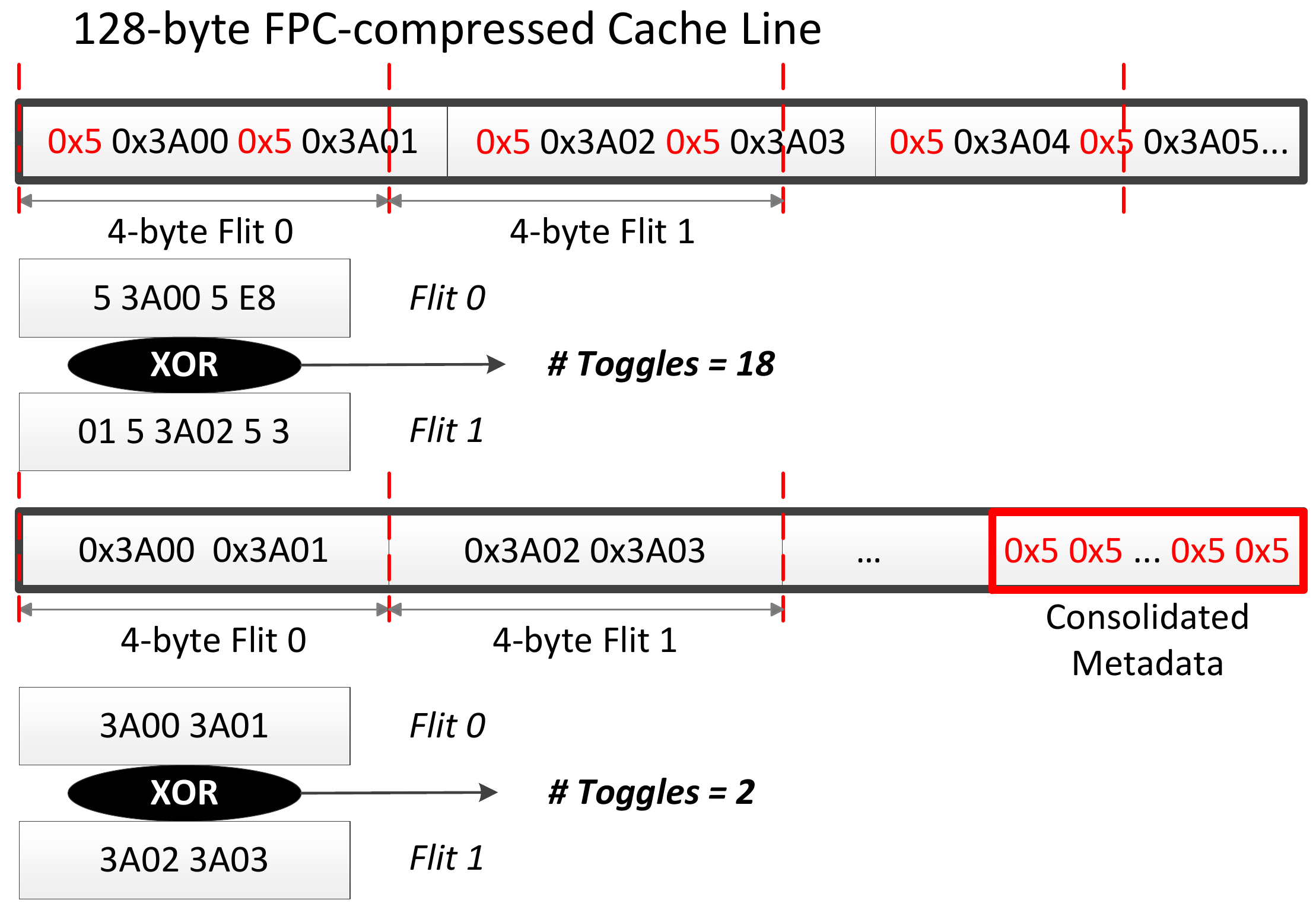}
\caption{Bit toggle count w/o and with Metadata Consolidation.}
\label{fig:mc}
\end{figure}

Figure~\ref{fig:mc} shows an example cache line compressed with the
FPC algorithm, with and without MC. We assume 4-byte flits. Without
MC, the bit toggle count between the first two flits is 18 (due to
per-word metadata insertion). With MC, the corresponding bit toggle
count is only 2, showing the effectiveness of MC in reducing bit
toggles.

\section{EC Architecture}
\label{toggles:sec:design}
\begin{figure}[h!] 
\centering
\includegraphics[width=0.9\textwidth]{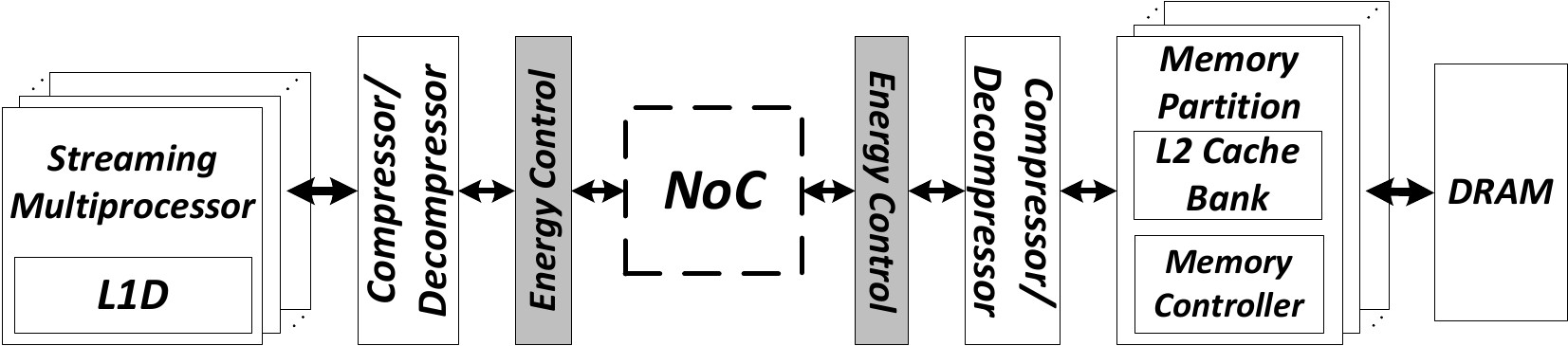}
\caption{System overview with interconnect compression and EC.}
\label{fig:system-icnt}
\end{figure}

\begin{figure}[h!]
\centering
\includegraphics[width=0.9\textwidth]{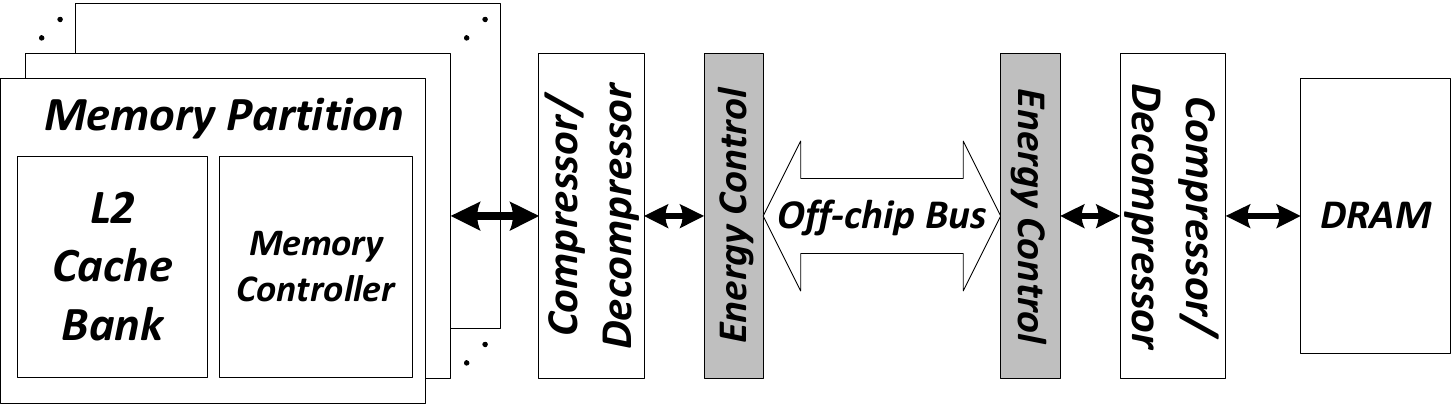}
\caption{System overview with off-chip bus compression and EC.}
\label{fig:system-DRAM}
\end{figure}

In this work, we assume a system where global on-chip network and main memory
communication channels are augmented with compressor and decompressor units as
described in Figure~\ref{fig:system-icnt} and Figure~\ref{fig:system-DRAM}.
While it is possible to store data in the compressed form as well (e.g., to
improve the capacity of on-chip caches~\cite{fvc,fpc,bdi,c-pack,dcc,sc2}), the
corresponding changes come with potentially significant hardware complexity
that we would like to avoid in our design.\ignore{In our system, the data
traffic coming in and out of the channel is attempted to be compressed with one
(or a few) compression algorithms.} We first attempt to compress the data
traffic coming in and out of the channel with one (or a few) compression
algorithms. The results of the compression, both the compressed cache line size
and data, are then forwarded to the Energy Control (EC) logic that is described
in detail in Section~\ref{sec:idea}. EC decides whether it is beneficial to
send data in the compressed or uncompressed form, after which the data is
transferred over the communication channel. It is then decompressed if needed
at the other end, and the data flow proceeds normally.  In the case of main memory
bus compression (Figure~\ref{fig:system-DRAM}), additional EC and
compressor/decompressor logic can be implemented in the already existing base-layer die
assuming stacked memory organization~\cite{hbm,hmc}, or in the additional  layer between DRAM and
the main memory bus. Alternatively, the data can be stored in the compressed form but without
any capacity benefits~\cite{GPUBandwidthCompression,memzip}.

%\gena{Figure 8 and 9 have missing border-lines on boxes - please fix}
%response: I don't see it on my machine in pdf.

\subsection{Toggle Computation for On-Chip Interconnect} 
%\evgeny{Talk about
%what are the changes needed to support toggle computation.  What are the
%overheads, e.g., latency, additional data to add to the flit headers}
As described in Section~\ref{sec:idea}, our proposed mechanism, EC,
aims to decrease the negative effect of data compression on bit toggling
while preserving most of the compression benefits. GPU on-chip communication is
performed via exchanging packets at a cache line size granularity.
But the physical width of the on-chip interconnect channels is usually
several times smaller than the size of a cache line (e.g.,
32-byte wide channels for 128-byte cache lines). As a result, the
communication packet is divided into multiple \emph{flits} that are stored at the 
transmission queue buffer before being transmitted over the communication
channel in a sequential manner. Our approach adds a simple bit toggle
computation logic that computes the bit toggle count across flits awaiting
transmission. 
This logic consists of a flit-wide array of XORs and a tree-adder
to compute the \emph{hamming distance}, the number of bits that are
different, between two flits.
We perform this computation for both compressed and uncompressed data, and
the results are then fed to the EC decision function (as
described in Figure~\ref{fig:ec-detailed}). 
This computation can be done sequentially while reusing the transition queue buffers to store
intermediate compressed or uncompressed flits, or in parallel with
the addition of some dedicated flit buffers (to reduce the latency overhead).
In this work we assume the second approach.

\subsection{Toggle Computation for DRAM} 
For modern DRAMs~\cite{jedec-gddr5,DDR4} the bit toggle definition is different from
the definition we used for on-chip interconnects. As we described in
Section~\ref{toggles:sec:background}, in the context of main memory bus what matters is
the number of zero bits per data transfer. This defines how we compute the toggle count
for DRAM transfers by simply counting the zero bits{\textemdash}which is known as the \emph{hamming weight} or the
\emph{population count} of the inverted value.  The difference in defining the toggle count also 
leads to the fact that the current toggle count does not depend on the previous data, which means that no
additional buffering is required to perform the computation. 

%\subsection{Metadata Consolidation} Different design considerations: sending
%first vs. sending last.  \evgeny{Discuss latency implications for ICNT}

\subsection{EC and Data Bus Inversion} 

Modern communication channels use different techniques
to minimize (and sometimes to maximize) the bit toggle count to reduce the energy 
consumption or/and preserve signal integrity. We now briefly summarize two major
techniques used in existing on-chip/off-chip interconnects: Data Bus Inversion
and Data Scrambling, and their effect on our proposed EC mechanism.

\subsubsection{Data Bus Inversion} 

Data Bus Inversion is an encoding technique proposed to reduce the power
consumption in data channels. Two commonly used DBI algorithms include \emph{Bus
invert coding}~\cite{dbi} and \emph{Limited-weight
coding}~\cite{limited-weight-codes1,limited-weight-codes2}. \emph{Bus invert coding} places an
upper-bound on the number
of bit flips while transmitting data along a channel. Consider a set of
\emph{N} bit lines transmitting data in parallel. If the Hamming distance
between the previous and current data value being transmitted exceeds \emph{N/2}, the
data is transmitted in the inverted form. This limits the number of bit flips to
\emph{N/2}. To preserve correctness, an additional bit line carries the
inverted status of each data tranmission. By reducing the number of bit flips,
\emph{Bus invert coding} reduces the switching power associated with charging
and discharging of bit lines. 

\emph{Limited weight coding} is a DBI technique that helps reduce power when one
of the two different bus states is more dissipative than the other. The
algorithm only observes the \emph{current} state of data. It decides to invert
or leave the data inverted based on the goal of minimizing either the number of \emph{zeros}
or \emph{ones} being transmitted. 

Implementing \emph{Bus invert coding} requires much the same circuitry for
toggle count determination in the proposed EC mechanism. Here,
hardware logic is required to compute the XOR between the different prior and
current data at a fixed granularity. The Hamming distance is then computed by
summing the number of 1's using a simple adder. Similar logic is required to
compute the toggle count for compressed versus uncompressed data in the Energy
Control mechanism. We expect that both EC and DBI can efficiently coexist. 
After compression is applied, we first apply DBI (to minimize the bit toggles), 
and after that we apply EC mechanism to evaluate the tradeoff between the compression ratio and
the bit toggle count. 

%\gena{Need to add here how EC and DBI co-exist. Because you just described DBI so far\ldots }

\subsubsection{Data Scrambling} 
To minimize the signal distortion, some modern
DRAM designs~\cite{ddr3-jedec,scrambling}  use a \emph{data scrambling} technique that
aims to minimize the running data disparity, i.e., the difference  between the
number of 0s and 1s, in the transmitted data. 
One way to ``randomize'' the bits is by XORing them with
a pseudo-random values generated at boot time~\cite{scrambling}. While techniques
like data scrambling can potentially decrease signal distortion, they also increase
the dynamic energy of DRAM data transfers. This approach also contradicts what
several designs aimed to achieve by using DBI for GDDR5~\cite{jedec-gddr5} and DDR4~\cite{DDR4}, since the bits become much
more random. In addition, using pseudo-random data scrambling techniques can be motivated by the existence of certain
pathological data patterns~\cite{scrambling}, where signal integrity requires much lower
operational frequency. At the same time, those patterns can usually be handled well
with data compression algorithms that can provide the appropriate data transformation to avoid repetitive failures at a certain frequency.
For the rest of this chapter, we assume GDDR5 memory without scrambling.

\subsection{Complexity Estimation }
\label{sec:overhead}
Toggle count computation is the main hardware addition introduced by the EC mechanism.
We modeled and synthesized the toggle-computational block in Verilog.
Our results show that the required logic can be performed in an energy-efficient way 
(4pJ per 128-byte cache line with 32-byte flits for 65nm process\footnote{This is significantly
lower than the corresponding energy for compression and decompression~\cite{memzip}.}). 
\ignore{
In our experiments,
we observe that even this overhead can be significantly reduced with simple
sampling techniques that monitor the decisions made with EC over time. 
More precisely, the decision made by EC stays very stable over time,
and hence we can avoid recomputing the toggle count for every transferred cache line.
}

\section{Methodology}
\label{toggles:sec:methodology}

In our work, we analyze two distinct groups of applications. First, a 
group of 221 applications from a major GPU vendor in the form
of memory traces with real application data. This group consists
of two subgroups: \emph{discrete} applications (e.g., HPC workloads, general-purpose
applications, physics etc.) and \emph{mobile} applications.
 As there is no existing simulator that can run these traces for cycle-accurate simulation,
we use them to demonstrate (i) the benefits of compression on a large pool
of existing applications operating on real data, and (ii) the existence
of the toggle count increase problem.
Second, we use 21 \emph{open-sourced} GPU computing applications derived
from CUDA SDK~\cite{sdk} (\emph{BFS, CONS, JPEG, LPS, MUM, RAY, SLA, TRA}),
Rodinia~\cite{rodinia} (\emph{hs, nw}), Mars~\cite{mars} (\emph{KM, MM, PVC, PVR, SS}), and
Lonestar~\cite{lonestar} (\emph{bfs, bh, mst, sp, sssp}) suites.

We evaluate the performance of our proposed mechanisms with the second group of applications using GPGPU-Sim
3.2.2~\cite{GPGPUSim} cycle-accurate simulator. Table~\ref{tab:meth} provides all the details of the
simulated system. Additionally, we use GPUWattch~\cite{gpuwattch} for energy analysis with proper
modifications to reflect bit-toggling effect.
 We run all applications to completion or 1 billion instructions
(whichever comes first). Our evaluation in Section~\ref{toggles:sec:results} demonstrates
detailed results for applications that exhibit at least 10\% bandwidth compressibility.

\textbf{Evaluated Metrics.} 
We present Instruction per Cycle (\emph{IPC}) as the primary performance
metric. In addition, we also use average bandwidth utilization defined as the
fraction of total DRAM cycles that the DRAM data bus is busy, and
\emph{compression ratio} defined as the effective bandwidth increase.
For both on-chip interconnect and DRAM we assume the highest possible compression ratio of
4.0. For on-chip interconnect, this is because we assume a flit size of 32 bytes for a 128-byte packet.
For DRAM, there are multiple ways of achieving the desired flexibility in data transfers:
(i) increasing the size of a cache line (from 128 bytes to 256 bytes), (ii) using sub-ranking as was 
proposed for DDR3 in MemZip~\cite{memzip}, (iii) transferring multiple compressed cache lines
instead of one uncompressed line as in LCP design~\cite{lcp-micro}, and (iv) any combination of 
the first three approaches. Existing GPUs (e.g., GeForce FX series) are known to support
4:1 data compression~\cite{maxwell}.   

%ratio of the number of DRAM bursts
%required to transfer data in the compressed vs. uncompressed form. 

\begin{table}[!t]
\vspace{-0.3cm}
\begin{scriptsize}
  \centering
    \begin{tabular}{ll}
        \toprule
System Overview           &  15 SMs, 32 threads/warp,  6 memory channels\\
        \cmidrule(rl){1-2}
Shader Core Config           &  1.4GHz, GTO scheduler~\cite{tor-micro12}, 2 schedulers/SM\\
        \cmidrule(rl){1-2}
Resources / SM     &  48 warps/SM, 32K registers, 32KB Shared Mem.\\
        \cmidrule(rl){1-2}
L1 Cache    &  16KB, 4-way associative, LRU   \\
        \cmidrule(rl){1-2} L2 Cache   &  768KB, 16-way associative, LRU  \\
        \cmidrule(rl){1-2} Interconnect   &  1 crossbar/direction (15 SMs, 6 MCs), 1.4GHz  \\
        \cmidrule(rl){1-2} Memory Model  &  177.4GB/s BW, 6 GDDR5 Memory Controllers,\\
& FR-FCFS scheduling, 16 banks/MC
  \\
        \cmidrule(rl){1-2}GDDR5 Timing~\cite{jedec-gddr5}  &  $t_{CL}=12, t_{RP}=12, t_{RC}=40, t_{RAS}=28,$\\
        &$t_{RCD}=12, t_{RRD}=6, t_{CLDR}=5, t_{WR}=12$
  \\%        \cmidrule(rl){1-2} Main Memory   &  2 partitions per controller (what does this mean?) 8? channels, FRFCFS,\\ & 8? banks-per-rank XX ranks per channel\\
  %      \cmidrule(rl){1-2} GDDR3 timing   & $(t_{CL}:t_{RP}:t_{RC}:t_{RAS}:t_{RCD}:t_{RRD}:t_{CLDR}:t_{WR}$ Fill in here~\cite{} \\
       % \cmidrule(rl){1-2}
%\multirow{2}[2]{*}{\centering Memory}              &  Timing: GDDR3 (8-8-8)~\cite{micron} \\
% & Organization: 1 channel, 1
 %rank-per-channel,\\ & 8 banks-per-rank, 8 KB row-buffer \\
        \bottomrule
    \end{tabular}%
\vspace{-0.1cm}
  \caption{Major Parameters of the Simulated Systems.}
  \label{tab:meth}%
\end{scriptsize}%
\vspace{-0.4cm}
\end{table}%

\section{Evaluation}
\label{toggles:sec:results}
%\subsection{Methodology} 
%We analyze 221 memory traces from compute (167) and graphics (54)
%application traces. We collect
%the information about the bit toggle count that reflects
%energy consumption and compression ratio that serves as a proxy
%for bandwidth consumption.
%\footnote{Unfortunately, there is no
%open-source cycle-accurate simulator that is capable
%of measuring performance for these applications.}
%Different encoding techniques (e.g., DBI~\cite{dbi} or DESC~\cite{desc}) can be
%applied to decrease the baseline bit toggle count of the data transfers. In our experiments,
%we found that the benefits of these techniques are largely orthogonal to whether or not
%data compression is applied. In our evaluation we use
%DBI as a part of the baseline for transferring both compressed and uncompressed data.

We present our results for two communication channels described
above: (i) off-chip DRAM bus and (ii) on-chip interconnect. 
 We exclude LZSS compression algorithm
from our detailed evaluation since its hardware implementation
is not practical with hundreds of cycles of
compression/decompression latency~\cite{MXT}.

\subsection{DRAM Bus Results}

\subsubsection{Effect on Toggles and Compression Ratio}
\begin{figure}[h!] \centering
%\begin{subfigure}[b]{0.24\textwidth}
%\includegraphics[width=\textwidth]{figures/TogglesEC.pdf}
%\caption{Discrete GPU applications.}
%\label{fig:tr-ec-compute}
%\end{subfigure}
%\begin{subfigure}[b]{0.24\textwidth}
%\includegraphics[width=\textwidth]{figures/TogglesEC-Mobile.pdf}
%\caption{Mobile GPU applications.}
%\label{fig:tr-ec-mobile}
%\end{subfigure}
\includegraphics[width=0.9\textwidth]{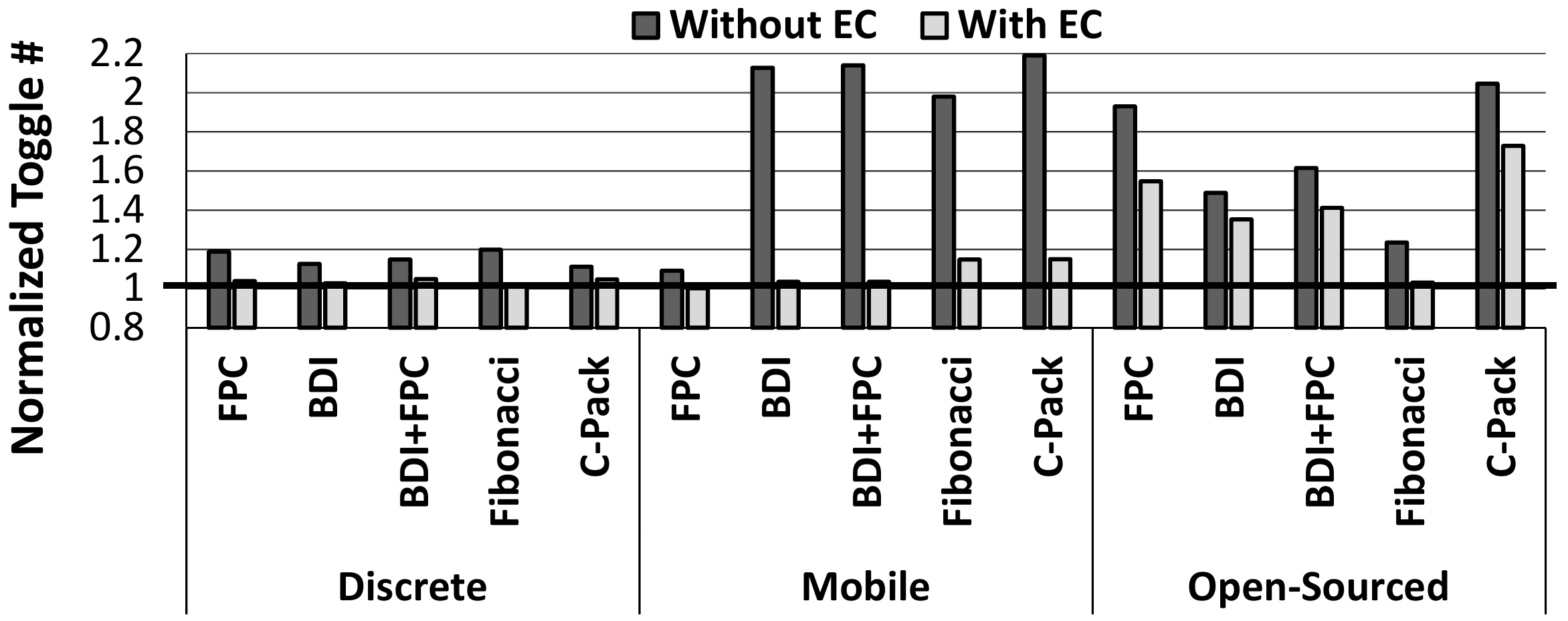}
\caption{Effect of Energy Control on the number of toggles on DRAM bus.}
\label{fig:tr-ec-all}
\end{figure}

\begin{figure}[h!]
\centering
%\begin{subfigure}[b]{0.24\textwidth}
%\includegraphics[width=\textwidth]{figures/CompRatioEC.pdf}
%\caption{Discrete GPU applications.}
%\label{fig:cr-ec-gpu}
%\end{subfigure}
%\begin{subfigure}[b]{0.24\textwidth}
%\includegraphics[width=\textwidth]{figures/CompRatioEC-Mobile.pdf}
%\caption{Mobile GPU applications.}
%\label{fig:cr-ec-mobile}
%\end{subfigure}
\includegraphics[width=0.9\textwidth]{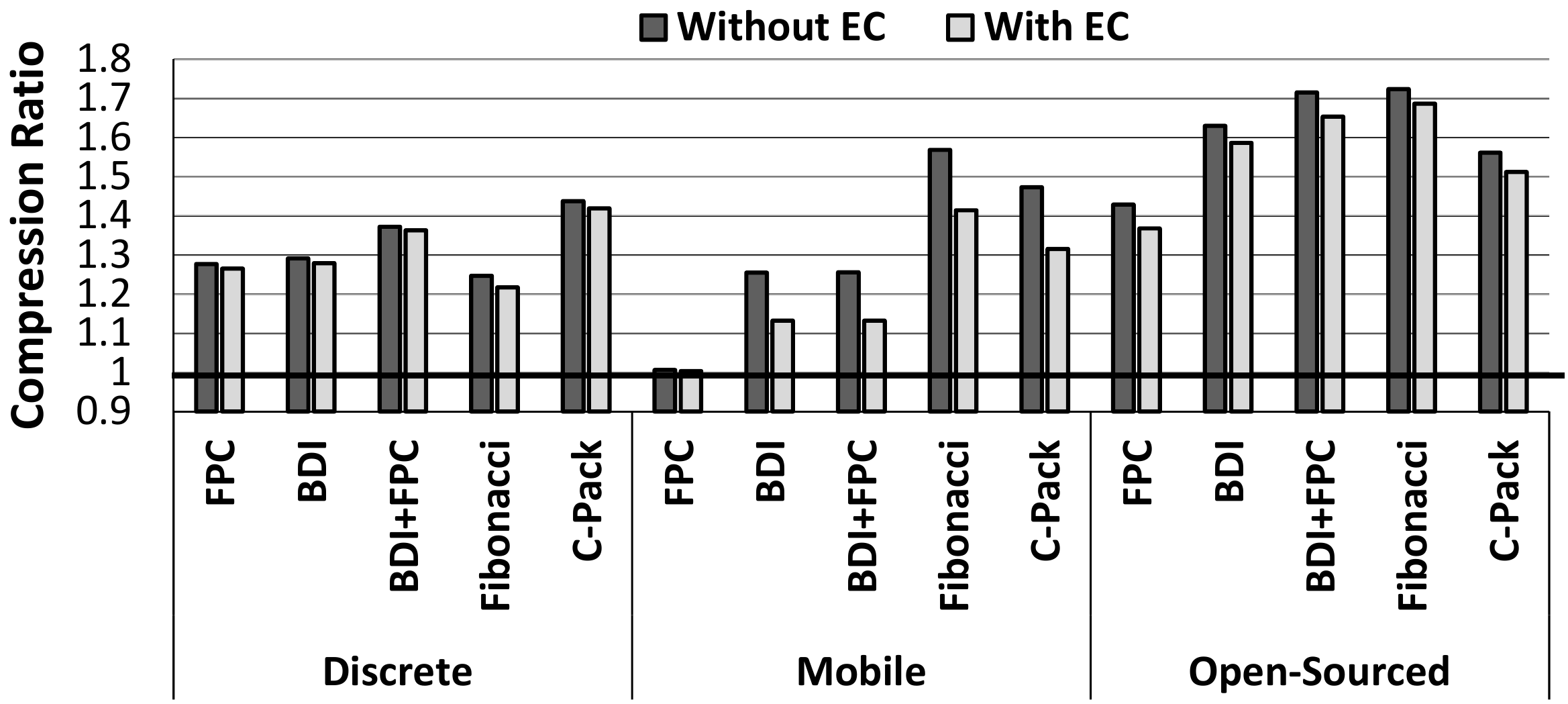}
\caption{Effective DRAM bandwidth increase for different applications.}
\label{fig:cr-ec-all}
\end{figure}

We analyze the effectiveness of the proposed EC optimization by
examining how it affects both the number of toggles
(Figure~\ref{fig:tr-ec-all}) and the compression ratio
(Figure~\ref{fig:cr-ec-all}) for five compression algorithms. 
 In both figures, results are averaged across all applications within the 
corresponding application subgroup and normalized
to the baseline design with no compression. Unless specified otherwise, we use the EC mechanism
with the decision function based on the $Energy \times Delay^2$ metric 
using our model from Section~\ref{sec:ec}.
We make two observations from these figures.

First, we observe that EC can effectively reduce the overhead in terms of
toggle count for both discrete and mobile GPU applications
(Figure~\ref{fig:tr-ec-all}). For discrete GPU applications, the toggle
reduction varies from 6\% to 16\% on average, and the toggle increase due to
compression is almost completely eliminated in the case of the Fibonacci
compression algorithm. For mobile GPU applications, the reduction is as high as
51\% on average for the BDI+FPC compression algorithm (more than 32$\times$
reduction in \emph{extra} bit toggles), with only a modest
reduction\footnote{Compression ratio reduces because EC decides to transfer
some compressible lines in the uncompressed form.} in compression ratio. 

Second, the reduction in compression ratio with EC is usually minimal.
For example, in discrete
GPU applications, this reduction for the BDI+FPC algorithm is only
0.7\% on average (Figure~\ref{fig:cr-ec-all}). For mobile and open-sourced GPU applications, the
reduction in compression ratio is more noticeable (e.g., 9.8\% on
average for Fibonacci with mobile applications), which is still a very attractive trade-off
since the 2.2$\times$ growth in the number of toggles is practically
eliminated. We conclude that EC offers an effective way to control
the energy efficiency of data compression for DRAM by applying it only when it
provides a high compression ratio with only a small increase in the
number of toggles.

\begin{figure}[h!]
\centering
\includegraphics[width=0.9\textwidth]{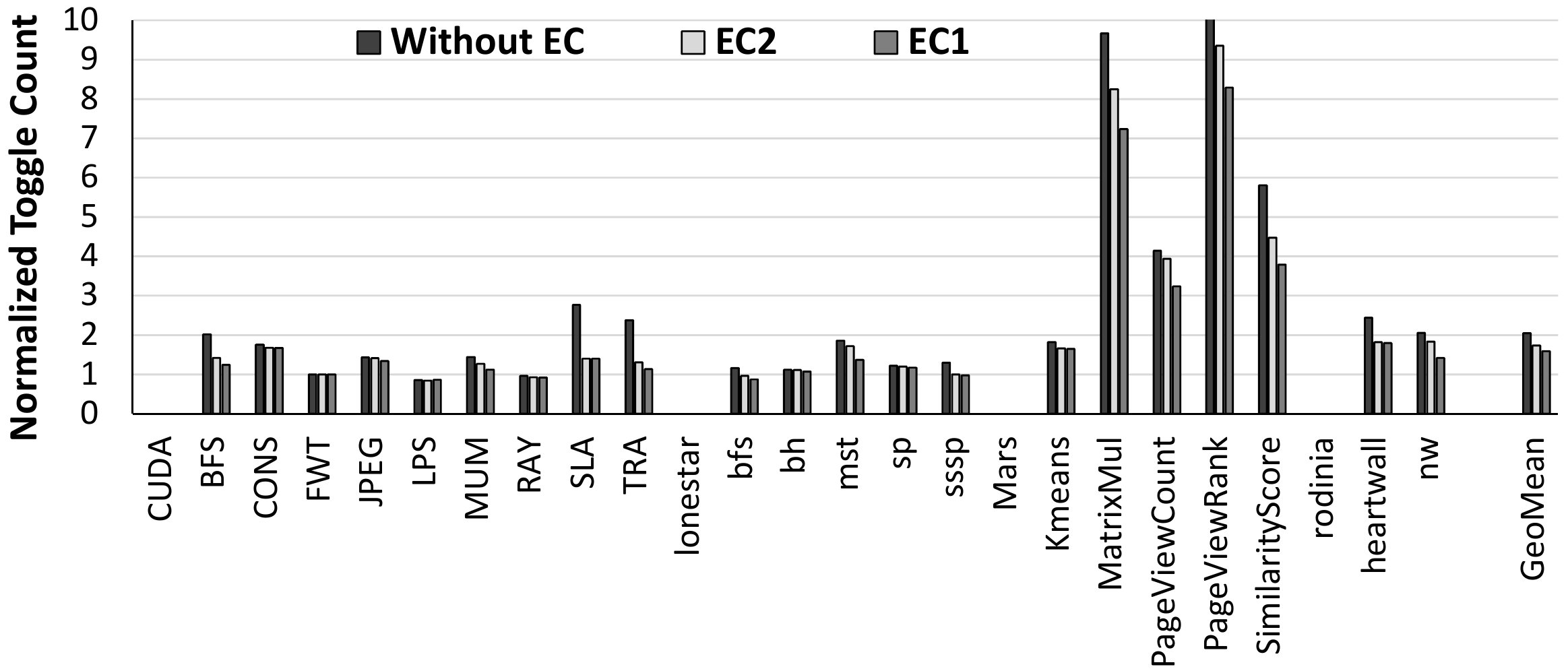}
\caption{Effect of Energy Control with C-Pack compression algorithm on the number of DRAM toggles.}
\label{fig:toggles-c-pack}
\end{figure}

\begin{figure}[h!]
\centering
\includegraphics[width=0.9\textwidth]{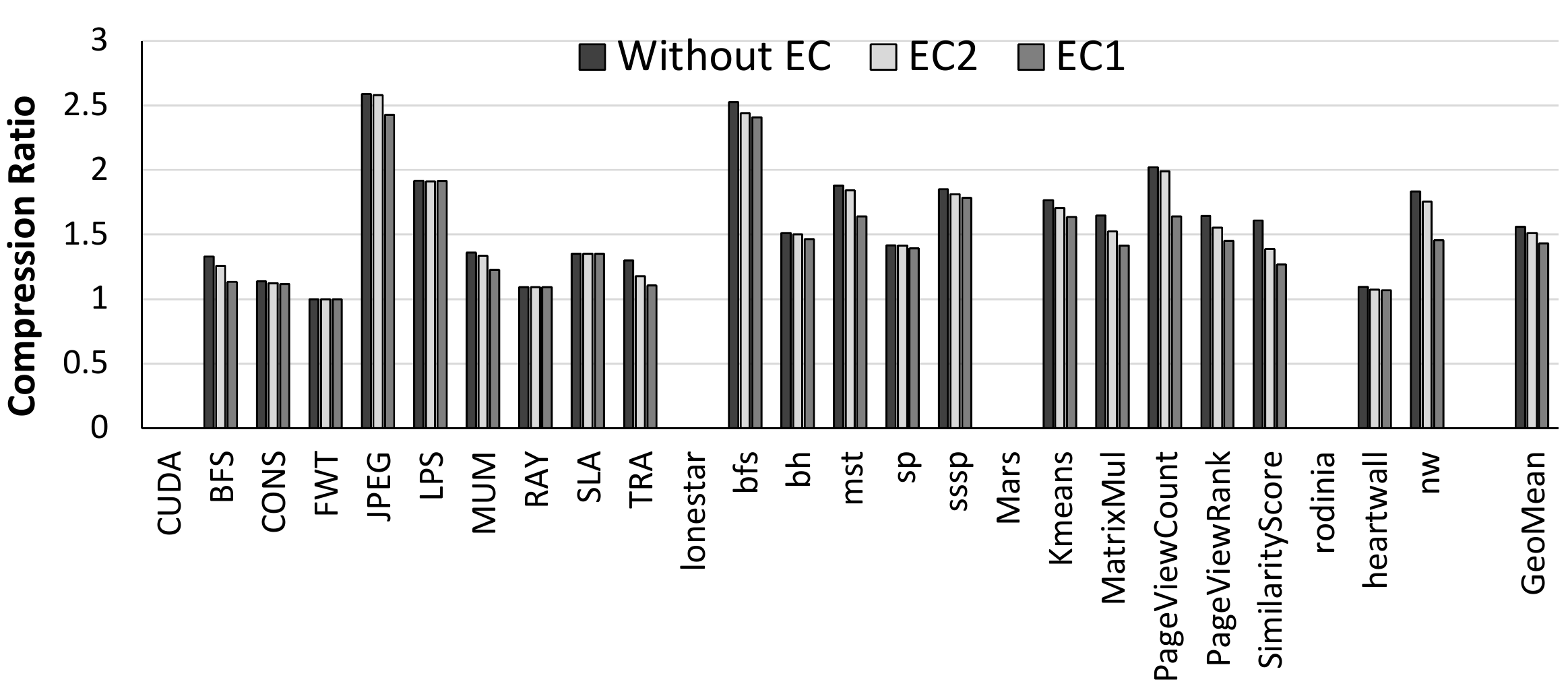}
\caption{Effective DRAM bandwidth increase with C-Pack algorithm.}
\label{fig:cr-c-pack}
\end{figure}

While the average numbers presented express the general
effect of the EC mechanism on both the number of toggles and compression ratio,
it is also interesting to see how the results vary for individual applications.
To perform this deeper analysis, we pick one compression algorithm
 (\emph{C-Pack}), and a single subgroup of applications (\emph{Open-Sourced}), 
and show the effect of compression with and without EC on the toggle count (Figure~\ref{fig:toggles-c-pack}) and compression
ratio (Figure~\ref{fig:cr-c-pack}). We also study two versions of the EC mechanism:
(i) \emph{EC1} which uses the $Energy \times Delay$ metric and (ii) \emph{EC2} which uses
the $Energy \times Delay^2$ metric. 
We make three major observations from these figures.

First, both the increase in bit toggle count and compression ratio vary significantly for different applications.
For example, \emph{bfs} from the Lonestar application suite has a very high compression
ratio of more than 2.5$\times$, but its increase in toggle count is relatively small (only 17\% 
for baseline C-Pack compression without EC mechanism). In contrast, \emph{PageViewRank} application
from the Mars application suite has more than 10$\times$ increase in toggles with 1.6$\times$ compression
ratio. This is because different data is affected differently by data compression. There can
be cases where the overall toggle count is lower than in the uncompressed baseline even without EC mechanism
(e.g., \emph{LPS}).

Second, for most of the applications in our workload pool, the proposed mechanisms (EC1 and EC2) can significantly reduce
the bit toggle count while retaining most of the benefits of compression. For example, for \emph{heartwall}
we reduce the bit toggle count with our EC2 mechanism from 2.5$\times$ to 1.8$\times$ by only sacrificing 8\% of the compression
ratio (from 1.83$\times$ to 1.75$\times$). This could significantly reduce the bit toggling energy overhead with C-Pack algorithm
while preserving most of the bandwidth (and hence potentially performance) benefits. 

Third, as expected, EC1 is more aggressive in disabling compression, because it weights bit toggles and compression
ratio equally in the trade-off, while in the EC2 mechanism, compression ratio has higher value (squared in the formula) than bit toggle count. Hence,
for many of our applications (e.g., \emph{bfs}, \emph{mst}, \emph{Kmeans}, \emph{nw}, etc.) we see a gradual reduction in toggles,
with corresponding small reduction in compression ratio, when moving from baseline to EC1 and then EC2. This means that depending on the application characteristics, we have
multiple options with varying aggressiveness to trade-off bit toggle count with compression ratio. As we will show in the next section, we can achieve these trade-offs with minimal effect on performance. 

\subsubsection{Effect on Performance}
\begin{figure}[h!]
\centering
\includegraphics[width=0.9\textwidth]{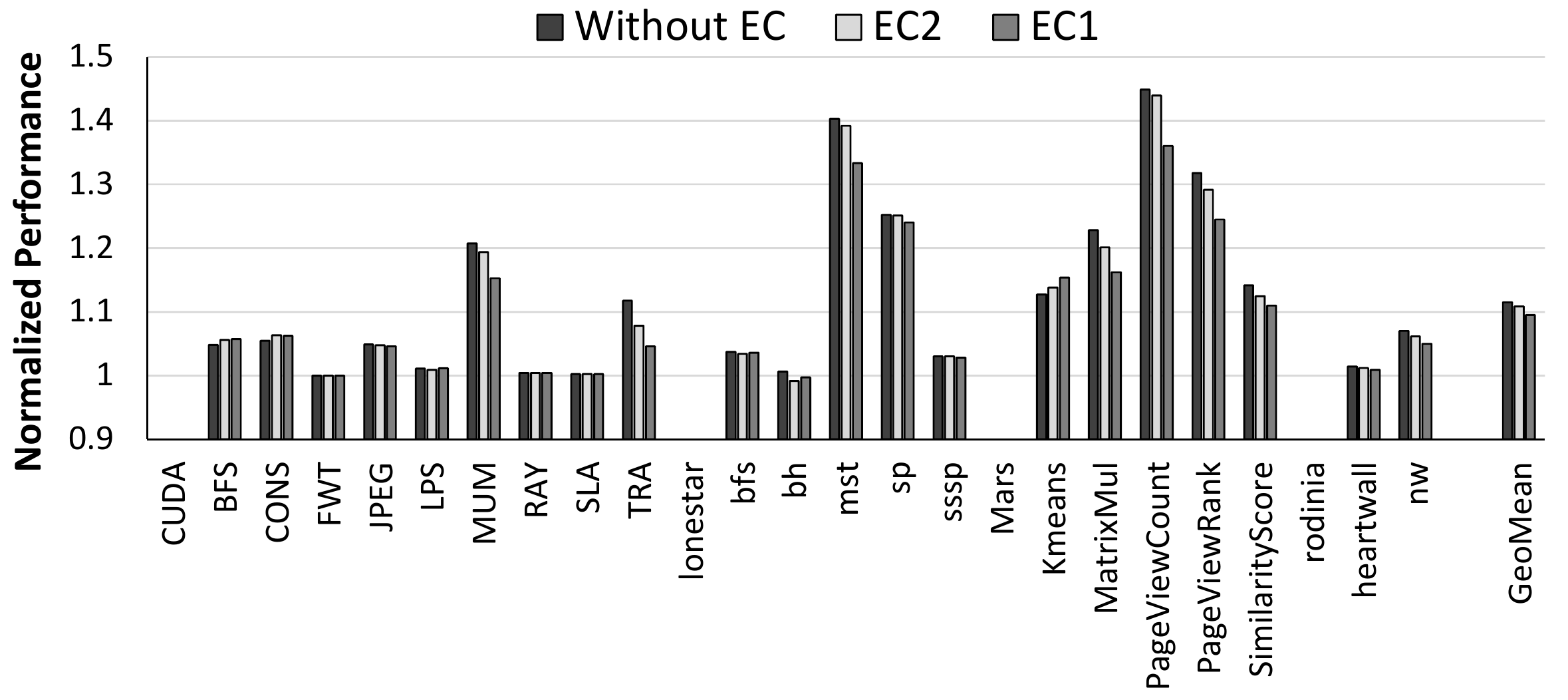}
\caption{Speedup with C-Pack compression algorithm.}
\label{fig:perf-c-pack}
\end{figure}

While previous results show that EC1 and EC2 mechanisms are very effective in
trading off bit toggle count with compression ratio, it is still important to
understand how much this trade-off ``costs'' in actual performance. This is
especially important for the DRAM, that is commonly one of the major
bottlenecks in GPU applications performance, and hence even a minor degradation
in compression ratio can potentially lead to a noticeable degradation in
performance and overall energy consumption.  Figure~\ref{fig:perf-c-pack} shows
this effect on performance for both EC1 and EC2 mechanisms in comparison to a
baseline employing compression with C-Pack. We make two observations here.

First, our proposed mechanisms (EC1 and EC2) usually have minimal negative
impact on the applications' performance. The baseline mechanism
(\emph{Without EC}) provides 11.5\% average performance improvement, while the
least aggressive EC2 mechanism reduces performance benefits by only 0.7\%, and
the EC1 mechanism - by 2.0\%. This is significantly smaller than the corresponding
loss in compression ratio (shown in Figure~\ref{fig:cr-c-pack}). The primary
reason is a successful trade-off between compression ratio,
toggles and performance. Both EC mechanisms consider current
DRAM bandwidth utilization, and only trade-off compression when it is unlikely to
hurt performance.

Second, while there are applications (e.g., \emph{MatrixMul}) where we could lose up
to 6\% performance using the most aggressive mechanism (EC1), this is
absolutely justified because we also reduce the bit toggle count from almost
10$\times$ to about 7$\times$. It is hard to avoid any degradation in
performance for such applications since they are severely bandwidth-limited,
and any loss in compression ratio is conspicuous in performance.  If such
performance degradation is unacceptable, then a less aggressive version of the
EC mechanism, EC2, can be used.
Overall, we conclude that our proposed mechanisms EC1 and EC2 are both very
effective in preserving most of the performance benefit that comes from data
compression while significantly reducing the negative effect of bit toggling
increase (and hence reducing the energy overhead).

\subsubsection{Effect on DRAM and System Energy}

\begin{figure}[h!]
\centering
\includegraphics[width=0.9\textwidth]{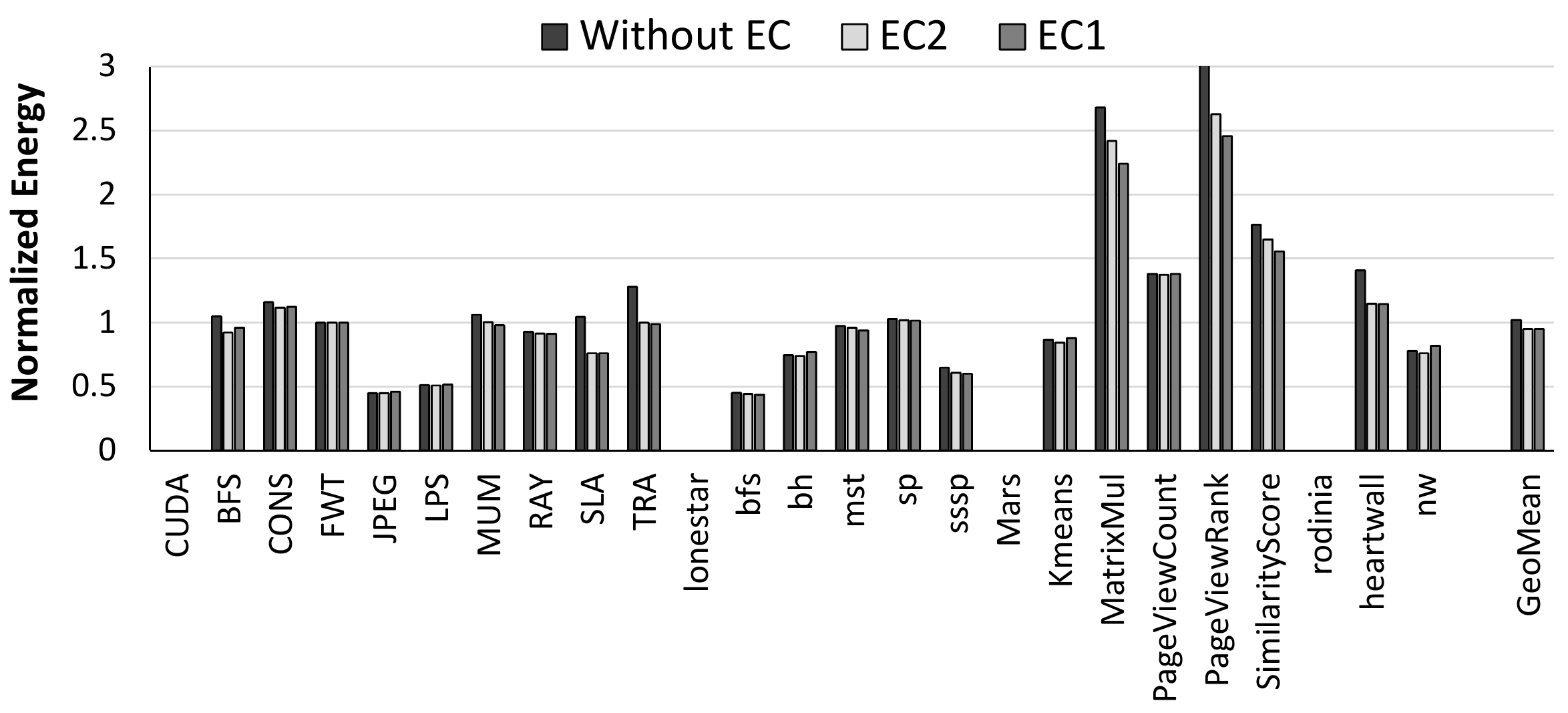}
\caption{Effect on the DRAM energy with C-Pack compression algorithm.}
\label{fig:energy-c-pack}
\end{figure}

Figure~\ref{fig:energy-c-pack} shows the effect of C-Pack compression algorithm
on the DRAM energy consumption with and without energy control (normalized to the
energy consumption of the uncompressed baseline). These results
include the overhead of the compression/decompression hardware~\cite{c-pack} and our
mechanism (Section~\ref{sec:overhead}).  and  We make two observations
from the figure. First, as expected, many applications significantly reduce
their DRAM energy consumption (e.g., \emph{SLA}, \emph{TRA}, \emph{heartwall}, \emph{nw}).
For example, for \emph{TRA}, the 28.1\% reduction in the DRAM energy (8.9\% reduction in the total energy) 
is the direct cause of the significant reduction in the bit toggle count (from
2.4$\times$ to 1.1$\times$ as shown in Figure~\ref{fig:toggles-c-pack}).  
Overall, the DRAM energy is reduced by 8.3\% for both EC1 and EC2.
As DRAM energy constitutes on average 28.8\% out of total system energy (ranging from
7.9\% to 58.3\%), and the decrease in performance is less than 1\%,
 this leads to a total system energy reduction of 2.1\% on average across applications using EC1/EC2 mechanisms.

Second, many applications that have significant growth in their bit toggle
count due to compression (e.g., \emph{MatrixMul} and \emph{PageViewRank}) are
also very sensitive to the available DRAM bandwidth. Therefore to
provide any energy savings for these applications, it is very important to
dynamically monitor their current bandwidth utilization. We observe that
without the integration of current bandwidth utilization metric into our mechanisms
(described in Section~\ref{sec:ec}), even a minor reduction in compression
ratio for these applications could lead to a severe degradation in performance,
and system energy. We conclude that our proposed mechanisms can efficiently trade off compression ratio and bit toggle count to improve both
the DRAM and overall system energy.

%\textbf{Total System Energy.}

\ignore{
\gena{ Need to rework this,
first I am not sure you need this, if you will mention it above.
}
Since there is a significant reduction in DRAM energy with almost no loss in compression ratio (and hence
the performance) the total system energy is reduced by 2.1\% when EC2 mechanism is used.

\gena{This below is not clear, it relates to DRAM energy and not system energy right?}

We also observe that no application has energy consumption higher than that of the uncompressed baseline
after EC mechanisms are applied, and there is only a few (e.g., \emph{bh}) where the system
energy increases slightly after EC-based mechanism is applied (due to some loss in useful 
compression ratio).
}

\subsection{On-Chip Interconnect Results}
\subsubsection{Effect on Toggles and Compression Ratio}
\begin{figure}[h!]
\centering
\includegraphics[width=0.9\textwidth]{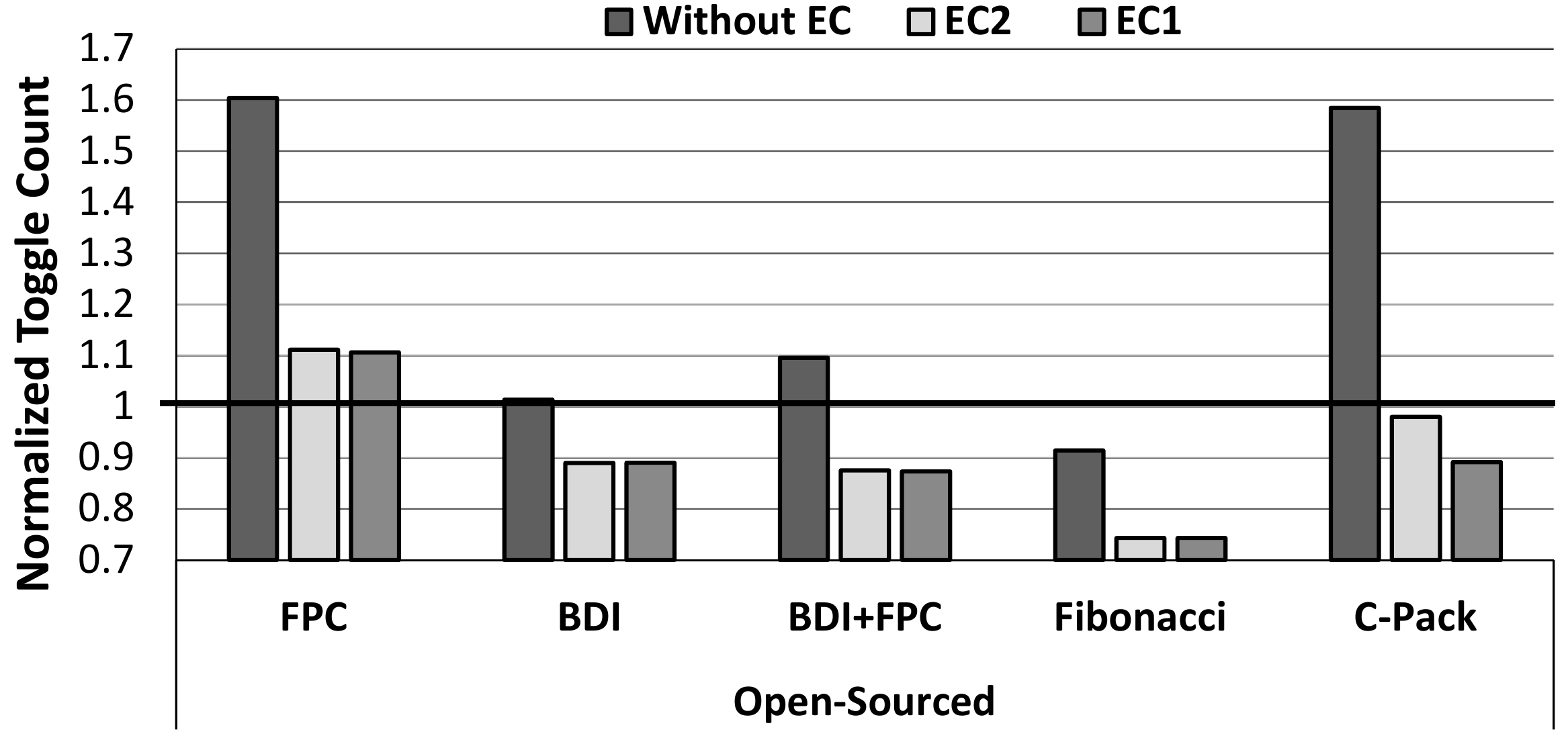}
\caption{Effect of Energy Control on the number of toggles in on-chip interconnect.}
\label{fig:toggles-icnt}
\end{figure}

\begin{figure}[h!] 
\centering
\includegraphics[width=0.9\textwidth]{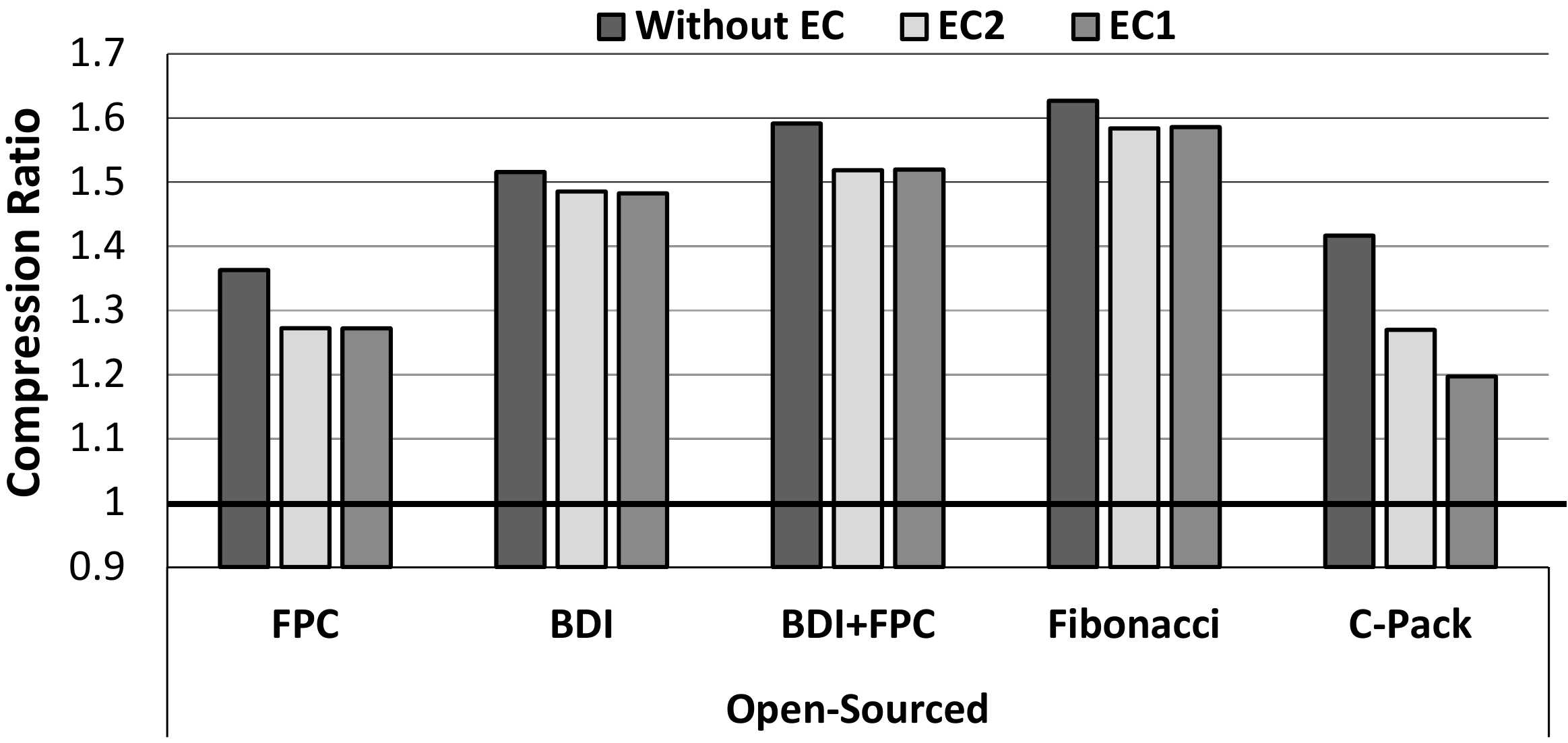}
\caption{Effect of Energy Control on compression ratio in on-chip interconnect.}
\label{fig:cr-icnt}
\end{figure}

Similar to the off-chip bus, we evaluate the effect of five compression
algorithms on toggle count and compression ratio for the on-chip interconnect
(Figure~\ref{fig:toggles-icnt} and Figure~\ref{fig:cr-icnt} correspondingly)
using GPGPU-sim and open-sourced applications as described in
Section~\ref{toggles:sec:methodology}.
We make three major observations from these figures.

First, the most noticeable difference when compared with the DRAM bus is
that the increase in bit toggle count is not as significant for all compression algorithms.
It still increases for all but one algorithm (\emph{Fibonacci}), but we observe
steep increases in bit toggle count (e.g., around 60\%)  
only for FPC and C-Pack algorithms. The reason for this behaviour is two fold.
First, the on-chip data working set is different from that of the off-chip
working set for some applications, and hence these data sets have different characteristics. Second, we
define \emph{bit toggles} differently for these two channels (see Section~\ref{toggles:sec:background}). 

Second, despite the variation in how different compression algorithms affect the bit toggle count, both of our proposed mechanisms are effective in reducing
the bit toggle count (e.g., from 1.6$\times$ to 0.9$\times$ with C-Pack). Moreover, both mechanisms, EC1 and EC2, preserve most of the compression
ratio achieved by C-Pack algorithm. Therefore, we conclude that our proposed mechanisms are effective in reducing bit toggles for both on-chip
interconnect and off-chip buses. 

Third, in contrast to our evaluation of the DRAM bus, our results with
interconnect show that for all but one algorithm (C-Pack), both EC1 and EC2 are almost equally effective
in reducing the bit toggle count while preserving the compression ratio. This means that in the case of on-chip interconnect, there is no need to use
more aggressive decision functions to trade-off bit toggles with compression
ratio, because the EC2 mechanism{\textemdash}the less aggressive of the
two{\textemdash}already provides most
of the benefits.

Finally, while the overall achieved compression ratio is slightly lower than in
case of DRAM, we still observe impressive compression ratios in on-chip interconnect, reaching up to 1.6$\times$ 
on average across all open-sourced applications.
While DRAM bandwidth traditionally is a primary performance bottleneck for
many applications, on-chip interconnect is usually designed such that its bandwidth 
will not be the primary performance limiter. Therefore the achieved
compression ratio in case of on-chip interconnect is expected to translate directly into
overall area and silicon cost reduction assuming fewer ports, wires and switches are required 
to provide the same effective bandwidth.
Alternatively, the compression ratio can be translated into lower power and energy assuming lower clock frequency can
be applied due to lower bandwidth demands from on-chip interconnect.

\subsubsection{Effect on Performance and Interconnect Energy}
\begin{figure}[h!]
\centering
\includegraphics[width=0.9\textwidth]{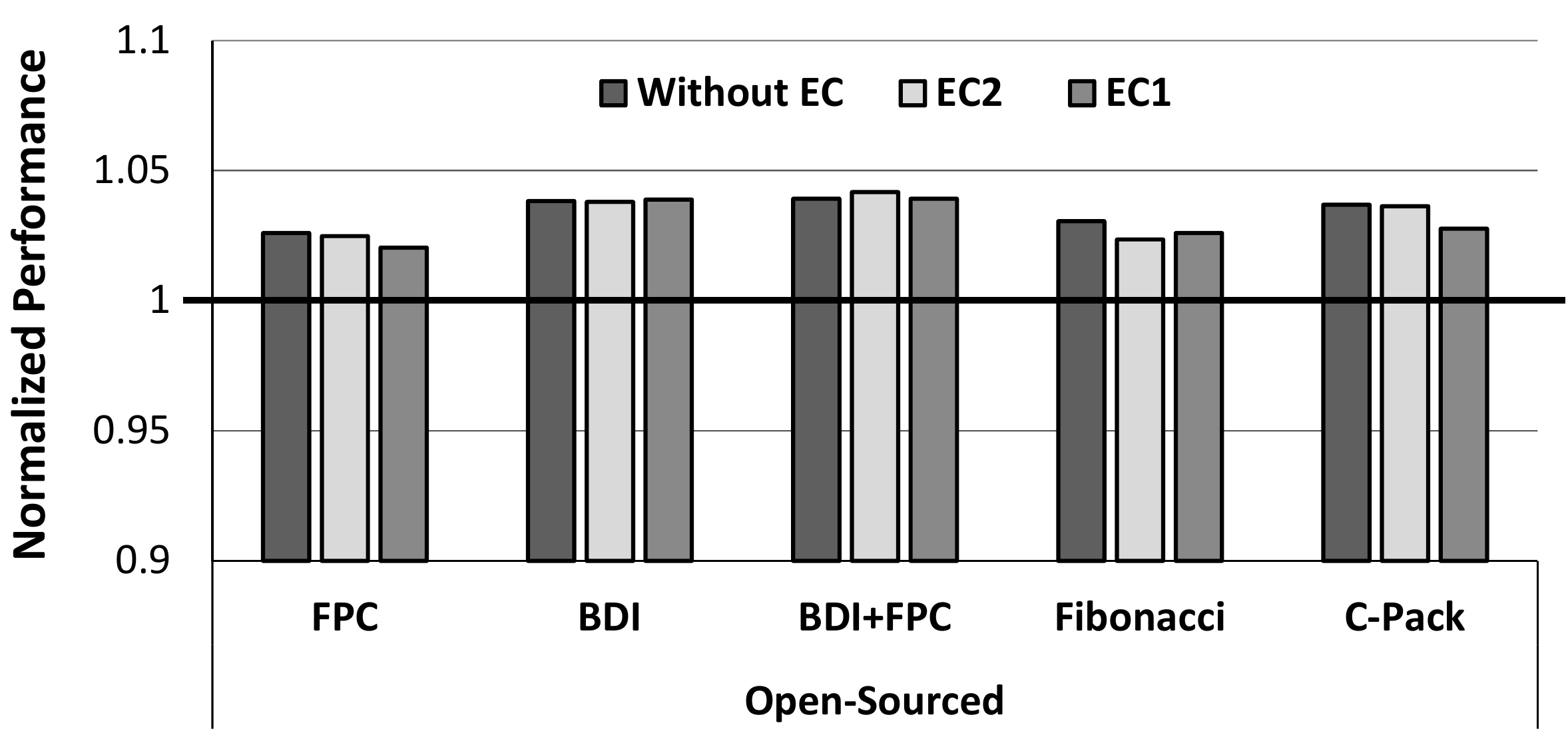}
\caption{Effect of Energy Control on performance when compression is applied to on-chip interconnect.}
\label{fig:perf-icnt}
\end{figure}

While it is clear that both EC1 and EC2 are effective in reducing the bit toggle count,
it is important to understand how they affect performance and interconnect energy in our simulated system.
Figure~\ref{fig:perf-icnt} shows the effect of both proposed techniques on performance (normalized
to the performance of the uncompressed baseline). The key takeaway from this figure is that for all
compression algorithms, both EC1 and EC2 are within less than 1\% of the performance of the designs
without the energy control mechanisms. There are two reasons for this. First,
both EC1 and EC2 are effective in deciding when compression is useful to improve performance and when
it is not. Second, the on-chip interconnect is less of a bottleneck in our example configuration than the off-chip bus, hence
disabling compression in some cases has smaller impact on the overall performance.

\begin{figure}[h!]
\centering
\includegraphics[width=0.9\textwidth]{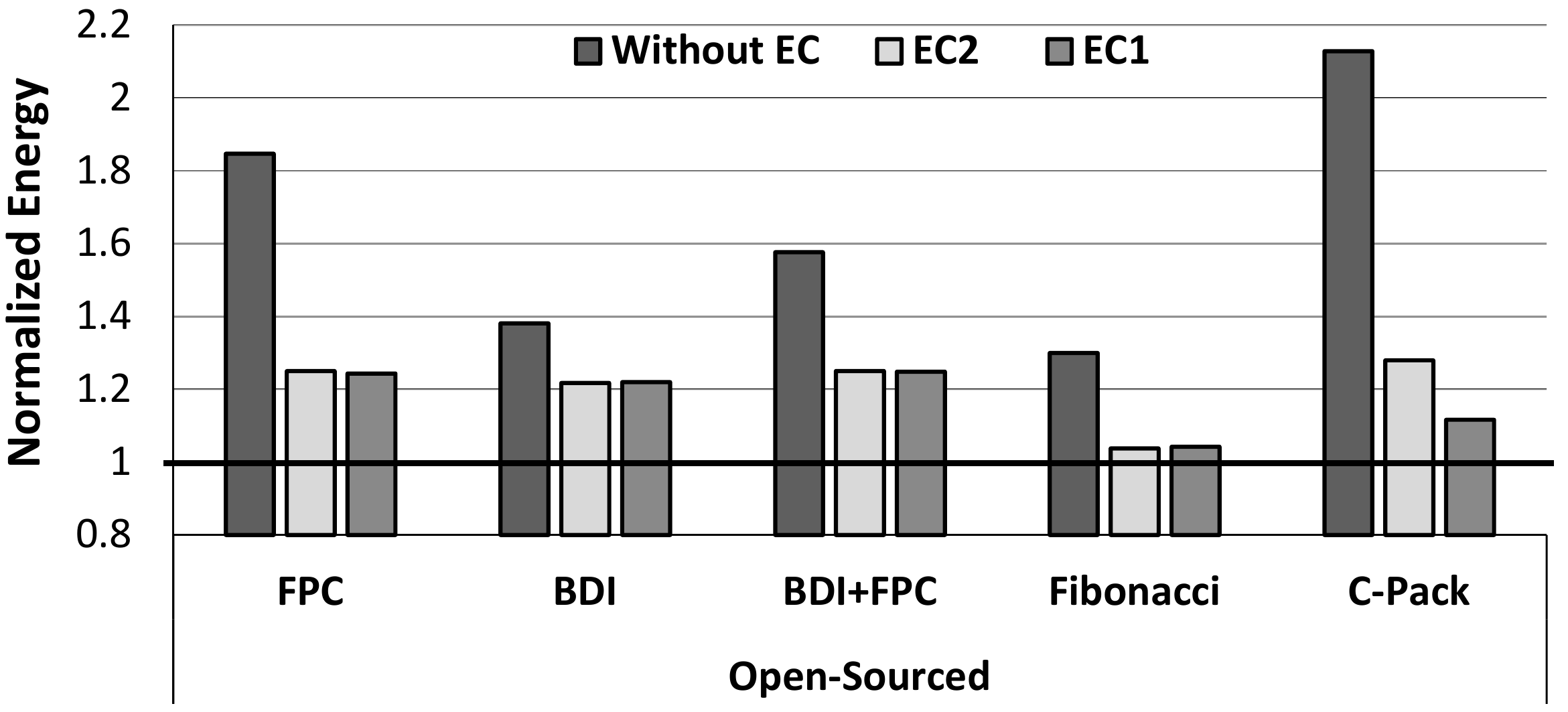}
\caption{Effect of Energy Control on on-chip interconnect energy.}
\label{fig:energy-icnt-icnt}
\end{figure}

Figure~\ref{fig:energy-icnt-icnt} shows the effect of data compression and bit toggling on the energy consumed by
the on-chip interconnect (results are normalized to the energy of the uncompressed interconnect). 
As expected, compression algorithms that have higher bit toggle count, have much higher
energy cost to support data compression, because bit toggling is the dominant
part of the on-chip interconnect energy consumption.
From this figure, we observe that our proposed mechanisms, EC1 and EC2, are both effective in reducing the energy
overhead. The most notable reduction is for \emph{C-Pack} algorithm, where we reduce the overhead from 2.1$\times$ to
just 1.1$\times$. 

Overall, we conclude that our mechanisms are effective in reducing the
energy overheads related to increased bit toggling due to compression, while
preserving most of the bandwidth and performance benefits achieved through compression.

\subsection{Effect of Metadata Consolidation}
%Finally, we report the effectiveness of Metadata Consolidation (MC) optimization mechanism. 
%Our results for Local (flit) Reordering
%(when applied on top of EC) show the significant potential in further reduction
%in bit toggles. For several compression algorithms (e.g., BDI, Fibonacci),
%Local Reordering reduced the number of toggles by up to 9.1\% on average,
%making it 7.3\% lower than in the uncompressed baseline (in case of Fibonacci).
%This technique can be applied on the uncompressed data as well, achieving a
%9.3\% toggle reduction on average. 
Metadata Consolidation (MC) is able to reduce the bit-level misalignment
for several compression algorithms (currently implemented for FPC and C-Pack
compression algorithms). We observe additional toggle reduction on the \emph{DRAM
bus} from applying MC (over EC2) of 3.2\% and 2.9\% for
FPC and C-Pack respectively across applications
in the discrete and mobile subgroups.  
Even though MC can mitigate some negative effects of bit-level
misalignment after compression, it is not effective in cases where data values
within the cache line are compressed to
different sizes. These variable sizes frequently lead to misalignment
at the byte granularity. While it is possible to insert some amount of padding into the compressed line to reduce the
misalignment, this would counteract the primary goal of
compression to minimize data size.  

\begin{figure}[h!]
\centering
\includegraphics[width=0.9\textwidth]{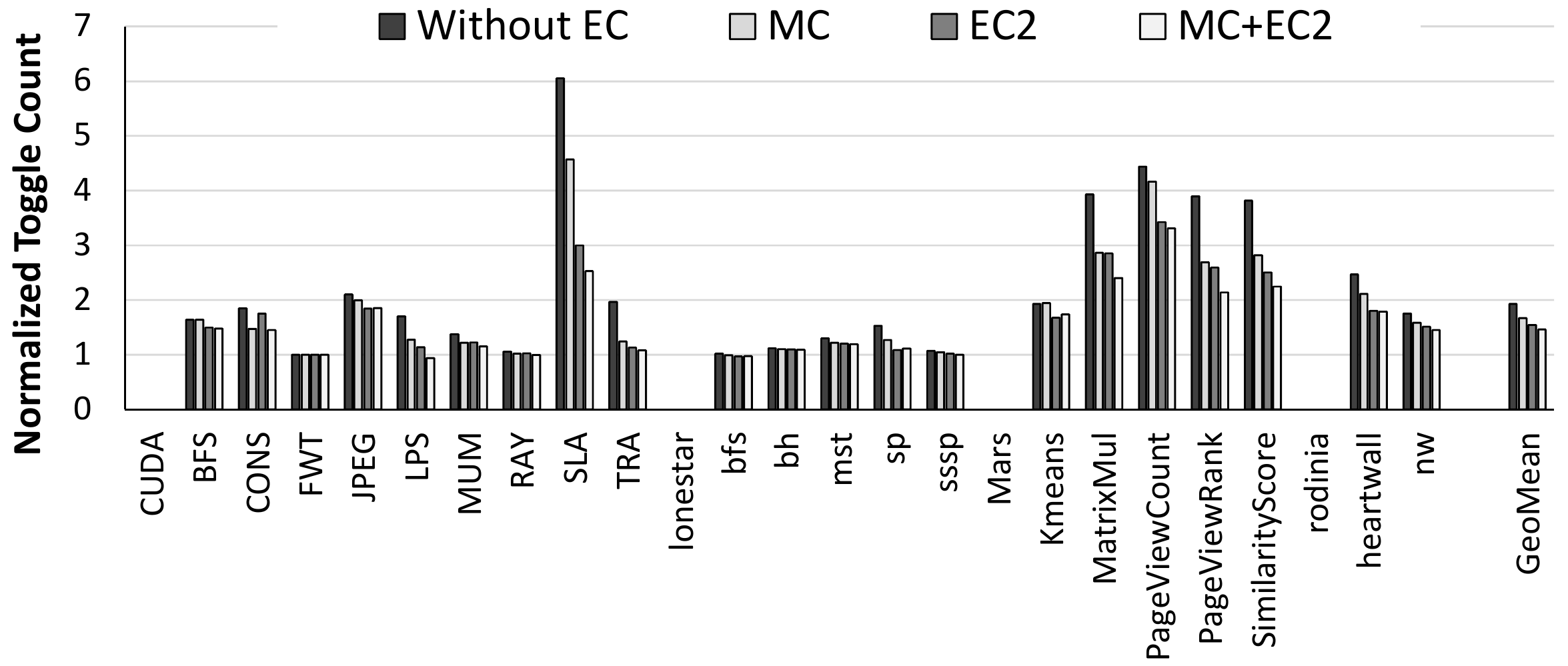}
\caption{Effect of Metadata Consolidation on DRAM bit toggle count with FPC compression algorithm.}
\label{fig:toggles-mc}
\end{figure}

We also conducted an experiment with open-sourced applications where
we compare the impact of MC and EC separately, as well as together,
for the FPC compression algorithm. We observe similar results with the
C-Pack compression algorithm.  Figure~\ref{fig:toggles-mc} lead to two
observations.  First, when EC is not employed, MC can substantially
reduce the bit toggle count, from 1.93$\times$ to 1.66$\times$ on
average.  Hence, in the case when the hardware changes related to EC
implementation are undesirable, MC can be used to avoid some of the
increase in the bit toggle count.  Second, when energy control is
employed (see \emph{EC2} and \emph{MC+EC2}), the additional reduction
in bit toggle count is relatively small. This means that EC2 mechanism
can cover most of the benefits that MC can provide.  In summary, we
conclude that MC mechanism can be effective in reducing the bit toggle
count when energy control is not used.  It does not require
significant hardware changes other than the minor modifications in the
compression algorithm itself. At the same time, in the presence of
energy control mechanism, the additional effect of MC in toggle
reduction is marginal.

\section{Related Work}
To the best of our knowledge, this is the first work that (i) identifies increased bit toggle count
in communication channels as a major drawback in enabling efficient data
compression in modern systems, (ii) evaluates the impact and causes for this
inefficiency in modern GPU architectures for different channels across multiple
compression algorithms, and (iii) proposes and extensively evaluates different mechanisms to
mitigate this effect to improve overall energy efficiency. We first discuss
prior works that propose more energy efficient designs for DRAM, interconnects
and mechanisms for energy efficient data communication in on-chip/off-chip
buses and other communication channels. 
We then discuss prior work that aims to address different challenges in
efficiently applying data compression. 

\textbf{Low Power DRAM and Interconnects.} A wide range of previous works
propose mechanisms and architectures to enable more energy-efficient operation
of DRAM. Examples of these proposals include activating fewer
bitlines~\cite{udipi-isca2010}, using shorter bitlines~\cite{lee-hpca2013},
more intelligent refresh policies~\cite{raidr, liu-asplos2011,
taku-islped1998, ahn-asscc2006, kim-patmos2000,Avatar,khan-sigmetrics2014}, dynamic voltage and frequency
scaling~\cite{david-icac11} and better management of data
placement~\cite{zhu-itherm2008,lin-islped2009,liu-hpca2011}. In the case of
interconnects, Balasubramonian et al.~\cite{balasubramonian-hpca2005} propose a
hybrid interconnect comprising wires with different latency, bandwidth, and
power characteristics for better performance and energy efficiency. Previous
works also propose different schemes to enable and exploit \emph{low-swing}
interconnects~\cite{zhang-1998,nuca,tlc} where reduced
voltage swings during signalling enables better energy efficiency. These works
do not consider energy efficiency in the context of data compression and are usually
data-oblivious, hence the proposed solutions can not alleviate the negative impact of increased toggle
rates with data compression.

\textbf{Energy Efficient Encoding Schemes.} \emph{Data Bus Inversion (DBI)} is
an encoding technique proposed to enable energy efficient data communication.
Widely used DBI algorithms include \emph{bus invert coding}~\cite{dbi} and
\emph{limited-weight coding}~\cite{limited-weight-codes1,limited-weight-codes2}
which selectively invert all the bits within a fixed granularity to either
reduce the number of bit flips along the communication channel or reduce the
frequency of either 0's or 1's when transmitting data.
Recently, \emph{DESC}~\cite{desc} was proposed in the context of on-chip interconnects to
reduce power consumption by representing information by the delay between two
consecutive pulses on a set of wires, thereby reducing the number of bit
toggles. Jacobvitz et al.~\cite{coset-coding} applied \emph{coset coding} to
reduce the number of bit flips while writing to memory by mapping each dataword
into a larger space of potential encodings. These encoding techniques do not
tackle the excessive bit toggle count generated by data compression and are
largely orthogonal to the our proposed mechanisms for toggle-aware data
compression.  

\ignore{\textbf{Low Power Interconnect and DRAM}

{\bf Low Power DRAM Architecture}\\ Reducing Wordline
Length~\cite{udipi-isca2010}\\ TL-DRAM~\cite{lee-hpca2013}\\
RowClone~\cite{seshadri-micro2013}\\

{\bf Low Power DRAM Feature in Industry}\\ JEDEC~\cite{jedec-lpdram}\\
MICRON~\cite{micron-lpdram}\\

{\bf 3D-Stacked DRAM with Low Channel Power}\\ 3D-DRAM~\cite{kang-isscc2009,
jeong-isscc2009, woo-hpca2010, harward-mwscas2011, loh-isca2008, loh-micro2009,
black-micro2006}\\

{\bf DRAM energy reduction with managing data placement and temperature}\\
\cite{zhu-itherm2008}\\ \cite{lin-islped2009} put overheated DRAMs to idle or
power-downed states to distribute temperature across all ranks.\\
\cite{liu-hpca2011} change memory access rates based on temperatures of
individual DRAM chips by optimizing cache replacement and page allocation
policies.\\

{\bf Reducing DRAM Refresh}\\ \cite{liu-isca2012, liu-asplos2011,
taku-islped1998, ahn-asscc2006, kim-patmos2000}\\

{\bf DVFS on DRAM}\\ DVFS on DRAM~\cite{david-icac11}\\

{\bf Low Power Interconnects}\\ Bus-invert coding~\cite{stan-vlsi1995}. There
are many bus-invert coding mechanisms.\\ Low-Swing
Interconnect~\cite{zhang-1998}\\ Cache using low-swing
wires~\cite{udipi-hipc2009, beckmann-2003}\\ Hybrid
Interconnect~\cite{balasubramonian-hpca2005}\\ DESC~\cite{desc}\\ }

\textbf{Efficient Data Compression.} Several prior
works~\cite{LinkCompression,CompressionPrefetching,GPUBandwidthCompression,
lcp-micro,memzip,MXT} study main memory and cache compression with several
different compression algorithms~\cite{fpc,bdi,c-pack,dcc,sc2}. These works
exploit the capacity and bandwidth benefits of data compression to enable
higher performance and energy efficiency. These prior works primarily tackle
improving compression ratios, reducing the performance/energy overheads of
processing data for compression/decompression, or propose more efficient
architectural designs to integrate data compression. These works address
different challenges in data compression and are orthogonal to our proposed
toggle-aware compression mechanisms. To the best of our knowledge, this is the
first work to study the energy implications of transferring compressed data
over different on-chip/off-chip channels.

\section{Summary}

We observe that data compression, while very effective in improving
bandwidth efficiency in GPUs, can greatly increase the bit toggle
count in the on-chip/off-chip interconnect.  Based on this new
observation, we develop two new {\em toggle-aware compression}
techniques to reduce bit toggle count while preserving most of the
bandwidth reduction benefits of compression.
%In this article, we analyze the potential benefits of data compression
%for a large spectrum of real GPU compute applications.  We show that
%bandwidth compression can be an effective technique in reducing the
%bandwidth consumption of modern GPUs.  At the same time, we make a new
%observation that there is an important problem related to bandwidth
%compression that must be addressed in the context of communication
%energy efficiency: a substantial growth in the number of bit-toggles
%when transferring compressed data. We provide a detailed analysis of
%the problem, and propose two toggle-aware mechanisms to address this
%new challenge: Energy Control and Metadata Consolidation. 
Our evaluations across six compression algorithms and 242 workloads 
show that these techniques are effective as they greatly reduce the bit
toggle count while retaining most of the bandwidth reduction
advantages of compression. We conclude that toggle-awareness is an
important consideration in data compression mechanisms for modern GPUs
(and likely CPUs as well), and encourage future work to develop new
solutions for it.

%This analysis proves the
%effectiveness of compression in reducing the bandwidth consumption in GPUs. 

%%% ONUR: I will go over conclusion again
%%% ONUR: Can we remove ``bandwidth'' form the title? How about just ``Toggle-Aware Compression for GPUs''
%%% Sounds very resonable and probably more effective

%gpekhime: done

%

\chapter{Putting It All Together}
In the previous chapters, we analyzed hardware-based data compression on a per layer basis;
i.e., as applied to only main memory, only cache, or only interconnect.
In this chapter, we focus on issues that arise when combining data compression applied
to multiple layers of the memory system at the same time in a single design.

%But at the same time, it is clear that some (if not all) of these ideas can be 
%combined in a single design with multiple layers of the memory hierarchy potentially
%having data in the compressed form.

%TODO: Explain while bandwidth is kept separately.
%In the context of modern CPUs, off-chip/on-chip bandwidth tends to be overprovisioned and 
%also hard to implement due to current DRAM restrictions.

In the context of modern GPUs, on-chip cache capacity is usually not the bottleneck. 
Instead, the bottleneck for most of our GPGPU applications is the off-chip bandwidth.
In addition, all of our GPU workloads have working set sizes that are too small to benefit from main memory
compression, and their compression ratios are very close to those of the corresponding off-chip
compression ratios (since most of the data has little reuse/locality and most of the
data in these GPGPU applications is frequently accessed only once). Hence
there is little benefit in separately evaluating main memory compression and bandwidth compression
for the GPGPU applications that were available to us.

%At the same time, such an evaluation can be done in the context of modern CPUs.
%In the next section, we will show the potential of one promising combination: 
%main memory compression (with some support for bandwidth compression) 

%together with cache compression.
Thus, the focus of this chapter is on combining cache compression and main memory compression for modern CPUs.

\section{Main Memory + Cache Compression}
We now show how main memory compression can be efficiently combined with cache compression with two compression
algorithms: FPC~\cite{fpc} and BDI~\cite{bdi}.

\subsection{Effect on Performance}
Main memory compression (including the LCP-based designs we introduced in Section 5) 
can improve performance in two major ways: 1) reducing
memory footprint can reduce long-latency disk accesses, 2) reducing memory
bandwidth requirements can enable less contention on the main memory bus, which is an
increasingly important bottleneck in systems. In our 
evaluations, we do not take into account the former benefit as we do not model
disk accesses (i.e., we assume that the uncompressed working set fits entirely
in memory). However, we do evaluate the performance improvement due to memory
bandwidth reduction (including our optimizations for compressing zero values). 
Evaluations using our LCP framework
show that the performance gains due to the bandwidth reduction more than
compensate for the slight increase in memory access latency due to memory
compression.
In contrast, cache compression (as we introduced it in Section 3) improves performance
by reducing the number of main memory accesses, which is also an important
bottleneck in many systems today.  

In our experiments, we compare eight
different schemes that employ compression either in the last-level cache, main
memory, or both. Table~\ref{table:schemes} describes the eight schemes.
Each scheme is named (X, Y) where X defines the cache compression mechanism
(if any) and Y defines the memory compression mechanism the scheme uses.
\begin{table}[h!]\small
  \centering
  \begin{tabular}{|l|l|l|}
    \hline
    \textbf{No.} & \textbf{Label} & \textbf{Description}\\
    \hline
    1 & (None, None) & Baseline with no compression\\
    \hline
    2 & (FPC, None) or FPC-Cache & LLC compression using FPC~\cite{fpc}\\
    \hline
    3 & (BDI, None) or BDI-Cache & LLC compression using BDI~\cite{bdi}\\
    \hline
    4 & (None, FPC) or FPC-Memory & Main memory compression (Ekman and Stenstrom~\cite{MMCompression})\\
    \hline
    5 & (None, LCP-BDI) or LCP-BDI & Main memory compression using LCP framework with BDI~\cite{lcp-micro}\\
    \hline
    6 & (FPC, FPC) &  Designs 2 and 4 combined\\
    \hline
    7 & (BDI, LCP-BDI) &  Designs 3 and 5 combined\\
    \hline
    8 & (BDI, LCP-BDI+FPC-Fixed) & Design 3 combined with LCP-framework using BDI+FPC-Fixed\\
    \hline
  \end{tabular}
  \caption{List of evaluated designs.}
  \label{table:schemes}
\end{table}

Figure~\ref{fig:IPC} shows the performance of single-core workloads using all
our evaluated designs, normalized to the baseline (None, None). We draw two
major conclusions from the figure.

\begin{figure}[h]
  \centering
  \includegraphics[width=0.99\textwidth]{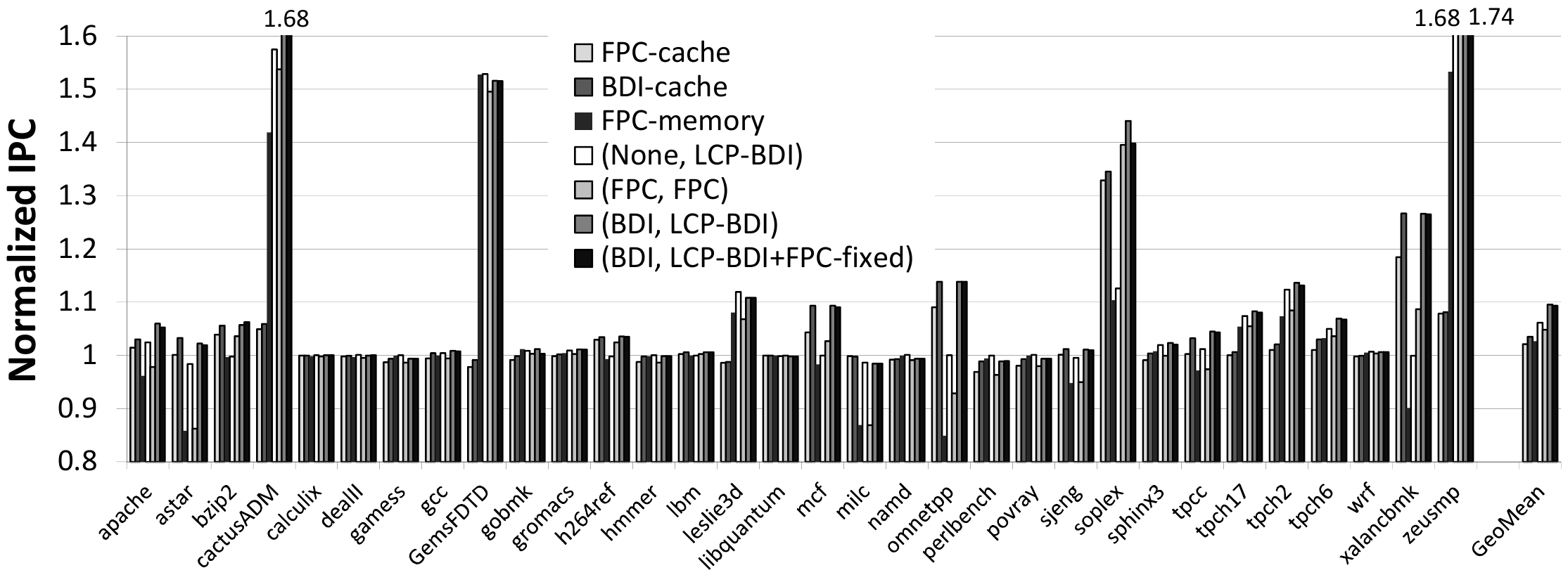}
  \caption{Performance comparison (IPC) of different compressed designs.}
  \label{fig:IPC}
\end{figure}

First, the performance improvement of combined LLC and DRAM
compression is greater than that of LLC-only or DRAM-only compression
alone.  For example, LCP-BDI improves performance by 6.1\%,
whereas (BDI, LCP-BDI) improves performance by 9.5\%.  Intuitively, this
is due to the orthogonality of the benefits provided by cache
compression (which retains more cache lines that otherwise would have been
evicted) and DRAM compression (which brings in more cache lines that would
otherwise have required separate memory transfers on the main memory
bus). We
conclude that main memory and cache compression frameworks integrate well and
complement each other.

Second, a high compression ratio does not always imply an improvement
in performance. For example, while GemsFDTD is an application with a
highly compressible working set in both the cache and DRAM, its
performance does not improve with LLC-only compression schemes (due
to the extra decompression latency), but improves significantly with
DRAM-only compression schemes. In
contrast, LLC-only compression is beneficial for
omnetpp, whereas DRAM-only compression is not. This difference across
applications can be explained by the difference in their memory access
patterns. We observe that when temporal locality is critical for the
performance of an application (e.g., omnetpp and xalancbmk), then
cache compression schemes are typically more helpful. On the other
hand, when applications have high spatial locality and less temporal
locality (e.g., GemsFDTD has an overwhelmingly streaming access pattern
with little reuse), they benefit significantly from the bandwidth
compression provided by the LCP-based schemes. Hence, if the goal is to
improve performance of a wide variety of applications, which may have a
mix of temporal and spatial locality, our results suggest that
employing both memory and cache compression using our LCP-based designs are the best
option. We conclude that combined LLC and DRAM compression that takes
advantage of our main memory compression framework improves the performance of a wide
variety of applications.

\subsection{Effect on Bus Bandwidth}
\label{sec:results-bandwidth}

When cache blocks and DRAM pages are compressed, the traffic between the LLC and DRAM can also be
compressed. This can have multiple positive effects: {\em i)} reduction in the
average latency of memory accesses, which can lead to improvement in the overall
system performance, {\em ii)} decrease in the bus energy consumption due to the
decrease in the number of transfers. 

%In a system, where DRAM pages are stored in the compressed form, there is an
%opportunity to compress the traffic between the LLC and DRAM. This can have
%multiple positive effects: 1) reduction in average latency of memory accesses,
%which can lead to improvement in the overall system performance, 2) decrease in
%the bus power consumption due to the decrease in the number of transfers. 

Figure~\ref{fig:bandwidth} shows the reduction in main memory
bandwidth between LLC and DRAM (in terms of bytes per kiloinstruction,
normalized to a system with no compression) using different
compression designs. Two major observations are in order.

\begin{figure}[h]
  \centering
  \includegraphics[width=0.99\textwidth]{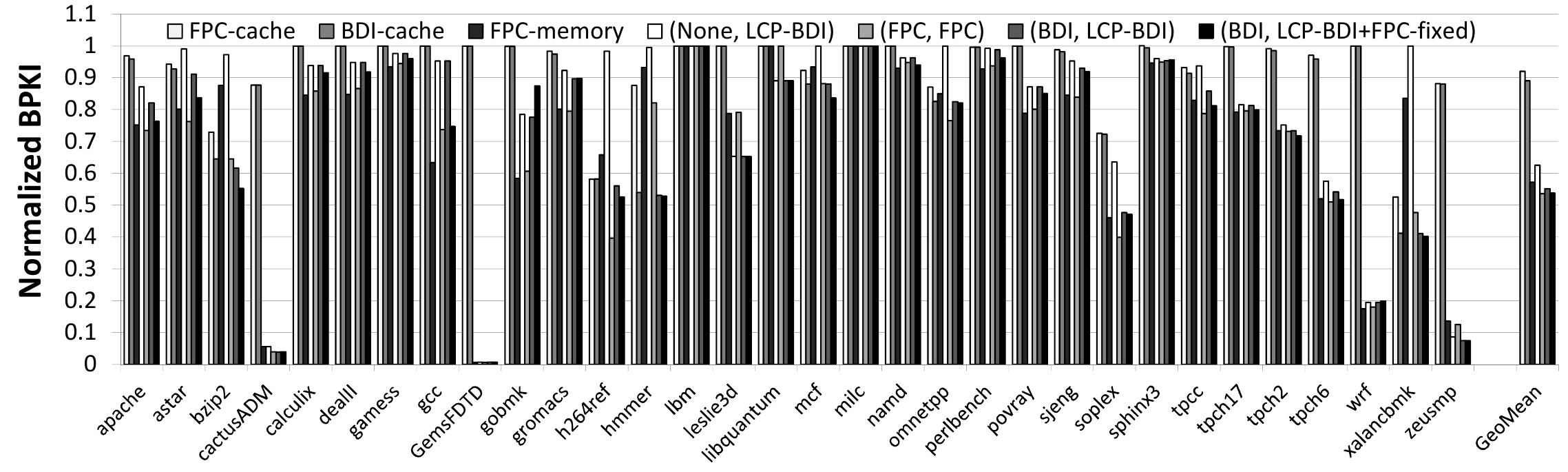}
  \caption{Effect of cache and main memory compression on memory bandwidth.}
  \label{fig:bandwidth}
\end{figure}

%Figure~\ref{fig:bandwidth} shows the total reduction in bandwidth between LLC
%and DRAM (in terms of bytes per kilo instuction, normalized to the design with
%no compression) with different compression designs. We can draw three major
%conclusions from this figure.

First, DRAM compression schemes are more effective in reducing
bandwidth usage than cache compression schemes. This is because
cache-only compression schemes reduce bandwidth consumption by
reducing the number of LLC misses but they cannot reduce the bandwidth
required to transfer a cache line from main memory. Overall, combined
cache-DRAM compression schemes such as (FPC, FPC) and (BDI,
LCP-BDI+FPC-fixed) decrease bandwidth consumption by more than 46\%, by
combining the reduction in both LLC misses and bandwidth required to
transfer each cache line.

%First, DRAM compression designs are more effective in compressing bandwidth than
%cache compression designs. This is because compression designs can only improve
%bandwidth indirectly by reducing the number of LLC misses.  Overall, FPC-FPC and
%BDI-BDI+FPC-fixed design can decrease bandwidth consumption by more than 46\%.

Second, there is a strong correlation between bandwidth compression
and performance improvement (Figure~\ref{fig:IPC}). Applications that
show a significant reduction in bandwidth consumption (e.g., GemsFDFD,
cactusADM, soplex, zeusmp, leslie3d, tpc*) also see large performance
improvements. There are some noticeable exceptions to this
observation, e.g., h264ref, wrf and bzip2. Although the memory bus
traffic is compressible in these applications, main memory bandwidth
is not the bottleneck for their performance.

%Second, there is a strong correlation between bandwidth compression and
%performance improvement (Figure~\ref{fig:IPC}).  Applications that have
%significant reduction in bandwidth consumption(GemsFDFD, cactusADM, soplex,
%zeusmp, leslie3d, tpc*) typically have significant improvement in performance as
%well. There are some noticable exclusions from this observation, e.g., h264ref,
%wrf and bzip2. The reason for this is that although their bus traffic is
%compressible, bandwidth is not the bottleneck for their performance.

\subsection{Effect on Energy}

By reducing the number of data transfers on the memory bus, a
compressed cache and main memory design also reduces the energy consumption of
the memory bus. Figure~\ref{fig:energy} shows the reduction in
consumed energy\footnote{\small Normalized to the energy of the baseline system
  with no compression.} by the main memory bus with different
compression designs. We observe that DRAM compression designs
outperform cache compression designs, and LCP-based designs provide
higher reductions than previous mechanisms for main memory
compression. The largest energy reduction, 33\% on average, is achieved
by combined cache compression and LCP-based main memory compression
mechanisms, i.e., (BDI, LCP-BDI) and (BDI, LCP-BDI+FPC-fixed).  Even
though we do not evaluate full system energy due to simulation
infrastructure limitations, such a large reduction in main memory bus
energy consumption can have a significant impact on the overall
system energy, especially for memory-bandwidth-intensive
applications. We conclude that our framework for main memory
compression can enable significant energy savings, especially when
compression is applied in both the last level cache and main memory. 

\begin{figure}[htb]
  \centering
  \includegraphics[width=0.99\textwidth]{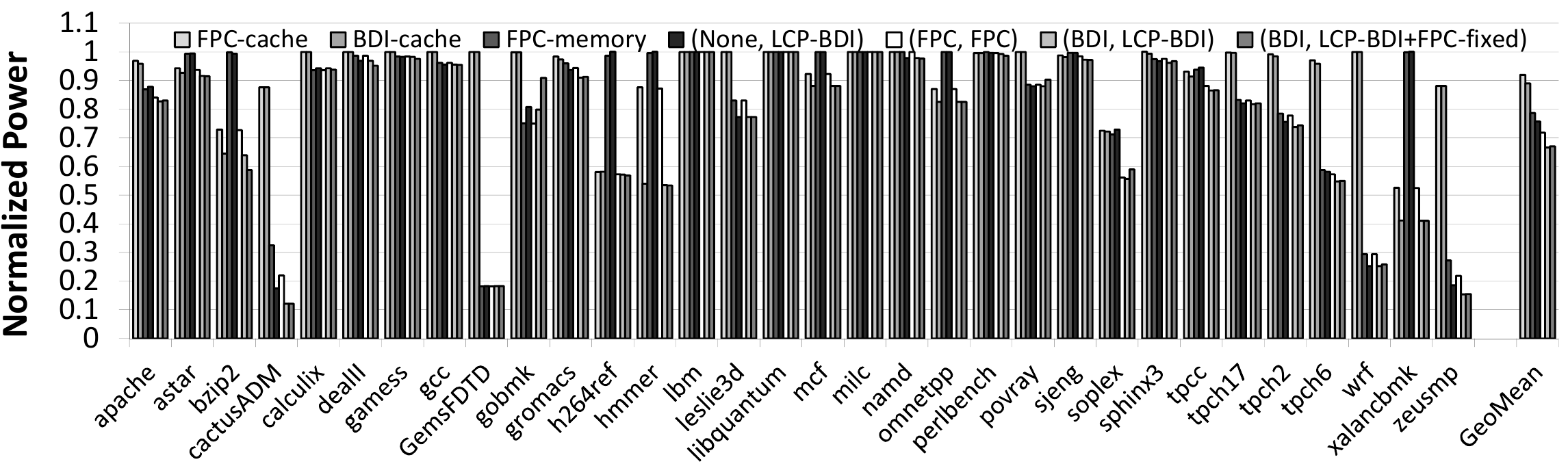}
  \caption{Effect of cache and main memory compression on DRAM bus energy.}
  \label{fig:energy}
\end{figure}

\chapter{Conclusions and Future Work}
Memory hierarchies play a significant role in the performance and energy
efficiency of many modern systems, from mobile devices to data centers and
supercomputers. Unfortunately, the limited resources of these memory
hierarchies are not always utilized efficiently. One of these sources of
inefficiency is redundancy in the data that is stored and transferred. We
observe that this redundancy can be efficiently explored using hardware-based
data compression. In Chapter 2, we described what are the key challenges against
making hardware-based data compression practical across major layers of the
memory hierarchy: caches, main memory, and on-chip/off-chip buses.

In this dissertation, we proposed three major sets of solution to make
hardware-based data compression efficient and practical in the context of 
all three layers of the memory
hierarchy. First, we observed that a simple and fast, yet efficient compression
algorithm can make data compression practical even for on-chip caches.  In
Chapter 3, we described such an algorithm, called \emph{Base-Delta-Immediate
Compression}, and a corresponding on-chip cache design to support data
compression. The performance benefits observed are on-par with the performance 
benefits of doubling the
cache size.  Then, in Chapter 4, we showed that compressed block size can be
sometimes indicative of data reuse and can be efficiently used as a new
dimension in cache management decisions. The performance benefits of our
proposed compression-aware mechanism which takes into account compressed block size
in making cache replacement and insertion decisions,
 results in performance on-par with that provided by doubling the cache
size.  Overall, both cache compression and compression-aware replacement policies using
compressed block size deliver performance on par with that of a conventional cache with 4$\times$ capacity.

Second, we proposed a new main memory compression framework, called
\emph{Linearly Compressed Pages (LCP)}, that can provide low-overhead support
for data compression in memory with different compression algorithms, to achieve
higher effective memory capacity (69\% on average) and higher off-chip
bandwidth (24\% on average).  LCP improves performance by 6\%/14\%/11\% for
single-/two-/four-core workloads, relative to a system without main memory
compression. 

Third, we observed that there is a high potential for bandwidth compression for
modern GPGPU applications. However, in order to realize this potential
in an energy efficient manner, a new
problem---the significant increase in bit flips (bit toggles) due to
compressed data transfers on the interconnect---needs to be
properly addressed.  This increase is so high that it can lead to a 2.1$\times$
average increase in the consumed energy by the on-chip communication channel.
We showed two major potential solutions to this problem, called \emph{Energy
Control} and \emph{Metadata Consolidation}, which can preserve most of the
benefits of compression without significant increase in energy consumption due
to the bit toggle problem. 

\section{Future Work Directions} This dissertation on data compression
significantly advances this subfield of computer architecture, but as it
commonly happens, also highlights some completely new problems and
opportunities. We conclude our dissertation by describing three such opportunities.

\subsection{Compiler-Assisted Data Compression} One problem is the dependence
of the existing compression algorithms on how the application data structures
are mapped to main memory and on-chip caches (as we show in Chapter 3).  For
example, if pointer-like values are allocated side by side, they have a higher
chance to be compressed well with BDI compression algorithm, but putting
together (e.g., in the same cache line) a pointer and a boolean value would
obviously lead to higher dynamic range, and hence lower compressibility. The latter
frequently happens when arrays or lists of structs are defined in the program
with different types mixed together.  For applications with such data types, we
want to allocate objects such that the spatial locality of similar-valued
members is preserved. More precisely, we would like to \emph{split} an object
up into respective members and  allocate space for those members based on what
kinds of values they hold. These decisions of splitting and allocation
may be made during compile time or
runtime, depending on the implementation. Compression ratio improves from
using members with similar
value-types that are \emph{pooled} (allocated) together and our preliminary
studies already show a significant potential of such an approach. We aim
to extend this idea to improve the compressibility of main memory pages that
suffer from mixing data of very different types.

\subsection{Data Compression for Non-Volatile Memories}

LCP~\cite{lcp-micro} main memory compression design was built on top of
commodity DRAM main memory, but data compression is fundamentally independent
of the technology that was used to build main memory. In our work, we aim to
investigate the potential of extending LCP to other emerging non-volatile
memory technologies (e.g., PCM~\cite{PCM,PCM2,PCM3,PCM4,PCM5,PCM6,PCM7}, 
STT-MRAM~\cite{STT-MRAM,STT-MRAM2}, RRAM~\cite{RRAM}) and hybrid
memory technologies (e.g.,~\cite{hb1,hb2,hb3,hb4,hb5}). We
expect that longer access/write latencies of these emerging memory technologies
will allow the system designs to use more aggressive compression algorithms, and hence the
capacity benefits of LCP-based designs can increase even further.

\subsection{New Efficient Representations for Big Data} Many modern
applications, such as machine learning applications, applications from
the bioinformatics field, modern databases etc., operate on data sets that
significantly exceed the available main memory. At the same time, these
applications do not always require the full precision or accuracy in computation, as
their input data are already significantly imprecise or noisy.  In our future work,
we would like to investigate the potential of partially replacing the accesses
to the huge data sets in these applications with the accesses to their much
smaller representations or signatures.  The key idea is to build a
lower-resolution representation of the data set, keep it up-to-date in main
memory, and refer to it when information to this data set is missing in the
main memory. We then dynamically monitor whether the application meets its
desired quality of output, and update the aggressiveness of our speculation
accordingly. Our related work in recovery-free value prediction
using approximate loads~\cite{rfvp-pact,rfvp-taco,rfvp-dt} hints that this
can be significant promise toward this direction of research.

\begin{comment} \section{Summary} In this dissertation, we showed that modern
memory hierarchies not always utilize their limited resources efficiently by
storing a lot of redundant bits of data. We proposed several techniques based
on the general idea of hardware-based data compression to avoid this redundancy
for (i) on-chip caches (Base-Delta-Immediate Compression and Compression-Aware
Management Policies), (ii) main memory (Linearly Compressed Pages), and (iii)
on-chip/off-chip bandwidth compression (Toggle-Aware bandwidth compression
through Energy Control mechanism).  As we showed in Section 8.1, the ideas
behind these mechanisms can be extended in many directions for new research in
this area of computer architecture and can also enable new mechanisms that
could improve the efficiency of modern memory hierarchies even further.
\end{comment}

\chapter*{Other Works of This Author}

I have been actively involved in research projects outside the scope of my thesis.

\textbf{Systems.} I worked on web search systems for mobile phones where users'
interest in certain trending events can be predicted and efficiently prefetched
to extend the phone's battery life~\cite{pockettrend}.  Previously, I also
worked on improving the compile time of existing compilers with machine
learning techniques that can predict which optimizations are actually useful for
performance~\cite{ml-compilers}.
 
\textbf{Main Memory.}
In collaboration with Vivek Seshadri, I proposed several ways of better utilizing
existing DRAM-based main memories: (i) fast bulk data operations like copying
and memory initialization using RowClone~\cite{rowclone}, and (ii) an enhanced
virtual memory framework that enables fine-grained memory management~\cite{overlays}. 
In collaboration with Donghyuk Lee, I worked on (i) reducing the
latency of existing DRAM memories~\cite{lee-hpca2015}, and (ii) increasing the
bandwidth available for existing (and future) 3D stacking
designs~\cite{smla}. In collaboration with
Hasan Hassan, I also worked on reducing DRAM latency by exploiting our new
observation that many DRAM rows can be accessed significantly faster since they
have sufficient amount of charge left~\cite{chargecache}. 
In collaboration with Kevin Chang, I investigated the potential of reducing
different DRAM timing parameters to decrease its latency and their effect
on the error rate~\cite{ChangKHGHLLPKM16}.

\textbf{GPUs.}
In collaboration with Nandita Vijaykumar, I worked on new ways of utilizing
existing GPU resources through flexible data compression~\cite{caba,caba-book}
and virtualization with oversubscription~\cite{proteus}.

\textbf{Bioinformatics.}
In collaboration with Hongyi Xin, I worked on new filters for alignment in
genome read mapping~\cite{shd}, and techniques to find the optimal seeds for a
particular read in the genome mapping process~\cite{oss}.

\textbf{Approximate Computing.}
Together with my collaborators from Georgia Tech, I worked on rollback-free
value prediction mechanisms for both CPUs~\cite{rfvp-pact} and
GPUs~\cite{rfvp-dt,rfvp-taco}.

\appendix

\backmatter

%\renewcommand{\baselinestretch}{1.0}\normalsize

% By default \bibsection is \chapter*, but we really want this to show
% up in the table of contents and pdf bookmarks.
\renewcommand{\bibsection}{\chapter{\bibname}}
\bibliographystyle{plain}
\bibliography{main}

\end{document}

%% Double space document for easy review:
%\renewcommand{\baselinestretch}{1.66}\normalsize
\renewcommand{\baselinestretch}{1.2}\normalsize

\appendix
\appendix
\chapter{I like Pie}

Yum!  Pie is good!  3.1415926535897932384, I adore thee!

%%% Local Variables: 
%%% mode: latex
%%% TeX-master: t
%%% End: 

%%% Local Variables: 
%%% mode: latex
%%% TeX-master: "main"
%%% End: 

\renewcommand{\baselinestretch}{1.0}\normalsize

\bibliographystyle{plain}
\bibliography{references}